\documentclass[sigconf,screen]{acmart}

\usepackage{xr-hyper} 
\usepackage{xr}

\providecommand{\ifnotsplit}[1]{{#1}}

\usepackage{etoolbox}

\usepackage{amsmath}
\usepackage{amsfonts}
\usepackage{amsthm}
\usepackage{comment}

\usepackage{tabularx}
\usepackage{multirow}

\usepackage{booktabs}   
\usepackage{subcaption} 

\usepackage[nomargin,inline,index,%
  status=final
]{fixme} 
\fxusetheme{colorsig}
\FXRegisterAuthor{fxAS}{anfxAS}{}
\FXRegisterAuthor{fxNY}{anfxNY}{}

\usepackage[inline]{enumitem}

\usepackage{graphicx}

\usepackage{multibib}
\newcites{app}{Additional references for the Appendix}

\usepackage{mdframed}

\usepackage{nicefrac}
\usepackage{mathpartir}

\usepackage{proof}

\usepackage{stmaryrd}

\usepackage{thmtools}
\usepackage{thm-restate}

\theoremstyle{definition}
\newtheorem{newdefinition}{Definition}[section]
\newtheorem{newexample}{Example}[section]
\newtheorem{newremark}{Remark}[section]
\newtheorem{newcorollary}{Corollary}[section]
\newtheorem{newlemma}{Lemma}[section]
\theoremstyle{plain}

\usepackage{tikz}
	\usetikzlibrary{trees,decorations,decorations.shapes,arrows,automata,positioning,plotmarks,
	shapes,backgrounds,petri}
\usepackage{array}
\usepackage{ifthen}
\usetikzlibrary{decorations,arrows,shapes,shapes.geometric,shapes.misc,shapes.multipart,decorations.pathreplacing,calc,positioning}
\usetikzlibrary{arrows.meta}
\usepackage{tcolorbox}

\usepackage{dashbox}

\usepackage{url}

\usepackage{xifthen}
\usepackage{xspace}

\usepackage{mathtools}

\usepackage{cleveref}

\usepackage{listings}

\definecolor{mygreen}{rgb}{0,0.6,0}
\definecolor{myblue}{rgb}{0.4,0.5,0}
\definecolor{mygray}{rgb}{0.5,0.5,0.5}
\definecolor{mymauve}{rgb}{0.58,0,0.82}

\AtBeginDocument{%
	\providecommand\BibTeX{{%
			\normalfont B\kern-0.5em{\scshape i\kern-0.25em b}\kern-0.8em\TeX}}}

\usepackage{pifont}%

\usepackage{pfsteps} 

\usepackage{wrapfig}

\setlist[description]{
    labelwidth=0pt,
    leftmargin=15pt,
    itemindent=\dimexpr-5pt-\labelsep\relax,
}

\usepackage{array,longtable}

\newcommand{\mathstyle}[1]{\ensuremath{\mathsf{#1}}\xspace}
\definecolor{ruleColor}{rgb}{0.1, 0.3, 0.1}%
\newcommand{\mathparfont}[1]{\ensuremath{\text{{\footnotesize\color{ruleColor}{[#1]}}}}\xspace}

\newcommand{\myparagraph}[1]{{\bf #1}}
\newcommand{\KPCom}[1]{#1}

\newcommand{\set}[1]{\ensuremath{\{#1\}}\xspace}
\newcommand{\setbar}{\ensuremath{\ \ |\ \ }\xspace}

\newcommand{\bnfis}{\ensuremath{::=}}
\newcommand{\bnfbar}{\ensuremath{\ \: |\ \: }}

\newcommand{\bind}[2]{\nicefrac{#2}{#1}}
\newcommand{\substenum}[1]{\mathord{{\color{black}\left\{{#1}\right\}}}}
\newcommand{\subst}[2]{\ensuremath{\substenum{\bind{#2}{#1}}}}

\newcommand{\function}[2]{\ensuremath{\mathsf{#1}\!\left(#2\right)}\xspace}

\newcommand{\powerset}[1]{\ensuremath{\mathcal{P}(#1)}\xspace}

\newcommand{\domain}[1]{\function{dom}{#1}}

\newcommand{\es}{\ensuremath{\emptyset}\xspace}

\newcommand{\tree}[4][] {
	\begin{array}{l}
		\ifthenelse {\equal {#1} {} }
		{}
		{
		#1
		\\
		}
		{\textstyle \frac{
			\begin{array}{cl}
			#2
			\end{array}
		}{
			\begin{array}{l}
			#3
			\end{array}
		}
		}
		\ifthenelse { \equal {#4} {} }
		{}
		{
		#4
		}
	\end{array}
}

\newcommand{\Tfrac}[2]{%
	\ooalign{%
		$\genfrac{}{}{1.2pt}1{#1}{#2}$\cr%
		$\color{white}\genfrac{}{}{.6pt}1{\phantom{#1}}{\phantom{#2}}$}%
}

\newcommand{\ttree}[4][] {
	\begin{array}{l}
		\ifthenelse {\equal{#1} {} }
		{}
		{
			#1
			\\
		}
		{\textstyle \Tfrac{
				\begin{array}{cl}
					#2
				\end{array}
			}{
				\begin{array}{l}
					#3
				\end{array}
			}
		}
		\ifthenelse{\equal{#4}{} }
		{}
		{
			\ #4
		}
	\end{array}
}

\newcommand{\mcmpst}{\text{MCMP}\xspace} 

\newcommand{\scmpst}{\text{SCMP}\xspace} 

\newcommand{\msmpst}{\text{MSMP}\xspace} %

\newcommand{\dmpst}{\text{DMP}\xspace} %

\newcommand{\smpst}{\text{SMP}\xspace} %

\newcommand{\mpst}{\text{MP}\xspace} 

\newcommand{\mcbs}{\text{MCBS}\xspace} 

\newcommand{\msbs}{\text{MSBS}\xspace}

\newcommand{\scbs}{\text{SCBS}\xspace}

\newcommand{\bs}{\text{BS}\xspace}

\newcommand{\cnvms}{\text{CMV}${}^+$\xspace}
\newcommand{\cnv}{\text{CMV}\xspace}
\newcommand{\lcnvms}{\text{LCMV}${}^+$\xspace}
\newcommand{\lcnv}{\text{LCMV}\xspace}

\newcommand{\SETB}[1]{\ensuremath{\mathbb{#1}}\xspace}
\newcommand{\Part}{\SETB{P}}

\newcommand{\parts}[1]{\function{pt}{#1}}
\newcommand{\partsf}{\mathstyle{pt}}

\newcommand{\ftv}[1]{\function{ftv}{#1}}
\newcommand{\ftvsf}{\mathstyle{ftv}}

\newcommand{\fv}[1]{\function{fv}{#1}}
\newcommand{\fvsf}{\mathstyle{fv}}

\newcommand{\Roles}{{\color{myblue}\ensuremath{\mathbb{R}}}\xspace}

\newcommand{\localise}[1]{\ensuremath{#1\!:}\xspace}

\newcommand{\LLfont}{\Delta}
\newcommand{\LL}{{\color{myblue}\ensuremath{\LLfont}}\xspace}
\newcommand{\LLd}{{\color{myblue}\ensuremath{\LLfont'}}\xspace}
\newcommand{\LLi}[1]{{\color{myblue}\ensuremath{\LLfont_{#1}}}\xspace}

\newcommand{\LLdd}{{\color{myblue}\ensuremath{\LLfont''}}\xspace}

\newcommand{\LLddd}{{\color{myblue}\ensuremath{\LLfont'''}}\xspace}

\newcommand{\Defont}{\Delta}
\newcommand{\De}{{\color{myblue}\ensuremath{\Defont}}\xspace}
\newcommand{\Ded}{{\color{myblue}\ensuremath{\Defont'}}\xspace}

\newcommand{\rolecolor}{mymauve}

\newcommand{\role}[1]{\ensuremath{{\mathsf{{\color{\rolecolor}#1}}}}\xspace}

\newcommand{\rolei}[2]{\ensuremath{{\color{\rolecolor}\mathsf{#1}_{#2}}}\xspace}
\newcommand{\roled}[1]{\ensuremath{{\color{\rolecolor}\mathsf{#1}'}}\xspace}

\newcommand{\AT}[2]{#1{:}#2}

\newcommand{\p}{\role{p}}

\newcommand{\pii}[1]{\rolei{\p}{#1}}
\newcommand{\q}{\role{q}}

\newcommand{\rr}{\role{r}}

\newcommand{\Labels}{{\SETB{L}\xspace}}

\newcommand{\UType}{\ensuremath{U}\xspace}

\newcommand{\UTypei}[1]{{\color{black}\ensuremath{U_{#1}}}\xspace}
\newcommand{\UTypeid}[1]{{\color{black}\ensuremath{U_{#1}'}}\xspace}
\newcommand{\UTyped}{{\color{black}\ensuremath{U'}}\xspace}

\newcommand{\type}[1]{\ensuremath{\textsf{\small #1}}\xspace}
\newcommand{\lbl}[1]{\ensuremath{\mathit{#1}}\xspace}
\newcommand{\lab}{\lbl{\ell}}
\newcommand{\labd}{\lbl{\ell'}}
\newcommand{\labi}[1]{\lbl{\ell_{#1}}}

\newcommand{\true}{\mathtt{tt}}
\newcommand{\false}{\mathtt{ff}}

\newcommand{\cond}[3]{\mathtt{if}\ {#1}\ \mathtt{then} \ {#2}\
  \mathtt{else} \ {#3}}

\newcommand{\nat}{\type{nat}}

\newcommand{\bool}{\type{bool}}

\newcommand{\lblmessage}[2]{\ensuremath{#1\langle{#2\rangle}}}

\newcommand{\beh}{\ensuremath{\mathsf{beh}}}

\newcommand{\guarded}[1]{\function{guarded}{#1}}
\newcommand{\guards}[2]{\mathstyle{guards}\ #1\ \mathstyle{in}\ #2}

\newcommand{\lcolor}{\color{blue}}

\newcommand{\Local}{{\lcolor\ensuremath{\SETB{T}}}\xspace}
\newcommand{\local}{{\lcolor\ensuremath{T}}\xspace}
\newcommand{\locald}{{\lcolor\ensuremath{T'}}\xspace}

\newcommand{\localdd}{{\lcolor\ensuremath{T''}}\xspace}

\newcommand{\locali}[1]{{\ensuremath{\lcolor{T_{#1}}}}\xspace}
\newcommand{\localid}[1]{{\lcolor\ensuremath{T'_{#1}}}\xspace}
\newcommand{\localidd}[1]{{\lcolor\ensuremath{T''_{#1}}}\xspace}
\newcommand{\localiddd}[1]{{\lcolor\ensuremath{T'''_{#1}}}\xspace}

\newcommand{\tseq}{{\lcolor{;}}}

\newcommand{\tinact}{{\lcolor\mathstyle{end}}\xspace}

\newcommand{\tout}[2]{{\ensuremath{#1 {\lcolor{\textbf{!}}} #2\tseq}}}
\newcommand{\ntout}[2]{{\ensuremath{#1 {\lcolor{\textbf{!}}} #2}}}
\newcommand{\toutlbl}[3]{\ensuremath{#1 {\lcolor{\textbf{!}}} \lblmessage{#2}{#3} \tseq}}
\newcommand{\ntoutlbl}[3]{\ensuremath{#1 {\lcolor{\textbf{!}}} \lblmessage{#2}{#3}}}

\newcommand{\tinp}[2]{{\ensuremath{#1 {\lcolor{\textbf{?}}} #2\tseq}}}
\newcommand{\ntinp}[2]{{\ensuremath{#1 {\lcolor{\textbf{?}}} #2}}}
\newcommand{\tinplbl}[3]{\ensuremath{#1 {\lcolor{\textbf{?}}} \lblmessage{#2}{#3} \tseq}}
\newcommand{\ntinplbl}[3]{\ensuremath{#1 {\lcolor{\textbf{?}}} \lblmessage{#2}{#3}}}

\newcommand{\tinout}[2]{{\ensuremath{#1 {\inoutmark} #2\tseq}}}

\newcommand{\tvar}[1]{\mathstyle{\lcolor{#1}}}
\newcommand{\trec}[1]{\ensuremath{{\lcolor{\mu \tvar{#1}.}}\xspace}}

\newcommand{\tinoutlbl}[3]{\ensuremath{#1 {\inoutmark} \lblmessage{#2}{#3} \tseq}}

\newcommand{\tinoutlbli}[4][i]{\ensuremath{\rolei{#2}{#1} {\inoutmarki{#1}} \lblmessage{#3_{#1}}{#4_{#1}}\tseq}}

\newcommand{\tor}{\ensuremath{{\lcolor{+}}}}

\newcommand{\outmark}{{\lcolor{\textbf{!}}}}
\newcommand{\inpmark}{{\lcolor{\textbf{?}}}}
\newcommand{\inoutmark}{{\lcolor{\mathbf{\dagger}}}}
\newcommand{\inoutmarki}[1]{{\lcolor{\mathbf{\dagger}_{#1}}}}

\newcommand{\Tor}[2]{\ensuremath{{\lcolor{\sum_{#1}}} #2}\xspace}

\newcommand{\prefix}[1]{\function{pre}{#1}}

\newcommand{\Act}{\mathstyle{Act}}
\newcommand{\LAct}{\mathstyle{Act_\mathit{L}}}

\newcommand{\act}[3]{{\ensuremath{#1#2 \mathbf{:} #3}\xspace}}
\newcommand{\actlbl}[4]{\ensuremath{#1#2 \mathbf{:} \lblmessage{#3}{#4}}\xspace}

\newcommand{\actinp}[3]{{\ensuremath{{#1}{#2}{\lcolor{\textbf{?}}}{#3}}\xspace}}
\newcommand{\actoutlbl}[4]{\ensuremath{{#1}{#2}{\lcolor{\textbf{!}}}{\lblmessage{#3}{#4}}}\xspace}
\newcommand{\actinplbl}[4]{\ensuremath{{#1}{#2}{\lcolor{\textbf{?}}}{\lblmessage{#3}{#4}}}\xspace}
\newcommand{\actoutinlbli}[5]{\ensuremath{{\rolei{#1}{#5}}{\rolei{#2}{#5}}}
{\lcolor{\textbf{$\dagger$}_{#5}}}{\lblmessage{{#3}_{#5}}{{#4}_{#5}}}\xspace}

\newcommand{\by}[1]{\ensuremath{{\xrightarrow{#1}}}\xspace}

\newcommand{\safe}[1]{\function{safe}{#1}}

\newcommand{\df}[1]{\function{dfree}{#1}}

\newcommand{\subt}{\ensuremath{\mathrel{\leqslant}}\xspace}%

\newcommand{\subtfont}[1]{\mathparfont{#1}}

\newcommand{\SEnd}{\subtfont{SEnd}}

\newcommand{\SSel}{\subtfont{SSl}}
\newcommand{\SBra}{\subtfont{SBr}}
\newcommand{\SChoice}{\subtfont{S$\Sigma$}}
\newcommand{\SRcL}{\subtfont{S$\mu$L}}
\newcommand{\SRcR}{\subtfont{S$\mu$R}}
\newcommand{\SSubset}{\subtfont{SSet}}

\newcommand{\PP}{\ensuremath{P}\xspace}

\newcommand{\seq}{\ensuremath{.}\xspace}

\newcommand{\inact}{\ensuremath{\mathbf{0}}\xspace}

\newcommand{\pvar}[1]{\mathstyle{#1}}
\newcommand{\rec}[1]{\ensuremath{\mu \pvar{#1}.}\xspace}

\newcommand{\Par}{\ensuremath{\,\,|\,\,}\xspace}
\newcommand{\por}{\ensuremath{+}\xspace}

\newcommand{\choice}[2]{\ensuremath{\sum_{#2} #1}\xspace}

\newcommand{\sendp}[1]{\function{subj}{#1}}

\newcommand{\outsymbol}{\ensuremath{{\lcolor{\textbf{!}}}}\xspace}
\newcommand{\inpsymbol}{\ensuremath{{\lcolor{\textbf{?}}}}\xspace}

\newcommand{\pprefix}{\ensuremath{\pi}\xspace}

\newcommand{\OR}{\ensuremath{+}\xspace}

\newcommand{\xx}{\ensuremath{x}\xspace}
\newcommand{\val}{\ensuremath{v}\xspace}

\newcommand{\scong}{\ensuremath{\equiv}\xspace}
\newcommand{\red}{\ensuremath{\longrightarrow}\xspace}

\newcommand{\RCondT}{\mathparfont{If-$\true$}}
\newcommand{\RCondF}{\mathparfont{If-$\false$}}
\newcommand{\RChoice}{\mathparfont{$\Sigma$}}

\newcommand{\RCong}{\mathparfont{Cong}}

\newcommand{\Ga}{\ensuremath{\Gamma}\xspace}

\newcommand{\ee}{\ensuremath{\emptyset}\xspace}

\newcommand{\types}{\ensuremath{\vdash}\xspace}

\newcommand{\as}{\ensuremath{\triangleright}\xspace}

\newcommand{\LRec}{\mathparfont{L$\mu$}}

\newcommand{\LChoice}{\mathparfont{Sum}}

\newcommand{\RPass}{\mathparfont{RPass}}

\newcommand{\TBool}{\mathparfont{Bool}}
\newcommand{\TNat}{\mathparfont{Nat}}
\newcommand{\TGround}{\mathparfont{Gr}}

\newcommand{\TInact}{\mathparfont{TInact}}

\newcommand{\TReq}{\mathparfont{TReq}}

\newcommand{\TSend}{\mathparfont{TSnd}}
\newcommand{\TRcv}{\mathparfont{TRcv}}

\newcommand{\TSum}{\mathparfont{T$\Sigma$}}

\newcommand{\TVar}{\mathparfont{TVar}}
\newcommand{\TRec}{\mathparfont{TRec}}
\newcommand{\TCond}{\mathparfont{TIf}}

\newcommand{\TSubs}{\mathparfont{T$\subt$}}

\newcommand{\TSessPlus}{\mathparfont{TSess$+$}}

\newcommand{\NY}[1]{#1}

\newcommand{\dkmargin}[2][] {
	\ifthenelse { \equal {#1} {} } {
		{\color{blue} $\bullet$}
	}{
		{\color{blue} $\bullet$ #1}
	}
	{
		\marginpar{{\color{blue} \footnotesize $\bullet$ #2 }}
	}
}

\definecolor{keywordcolour}{rgb}{0.5,0,0.35}
\definecolor{greencomments}{rgb}{0,0.5,0}
\definecolor{codegengray}{rgb}{0.92, 0.92, 0.92}
\lstdefinelanguage{Scribble}{
	basicstyle=\small\ttfamily,
	stringstyle=\color{Blue},
	keywordstyle=\color{keywordcolour},
	showstringspaces=false,
	keywords={and,as,at,by,catches,choice,continue,do,from,global,import,instantiates,interruptible,local,module,or,par,protocol,rec,role,sig,throws,to,type,with,int,aux,nested,new, calls, invite, accept, end, mchoice},
	morestring=[b]",
	morestring=[b]',
	morecomment=[l][\color{greencomments}]{//},
	literate={->}{{${\rightarrow\ }$}}1 {>}{{$>$}}1 {<}{{$<$}}1 {<=}{{$\leq$}}1 {>=}{{$\geq$}}1 {&&}{{$\land$}}1 {||}{{$\lor$}}1 {!=}{{$\neq$}}1 {=}{{$=$}}1,
	mathescape=true
}
\definecolor{gokeyword1}{rgb}{0.043, 0.066, 0.717}
\definecolor{gokeyword2}{rgb}{0.109, 0.490, 0.145}
\definecolor{gocomments}{rgb}{0.654, 0.352, 0.086} 
\definecolor{pkgaccesscolour}{rgb}{0, 0.643, 0.701} 
\definecolor{gostrings}{rgb}{0.384, 0.737, 0.003}
\lstdefinelanguage{Golang}{
	morekeywords=[1]{package,import,func,type,struct,return,defer,panic,%
		recover,select,var,const,iota,interface,},%
	morekeywords=[2]{string,uint,uint8,uint16,uint32,uint64,int,int8,int16,%
		int32,int64,bool,float32,float64,complex64,complex128,byte,rune,uintptr,%
		error,chan,},%
	morekeywords=[3]{map,slice,make,new,nil,len,cap,copy,close,true,false,%
		delete,append,real,imag,complex,},%
	morekeywords=[4]{for,break,continue,range,go,goto,switch,case,fallthrough,if,%
		else,default,},%
	morekeywords=[5]{Printf,Error,Print,},%
	morekeywords=[6]{channels, messages, sync, fibonacci, invitations, protocol,%
		fibonacci_2, roles, callbacks, results, boundedfib_2, boundedfib, fmt, dyntg,%
		dyntg_2, bfib_2},%
	morekeywords=[7]{Fibonacci_F2, StartFib2_From_Start, To_Fib_F2_Env, Fib_F2, ResultFrom_Fib_F2,%
		Done, BoundedFib_F2, Fib2_To_F3, End_From_F3, To_BoundedFib_F1_Env, BoundedFib_F1,%
		ResultFrom_BoundedFib_F1, BFib_F1, Worker, Master, Client, Println, ResultFrom_BFib_F1,%
		Req_To_W, DynTG_SendChannels, To_DynTG_S_Env, DynTG_S, ResultFrom_DynTG_S,%
		Resp_From_W, Pkt_From_Processing, Pkt_To_Output1, Pkt_To_Output2, Pkt_From_Processing_2,%
		Pkt_From_Processing_3, hash},%
	morekeywords=[8]{WaitGroup, F2_Chan, Fibonacci_F2_InviteChan, Fibonacci_F2_Env,%
		F2_Result,BoundedFib_F2_InviteChan, BoundedFib_F2_Env, S_Chan, DynTG_S_InviteChan,%
		DynTG_S_Env, S_Result, BFib_F2_InviteChan, BFib_F2_Env, DynTG_RoleSetupChan, DynTG_InvSetupChan,%
		Routing_Result, Routing_Env, Routing_Chan},%
	keywordstyle=[1]\color{gokeyword1}\ttfamily\bfseries,
	keywordstyle=[2]\color{gokeyword2}\ttfamily\bfseries,
	keywordstyle=[3]\color{keywordcolour},
	keywordstyle=[4]\color{gokeyword1}\ttfamily\bfseries,
	keywordstyle=[5]\color{black},
	keywordstyle=[6]\color{gokeyword2},
	keywordstyle=[7]\color{keywordcolour},
	keywordstyle=[8]\color{pkgaccesscolour},
	sensitive=true,%
	morecomment=[l][\color{gocomments}]{//},%
	morecomment=[s][\color{gocomments}]{/*}{*/},%
	morestring=[b]',%
	morestring=[b]",%
	morestring=[s]{`}{`},%
	stringstyle=\color{gostrings},%
	mathescape=true,
	escapeinside={@}{@}
}

\lstdefinelanguage{raw}{
	morekeywords={},
	otherkeywords={},
}
\newcommand{\enodei}[1]{\role{#1}}

\newcommand{\env}{\role{station}}

\newcommand{\leader}{\lbl{leader}}
\newcommand{\elect}{\lbl{elect}}
\newcommand{\delete}{\lbl{del}}

\newcommand{\mcfont}[1]{\ensuremath{\texttt{\small #1}}\xspace}

\newcommand{\id}{\mcfont{id}}

\newcommand{\ruleLEll}{\mathparfont{L-$\ell$}}

\newif\ifdraft%
  \draftfalse%
  \drafttrue%

\newtoggle{techreport}%
\toggletrue{techreport}%

\newcommand{\ifempty}[3]{%
  \ifthenelse{\isempty{#1}}{#2}{#3}%
}%

\newcommand{\cinference}[3]{
	\begin{array}{ll}
	\infer=
	{
		#2
	}{
		#1
	}
	\end{array}\!\!{#3}
}%

\definecolor{citrine}{rgb}{0.89, 0.82, 0.04}
\newcommand{\highlight}[1]{{\setlength{\fboxsep}{1pt}\colorbox{citrine!30}{$\displaystyle
      #1$}}}

\newcommand{\dual}[1]{\overline{#1}}

\newcommand{\caseof}[1]{  \resetpfcounter\textbf{Case : } #1}

\newcommand{\pfitem}[2][]{
	\ifthenelse{\equal{#1}{}}{}{(#1)} $#2$ \\
}

\newcommand{\proofstep}[3][]{\ifthenelse{\equal{$#1$}{}}{}{(#1)} $#3$ \hfill #2\\}

\newcommand{\proofref}[1]{($#1$)}

\newcommand{\wfjudgement}[1]{\vdash {#1}}

\definecolor{modalColour}{rgb}{0.2, 0.4, 0.2}%

\newcommand{\M}{\ensuremath{M}}
\newcommand{\pa}[2]{#1 \triangleleft  #2}
\newcommand{\soutprlbl}[3]{\ensuremath{#1\outsymbol#2\langle#3\rangle}}
\newcommand{\sinprlbl}[3]{\ensuremath{#1\inpsymbol#2(#3)}}
\newcommand{\soutlbl}[4]{\ensuremath{#1\outsymbol#2\langle#3\rangle}.{#4}}
\newcommand{\sinplbl}[4]{\ensuremath{#1\inpsymbol#2(#3).{#4}}}

\newcommand{\ssout}[2]{\ensuremath{#1\outsymbol#2}}
\newcommand{\ssinp}[2]{\ensuremath{#1\inpsymbol#2}}

\makeatletter
\newcommand{\superimpose}[3][\mathord]{#1{\mathpalette\superimpose@{{#2}{#3}}}}
\newcommand{\superimpose@}[2]{\superimpose@@{#1}#2}
\newcommand{\superimpose@@}[3]{%
    \ooalign{%
        \hfil$\m@th#1#2$\hfil\cr
        \hfil$\m@th#1#3$\hfil\cr
        }%
    }
\makeatother

\newcommand{\ie}{i.e.,\xspace}
\newcommand{\wrt}{w.r.t.\ }
\newcommand{\eg}{e.g.\ }
\newcommand{\Set}[1]{{\left\lbrace #1 \right\rbrace}}
\newcommand{\Subst}[2]{\nicefrac{#1}{#2}}

\newcommand{\Orbit}[2]{\mathit{O}_{#1}{\left( #2 \right)}}

\newcommand{\patternM}{\textnormal{\textbf{\textsf{M}}}\xspace}
\newcommand{\patternStar}{\ensuremath{\star}\xspace}
\newcommand{\PM}{M^{\patternM}}
\newcommand{\PS}{M^{\patternStar}}

\newcommand{\proc}{\mathcal{P}}
\newcommand{\indexUntyped}{\mathsf{ut}}
\newcommand{\indexCMVmix}{\mathsf{CMV}^+}
\newcommand{\procCMVmixUntyped}{\proc_{\indexCMVmix}^{\indexUntyped}}
\newcommand{\procCMVmix}{\proc_{\indexCMVmix}}
\newcommand{\stepCMVmix}{\red}

\newcommand{\linCMVmix}{\mathsf{lin}}
\newcommand{\unCMVmix}{\mathsf{un}}
\newcommand{\ChoiceCMVmix}[3]{#1\,#2\,#3}
\newcommand{\ResCMVmix}[3]{{\left( \nu #1 #2 \right)} #3}
\newcommand{\BranchCMVmix}[4]{#1{#2}#3.#4}
\newcommand{\OutCMVmix}[3]{#1!#2.#3}
\newcommand{\InpCMVmix}[3]{#1?#2.#3}
\newcommand{\inactCMVmix}{\mathbf{0}}
\newcommand{\ConditionalCMVmix}[3]{\mathsf{if} \, #1 \, \mathsf{then} \, #2 \, \mathsf{else} \, #3}

\newcommand{\ruleRIfTCMVmix}{\textsc{(R-IfT$ _{\indexCMVmix} $)}\xspace}
\newcommand{\ruleRIfFCMVmix}{\textsc{(R-IfF$ _{\indexCMVmix} $)}\xspace}
\newcommand{\ruleRLinLinCMVmix}{\textsc{(R-LinLin$ _{\indexCMVmix} $)}\xspace}
\newcommand{\ruleRLinUnCMVmix}{\textsc{(R-LinUn$ _{\indexCMVmix} $)}\xspace}
\newcommand{\ruleRUnLinCMVmix}{\textsc{(R-UnLin$ _{\indexCMVmix} $)}\xspace}
\newcommand{\ruleRUnUnCMVmix}{\textsc{(R-UnUn$ _{\indexCMVmix} $)}\xspace}
\newcommand{\ruleRResCMVmix}{\textsc{(R-Res$ _{\indexCMVmix} $)}\xspace}
\newcommand{\ruleRParCMVmix}{\textsc{(R-Par$ _{\indexCMVmix} $)}\xspace}
\newcommand{\ruleRStructCMVmix}{\textsc{(R-Struct$ _{\indexCMVmix} $)}\xspace}

\newcommand{\ChoiceTCMVmix}[3]{#1{#2}#3}
\newcommand{\IntCMVmix}[2]{#1{\oplus}#2}
\newcommand{\ExtCMVmix}[2]{#1{\&}#2}
\newcommand{\finCMVmix}{\mathsf{end}}
\newcommand{\RecCMVmix}[2]{\mu #1.#2}

\newcommand{\Dual}[2]{#1 \, \bot \, #2}
\newcommand{\At}[2]{#1:#2}
\newcommand{\Subtype}[2]{#1 \, <: \, #2}
\newcommand{\UnT}[1]{#1 \, \unCMVmix}

\newcommand{\ruleTUnitCMVmix}{\textsc{(T-Unit$ _{\indexCMVmix} $)}\xspace}
\newcommand{\ruleTTrueCMVmix}{\textsc{(T-True$ _{\indexCMVmix} $)}\xspace}
\newcommand{\ruleTFalseCMVmix}{\textsc{(T-False$ _{\indexCMVmix} $)}\xspace}
\newcommand{\ruleTVarCMVmix}{\textsc{(T-Var$ _{\indexCMVmix}
    $)}\xspace}
\newcommand{\ruleTSubCMVmix}{\textsc{(T-Sub$ _{\indexCMVmix} $)}\xspace}
\newcommand{\ruleTOutCMVmix}{\textsc{(T-Out$ _{\indexCMVmix} $)}\xspace}
\newcommand{\ruleTInCMVmix}{\textsc{(T-In$ _{\indexCMVmix} $)}\xspace}
\newcommand{\ruleTInactCMVmix}{\textsc{(T-Inact$ _{\indexCMVmix} $)}\xspace}
\newcommand{\ruleTParCMVmix}{\textsc{(T-Par$ _{\indexCMVmix} $)}\xspace}
\newcommand{\ruleTIfCMVmix}{\textsc{(T-If$ _{\indexCMVmix} $)}\xspace}
\newcommand{\ruleTResCMVmix}{\textsc{(T-Res$ _{\indexCMVmix} $)}\xspace}
\newcommand{\ruleTChoiceCMVmix}{\textsc{(T-Choice$ _{\indexCMVmix} $)}\xspace}

\newcommand{\scCMVmix}{\equiv}

\newcommand{\indexCMV}{\mathsf{CMV}}
\newcommand{\procCMVUntyped}{\proc_{\indexCMV}^{\indexUntyped}}

\newcommand{\stepCMV}{\red}

\newcommand{\linCMV}{\mathsf{lin}}
\newcommand{\unCMV}{\mathsf{un}}
\newcommand{\ResCMV}[3]{{\left( \nu #1 #2 \right)} #3}
\newcommand{\OutCMV}[3]{#1!#2.#3}
\newcommand{\InpCMV}[4]{#1 \, #2?#3.#4}
\newcommand{\inactCMV}{\mathbf{0}}
\newcommand{\ConditionalCMV}[3]{\mathsf{if} \, #1 \, \mathsf{then} \, #2 \, \mathsf{else} \, #3}
\newcommand{\SelCMV}[3]{#1 \triangleleft #2.#3}
\newcommand{\BranCMV}[2]{#1 \triangleright #2}
\newcommand{\BranchCMV}[2]{#1 : #2}

\newcommand{\ruleRLinComCMV}{\textsc{(R-LinCom$ _{\indexCMV} $)}\xspace}
\newcommand{\ruleRUnComCMV}{\textsc{(R-UnCom$ _{\indexCMV} $)}\xspace}
\newcommand{\ruleRCaseCMV}{\textsc{(R-Case$ _{\indexCMV} $)}\xspace}

\newcommand{\ComTCMV}[4]{#1 \, #2#3.#4}
\newcommand{\ChoiceTCMV}[3]{#1{#2}#3}
\newcommand{\RecCMV}{\@ifstar\RecCMVStar\RecCMVNoStar}
\newcommand{\RecCMVStar}[2]{{\left( \RecCMVNoStar{#1}{#2} \right)}}
\newcommand{\RecCMVNoStar}[2]{\mu #1.#2}

\newcommand{\ruleTOutCMV}{\textsc{(T-OutC$ _{\indexCMV} $)}\xspace}
\newcommand{\ruleTInCMV}{\textsc{(T-InC$ _{\indexCMV} $)}\xspace}
\newcommand{\ruleTBranchCMV}{\textsc{(T-Branch$ _{\indexCMV} $)}\xspace}
\newcommand{\ruleTSelCMV}{\textsc{(T-Sel$ _{\indexCMV} $)}\xspace}

\newcommand{\noRed}{\centernot\red}

\newcommand{\indexSource}{\operatorname{S}}
\newcommand{\indexTarget}{\operatorname{T}}

\newcommand{\ArbitraryEncoding}[1]{\left\llbracket #1 \right\rrbracket}
\newcommand{\arbitraryEncoding}{\ArbitraryEncoding{\cdot}}
\newcommand{\TypeEncoding}[1]{\llparenthesis {#1} \rrparenthesis}

\newcommand{\context}{\mathcal{C}}
\newcommand{\Context}[3]{\context^{#1}_{#2}\!\left( #3 \right)}
\newcommand{\hole}{[\cdot]}

\newcommand{\success}{\checkmark}
\newcommand{\hasSuccess}{\downarrow_{\success}}
\newcommand{\reachSuccess}{\Downarrow_{\diamond\success}}
\newcommand{\reachSuccessFin}{\Downarrow_{\square\diamond\success}}
\newcommand{\HasBarb}[1]{\downarrow_{#1}}
\newcommand{\WeakBarb}[1]{\Downarrow_{#1}}

\newcommand{\redArrow}{\begin{tikzpicture}
	\path[-stealth, thick, color=red] (0, 0) edge (0.5, 0);
	\draw[decorate, color=red, decoration={crosses, shape size=1.5mm, pre=, pre length=0.2cm, post=, post length=0.2cm}] (0, 0) to (0.5, 0);
	\node (dummy) at (0.25, 0) {};
\end{tikzpicture}\xspace}
\newcommand{\blueArrow}{\begin{tikzpicture}
	\path[-stealth, color=blue, thick] (0, 0) edge (0.5, 0);
	\node (dummy) at (0.25, 0) {};
\end{tikzpicture}\xspace}
\newcommand{\blackArrow}{\begin{tikzpicture}
	\path[-stealth, thick] (0, 0) edge (0.5, 0);
	\node (dummy) at (0.25, 0) {};
\end{tikzpicture}\xspace}
\newcommand{\greenArrow}{\begin{tikzpicture}
	\path[-stealth, color=green, thick] (0, 0) edge (0.5, 0);
	\node (dummy) at (0.25, 0) {};
\end{tikzpicture}\xspace}
\newcommand{\GreyBubble}[1]{\begin{tikzpicture}[baseline=(A.base)]
	\node[rounded corners, rectangle, fill=black!10] (A) at (0, 0) {#1};
\end{tikzpicture}\xspace}

\providecommand{\ifnotsplit}[1]{{#1}}
\externalcitedocument{full}

\AtBeginDocument{%
  \providecommand\BibTeX{{%
    Bib\TeX}}}

\setcopyright{acmlicensed}
\copyrightyear{2018}
\acmYear{2018}
\acmDOI{XXXXXXX.XXXXXXX}

\acmConference[LICS '24]{Make sure to enter the correct
  conference title from your rights confirmation emai}{June 03--05,
  2018}{Woodstock, NY}
\acmISBN{978-1-4503-XXXX-X/18/06}

\copyrightyear{2024}
\acmYear{2024}
\setcopyright{rightsretained}
\acmConference[LICS '24]{39th Annual ACM/IEEE Symposium on Logic in Computer Science}{July 8--11, 2024}{Tallinn, Estonia}
\acmBooktitle{39th Annual ACM/IEEE Symposium on Logic in Computer Science (LICS '24), July 8--11, 2024, Tallinn, Estonia}\acmDOI{10.1145/3661814.3662085}
\acmISBN{979-8-4007-0660-8/24/07}

\begin{document}

\title{Separation and Encodability in Mixed Choice Multiparty Sessions
(Full Version)}

\author{Kirstin Peters}
\email{kirstin.peters@uni-a.de}
\orcid{0000-0002-4281-0074}
\affiliation{%
  \institution{Augsburg University}
  \city{Augsburg}
  \country{Germany}
}

\author{Nobuko Yoshida}
\email{nobuko.yoshida@cs.ox.ac.uk}
\orcid{0000-0002-3925-8557}
\affiliation{%
  \institution{University of Oxford}
  \city{Oxford}
  \country{United Kingdom}
}

\begin{abstract}
Multiparty session types (\mpst) are a type discipline for enforcing the
structured, 
deadlock-free communication of concurrent and message-passing programs. 
Traditional \mpst have a limited form of choice in which 
alternative communication possibilities are offered by a single participant and selected by another. 
\emph{Mixed choice multiparty session types} (\mcmpst) extend the choice
construct to include both selections and offers in the same choice. 
This paper first proposes a general typing system for a mixed choice
synchronous multiparty session calculus, and prove  
type soundness, communication safety, and deadlock-freedom. 

Next we compare expressiveness of nine subcalcli of \mcmpst-calculus
by examining their \emph{encodability} (there exists 
a \emph{good} encoding from one to another) 
and \emph{separation} (there exists \emph{no} 
good encoding from one calculus to another). 
We prove 8 new encodablity results 
and 20 new separation results. 
In summary, \mcmpst 
is strictly more expressive than 
classical multiparty sessions (\mpst) in 
\cite{DBLP:journals/jlp/GhilezanJPSY19} and 
mixed choice in mixed sessions in \cite{CASAL202223}.
This contrasts to the results proven in \cite{CASAL202223,PY2022}
where mixed sessions \cite{CASAL202223} do not add 
any expressiveness to non-mixed fundamental sessions 
in \cite{VASCONCELOS201252},
shedding a light on expressiveness of multiparty mixed choice. 
\end{abstract}

\ccsdesc[500]{Theory of computation~Process calculi}
\ccsdesc[500]{Theory of computation~Parallel computing models}
\ccsdesc[500]{Theory of computation~Distributed computing models}
\ccsdesc[100]{Software and its engineering~Concurrent programming languages}

\keywords{Session Types, Mixed Choice, Concurrency, Pi-Calculus, Typing System, Protocols, Expressiveness} 

\maketitle

\section{Introduction}
\label{sec:intro}
\emph{Mixed choice}, which allows non-deterministic
choice between enabled inputs or outputs, 
has been used to represent mutual exclusion such as 
semaphores and concurrent scheduling algorithms in communicating 
systems \cite{CCS}. 
Mixed choice offers the ability to rule out
alternative options, i.e., discard inputs by selecting an output in
the same choice, and vice versa. 
In concurrent and message-passing programming 
languages, there has been interest 
in including and efficiently implementing mixed
choice, as exemplified by 
Concurrent ML
\cite{CML,RRY09}, and
more recently by Go \cite{golang} (where choice is
synchronous by default). 
In Esterel \cite{BERRY199287} 
and Facile \cite{facile96},    
 mixed choice is used 
as a key construct 
to lump all 
IO-synchronisations among parallel processes
as a single choice.

This paper shows that 
an introduction of mixed choice 
in the behavioural type 
theory based on protocols, \emph{multiparty session types}
\cite{HYC08} (\mpst), 
not only offers more safe and deadlock-free processes, 
but also gains \emph{expressiveness} \cite{DBLP:journals/iandc/Gorla10, DBLP:journals/corr/PetersG15},  which was not the case in 
binary (two-party) mixed sessions by \citet{CASAL202223}. 

\myparagraph{Two Party Mixed Choice Sessions.\ }
Session types \cite{THK,honda.vasconcelos.kubo:language-primitives,yoshida.vasconcelos:language-primitives} 
govern communication behaviours 
of concurrent programs, ensuring
\emph{type error freedom} and  
\emph{communication safety} 
(no mismatch between sent and expected data types).
The shape of session types originated in Linear Logic 
\cite{HondaK:typdyi,GirardJY:linlog},
where choices are \emph{separated} (not mixed) and 
\emph{binary} (between two participants). Such choices are 
either a sum of inputs (\emph{external} choice) or 
of outputs (\emph{internal} choice). 

Using session process notation, we can write
\emph{external choice} and 
\emph{internal choice} processes as:\\[1mm] 
\centerline{$P_{\&} \ = \
\sinprlbl{s}{\lab_1}{x_1}.P_1 +
\sinprlbl{s}{\lab_2}{x_2}.P_2$}\\[1mm]
\centerline{
$Q_{\oplus} = \text{if} \ v \ \text{then} \  \soutprlbl{s}{\lab_1}{v_1}.Q_1
\ \text{else} \ 
\soutprlbl{s}{\lab_2}{v_2}.Q_2
$}\\[1mm] 
Here 
$\outmark$ denotes output, 
$\inpmark$ denotes input, and 
$\ell$ is a \emph{label}
used for matching. 
The input process $\sinprlbl{s}{\lab_1}{x_1}.P_1$ 
indicates that the recipient at channel $s$ expects to
receive a value with label $\ell_1$, after which it will continue with behaviour 
$P_1\subst{v_1}{x_1}$.
The output 
$\soutprlbl{s}{\lab_1}{v_1}.Q_1$ 
selects 
label $\ell_1$, sending value $v_1$ and continues as $Q_1$.

A natural next step is the extension of separate 
binary choice to  
\emph{mixed choice} 
(a mixture of synchronous input and outputs in a single choice), 
making it \emph{non-deterministic}, e.g.,
a process waits for 
an input on label $\ell_1$, or can
\emph{non-deterministically} select to output $\ell_2$: \\[1mm]
\centerline{
$P_{+} \ = \
\sinprlbl{s}{\lab_1}{x_1}.P_1 +
\soutprlbl{s}{\lab_2}{v_2}.P_2
\quad 
Q_{+} = \soutprlbl{s}{\lab_1}{v_1}.Q_1 +
\sinprlbl{s}{\lab_2}{x_2}.Q_2$
}\\[1mm]
where 
a parallel composition of $P_{+}$ and $Q_{+}$ synchronises (reduces) either to 
$P_1\subst{v_1}{x_1} \Par Q_1$ or 
$P_2 \Par Q_2\subst{v_2}{x_2}$. 
\NY{This mixed choice behaviour follows the standard 
CCS semantics \cite{CCS}: an output action always chooses a receive action at a choice in another process, and they are synchronised together.} 
Recently, 
\citet{CASAL202223} 
studied this extension (denoted by \cnvms), and proved its type safety. 

\myparagraph{Expressiveness. }
A standard method to compare the abstract \emph{expressive power} of process calculi is to analyse the existence of encodings, \ie translations from one calculus into another.
To rule out trivial or meaningless encodings, they are augmented with a set of quality criteria.
An \emph{encodability} result, \ie an encoding from a \emph{source calculus} into a \emph{target calculus} that satisfies relevant criteria, shows that the target can mimic or \emph{emulate} the behaviours of the source---the target is at least as expressive as the source.
Conversely, a \emph{separation} result, \ie the proof that there is no encoding satisfying the considered criteria, shows that some behaviours of the source cannot be emulated by the target.
The combination of encodability and separation results can build hierarchies of the analysed calculi, if these results are based on the same set of criteria. 
Encodings become stronger by using a stronger criteria, whereas separations become stronger by using weaker criteria.
Sets of criteria that are well suited for encodability and separation were discussed \eg in \cite{Palamidessi03, DBLP:journals/entcs/Parrow08, gorla09, DBLP:journals/iandc/Gorla10, DBLP:journals/tcs/FuL10, DBLP:journals/corr/abs-1208-2750, peters12, PetersN12, DBLP:conf/esop/PetersNG13, DBLP:journals/corr/PetersG15, fu16, glabbeek18, peters19}.
By following in particular \cite{DBLP:journals/iandc/Gorla10, DBLP:conf/esop/PetersNG13}, we consider the criteria \emph{compositionality} (the encoding is a function on the operators of the source), \emph{name invariance} (the encoding treats all names of the source in the same way), sound and complete \emph{operational correspondence} (the encoding preserves and reflects the behaviours of the source), \emph{divergence reflection} (the target diverges only if the source diverges), \emph{success-sensitiveness} (reachability of success is preserved and reflected), and \emph{distributability preservation} (the target has the same degree of distribution as the source).
Encodings that satisfy these criteria are denoted as \emph{good} encodings.

Unfortunately, \citet{PY2022} have shown that 
mixed sessions (\cnvms) proposed by \citet{CASAL202223} do \emph{not} add  
any expressive power 
to the binary session calculus (denoted by \cnv)  \cite{VASCONCELOS201252}  
which has the standard branching and selection constructs (i.e., separate 
choices). 
This is perhaps surprising as it is against a landmark result on expressiveness  by \citet{Palamidessi03}---the $ \pi $-calculus \cite{short:MilnerR:calmp1} with mixed choice is strictly more expressive than the $ \pi $-calculus without choice.
An open question is under which circumstances, mixed choice in session types
is strictly more expressive. 

\begin{figure}
	\begin{tikzpicture}[node distance=3cm, auto]
		\node[scale=2] (star) at (2.75, 0.5) {\ \patternStar};
		\draw[dashed] (-2, 0.5) -- (2, 0.5) -- (3, 1);
		\draw[dashed] (2, 0.5) -- (3, 0);
		\node[scale=1.2] (M) at (-2.75, -0.5) {\patternM\quad};
		\draw[dashed] (2, -0.5) -- (-2, -0.5) -- (-3, -1);
		\draw[dashed] (-2, -0.5) -- (-3, 0);
		\node (mix) at (0, 0.8) {$ \pi $};
		\node (sep) at (-2, 0) {$ \pi_{\mathsf{s}} $};
		\node (asyn) at (-1.25, 0) {$ \pi_{\mathsf{a}} $};
		\node (ma) at (-0.3, 0) {$ \mathsf{MA} $};
		\node (CMV) at (0.75, 0) {\cnv};
		\node (CMVmix) at (2, 0) {\cnvms};
		\node (join) at (-0.75, -0.8) {$ \mathsf{Join} $};
		\node (mau) at (0.75, -0.8) {$\mathsf{MA}_{\mathsf{u}} $};
	\end{tikzpicture}
	\caption{Hierarchy of $ \pi $-like calculi \cite{PY2022} \label{fig:hierarchyPi}}
\end{figure}
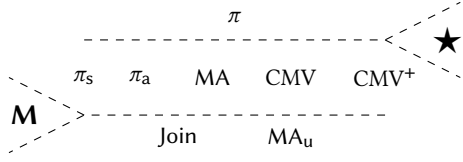

\myparagraph{Expressiveness Hierarchy with Binary Mixed Sessions.\ }
To explain the above open problem more precisely, we first 
present a hierarchy of $ \pi $-like calculi from \cite{PY2022}
in Figure~\ref{fig:hierarchyPi}.

\noindent
\begin{minipage}{\columnwidth}
	\begin{wrapfigure}{R}{0.25\columnwidth}
		\hspace*{-2em}
		\scalebox{0.65}{
		\tikzstyle{place}=[circle,draw=black,thick,minimum size=5mm]
		\tikzstyle{transition}=[rectangle,draw=black,thick,minimum size=5mm]
		\begin{tikzpicture}
			\node[place,tokens=1]	(p) at (0.75, 1.5) {};
			\node[place,tokens=1]	(q) at (2.25, 1.5) {};
			\node[transition]		(a) at (0, 0) {$ a $};
			\node[transition]		(b) at (1.5, 0) {$ b $};
			\node[transition]		(c) at (3, 0) {$ c $};
			\draw[-latex] (p) -- (a);
			\draw[-latex] (p) -- (b);
			\draw[-latex] (q) -- (b);
			\draw[-latex] (q) -- (c);
		\end{tikzpicture}
		}
	\end{wrapfigure}
\smallskip

The hierarchy orders calculi along their ability to express the \emph{synchronisation patterns} \patternM and \patternStar.
An \patternM (see \cite{Glabbeek2008, glabbeekGoltzSchicke12}), as visualised on the right, describes a Petri net that consists of two parallel transitions ($ a $ and $ c $) and one transition ($ b $) that is in conflict with both of the former.
	In other words, it describes a situation where either two parts of the net can proceed independently or they synchronise to perform a single transition together.

\smallskip

\end{minipage}

As stated by Glabbeek et al. \cite{Glabbeek2008, glabbeekGoltzSchicke12}, a Petri net specification can be implemented in an asynchronous, fully distributed setting iff it does not contain a fully reachable pure \patternM.
They also present a description of a fully reachable pure \patternM as conditions on a state in a step transition system, which allows us to directly use this pattern to reason about process calculi 
as shown in \cite{petersSchickeNestmann11, peters12, DBLP:conf/esop/PetersNG13}. 

\noindent
\begin{minipage}{\columnwidth}
	\begin{wrapfigure}{R}{0.25\columnwidth}
		\vspace{-1em}
		\hspace{-2em}
		\scalebox{0.6}{
		\tikzstyle{place}=[circle,draw=black,thick,minimum size=5mm]
		\tikzstyle{transition}=[rectangle,draw=black,thick,minimum size=5mm]
		\begin{tikzpicture}
			\foreach \x/\xtext in {1/e,2/d,3/c,4/b,5/a}
	        {
	            \path (360*\x/5+125:0.8) node[transition] (\xtext) {$\xtext$};
	            \path (360*\x/5-55:1.75) node[place,tokens=1] (p\x) {};
	        }
	        \draw[-latex] (p2) -- (a);
	        \draw[-latex] (p2) -- (b);
	        \draw[-latex] (p1) -- (b);
	        \draw[-latex] (p1) -- (c);
	        \draw[-latex] (p5) -- (c);
	        \draw[-latex] (p5) -- (d);
	        \draw[-latex] (p4) -- (d);
	        \draw[-latex] (p4) -- (e);
	        \draw[-latex] (p3) -- (e);
	        \draw[-latex] (p3) -- (a);
		\end{tikzpicture}
		}
		\vspace{-1em}
	\end{wrapfigure}

\smallskip

Further, Peters et al.~\cite{peters12, DBLP:conf/esop/PetersNG13} introduce the synchronisation pattern \patternStar, 
which is a chain of conflicting and distributable steps as they occur in an \patternM that build a circle of odd length.
	As visualised on the right, there is \eg one \patternM consisting of the transitions $ a $, $ b $, and $ c $ with their corresponding two places.
	Another \patternM is build by the transitions $ b $, $ c $, and $ d $ with their corresponding two places and so on.

\smallskip
\end{minipage}

The \patternM captures only a small amount of synchronisation, whereas the \patternStar requires considerably more synchronisation in the calculus.
\emph{Asynchronously distributed calculi} that cannot express \patternM nor \patternStar such as the $ \mathsf{Join} $-calculus from \cite{fournet.gonthier:join} are placed in the bottom layer of Figure~\ref{fig:hierarchyPi}.
\emph{Synchronously distributed calculi} that can express \patternStar and thus also \patternM such as the $ \pi $-calculus with mixed choice ($ \pi $) are placed in the top layer.
The middle layer consists of the calculi that can 
express \patternM but not \patternStar. 
The asynchronous $\pi$-calculus without choice \cite{HondaTokoro91,Boudol92} 
($\pi_{\mathsf{a}}$),
the $ \pi $-calculus 
with separate choice \cite{Nestmann00}
($\pi_{\mathsf{s}}$),
and mobile ambients \cite{CardelliL:mobamb} ($ \mathsf{MA} $)
are placed in the middle---unless it is restricted to unique ambient names ($ \mathsf{MA}_{\mathsf{u}} $) \cite{petersNestmannIC20}.
That \cnv and \cnvms are also placed in the middle layer was proven in \cite{PY2022}.
Hence, they are \emph{strictly less expressive} than $ \pi $.
This completes Figure~\ref{fig:hierarchyPi}.   

\myparagraph{Mixed Choice Multiparty Session Types.}
In the presence of 
mixed choice, 
the extension from two parties to \emph{more than two} parties
gives us strictly more expressive power to session types and is placed in the top layer. 
This is because 
not only inputs and outputs but also 
\emph{destinations of messages} represented by \emph{participants}
can be mixed in one single choice.

Consider three participants
$\role{a}$, $\role{b}$, and $\role{c}$, and  
assume $P_{\role{a}}$, $P_{\role{b}}$ and $P_{\role{c}}$ below:\\[1mm]
\centerline{
$P_{\role{a}} = 
(\tout{\role{b}}{\ell_1}{
 \ntout{\role{c}}{\ell_2}{}})
+
\ntinp{\role{b}}{\ell_1}{}
+
\ntinp{\role{c}}{\ell_2}{}
\quad
P_{\role{b}} = 
\ntinp{\role{a}}{\ell_1}{}
+
(\tout{\role{a}}{\ell_1}{\ntout{\role{c}}{\ell_2}{}})
+
\ntinp{\role{c}}{\ell_2}{}$}\\[1mm]
\centerline{
$P_{\role{c}} = 
\ntinp{\role{a}}{\ell_2}{}
+
\ntinp{\role{b}}{\ell_2}{}
+
(\tout{\role{a}}{\ell_2}{\ntout{\role{b}}{\ell_2}{}})$}\\[1mm]
where we omit the payload and nil processes. In multiparty sessions,
each process who plays role $\p$
is represented as $\pa{\p}{\PP}$. Assuming, e.g.,  
$\ntout{\role{b}}{\ell_1}$ in $P_{\role{a}}$ matches with
$\ntinp{\role{a}}{\ell_1}$ in $P_{\role{b}}$,
their parallel composition 
$(\pa{\role{a}}{\PP_{\role{a}}}
\Par 
\pa{\role{b}}{\PP_{\role{b}}}
\Par 
\pa{\role{c}}{\PP_{\role{c}}})$ 
non-deterministically leads 
to several possible states such as 
$\pa{\role{a}}{\ntout{\role{c}}{\ell_2}{}}
\Par 
\pa{\role{c}}{\PP_{\role{c}}}$ 
or 
$\pa{\role{b}}{\PP_{\role{b}}}
\Par 
\pa{\role{c}}{\ntout{\role{b}}{\ell_2}{}}$.  
We can observe that the choice behaviours of processes are 
\emph{distributed} to and from two distinct participants.

\myparagraph{Expressiveness in Mixed Choice Multiparty Sessions}
After introducing the typing system of \mcmpst, we study the expressive power introduced by multiparty mixed choice, \ie determine in which layer of Figure~\ref{fig:hierarchyPi}, \mcmpst is situated.
A precise answer to this question was not as simple as expected.
To more clearly understand the causes of an increase in expressiveness, 
we first restrict our attention to a \emph{single multiparty session} 
which has neither shared names, name passings, 
session delegation nor interleaved sessions, and   
consider subcalculi of \mcmpst. 
By limiting the form of the choices and number of participants, \mcmpst includes \emph{9 subcalculi} (including itself).
For example, we can define
\emph{separated choice from/to multiple participants} (\scmpst),  
e.g.,
\\[1mm]  
\centerline{
$P_{\text{\scmpst}}=\tout{\role{a}}{\ell_1}{\tout{\role{b}}{\ell_1}} 
+
\tout{\role{a}}{\ell_2}{\tinp{\role{c}}{\ell_2}}
+ 
\tout{\role{b}}{\ell_3}{\tinp{\role{c}}{\ell_2}}
+
\tout{\role{b}}{\ell_4}{\tinp{\role{a}}{\ell_2}}$
}\\[1mm]
and \emph{directed mixed choice} (\dmpst) 
where choice is mixed but from/to the same participant,  
e.g.,  
\\[1mm]  
\centerline{
$P_{\text{\dmpst}}
=\tout{\role{a}}{\ell_1}{\tout{\role{b}}{\ell_1}} 
+
\ntinp{\role{a}}{\ell_2}{\tinp{\role{c}}{\ell_2}}
+
\tout{\role{a}}{\ell_3}{\tout{\role{b}}{\ell_3}}$
}\\[1mm]
This makes the expressiveness analysis 
very subtle and comprehensive:  
we prove 20 new separation results and 8 
new encodability results among 
9 subcalculi of \mcmpst and 4 variants of \cnvms\ 
\cite{CASAL202223}. 
Our observation is, 
by allowing a \emph{combination of different 
communication patterns in a single choice}, we can cross the boundary of expressiveness. For example, \mcmpst is strictly more expressive than \scmpst 
and can express the pattern \patternStar; and 
\scmpst is strictly more expressive than \dmpst and can 
express pattern \patternM; 
but \dmpst and \mpst have the same expressive power and belong to 
the bottom. 

The highlight is, in spite of the fact that \cnvms enables creating an infinite number of interleaved binary mixed sessions with unrestricted names, \mcmpst 
(a single mixed choice multiparty session) cannot be emulated by \cnvms.
In our results, properties such as \emph{communication safety}  and \emph{deadlock-freedom} \NY{(if a session terminates, then all participants terminate, completing all actions)} ensured by the \mcmpst typing system play the key role to prove the encodability results. 

\myparagraph{Contributions.\ } 
\begin{enumerate*}[=leftmargin]
\item We first introduce the calculus, types and a typing system of 
\emph{mixed choice multiparty sessions} (\mcmpst) which subsume  
the types and typability of the classical synchronous multiparty sessions  
(\mpst). We follow a general \mpst typing system in \cite{less_is_more} which does not require global types. 
We then prove 
\emph{communication safety} and \emph{deadlock-freedom} of typable sessions.
\item \KPCom{We analyse the expressive power of mixed choice in \mcmpst showing that it is strictly more expressive than choice in \mpst.
Therefore we use 8 subcalculi of \mcmpst deliberately chosen to explain the features of choice that raise expressiveness and compare to 4 variants of \cnvms.}
We prove 8 encodability and 20 separation results among the 13 calculi, introducing 8 new \emph{good} encodings. 
\end{enumerate*}
The omitted proofs and more details can be found in Appendix.

\section{Mixed Choice Multiparty Session $\pi$-Calculus (\mcmpst-calculus)}
\label{sec:calculus}
\begin{figure}
$
\begin{array}{cl}
\quad \pa{\p}{(\sinprlbl{\role{q}}{\lab}{x} \seq\PP + R_1)} 
\Par 
\pa{\q}{(\soutprlbl{\role{p}}{\lab}{v} \seq Q + R_2)} \Par \M\\ 
\red 
\pa{\p}{\PP\subst{v}{x}} \Par \pa{\q}{Q} \Par \M 
& 
\RChoice
\\[1mm]
	\pa{\p}{\cond{\ \true\ }{\PP_1}{\PP_2}}\Par \M 
	\red
	\pa{\p}{\PP_1}\Par \M  
	& \RCondT\\
	\pa{\p}{\cond{\ \false\ }{\PP_1}{\PP_2}}\Par \M
	\red
	\pa{\p}{\PP_2} \Par \M
	& \RCondF\\
		\M \scong \M_1 \red \M_2 \scong \M'
\quad \Rightarrow \quad 
		\M \red \M'
& \RCong
	\end{array}
	$\\
	\rule{0.45\textwidth}{0.3mm}\\
	$
\hspace*{-0.3cm}	\begin{array}{c}
P \scong_\alpha Q \Rightarrow P\equiv Q \quad 
\rec{X} \PP \scong \PP \subst{\rec{X} \PP}{\pvar{X}} 
\quad 
\M \scong_\alpha \M' \Rightarrow \M \equiv \M'
\\
\PP \equiv Q \Rightarrow \pa{\p}{\PP} \Par \M \equiv \pa{\p}{Q}\Par \M
\quad \pa{\p}{\inact} \Par \M \equiv \M\\
\M \Par \M' \equiv \M'\Par \M \quad 
(\M \Par \M')\Par \M''  \equiv \M \Par (\M'\Par \M'')
\\[-3mm]
\end{array}
$
\caption{Reduction and structure congruence of \mcmpst
\label{fig:reduction}\label{def:scong}}
\end{figure}

This section introduces the syntax and operational semantics
for the \emph{mixed choice multiparty session calculus}
(\mcmpst-calculus), then define its subcalculi. 

\subsection{Syntax of \mcmpst-Calculus and its Family} 
\label{subsec:syntax}
The syntax of the \mcmpst-calculus follows the simplest synchronous multiparty 
session calculus \cite{YG2020,DBLP:journals/jlp/GhilezanJPSY19} \NY{(which 
consists of only a single multiparty session without session delegations)}, 
extended with nondeterministic mixed choices.

\begin{newdefinition}[syntax]
\label{def:syntax}
Assume a set of \emph{participants} $\Part$ $(\p, \q, \rr, \dots)$
and a set of \emph{labels} $\Labels$ $(\lab, \lab', \dots)$.  
\emph{Values} contain either \emph{variables} 
$(x, y, z,\ldots)$ or constants; 
$\pprefix, \pprefix', \dots$ denote \emph{prefixes}; 
$\pvar{X}, \pvar{Y}, \dots$ denote \emph{process variables}; 
$P, Q, \dots$ denote \emph{processes} and  
$\M, \M', \dots$ denote \emph{multiparty sessions} (often called 
\emph{sessions}). 
\\[1mm]
\centerline{$
\begin{array}{r@{~}c@{~}ll}
\val	&\bnfis&	
x,y,z,\dots
\bnfbar 
1, 2,\dots 
\bnfbar \true,                                                       \false
		& \text{\footnotesize{(variables, numbers, booleans)}}\\
\pprefix &\bnfis& 
				\soutprlbl{\p}{\lab}{v} \bnfbar 
				\sinprlbl{\p}{\lab}{x}
		& \text{\footnotesize{(output prefix, input prefix)}}
		\\ 
		\PP		&\bnfis&	\inact 
				\bnfbar \pvar{X} 
				\bnfbar \rec{X} \PP 
				& \text{\footnotesize{(nil, proc var, 
			recursion)}}\\
				& \bnfbar& 	\choice{\pprefix_i \seq
                                  \PP_i}{i \in I}
				& \text{\footnotesize{(mixed choice)}}
\\
		& \bnfbar & \cond{v}{\PP}{\PP}
				& \text{\footnotesize{(conditional)}}\\
\M				&\bnfis& \pa{\p}{\PP}
\bnfbar \M \Par \M 
				& \text{\footnotesize{(multiparty session, parallel)}}
\end{array}
$}
\end{newdefinition}
\noindent Output prefix $\soutprlbl{\p}{\lab}{v}$ 
which selects label $\lab$ at participant \p
by sending value $v$; 
and the matching input prefix, 
$\sinprlbl{\p}{\lab}{x}$ which 
receives a value with label \lab from participant \p  
and substitutes the value as
variable $x$.
We often omit values and variables 
($\p\outsymbol{\lab}$/$\p\inpsymbol{\lab}$) and labels for a
singleton prefix ($\soutprlbl{\p}{}{v}$/$\sinprlbl{\p}{}{x}$).  

Process terms include a 
\emph{nil}, \inact,
process variables and 
\emph{recursions} \rec{X}{\PP} where \pvar{X} is a binder.  
We assume $\PP$ is guarded \cite[\S~2]{DBLP:journals/jlp/GhilezanJPSY19}, i.e.,$\rec{X}{\pvar{X}}$ is not allowed. 
The nondeterministic \emph{mixed choice}
$\choice{\pprefix_i \seq \PP_i}{i \in I}$ 
($I\not=\emptyset$) is a sum of prefixed processes. 
Conditional $\cond{v}{\PP_1}{\PP_2}$ is standard. 
We assume standard $\alpha$-conversion, 
capture-avoiding substitution, $\PP \subst{\val}{\xx}$ 
and $\PP \subst{Q}{\pvar{X}}$; and use function $\fvsf(P)$ 
to denote free variables in $P$. We often omit 
$\inact$.   

A \emph{multiparty session} is a parallel composition of 
a participant process (denoted by $\pa{\p}{\PP}$) where 
process $P$ plays the role of participant $\p$, and can interact with 
other processes playing other roles in $\M$. 
We assume all participants in $\M$ are different.

We sometimes write $\pi_1.\PP_1 \por \cdots \por \pi_n.\PP_n$ for
$\choice{\pi_i.\PP_i}{i\in I}$ and 
$\Pi_{i \in I} \pa{\rolei{p}{i}}{\PP_i}$ for 
$\pa{\rolei{p}{1}}{\PP_1} \Par \cdots \Par \pa{\rolei{p}{n}}{\PP_n}$
with $I=\{1,..,n\}$. We omit $\Pi$ if $I$ is a singleton, i.e., 
we write $\pa{\rolei{p}{0}}{\PP_0}$ for 
$\Pi_{i \in \{0\}} \pa{\rolei{p}{i}}{\PP_i}$. Similarly for 
$\Sigma$. 
We often use $P+Q$ to denote
$\choice{\pi_i.\PP_i}{i \in \{1,..,n\}}$
with $P=\choice{\pi_i.\PP_i}{i \in \{1,..,k\}}$ and
$Q =\choice{\pi_i.\PP_i}{i \in \{k+1,..,n\}}$, and assume 
commutativity and associativity of $+$. 
We also use the nested choices 
$\choice{\choice{\pi_{ij}.\PP_{ij}}{i\in I_j}}{j\in J}$ with $J=\{1..n\}$ 
to denote $\choice{\pi_{ij}.\PP_{ij}}{i\in I_1}+\cdots \choice{\pi_{ij}.\PP_{ij}}{i\in I_n}$.

We next define subcalculi of \mcmpst which are used in the paper. 
\begin{newdefinition}[Family of \mcmpst]
\label{def:family} 
\strut
\begin{itemize}[leftmargin=*]
\item {\bf\msmpst:} \emph{Multiparty Mixed Separate Choice per Participant}, which is defined replacing $\choice{\pprefix_i \seq \PP_i}{i \in I}$ by  
$\choice{\choice {\soutprlbl{\rolei{p}{i}}{\lab_{ij}}{v_{ij}} \seq \PP_{ij}}{j \in J_i}}{i \in I} + 
\choice{\choice{\sinprlbl{\rolei{q}{k}}{\lab_{kh}}{v_{kh}} \seq \PP_{kh}}{h\in H_k}}{k \in K}$ where 
$\set{\rolei{p}{i}}_{i\in I}\cap \set{\rolei{q}{k}}_{k\in K}=\emptyset$ and 
$\cup_{i\in I}J_i \cup \cup_{k\in K}H_k\neq \emptyset$, 
i.e., the choices to/from each participant is either outputs or inputs.  

\item {\bf\scmpst:} \emph{Separate Choice Multiparty Session} 
where we have a sum of inputs 
$\choice{\sinprlbl{\rolei{p}{i}}{\lab_i}{x_i} \seq\PP_i}{i \in I}$ 
or a sum of outputs 
$\choice{\soutlbl{\rolei{p}{i}}{\lab_i}{x_i} \PP_i}{i \in I}$
with $I\neq \emptyset$. 

\item {\bf\dmpst:} \emph{Directed Mixed Choice Multiparty Session}
  where we have a mixed choice but from/to the same participant, 
i.e., a mixed choice
$\choice{\sinprlbl{\rolei{p}{}}{\lab_i}{x_i} \seq\PP_i}{i \in I}+
\choice{\soutlbl{\rolei{p}{}}{\lab_j}{v_j}\PP_j}{j \in J}$ 
with 
$I\cup J\neq \emptyset$.     
 
\item {\bf\smpst:} \emph{Separate Choice Multiparty Session} 
where we have a sum of inputs 
$\choice{\sinprlbl{\rolei{p}{}}{\lab_i}{x_i} \seq\PP_i}{i \in I}$ 
or a sum of outputs 
$\choice{\soutlbl{\rolei{p}{}}{\lab_i}{x_i} \PP_i}{i \in I}$
with $I\neq \emptyset$ from the same participant $\role{p}$. 

\item {\bf\mpst:} \emph{Multiparty Session} 
in \cite{YG2020,DBLP:journals/jlp/GhilezanJPSY19} where  
we only have a sum of inputs from the same participant 
$\choice{\sinprlbl{\rolei{p}{}}{\lab_i}{x_i} \seq\PP_i}{i \in I}$ 
with 
$I\neq \emptyset$ and  
a single selection $\soutprlbl{\role{p}}{\lab}{v} \seq\PP$.

\item {\bf\mcbs:} \emph{Mixed Choice Binary Session} 
where $\mcmpst$ is limited to two parties only. 

\item {\bf \scbs:} \emph{Separate Choice Binary Session} where 
\scmpst is limited to two parties only.  
The binary version of \msmpst is syntactically same as \scbs. 

\item {\bf\bs:} \emph{Binary Session} where 
\mpst is limited to two parties only. 
\end{itemize}
\end{newdefinition}
Note that
\mcmpst $\supset$ \msmpst $\supset$ \scmpst  $\supset$ \smpst $\supset$ \mpst; and \msmpst $\supset$ \dmpst $\supset$ \smpst;   
\mcbs $\supset$ \msbs ($=$ \scbs) $\supset$ \bs; and  
\dmpst $\supset$ \mcbs; \scmpst $\supset$ \scbs; 
and \mpst $\supset$ \bs where $\supset$ indicates a strict 
inclusion. Figure~\ref{fig:hierarchy} shows these set inclusions. 

\begin{newexample}[A Family of \mcmpst]
	\label{ex:family}
We list examples of each calculi from syntactically larger ones.  
Consider:\\[1mm]
{
\begin{tabular}{llll}
\mcmpst & 
$P_1= \ssout{\enodei{a}}{\ell}.\ssinp{\enodei{b}}{\ell} \OR 
\ssout{\enodei{b}}{\ell}.\ssout{\enodei{c}}{\ell} \OR 
\ssinp{\enodei{a}}{\ell}.\ssinp{\enodei{a}}{\ell}$\\
\msmpst & 
$P_2= \ssinp{\enodei{a}}{\ell}.\ssout{\enodei{b}}{\ell} \OR 
\ssout{\enodei{b}}{\ell}.\ssinp{\enodei{c}}{\ell} \OR 
\ssinp{\enodei{c}}{\ell}.\ssout{\enodei{a}}{\ell}$ 
\\
\scmpst & 
$P_3= \ssinp{\enodei{a}}{\ell}.\ssout{\enodei{b}}{\ell} \OR 
\ssinp{\enodei{b}}{\ell}.\ssinp{\enodei{c}}{\ell} \OR 
\ssinp{\enodei{c}}{\ell}.\ssinp{\enodei{a}}{\ell}$\\
\dmpst & 
$P_4= \ssout{\enodei{a}}{\ell_1}.\ssout{\enodei{b}}{\ell} \OR 
\ssout{\enodei{a}}{\ell_2}.\ssinp{\enodei{c}}{\ell} \OR 
\ssinp{\enodei{a}}{\ell}.\ssinp{\enodei{a}}{\ell}$\\
\smpst & 
$P_5= \ssout{\enodei{a}}{\ell_1}.\ssout{\enodei{b}}{\ell} \OR 
\ssout{\enodei{a}}{\ell_2}.\ssinp{\enodei{c}}{\ell} \OR 
\ssout{\enodei{a}}{\ell_3}.\ssinp{\enodei{a}}{\ell}$\\
\mpst & 
$P_6= \ssinp{\enodei{a}}{\ell_1}.\ssout{\enodei{b}}{\ell} \OR 
\ssinp{\enodei{a}}{\ell_2}.\ssinp{\enodei{c}}{\ell} \OR 
\ssinp{\enodei{a}}{\ell_3}.\ssout{\enodei{a}}{\ell}$\\
\mpst & 
$P_7= \ssinp{\enodei{a}}{\ell_1}.\ssout{\enodei{b}}{\ell_1} \OR 
\ssinp{\enodei{a}}{\ell_2}.\ssinp{\enodei{b}}{\ell_2}$\\
\mcbs & 
$P_8= \ssinp{\enodei{a}}{\ell_1}.\ssout{\enodei{a}}{\ell_1} \OR 
\ssout{\enodei{a}}{\ell_2}.\ssinp{\enodei{a}}{\ell_3}$\\
\scbs & 
$P_9= \ssout{\enodei{a}}{\ell_1}.\ssout{\enodei{a}}{\ell_1} \OR 
\ssout{\enodei{a}}{\ell_2}.\ssinp{\enodei{a}}{\ell_2}$\\
\bs & 
$P_{10} = \rec{X}{(\ssinp{\enodei{a}}{\ell_1}.\ssout{\enodei{a}}{\ell_1} \OR
\ssinp{\enodei{a}}{\ell_1}.\ssout{\enodei{a}}{\ell_2} \OR 
\ssinp{\enodei{a}}{\ell_2}.\ssinp{\enodei{a}}{\ell_2}.\pvar{X})}$\\
Untypable  & 
$P_{11} = \rec{X}{(\ssinp{\enodei{a}}{\ell_1}.\ssinp{\enodei{a}}{\ell_1} \OR
\ssinp{\enodei{a}}{\ell_1}.\ssinp{\enodei{a}}{\ell_2} \OR 
\ssinp{\enodei{a}}{\ell_2}.\ssinp{\enodei{a}}{\ell_2}.\pvar{X})}$\\
\end{tabular}
}\\[1mm]
The table is read as follows: 
$P_1$ is \mcmpst but not \msmpst; $P_2$ is \msmpst (hence \mcmpst) but 
neither \scmpst nor \dmpst; $P_3$ is \scmpst but neither \dmpst nor \smpst; 
$P_4$ is \dmpst but neither \scmpst nor \smpst; 
$P_5$ is \smpst but not \mpst; 
$P_6$ and 
$P_7$ are \mpst but not \mcbs; $P_8$ is \mcbs but not \mpst;  
$P_9$ is \scbs but not \mpst; $P_{10}$ is \bs. 
All processes except $P_{11}$ are typable with appropriate types by rules defined in Definition~\ref{def:typingsystem}. 
Notice that 
$P_{10}$ holds the two inputs from $\role{a}$ with the same label $\ell_1$ 
in the choice with the different outputs ($\ell_1$ and $\ell_2$) to $\role{a}$, but it will be typable using subtying and rule \TSum in Definition~\ref{def:typingsystem}. 
\end{newexample}
\subsection{Reduction Semantics of \mcmpst-Calculus}
The reduction and structural congruence rules 
are defined in Figure~\ref{fig:reduction}.
\NY{Rule \RChoice represents mixed choice communication:
it non-deterministically chooses a pair of an output and 
an input with the same label $\lab$, 
and at the same time, value $v$ is passed from the sender \q to receiver \p.}
Rule \RCong closes 
under structural congruence.

\NY{
As an example of reductions, let us define  $\M$ as:} \\[1mm]
\centerline{
$\pa{\p}{(\sinprlbl{\role{q}}{\lab_1}{x} \seq\PP_1 + 
\soutprlbl{\role{q}}{\lab_2}{v_2} \seq\PP_2 )} 
\Par 
\pa{\q}{(\soutprlbl{\role{p}}{\lab_1}{v_1} \seq Q_1 + 
\sinprlbl{\role{p}}{\lab_2}{y}\seq Q_2)}
$
}\\[1mm]
\NY{Then by \RChoice, we have:} \\[1mm]
\centerline{
$\M \red \pa{\p}{\PP_1\subst{v_1}{x}} \Par \pa{\q}{Q_1}$ or 
$\M \red \pa{\p}{\PP_2} \Par \pa{\q}{Q_2\subst{v_2}{y}}$
}\\[1mm]
We define a multistep as 
$\red^\ast=(\equiv\cup \red)^\ast$ (zero or more steps) and 
$\red^+$ (one or more steps).
Let $ \red^{\omega} $ denote an infinite sequence of steps.
We call a term \emph{convergent} if it does not have any infinite sequence of steps; and 
write $ M \noRed $ if there is no $ M' $ such that $ M \red M' $.

To showcase our theory, we start with 
a \emph{leader election protocol} (which is used similar to \cite{Palamidessi03} for a separation result 
in \S~\ref{sec:hierarchy}).

\begin{newexample}[Leader election protocol]
	\label{ex:leaderElection-process}
	Consider a protocol that involves five participants (\enodei{a}, \enodei{b}, \enodei{c}, \enodei{d}, and \enodei{e}), interacting in two stages to elect a leader.

\smallskip

	\noindent
	\begin{minipage}{\columnwidth}
		\begin{wrapfigure}{R}{0.25\columnwidth}
			\begin{tikzpicture}[bend angle=22]
				\foreach \v/\x/\y in {2/1/a,1/2/b,5/3/c,4/4/d,3/5/e}
		        {
		            \path (360*\v/5-55:0.75) node[draw, inner sep=2pt, minimum size=3pt] (p\x) {$ \enodei{\y} $};
		        }
		        \draw[-latex, color=blue] (p1) edge [bend right] (p5);
		        \draw[-latex, color=blue] (p2) edge [bend right] (p1);
		        \draw[-latex, color=blue] (p3) edge [bend right] (p2);
		        \draw[-latex, color=blue] (p4) edge [bend right] (p3);
		        \draw[-latex, color=blue] (p5) edge [bend right] (p4);
		        \draw[-latex, color=red] (p1) edge (p3);
		        \draw[-latex, color=red] (p2) edge (p4);
		        \draw[-latex, color=red] (p3) edge (p5);
		        \draw[-latex, color=red] (p4) edge (p1);
		        \draw[-latex, color=red] (p5) edge (p2);
			\end{tikzpicture}
		\end{wrapfigure}

\smallskip

\indent 
In the first stage (depicted by the \textcolor{blue}{blue circle}) two times a process \enodei{x} asks a process \enodei{y} to become leader by sending $ \ssout{\enodei{y}}{\leader} $ that is accepted by \enodei{y} via $ \ssinp{\enodei{x}}{\leader} $.
		The respective two receivers of the first stage continue in the same way asking each other in a second stage (depicted as \textcolor{red}{red star}).
		The receiver in the second stage finally announces its election as leader by sending $ \ssout{\enodei{station}}{\elect} $ to external \enodei{station}, where 
$ \mathsf{Station} = \pa{\env}{P_\env}$ 
with $P_\env =
{\choice{\left( \ssinp{\enodei{i}}{\elect} \seq \choice{\ssout{\enodei{i}}{\delete} \seq \inact}{\enodei{i} \in \set{\enodei{a}, \enodei{b}, \enodei{c}, \enodei{d},\enodei{e}}} \right)}{\enodei{i} \in \set{\enodei{a}, \enodei{b}, \enodei{c}, \enodei{d},\enodei{e}}}} $.
	\end{minipage}
	\begin{align*}
		\mathsf{Election} &= \pa{\role{a}}{P_{\enodei{a}}} \Par \pa{\role{b}}{P_{\enodei{b}}} \Par \pa{\role{c}}{P_{\enodei{c}}} \Par \pa{\role{d}}{P_{\enodei{d}}} \Par \pa{\role{e}}{P_{\enodei{e}}}\\
		P_{\enodei{a}} &= 
		(\ssout{\enodei{e}}{\leader}. \inact \\
		& \quad\quad \OR \ssinp{\enodei{b}}{\leader}. 
		\left( \ssout{\enodei{c}}{\leader}. \inact \OR \ssinp{\enodei{d}}{\leader}. \ssout{\env}{\elect}. \inact \right)\\
		& \quad\quad \OR \ssinp{\env}{\delete}. \inact )\\
		\sigma &= \left[ \enodei{a} \mapsto \enodei{b}, \enodei{b} \mapsto \enodei{c}, \enodei{c} \mapsto \enodei{d}, \enodei{d} \mapsto \enodei{e}, \enodei{e} \mapsto \enodei{a} \right]\\
		P_{\enodei{b}} &= P_{\enodei{a}}\sigma \quad P_{\enodei{c}} = P_{\enodei{b}}\sigma \quad P_{\enodei{d}} = P_{\enodei{c}}\sigma \quad P_{\enodei{e}} = P_{\enodei{d}}\sigma
	\end{align*}
	The complete multiparty session is $ \mathsf{Election} \Par \mathsf{Station} $.
	The participants in $ \mathsf{Election} $ are symmetric \wrt $ \sigma$, for instance
	$ P_{\enodei{d}} = P_{\enodei{c}}\sigma = $
	$ \hspace*{1.2em} ( \ssout{\enodei{c}}{\leader}. \inact \OR \ssinp{\enodei{e}}{\leader}. \left( \ssout{\enodei{a}}{\leader}. \inact \OR \ssinp{\enodei{b}}{\leader}. \ssout{\env}{\elect}. \inact \right) \hspace*{1.2em} $
	$ \OR \; \ssinp{\env}{\delete}. \inact) $. 
	After the winner is announced to $ \env $, it garbage collects the process $ \enodei{z} $ that did not participate in the first stage by sending $ \ssout{\enodei{z}}{\delete} $ and then terminates itself.
\smallskip

\end{newexample}

\section{Mixed Choice Synchronous Multiparty Session Types}
\label{sec:mcmpst}
This section introduces the syntax 
of \mcmpst types,
which describe the interactions between
	{\em participants} at the end-point level. 
The syntax is based on \cite{DBLP:journals/jlp/GhilezanJPSY19}, 
extending to mixed choice. 

\begin{newdefinition}[\mcmpst types and local contexts]
	\label{def:syntax}
	Assume a set of participants $\Part$ $(\p, \q, \rr, \dots)$
	and a set of labels $\Labels$ $(\lab, \lab', \dots)$.
	The set of {\em local types}, \Local with $\local \in \Local$, are defined as:
	\\[1mm]
	\centerline{$
	\begin{array}{r@{~}c@{~}ll}
		\UType & \bnfis & \nat \bnfbar \bool  
		& \text{\footnotesize (numeric, boolean)} 
		\\ 
		\inoutmark & \bnfis & {\lcolor{\textbf{!}}} \bnfbar {\lcolor{\textbf{?}}}
		& 
		\text{\footnotesize (message send, message receive)} 
		\\
		\local     & \bnfis & \tinact \bnfbar \tvar{t}
		\bnfbar
		\trec{t} \local & 
		\text{\footnotesize (inaction, recursion variable and type)}\\
		& \bnfbar	& \Tor{i\in I}{\tinoutlbli{p}{\lab}{\UType} \locali{i}}
		& 
	\text{\footnotesize (mixed choice)}\\
\LL & \bnfis &  \ee \bnfbar \LL, \p{:}\local
& \text{\footnotesize (local contexts)}
\end{array}
$}
\end{newdefinition}
\emph{Payload type} $\UType$ ranges over
ground types (\nat, \bool). 
Termination is represented by \tinact.  
Recursive types are  
$\trec{t} \local$,
with \tvar{t} as the recursive variable.
We assume standard capture-avoiding
substitution and assume that
recursive types are \emph{guarded},  
e.g.,~for 
type $\trec{t_{\mathit{1}}}{\cdots\trec{t_{\mathit{n}}}{\tvar{t}}}$
is not allowed. 
We take the equirecursive view, i.e. $\trec{t} \local$ is identified 
with $\local \subst{\trec{t} \local}{\tvar{t}}$, 
see \cite[Notation 3.5]{DBLP:journals/jlp/GhilezanJPSY19}. 
$\ftv{\local}$ denotes
a set of free type variables in $\local$ 
and
$\local$ is \emph{closed}
if $\ftv{\local}=\emptyset$.

\emph{Mixed choice} type
$\Tor{i\in I}{\tinoutlbli{p}{\lab}{\UType} \locali{i}}$ 
enables choice between a non-empty collection of input or output local types,
$\tinoutlbli{p}{\lab}{\UType} \locali{i}$. 
Each type denotes sending (${\lcolor{\textbf{!}}}$) or receiving 
($\lcolor{\textbf{?}}$) a message of label $\lab_i$ 
with a payload type
$\UType_i$ 
from or to some different participant $\rolei{\p}{i}$
and continues as $\locali{i}$.
We often abbreviate  
input and output types as 
$\tinoutlbl{\role{p}}{}{U}\local$ 
or 
$\tinout{\role{p}}{\ell}\local$ 
if a label or payload is not important;
and omit trailing \tinact types. 
Similarly with processes, 
we omit $\Tor{}{}$ 
when $I$ is a singleton,
$\Tor{i \in \{1\}}{\tinoutlbl{\rolei{\p}{i}}{\lab_i}{\UType_i} \locali{i}} =
	\tinoutlbl{\rolei{\p}{1}}{\lab_1}{\UType_1}\locali{1}$.  
We abbreviate $\tinoutlbl{\rolei{\p}{1}}{\lab_1}{\UType_1} \locali{1}\tor\cdots \tor\tinoutlbl{\rolei{\p}{n}}{\lab_n}{\UType_n} \locali{n}$ 
to denote $\Tor{i \in \{1,..,n\}}{\tinoutlbl{\rolei{\p}{i}}{\lab_i}{\UType_i}} \locali{i}$ and  write 
$\Tor{j \in J}{\Tor{i \in I_j}{\tinoutlbl{\rolei{\p}{ij}}{\lab_{ij}}{\UType_{ij}}\locali{ij}}}$ 
with $J=\{1..n\}$ 
for\\ 
$\Tor{i \in I_1}{\tinoutlbl{\rolei{\p}{i1}}{\lab_{i1}}{\UType_{i1}}\locali{i1}}\tor\cdots \tor
\Tor{i \in I_n}{\tinoutlbl{\rolei{\p}{in}}{\lab_{in}}{\UType_{in}} \locali{in}}$. 
A choice is commutative and associative
with $\tor$. 
Function \parts{\local} returns a
set of \emph{participants} in \local defined as: 
$\parts{\tinact}=\parts{\tvar{t}}=\es$; 
$\parts{\trec{t}{\local}} = \parts{\local}$; 
and $\parts{\Tor{i \in I}{\tinoutlbli{p}{\lab}{\UType} \locali{i}}} = \set{\rolei{p}{i}}_{i \in I} \cup \bigcup_{i \in I} \parts{\locali{i}}$; 
and 
function $\prefix{\local}$
returns \emph{prefixes}  of \local defined as: 
$\prefix{\tinact}=\prefix{\tvar{t}}=\es$; 
$\prefix{\trec{t}{\local}} = \prefix{\local}$; 
and $\prefix{\Tor{i \in I}{\tinoutlbli{p}{\lab}{\UType} \locali{i}}} = 
\set{\rolei{p}{i}\inoutmarki{i}}_{i \in I}$. 
 
We define the \emph{duality} function as $\dual{\outsymbol}=\inpsymbol$ and
$\dual{\inpsymbol}=\outsymbol$.

Type
$\Tor{i\in I}{\tinoutlbli{p}{\lab}{\UType} \locali{i}}$ 
is \emph{well-formed} 
if:\\[1mm]
\centerline{
	\ruleLEll\quad 
	$(\forall i {\ne} j{\in} I.~
\pii{i}\inoutmarki{i}=\pii{j}\inoutmarki{j}
\implies
		      \lab_i \ne \lab_j)$
}\\[1mm]
i.e.,
for a mixed choice, any matching choice prefixes must
have distinguishable labels.
We write $\wfjudgement{\local}$  
if all choices in $\local$ are well-formed.  
\emph{Hereafter we assume all types are well-formed.}
Note that typable mixed choice
processes do not have the same well-formedness requirement as local types,
allowing non-deterministic process choice using the same label
(see $P_{10}$ in Example~\ref{ex:family} and Example~\ref{ex:choiceprocess}).

A \emph{local context} $\LL$ abstracts the behaviour of 
a set of participants where we assume for all $\role{p}\in \domain{\LL}$, 
$\ftv{\LL}=\emptyset$.


\subsection{Subtyping of \mcmpst}
\label{app:subtyping}
We define the subtyping relation for  mixed choice local types, which
subsumes the standard 
subtyping~\cite{DemangeonH11,carbone.honda.yoshida:esop07,ChenDSY17}.
\begin{newdefinition}[Subtyping]
\label{def:subtyping}  
	The subtyping relation $\subt$ 
	is \emph{coinductively} defined by:
	\\
	\centerline{$\small
			\begin{array}{c}
				%
                                \tinact \subt \tinact\ \SEnd
				\\[2mm] 
				\cinference{
					\forall i {\in} I, \locali{i} \subt \localid{i}
				}{
					\Tor{i \in I }{\toutlbl{\p}{\lab_i}{\UTypei{i}} \locali{i}} \subt
					\Tor{i \in I\cup J}{\toutlbl{\p}{\lab_i}{\UTypei{i}} \localid{i}}
				}{\SSel}
					\\[4mm]
				\cinference{
					\forall i {\in} I, \locali{i} \subt \localid{i}
				}{
					\Tor{i \in I\cup J}{\tinplbl{\p}{\lab_i}{\UTypei{i}} \locali{i}} \subt
					\Tor{i \in I }{\tinplbl{\p}{\lab_i}{\UTypei{i}} \localid{i}}
				}{\SBra} \\[4mm]
				\cinference{
				\forall k {\in} K,
				\locali{k}
				\subt
				\localid{k}
				 &
				\forall i{\ne}j{\in}K,
				\prefix{\locali{i}}
				\cap
				\prefix{\locali{j}}
				=
				\emptyset
				}{
				\Tor{k\in K}{\locali{k}}
				\subt
				\Tor{k\in K}{ \localid{k}}
				}{\SChoice}
				\\[4mm]
				\cinference{
					\locali{1}\subst{\trec{t}{\locali{1}}}{\tvar{t}} \subt \locali{2}
				}{
					\trec{t}{\locali{1}} \subt \locali{2}
				}{\SRcL}
				\cinference{
					\locali{1} \subt \locali{2}\subst{\trec{t}{\locali{2}}}{\tvar{t}}
				}{
					\locali{1} \subt \trec{t}{\locali{2}}
				}{\SRcR}
			\end{array}
		        $}\\[1mm]
\noindent
	We write $\LLi{1} {\subt} \LLi{2}$ iff
	for all $\role{p}\in \domain{\LLi{1}} \cap \domain{\LLi{2}}$, 
	$\LLi{1}(\p) {\subt} \LLi{2}(\p)$
        and for all  
$\role{p}\in \domain{\LLi{1}} \setminus \domain{\LLi{2}}$ and  
$\role{q}\in \domain{\LLi{2}} \setminus \domain{\LLi{1}}$, 
$\LLi{1}(\role{p})=\LLi{2}(\role{q})=\tinact$. 
\end{newdefinition}
\noindent
The above subtyping rules without \SChoice are the standard
from \cite{DemangeonH11,carbone.honda.yoshida:esop07,ChenDSY17}:
a smaller type has smaller internal
choices 
\SSel; larger external sums are
smaller 
\SBra.
The subtyping of a mixed choice which combines
selection and branching (the premise
is inferred by either \SSel or \SBra)
is invariant \SChoice. The side condition
given by $\prefix{\local}$ in \SChoice
ensures no overlap with \SSel and 
\SBra.

\begin{newexample}[Mixed choice subtyping]
	\label{ex:subtyping}

	The mixed choice subtype judgement
	is given by \SChoice. This rule partitions each mixed-choice term
	into a sum of non-mixed choices, i.e., 	$\left(\ntinp{\p}{\lab_1}
		\tor
		\ntinp{\p}{\lab_2}\right)
		\tor
		\left(\ntout{\p}{\lab_3} \right)
		\subt
		\left(\ntinp{\p}{\lab_1}\right)
		\tor
		\left(\ntout{\p}{\lab_3}\tor \ntout{\p}{\lab_4}\right)$.
	Standard subtyping rules (\SSel and \SBra) are then applied pair-wise
	to each non-mixed choice.
	\vspace{-.2cm}
	\[\small
	 	\infer=[\SChoice]{
	 	\tinp{\p}{\lab_1} \locali{1}
	 	\tor
	 	\tinp{\p}{\lab_2}\locali{2}
	 	\tor
		\tout{\p}{\lab_3}\locali{3}
	 	\subt
	 	\tinp{\p}{\lab_1} \locali{1}
	 	\tor
	 	\tout{\p}{\lab_3}\locali{3}
	 	\tor
		\tout{\p}{\lab_4}\locali{4}
	 	}{
	 	\cinference{
	 		\vdots
	 	}{
	 		\tinp{\p}{\lab_1} \locali{1}
	 		\tor
	 		\tinp{\p}{\lab_2}\locali{2}
	 		\subt
	 		\tinp{\p}{\lab_1} \locali{1}
	 	}{\SBra}
		\quad
	 	\cinference{
	 		\vdots
	 	}{
	 		\tout{\p}{\lab_3}\locali{3}
	 		\subt
	 		\tout{\p}{\lab_3}\locali{3}
                        \tor 
	 		\tout{\p}{\lab_4}\locali{4}
	 	}{\SSel}
	 	}
	 \]
We can also mix different participants such as:
$\left(\ntinp{\p}{\lab_1}
		\tor
		\ntinp{\p}{\lab_2}\right)
		\tor
		\left(\ntout{\q}{\lab_3} \right)
		\subt
		\left(\ntinp{\p}{\lab_1}\right)
		\tor
		\left(\ntout{\q}{\lab_3}\tor \ntout{\q}{\lab_4}\right)$.
\end{newexample}
\begin{restatable}[Subtyping]{newproposition}{propmergingandsubtyping}
	\label{prop:subtyping}
        		      Suppose $\wfjudgement{\locali{i}}$ ($i=1,2$). 
	\begin{enumerate*}
		\item[(a)] $\locali{1} \subt \locali{2}$ is a
		      preorder; and
		\item[(b)] $\LLi{1} \subt \LLi{2}$ is a preorder.
	\end{enumerate*}
\end{restatable}
\begin{proof}
Induction on delivation of $\locali{1} \subt \locali{2}$. 
\end{proof}

\begin{newremark}[Subtyping]
\label{rem:subtyping}
Our subtyping relation subsumes the standard branching-selection 
subtyping relation by omitting \SChoice. This inclusion is important for 
proving the expressiveness results.  If 
we replace \SChoice, \SBra and \SSel by the following 
simpler rule \SSubset:\\[1mm] 
\centerline{$\small
			\cinference{
					\forall i {\in} I, \locali{i} \subt \localid{i}
				}{
					\Tor{i \in I}{\tinoutlbli{p}{\lab}{\UType} \locali{i}} \subt
\Tor{i \in I\cup J}{\tinoutlbli{p}{\lab}{\UType}\localid{i}}}{\SSubset}$}
\\[1mm]
then 
Lemma~\ref{lem:subtypingproperties}(\ref{lem:subtypingproperties:df})
does \emph{not} hold. The lemma 
is crucial for proving \emph{deadlock-freedom} 
of the typed processes (Theorem~\ref{thm:deadlockfree}), 
see Remark~\ref{rem:dfree}(2).
\end{newremark}

%
\subsection{Labelled Transition System (LTS) of Types and Contexts}
\label{subsec:LTS:types}
This subsection defines the LTS of types and contexts. 
The behavioural properties (safety and deadlock-freedom) of local contexts 
defined by the LTS 
are used to prove 
the main theorems of typed processes in \S~\ref{sec:mcmp:properties}. 
\begin{description}
	\item[Local Types:]
      The set of actions of local types is defined as
      $\LAct = \set{\actoutlbl{\p}{\q}{\lab}{\UType},\actinplbl{\p}{\q}{\lab}{\UType} \setbar \p,
			      \q \in \Part, \lab\in \Labels}$ with
                            $\lambda \in \LAct$; and 
the set of local contexts is defined as $\Roles=\set{\LL}$.  
	      The transition relation
	      $\by{\lambda} \subseteq \Roles \times \LAct \times \Roles$
	      is defined by \LChoice and \LRec in
              Figure~\ref{fig:type_semantics}. 
	    \item[Local Contexts:]
              The set of actions of local contexts is defined as
              $\Act = \set{\actlbl{\p}{\q}{\lab}{\UType} \setbar \p,
			      \q \in \Part, \lab\in \Labels}$.
	      The transition relation
	      $\by{\actlbl{\p}{\q}{\lab}{\UType}} \subseteq \powerset{\Roles} \times \Act \times \powerset{\Roles}$
	      is defined by \RPass
	      in Figure~\ref{fig:type_semantics}.
\end{description}
%
\begin{figure}
	$
		\begin{array}{c}
\begin{array}{c}
  \Tor{i \in I}{\tinoutlbli{p}{\lab}{\UType} \locali{i}}
  \by{\actoutinlbli{p}{q}{\ell}{\UType}{j}}
  \locali{j} \ \ (j\in I)
\quad                             {\LChoice}\\
			\tree{
				\localise{\p} \local \subst{\trec{t}{\local}}{\tvar{t}} \by{\lambda} \localise{\p} \locald
			}{
				\localise{\p} \trec{t}{\local} \by{\lambda} \localise{\p} \locald
			}{\LRec}
			\end{array}
\\[2mm]
			\tree {
				\localise{\p} \locali{1} \by{\actoutlbl{\p}{\q}{\lab}{\UTypei{}}} \localise{\p} \localid{1}
				\quad
				\localise{\q} \locali{2} \by{\actinplbl{\q}{\p}{\lab}{\UTypei{}}} \localise{\q} \localid{2}
			}{
				\LL, \localise{\p} \locali{1}, \localise{\q} \locali{2}
				\by{\actlbl{\p}{\q}{\lab}{\UTypei{}}}
				\LL, \localise{\p} \localid{1}, \localise{\q} \localid{2}
			}{\RPass}
		\end{array}
$
\caption{Labelled transition systems of types and contexts \label{fig:type_semantics}}
\end{figure}

\noindent
Notice that the LTS is defined between closed types. 
The semantics for local types and contexts follow
the standard concurrency semantics \cite{CCS}.
Label \actoutlbl{\p}{\q}{\lab}{\UType} denotes that participant
\p may send a message with label \lab of type \UType to participant \q.
Dually, label \actinplbl{\p}{\q}{\lab}{\UType} denotes that participant \p
may receive a message with label \lab of type \UType from participant \q.
Rule \LChoice chooses one of choices and 
rule \LRec is standard.
Rule $\RPass$ states that if two roles exhibit dual local labels,
they can synchronise and perform action \actlbl{\p}{\q}{\lab}{\UType}.
%

\subsection{Properties for Mixed Choice Multiparty Session Types}
\label{subsec:types:properties}
\begin{newdefinition}[Local context reductions]\label{def:config}
	We write $\LL \red{} \LLd$ if
	$\LL \by{\actlbl{\p}{\q}{\lab}{\UType}} \LLd$;
	$\LL \red^+ \LLd$ for its transitive closure;
	$\LL \red^\ast \LLd$ for either $\LL =\LLd$ or 
	$\LL \red^+ \LLd$; and 
    $\LL\, \by{\actlbl{\p}{\q}{\lab}{\UType}}$ if 
$\exists \LLd. \ \LL\, \by{\actlbl{\p}{\q}{\lab}{\UType}}\, \LLd$.
\end{newdefinition}

To prove the subject reduction theorem,
we first introduce the \emph{safety} property from \cite[Definition 4.1]{less_is_more}. It states 
that there is no communication mismatch. This property is 
used to prove the subject reduction theorem and communication safety. 


\begin{newdefinition}[Safety property]
  \label{def:session_safety}
  Co-inductive property $\varphi$
  is a \emph{safety} property of local context 
  $\LL$ if and only if for all
$\left\{\localise{\p} \locali{1}, \localise{\q}
  \locali{2} \right\} \subseteq 
  \LL\in \varphi$, if 
		$\localise{\p} \locali{1}
                \by{\actoutlbl{\p}{\q}{\lab}{\UType}}$
		and
		$\localise{\q} \locali{2}
                \by{\actinplbl{\q}{\p}{\labd}{\UTyped}}$,  
				then $\LL 
                                \by{\actlbl{\p}{\q}{\lab}{\UType}} \LLd$
                                and $\LLd\in \varphi$.
                                The largest safety property 
                                is a union of all safety 
                                properties. We say $\LL$ is \emph{safe} 
                                and write $\safe{\LL}$ 
if $\LL \in \varphi$ and $\varphi$ is a safety property.
\end{newdefinition}

\noindent
The safety property says that 
if the output and input 
actions are ready each other,  they can synchronise
with the label provided by the output $\lab$ and they have the same
payload types ($\UType = \UTyped$) (note that the base types
do not have subtyping),  
and it is preserved after a step. \NY{Notice that the safety is \emph{not} symmetric--the label $\lab$ and type $U$ of the selection are always taken, while \emph{some} label of a branching needs to be reducible.}   


Next we define 
\emph{deadlock-freedom} following 
\cite[Figure 5(2)]{less_is_more}. 
It states that if typing context $\LL$ 
terminates, all participants typed by $\LL$ 
terminate \NY{as nil processes (typed by $\tinact$)}. 

\begin{newdefinition}[Deadlock-freedom] 
\label{def:dfree_live} 
Local context \LL is 
				{\em deadlock-free}  
				if $\LL \red^*\LLd \noRed$, then 
				$\forall \p\in \domain{\LLd}.\,
                                \LLd(\p) = \tinact$. 
We denote \df{\LL} if $\LL$ is deadlock-free. 
\end{newdefinition}

The following is the key lemma to ensure communication safety 
and deadlock-freedom of 
typed sessions. 

\begin{restatable}[Subtyping and 
    properties]{newlemma}{lemmasubtypingproperties} 
  \label{lem:subtypingproperties}
	$ $
	\begin{enumerate}
		\item	\label{lem:subtypingproperties:safe}
				If $\LL\subt \LLd$ and $\safe{\LLd}$, then 
				\begin{enumerate*}
					\item	$\safe{\LL}$.
					\item	If $\LL \red \LLdd$, then there exists \LLddd such that $\LLd \red \LLddd$ and $\LLdd \subt \LLddd$ and \safe{\LLddd}.
				\end{enumerate*}

		\item	\label{lem:subtypingproperties:df}
				If $\LL\subt \LLd$, and $\safe{\LLd}$ and $\df{\LLd}$, then 
					\begin{enumerate*}
		\item $\df{\LL}$.
		\item If $\LL \red \LLdd$, then there exists \LLddd such that $\LLd \red \LLddd$ and $\LLdd \subt \LLddd$ and \df{\LLddd}.
					\end{enumerate*}
\item Checking \safe{\LL} and \df{\LL} is decidable. 
	\end{enumerate}
\end{restatable}
\begin{proof}
(1,2) Induction on $\LL$ and $\LL\red \LLd$; and (3) Similar with 
\cite[Appendix K]{SY2019Technical}.
\end{proof}  

\begin{newremark}[Deadlock-freedom]
\label{rem:dfree}
\begin{enumerate*}
\item \label{item:dfsafe}
Let $\LL = \p: \ntout{\q}{\lab_1} \tor \ntout{\q}{\lab_2}, \q:
\ntinp{\p}{\lab_1}$. Then we can check 
\df{\LL} \NY{since the only possible transition is 
$\LL\by{\act{\p}{\q}{\lab_1}}\LLd$.} But, for $\LLd = \p: \ntout{\q}{\lab_2}, \q:
\ntinp{\p}{\lab_1}$ such that $\LLd \subt \LL$, we have 
$\neg\df{\LLd}$. Hence without the $\safe{\LL}$ assumption, 
$\df{\LL}$ is not preserved by subtyping. \\

\item \label{item:dfsafe}
Assume 
$\LLi{2} = \p: \ntout{\q}{\lab_1} \tor \ntout{\q}{\lab_2}, \q:
\ntinp{\p}{\lab_1} \tor \ntinp{\p}{\lab_2}$. 
Then $\safe{\LLi{2}}$ and $\df{\LLi{2}}$. 
Suppose we apply a wrong subtyping rule \SSubset in Remark~\ref{rem:subtyping}
to obtain  
$\LLi{3} = \p: \ntout{\q}{\lab_1}, \q:
\ntinp{\p}{\lab_2}$. Then $\neg\safe{\LLi{3}}$ and 
$\neg\df{\LLi{3}}$, i.e.~neither safety nor deadlock-freedom 
is preserved. 
\end{enumerate*}
\end{newremark}


\begin{restatable}[Local context properties]{newproposition}{propconfigurationproperties} \label{pro:conf_properties}
$ $
\begin{enumerate*}
    \item 
      \df{\LL}$\not\Longrightarrow$ \safe{\LL}; and
    \item 
      \safe{\LL} $\not\Longrightarrow$ \df{\LL}.
\end{enumerate*}      
\end{restatable}
\begin{proof} 
\begin{enumerate*}[leftmargin=*]
\item Let $\LLi{3} = \p: \trec{t}{\tout{\q}{\lab_1} \tvar{t}}, \q: \trec{t}{\tinp{\p}{\lab_1} \tvar{t}}, \rr: \ntoutlbl{\roled{r}}{\lab_2}{\bool},$\\$\roled{r}: \ntinplbl{\rr}{\lab_2}{\nat}$.
  Then 
				 \LLi{3} is deadlock-free because 
				$\LLi{3} \by{\act{\p}{\q}{\lab_1}} \LLi{3}$. 
				But \LLi{3}
				is not safe because $\nat\not =
                                \bool$. 
				\item	Let
						$\LLi{2} = \p: \ntout{\q}{\lab}$. 
						Then \LLi{2}
                                                is vacuously safe
                                                following
                                                Definition~\ref{def:session_safety},
                                                but not
                                                deadlock-free because  
						$\LLi{2}\noRed{}$
                                                and
                                                ${\LLi{2}}(\p)\not=\tinact$.
		\end{enumerate*}
\end{proof}

\section{Typing System of \mcmpst-Calculus}
\label{subsec:typing}
%
We introduce the typing system, extending the system in
\cite{DBLP:journals/jlp/GhilezanJPSY19} and 
highlight the three key rules which differ from
\cite{DBLP:journals/jlp/GhilezanJPSY19}.

%

\begin{newdefinition}[Typing system]
  \label{def:typingsystem}
\label{def:typingcontext}
We define a \emph{shared context} as:\\[1mm]
\centerline{
$\Ga	\bnfis	\ee \bnfbar \Ga, \pvar{X}: \local \bnfbar 
\Ga, x{:} \nat \bnfbar  
\Ga, x{:} \bool$}\\[1mm]  
We write $\Ga,\Ga'$ if $\domain{\Ga}\cap \domain{\Ga'}=\emptyset$ and 
use the notation $\Ga(x)$ or $\Ga(\pvar{X})$ to denote 
a constant type of $x$ and local type of \pvar{X}. 
%
%
%
%
%
%
  Figure~\ref{fig:typing} defines the three
  judgements:\\[1mm]
\centerline{
    $\Ga \types v \as \UType\quad \quad
    \Ga \types \PP \as \local\quad \quad
    \types \M \as \De$
  }\\[1mm]
  where judgement $\Ga \types v \as \UType$ 
	is read as `context $\Ga$ types value $v$ with type $\UType$'
	and judgement $\Ga \types \PP \as \local$
	as `context \Ga types process \PP with local 
	type $\local$'. Judgement 
$\types \M \as \De$ is read as `multiparty session $\M$ is typed with 
       local context $\De$'.
\end{newdefinition}
%


\noindent We highlight 
three rules in Figure~\ref{fig:typing}. \NY{These rules are different
from the rules for \mpst in 
\cite{DBLP:journals/jlp/GhilezanJPSY19}.} 
Rule \TInact assigns a 
nil by $\tinact$.
Rules \TSend and  \TRcv 
type the output and input
with the send and input types, respectively.
Notice that we usually type the input branching instead of a single
receiver as given by \TRcv, but our choice 
rule allows to type both input and output are mixed, subsuming the
standard input branching rule (\mathparfont{T-INPUT-CHOICE} in \cite{DBLP:journals/jlp/GhilezanJPSY19}).
Typing recursion and conditional is standard by \TVar, \TRec
and \TCond. Rule \TSubs is a subsumption rule.  

%

Typing of mixed choice is given by \TSum. 
We motivate the double sum in the conclusion by 
considering Example~\ref{ex:choiceprocess}(1) below, 
in which two branches ($\ssout{\q}{\lab_1}.Q$ and $\ssout{\q}{\lab_1}.R$)
have a common prefix and both $Q$ and $R$ are 
typed by the same local type, 
while the other two branches are typed independently. 
The outer sum, then, groups 
similarly prefixed branches
($R_i=\choice{\pprefix_i \seq{\PP_j}}{j \in
J_i}$); each branch in $R_i$ is then typed by a 
common $\tinoutlbli{p}{\lab}{\UType} \locali{i}$;
finally, the full term $\choice{R_i}{i\in I}$ is typed as 
$\Tor{i\in I}{\tinoutlbli{p}{\lab}{\UType} \locali{i}}$.

Multiparty session $\M$ is typed by rule
\TSessPlus which assumes a local context which consists of 
types of all participants is safe  
\cite{less_is_more} to guarantee the subject reduction. 

\begin{figure}
\hspace{-7mm}
	$\begin{array}{c}		
			\Ga, x{:}\nat \types x \as \nat
			\ \ \TGround\\[1mm]
			\Ga \types 1, 2, .. \as \nat
			\ \ \TNat\quad
			\Ga \types \true, \false \as \bool
 			\ \ \TBool\\[1mm]
			{\Ga, \pvar{X}: \local \types \pvar{X} \as \local}
			\ \ \TVar 
\ \ \ 
				\Ga \types \inact \as \tinact
\ \ \TInact\\[1mm]
			%
			%
			\tree {
				\Ga \types \PP \as \local
				\quad
				\Ga \types v \as U
			}{
	\Ga \types \soutprlbl{\p}{\lab}{v}.\PP \as 
\toutlbl{\p}{\lab}{U} \local
			}{\TSend}
			\\[3mm]
			\tree {
				\Ga, x: U \types \PP \as \local
			}{
		\Ga \types \sinprlbl{\p}{\lab}{\xx}. \PP 
                \as \tinplbl{\p}{\lab}{U} \local}{\highlight{\TRcv}}
			\\[3mm]
			%
			%
			\tree {
				\Ga \types v \as \bool \quad
				\Ga \types \PP_i \as \local \quad i \in \set{1,2}
			}
			{
				\Ga \types \cond{v}{\PP_1}{\PP_2}\as \local
			}
			{\TCond}
			\\[3mm]
                        \tree {
{
     \forall i \in  I, \forall j \in J_i.\ (\Ga \types \pprefix_i
    \seq{\PP_j} \as \tinoutlbli{p}{\lab}{\UType} \locali{i})}
			}{
{\Ga \types \choice{\choice{\pprefix_i \seq{\PP_j}}{j \in J_i}}{i\in I} 
\as \Tor{i\in I}{\tinoutlbli{p}{\lab}{\UType} \locali{i}}}
			}{\highlight{\TSum}} \\[3mm]
			\tree {
				\Ga, \pvar{X}: \local \types \PP \as \local
			}{
				\Ga \types \rec{X}{\PP} \as \local
			}{\TRec}
			\ \ \ 
			\tree{	\Ga \types \PP \as \local\quad  
				\local \subt \locald
			}{
\Ga \types \PP \as \locald
			}{\TSubs}
			\\[3mm]
			\tree {
\forall i\in I \ \types \PP_i \as \locali{i}
\quad \safe{\{ \rolei{p}{i}{:}\locali{i}\}_{i\in I}}
                        }{
			{\types \Pi_{i \in I} \ \pa{\rolei{p}{i}}{\PP_i}
                        \as    \{ \rolei{p}{i}{:}\locali{i}\}_{i\in I}}
                        }{\highlight{\TSessPlus}}
		\end{array}
$
\caption{
   \label{fig:typing}
   Typing rules for \mcmpst-calculus. 
}
\end{figure}

\begin{newexample}[Typed and untyped mixed choice processes]
\label{ex:choiceprocess}
\label{ex:choicederivation}
\strut
	\begin{enumerate}[leftmargin=*]
		\item	\label{ex:sum1}
				We explain how we can type a process with duplicated labels with a single label type. 
				Consider:
				\\[1mm]
				\centerline{
				$
				P\ = \ 	\ssout{\q}{\lab_1}.Q 
				\OR	\ssout{\q}{\lab_1}.R
				\OR	\ssinp{\q}{\lab_2}.\inact 
				\OR	\ssinp{\rr}{\lab_2}.\inact
				$}\\[1mm]
				Assuming $Q$ and $R$ have the same type $\local$, we 
				first type:
\\[1mm]
\centerline{
				$
					\types \ssout{\q}{\lab_1}. Q \OR \ssout{\q}{\lab_1}. R \as \tout{\q}{\lab_1} \local
					\quad
		\types \ssinp{\q}{\lab_2}. \inact \as 
\ntinp{\q}{\lab_2}
					\quad
		\types \ssinp{\rr}{\lab_2}. \inact \as 
\ntinp{\rr}{\lab_2}$}
\\[1mm]
				and then use rule \TSum to obtain:
$
			\es \types \PP \as
			\tout{\q}{\lab_1} \local \tor \ntinp{\q}{\lab_2} \tor \ntinp{\rr}{\lab_2}$.
				Note that a single local type can type two choices where $Q$ and
				$R$ might have different behaviours but have the same
				type:
\\[1mm]
\centerline{{$Q = \soutlbl{\q}{\labd}{5} \inact$}
\, and \,
{$R = \soutlbl{\q}{\labd}{105} \inact$}
}\\[1mm]
with
{$\types Q \as \ntoutlbl{\q}{\labd}{\nat}$}
and {$\types R \as \ntoutlbl{\q}{\labd}{\nat}$}.
		\item	\label{ex:sum2}
A combination of 
\TSum and subsumption \TSubs makes 
a process with duplicated labels with different output labels typable.  
Consider:\\ 
\centerline{{$Q_1 = \soutlbl{\role{a}}{\lab_1}{5} \inact$ and 
$Q_2 = \soutlbl{\role{a}}{\lab_2}{\true}\inact$}}
\\[1mm]
and let $\local=\ntoutlbl{\role{a}}{\lab_1}{\nat}+\ntoutlbl{\role{a}}{\lab_2}{\bool}$. Then $\es \types Q_1 \as \local$ and 
$\es \types Q_2 \as \local$ by \TSend and \TSubs. 
Let $R=\sinplbl{\role{a}}{\lab}{x}Q_1 + \sinplbl{\role{a}}{\lab}{x}Q_2$. 
Then by \TSum, $R$ is typable. Similarly,  
$P_{10}$ in Example~\ref{ex:family} is typable, but $P_{11}$ is not typable 
as the input choice types are co-variant. 
		\item	\label{ex:nonprojone}
				Session $\M_1$ 
				is typable 
but its reduction causes a \emph{deadlock}:\\
				\centerline{$ 
					\M_1 =		\pa{\p}
(\ssout{\q}{\ell}. \inact \tor \ssout{\role{r}}{\ell}. \inact)
					\Par	\pa{\q}\ssinp{\p}{\lab}. \inact
					\Par	\pa{\role{r}}\ssinp{\p}{\lab}. \inact
				$}
		\item	\label{ex:nonprojtwo}
				Session $M_2$
causes a type-error when reducing and is not generated from any typable session:
				\\[1mm]
{$
\M_2	=	\pa{\q}(\soutlbl{\p}{\lab}{7}\inact \OR
							\soutlbl{\p}{\lab}{\true}\inact)
	\Par	\pa{\p}\sinplbl{\q}{\lab}{x}{\cond{x}{P_1}{P_2}}$}
				\\[1mm]
$\M_2$ is untypable since the corresponding type 
$\ntoutlbl{\q}{\lab}{\nat} \tor \ntoutlbl{\q}{\lab}{\bool}$ 
is not well-formed by \ruleLEll in \S~\ref{sec:mcmpst}.
		\item For Example~\ref{ex:leaderElection-process}, we have\\[1mm]
			\centerline{$ \begin{array}{l}
				\types P_{\env} \as \Tor{\enodei{i} \in \set{\enodei{a}, \enodei{b}, \enodei{c}, \enodei{d},\enodei{e}}}{\tinp{\enodei{i}}{\elect}\Tor{\enodei{i} \in \set{\enodei{a}, \enodei{b}, \enodei{c}, \enodei{d},\enodei{e}}}{\tout{\enodei{i}}{\delete}\tinact}}
			\end{array} $}\\[1mm]
			and $ \types P_{\role{a}} \as T_{\role{a}} $, where $ T_{\role{a}} $ is\\[1mm]
			\centerline{$ \begin{array}{l}
				\tout{\enodei{e}}{\leader}\tinact \; \tor\\
				\tinp{\enodei{b}}{\leader}\left( \tout{\enodei{c}}{\leader}\tinact \; \tor \; \tinp{\enodei{d}}{\leader}\tout{\env}{\elect}\tinact \right) \; \tor\\
				\tinp{\env}{\delete}\tinact
			\end{array} $}\\[1mm]
			and $ \types P_{\role{b}} \as T_{\role{b}} $ with $ T_{\role{b}} = T_{\role{a}}\sigma $, \ldots, $ \types P_{\role{e}} \as T_{\role{e}} $ with $ T_{\role{e}} = T_{\role{d}}\sigma $.
\end{enumerate}
\end{newexample}


\section{Properties of Typed \mcmpst-Calculus}
\label{sec:mcmp:properties}
This section proves the main properties of the typed calculus,
starting from the subject reduction theorem. 
\label{subsec:mcmp:properties}

\begin{restatable}[Subject Congruence]{newlemma}{lemsubjectcongruence}
\em
	\label{app:lem:subject_congruence}
\begin{enumerate*}
\item 
	Assume $\Ga \types \PP \as \local$.
	If  $\PP \equiv Q$, then 
$\Ga \types Q \as \local$.
\item 
	Assume $\types \M \as \De$.
	If  $\M \equiv \M'$, then 
$\types  \M' \as \De$.
\end{enumerate*}
\end{restatable}
\vspace*{-2mm}
\begin{restatable}[Subject Reduction]{newthm}{thmsubjectreduction}
	\label{cor:subject_reduction}
	\label{th:subject_reduction}
Assume $\types \M \as \LL$. 
If $\M \red \M'$, then there exists 
such that $\types \M' \as \LLd$
and $\LL \red^\ast \LLd$.
\end{restatable}
\begin{proof}
We use the lemma that safety property of local contexts 
is closed under subtyping (Lemma~\ref{lem:subtypingproperties}). 
We then use the closure under the structure congruence
(Lemma~\ref{app:lem:subject_congruence}). 
\end{proof}

A consequence of Theorem~\ref{th:subject_reduction}
is that a well-typed process never reduces to an error state. 
In \mcmpst, the error definition needs to consider all possible
synchronisations among multiple parallel processes
(see Example~\ref{ex:sesserror}). 

\begin{newdefinition}[Session error]
	\label{def:sessionerror}
	A session 
	$\M$ is 
	a \emph{label error session} if:\\[1mm] 
\centerline{
$\M \equiv
\pa{\p}{\choice{\pi_{i}.\PP_{i}}{i \in  I}} 
\Par \pa{\q}{\choice{\pi_{j}.Q_{j}}{j \in  J}} \Par \M'
$}\\[1mm]
where if there exists $\pi_i=\soutprlbl{\q}{\lab}{v}$ 
with $i \in I$, then 
$\forall k\in j$ such that $\pi_k=\ssinp{\p}{\lab_k}{(x_k)}$, 
we have $\lab_{k} \not = \lab$ 
(i.e., all input labelled processes are unmatched).  
A session $\M$ is a \emph{value error process} if:\\
\centerline{
$\M\equiv \pa{\role{p}}{\cond{v}{\PP_1}{\PP_2}\Par \M'}
\quad \text{\ with \ } v\not\in \{\true,\false\}$}\\[1mm]
We call $\M$ is a \emph{session error} if it is either label or
value error session. 
\end{newdefinition}

\begin{newexample}[Label error session]
\label{ex:sesserror}
	A {\em label error session} is a session that
	contains a pair of input and output with dual participants,
	but does not have a correct labelled 
	redex. For example, session\\[1mm]
	\centerline{$\small
	\M = \pa{\p}{(\ssout{\q}{\lab_1}. \inact \OR
          \ssinp{\rr}{\lab_2}.\inact)} \Par
          \pa{\q}{\ssinp{\p}{\lab_2}. \inact} \Par 
          \pa{\rr}{\ssout{\p}{\lab_2}. \inact}
	$}\\[1mm]
	is a session error because it has an 
active redex ($\ssout{\q}{\lab_1}$ and $\ssinp{\p}{\lab_2}$) that
        is mismatched. $\M$ has a type context:
	$\De = {\p}: (\ntout{\q}{\lab_1} \tor \ntinp{\rr}{\lab_2}), {\q}: \ntout{\p}{\lab_2}, {\rr}: \ntout{\p}{\lab_2}
	$ and $\De$ is not safe. 
        Note that session $\M'$ defined below  is \emph{not} 
a session error. \\[1mm]
\centerline{$\small
\M' = \pa{\p}{(\ssinp{\q}{\lab_1}. \inact \OR
          \ssinp{\q}{\lab_2}.\inact)} \Par
          \pa{\q}{\ssout{\p}{\lab_2}. \inact} 
$}\\[1mm]
as in the standard (multiparty) session types
\cite{less_is_more,ChenDSY17}.
\end{newexample}
From Theorem~\ref{th:subject_reduction}
	and the fact that error sessions are untypable by a safe
        context, we have:
\begin{newcorollary}[Communication Safety]
	\label{cor:typesafety}
	Assume 
	$\types \M \as \Delta$. 
	For all $\M'$, such that $\M \red^* \M'$, $\M'$ is 
        not a session error. 
\end{newcorollary}


\label{sub:deadlockfree}
Deadlock-freedom (from \cite[Definition 5.1]{less_is_more})
states that a process either terminates, completing
all actions, or makes progress. 

\begin{newdefinition}[Deadlock-freedom]
\label{def:properties}
Session $\M$ is \emph{deadlock-free} iff for all $\M'$ such that 
$\M \red^\ast \M'$ either 
\begin{enumerate*}
	\item $\M' \noRed$ and $\M' \equiv \pa{\p}{\inact}$ for
          some $\p$, or 
	\item there exists $\M''$ such that $\M'\red \M''$.
\end{enumerate*} 
\end{newdefinition}
\begin{restatable}[Deadlock-freedom]{newthm}{thmdeadlockfreedom}
\label{thm:deadlockfree} 
Assume $\types \M \as \Delta$ and $\df{\Delta}$.
Then $\M$ is 
deadlock-free. 
\end{restatable}
\begin{proof}
Assume $\M\red^* \M' \noRed$ and 
$\types \M' \as \LLd$. Then by Theorem~\ref{th:subject_reduction}, 
there exists $\LLd$ such that 
$\LL\red^* \LLd\noRed$ and 
$\types \M' \as \LLd$. 
By the definition of $\df{\Delta}$, 
$\LLd = \{\rolei{p}{i}:\tinact\}_{i\in I}$. 
Hence $\M'=\Pi_{i\in I} \ \pa{\rolei{p}{i}}{\inact}$ and  
$\M'\equiv \pa{\rolei{p}{k}}{\inact}$ for some $k \in I$
by definition of $\equiv$. 
\end{proof} 

\begin{newremark}[Properties]
\label{rem:properties}
\NY{
In this paper, we follow a \emph{general typing system} in 
\cite{less_is_more} where a property $\varphi$
of a session $\M$ is guranteed by checking 
the same propery $\varphi$ of its typing context $\LL$ \cite[Definition 5.1, Theorem 5.15]{less_is_more}. 
Hence if we replace $\df{\LL}$ in Theorem~\ref{thm:deadlockfree} 
to a liveness property in \cite{GPPSY2021}, 
then a typed session $\M$ can gurantee a liveness property.  
We selected deadlock-freedom since it is used to prove 
all the encodability results (see \S~\ref{sec:hierarchy}). 
This methodology from \cite{less_is_more} is often called \emph{bottom-up}. 
The classical \mpst \cite{HYC2016,YG2020,CDPY2015} takes 
the \emph{top-down} approach 
where the user first writes a \emph{global type} as a protocol specification. 
See \S~\ref{sec:related} for further discussions.
}
\end{newremark}

\section{Hierarchy of Expressiveness of Session Calculi}
\label{sec:hierarchy}
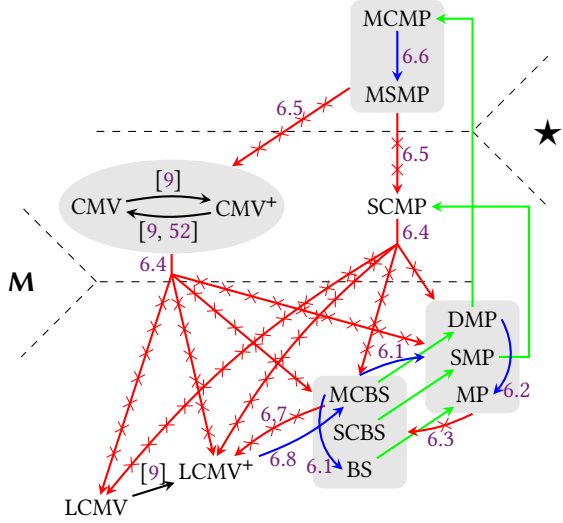
\begin{figure}
	\centering
	\begin{tikzpicture}[bend angle=10, inner sep=0.5mm]
		\node[scale=2] (star) at (3.5, 1) {\patternStar};
		\draw[dashed] (-2.5, 1) -- (2.5, 1) -- (3.5, 2);
		\draw[dashed] (2.5, 1) -- (3.5, 0);
		\node[scale=1.2] (M) at (-3.5, -1) {\patternM};
		\draw[dashed] (2.5, -1) -- (-2.5, -1) -- (-3.5, -2);
		\draw[dashed] (-2.5, -1) -- (-3.5, 0);
		\node[rounded corners, rectangle, fill=black!10, minimum width=1.25cm, minimum height=1.5cm] (A) at (1.5, 2) {};
		\node (MCMP) at (1.5, 2.5) {\mcmpst};
		\node (MSMP) at (1.5, 1.5) {\msmpst};
		\node (SCMP) at (1.5, 0) {\scmpst};
		\node[rounded corners, ellipse, fill=black!10, minimum width=3cm, minimum height=1.25cm] (B) at (-1.5, 0) {};
		\node (CMV+) at (-0.5, 0) {\cnvms};
		\node (CNV) at (-2.5, 0) {\cnv};
		\node[rounded corners, rectangle, fill=black!10, minimum width=1.25cm, minimum height=1.5cm] (C) at (2.5, -2) {};
		\node (MMP) at (2.5, -1.5) {\dmpst};
		\node (SMP) at (2.5, -2) {\smpst};
		\node (MP) at (2.5, -2.5) {\mpst};
		\node[rounded corners, rectangle, fill=black!10, minimum width=1.25cm, minimum height=1.5cm] (D) at (1, -3) {};
		\node (MCBS) at (1, -2.5) {\mcbs};
		\node (SCBS) at (1, -3) {\scbs};
		\node (BS) at (1, -3.5) {\bs};
		\node (LCMV+) at (-0.9, -3.5) {\lcnvms};
		\node (LCNV) at (-2.5, -4) {\lcnv};
		\draw[-stealth, color=green, thick] (MMP) to (2.5, 2.5) to (MCMP);
		\draw[-stealth, color=green, thick] (SMP) to (3.25, -2) to (3.25, 0) to (SCMP);
		\draw[-stealth, color=green, thick] (SCBS) to (SMP);
		\draw[-stealth, color=green, thick] (MCBS) to (MMP);
		\draw[-stealth, color=green, thick] (BS) to (MP);
		\path[-stealth, color=red, thick] (A) edge node[near start, left] {\ref{thm:separateMCMP}\hspace*{0.5em}} (B);
		\draw[color=red, decorate, decoration={crosses, shape size=1.5mm, pre=, pre length=0.4cm, post=, post length=0.25cm}] (A) to (B);
		\path[-stealth, color=red, thick, bend angle=20] (A) edge node[right] {\ref{thm:separateMCMP}} (SCMP);
		\draw[color=red, bend angle=20, decorate, decoration={crosses, shape size=1.5mm, pre=, pre length=0.4cm, post=, post length=0.25cm}] (A) to (SCMP);
		\path[color=red, thick] (B) edge node[left] {\ref{thm:separateSCMP}} (-1.5, -0.9);
		\path[-stealth, color=red, thick] (-1.5, -0.9) edge (LCNV);
		\draw[color=red, decorate, decoration={crosses, shape size=1.5mm, pre=, pre length=0.4cm, post=, post length=0.25cm}] (-1.5, -0.9) to (LCNV);
		\path[-stealth, color=red, thick] (-1.5, -0.9) edge (LCMV+);
		\draw[color=red, decorate, decoration={crosses, shape size=1.5mm, pre=, pre length=0.4cm, post=, post length=0.25cm}] (-1.5, -0.9) to (LCMV+);
		\path[-stealth, color=red, thick] (-1.5, -0.9) edge (C);
		\draw[color=red, decorate, decoration={crosses, shape size=1.5mm, pre=, pre length=0.4cm, post=, post length=0.25cm}] (-1.5, -0.9) to (C);
		\path[-stealth, color=red, thick] (-1.5, -0.9) edge (D);
		\draw[color=red, decorate, decoration={crosses, shape size=1.5mm, pre=, pre length=0.4cm, post=, post length=0.25cm}] (-1.5, -0.9) to (D);
		\path[color=red, thick] (SCMP) edge node[right] {\ref{thm:separateSCMP}} (1.5, -0.5);
		\path[-stealth, color=red, thick] (1.5, -0.5) edge[bend right] (LCNV);
		\draw[color=red, decorate, decoration={crosses, shape size=1.5mm, pre=, pre length=0.4cm, post=, post length=0.25cm}] (1.5, -0.5) to[bend right] (LCNV);
		\path[-stealth, color=red, thick] (1.5, -0.5) edge[bend right] (LCMV+.north);
		\draw[color=red, decorate, decoration={crosses, shape size=1.5mm, pre=, pre length=0.4cm, post=, post length=0.25cm}] (1.5, -0.5) to[bend right] (LCMV+.north);
		\path[-stealth, color=red, thick] (1.5, -0.5) edge (C);
		\draw[color=red, decorate, decoration={crosses, shape size=1.5mm, pre=, pre length=0.4cm, post=, post length=0.25cm}] (1.5, -0.5) to (C);
		\path[-stealth, color=red, thick] (1.5, -0.5) edge (D.north);
		\draw[color=red, decorate, decoration={crosses, shape size=1.5mm, pre=, pre length=0.4cm, post=, post length=0.25cm}] (1.5, -0.5) to (D.north);
		\path[-stealth, color=red, thick] (C.south) edge[bend left] node[below] {\ref{thm:MpstIntoBs}} (D.east);
		\draw[color=red, decorate, decoration={crosses, shape size=1.5mm, pre=, pre length=0.4cm, post=, post length=0.25cm}] (C.south) to[bend left] (D.east);
		\path[-stealth, color=red, thick] (MCBS) edge[bend right] node[above] {\ref{thm:separateMCBS}\hspace*{0.25em}} (LCMV+);
		\draw[color=red, decorate, decoration={crosses, shape size=1.5mm, pre=, pre length=0.4cm, post=, post length=0.25cm}] (MCBS) to[bend right] (LCMV+);
		\path[-stealth, thick] (CNV) edge[bend left] node[above] {\citep{CASAL202223}} (CMV+);
		\path[-stealth, thick] (CMV+) edge[bend left] node[below] {\citep{CASAL202223, PY2022}} (CNV);
		\path[-stealth, thick] (LCNV) edge node[above] {\citep{CASAL202223}} (LCMV+);
		\path[-stealth, color=blue, thick] (MCMP) edge node[right] {\ref{thm:MCMP}} (MSMP);
		\path[-stealth, color=blue, thick, bend angle=35] (MMP.east) edge[bend left] node[near end, below] {\hspace*{1em}\ref{thm:MP}} (MP.east);
		\path[-stealth, color=blue, thick, bend angle=35] (MCBS.west) edge[bend right] node[near end, below] {\ref{thm:binarySessions}\hspace*{0.75em}} (BS.west);
		\path[-stealth, color=blue, thick] (D.north) edge[bend left] node[above] {\ref{cor:BsIntoMpst}} (C.west);
		\path[-stealth, color=blue, thick] (LCMV+) edge[bend right] node[near start, below] {\ref{thm:LcnvmsIntoMcbs}} (MCBS);
	\end{tikzpicture}
	\caption{Hierarchy of Session Calculi\label{fig:hierarchy}}
\end{figure}

We analyse the expressive power of the mixed choice construct in the \mcmpst-calculus by several separation and encodability results.
In all subcalculi of \mcmpst, we assume that they are typed and deadlock-free.
The hierarchy of expressiveness of the 9 subcalculi and 4 variants of \cnvms \cite{CASAL202223} is given in Figure~\ref{fig:hierarchy}.
We label arrows with a reference to the respective result.
More precisely: 
\begin{itemize}[leftmargin=*]
	\item $ L_1 $ \redArrow $ L_2 $: There is no good encoding from $ L_1 $ into $ L_2 $.
	\item $ L_1 $ \blueArrow $ L_2 $: There is a good encoding from $ L_1 $ into $ L_2 $ and the arrow presents a new encodability result.
	\item $ L_1 $ \blackArrow $ L_2 $: There is a good encoding from $ L_1 $ into $ L_2 $ that has been proven in the literature.
\end{itemize}
If $ L_1 \subset L_2 $, then identity is a good encoding from $ L_1 $ into $ L_2 $.
If $ L_1 $ is placed straight below $ L_2 $ then $ L_1 \subset L_2 $.
Additional inclusions are depicted by \greenArrow (e.g. \smpst \greenArrow \scmpst since $ \smpst \subset\scmpst $).

A \GreyBubble{grey shape} captures calculi of the same expressive power, \ie there is a good encoding between any pair of calculi in the same shape (e.g., \mcmpst and \msmpst).
Similarly, arrows from (or into) a shape are understood as an arrow from (or into) each calculus of the shape (e.g. from $ L_1\in \set{\mcbs, \scbs, \bs} $ into $ L_2 \in \set{\dmpst, \smpst, \mpst} $).

\myparagraph{Encodability Criteria. \ }
We combine the encodability criteria from \cite{DBLP:journals/iandc/Gorla10} and \cite{DBLP:conf/esop/PetersNG13}.
Two steps are in \emph{conflict}, if performing one step disables the other step, \ie if both reduce the same choice.
Otherwise they are \emph{distributable}.
$ M $ is distributable into $ M_1, \ldots, M_n $ if and only if we have $ M \equiv M_1 \Par \ldots \Par M_n $, where $ \equiv $ does not unfold any recursions.
We add $ \success $ (\emph{successful termination}) to all calculi in addition to $ \inact $ and type it in the same way as $ \inact $.
$ M\reachSuccess $ if $ M \red^\ast M' $ and $ M' $ has an unguarded occurrence of $ \success $.
Moreover, let the equivalence $ \asymp $ be a success respecting (weak) reduction bisimulation, \ie if $ M_1 \asymp M_2 $ then (1)~$ M_1 \red M_1' $ implies $ M_2 \red^\ast M_2' $ and $ M_1' \asymp M_2' $, (2)~$ M_2 \red M_2' $ implies $ M_1 \red^\ast M_1' $ and $ M_1' \asymp M_2' $, and (3)~$ M_1\reachSuccess $ iff $ M_2\reachSuccess $.

We consider an encoding $ \arbitraryEncoding $ to be \emph{good} if it is
\begin{description}
\item[compositional:] The translation of an operator $ \mathbf{op} $
  is a function $ \context^{\mathbb{P}}_{\mathbf{op}} $ on the
  translations of the subterms of the operator, \ie $
  \ArbitraryEncoding{{\mathbf{op}}\left( M_1, \ldots, M_n \right)} =
  \Context{\mathbb{P}}{\mathbf{op}}{\ArbitraryEncoding{M_1},
    \ldots, \ArbitraryEncoding{M_n}} $ for all $ M_1, \ldots, M_n $
  with $\mathbb{P}= \parts{M_1} \cup \ldots \cup \parts{M_n}$.

	\item[name invariant:] For every $ M $ and every substitution $ \sigma $, it holds that $ \ArbitraryEncoding{M\sigma} \asymp \ArbitraryEncoding{M}\sigma $.
	\item[operationally complete:] For all $ M \red M' $, $ \ArbitraryEncoding{M} \red \asymp \ArbitraryEncoding{M'} $.
	\item[operationally sound:] For all $ \ArbitraryEncoding{M} \red N $, there is an $ M' $ such that $ M \red M' $ and $ N \red \asymp \ArbitraryEncoding{M'} $.
	\item[divergence reflecting:] For every $ M $, $ \ArbitraryEncoding{M} \red^{\omega} $ implies $ M \red^{\omega} $.
	\item[success sensitive:] For every $ M $, $ M \reachSuccess $ iff $ \ArbitraryEncoding{M} \reachSuccess $.
	\item[distributability preserving:] For every $ M $ and for all $ M_1, \ldots, M_n $ that are distributable within $ M $ there are some $ N_1, \ldots, N_n $ that are distributable within $ \ArbitraryEncoding{M} $ s.t.\ $ N_i \asymp \ArbitraryEncoding{M_i} $ for all $ 1 \leq i \leq n $.
\end{description}
Operational correspondence is the combination of operational \emph{completeness} and \emph{soundness}.

\myparagraph{Hierarchy. \ }
We summarise the keys to the presented separation and encodability results, \ie the counterexample for separation and the interesting parts of the encoding functions, in Figure~\ref{fig:KeysExpressiveness} and explain them below.
The parts of encodings given in Figure~\ref{fig:KeysExpressiveness} always present the encoding of a participant \p.

\begin{figure*}
	\renewcommand{\arraystretch}{1.2}
	\arraycolsep2pt
	\begin{tabular}{|l|l|c|}
		\hline
		1 &
		$ \begin{array}{l}
			\scbs \blueArrow \bs\\
			\smpst \blueArrow \mpst
		\end{array} $ &
		$ \begin{array}{rcl}
			\ArbitraryEncoding{\choice{\soutprlbl{\q}{\lab_i}{v_i}\seq\PP_i}{i \in I}} &=& \choice{\colorbox{green!20}{\ssinp{\q}{\lbl{enc_o}}}\seq\soutprlbl{\q}{\lab_i}{v_i}\seq\ArbitraryEncoding{\PP_i}}{i \in I}\\
			\ArbitraryEncoding{\choice{\sinprlbl{\q}{\lab_j}{x_j}\seq\PP_j}{j \in J}} &=& \colorbox{green!20}{\ssout{\q}{\lbl{enc_o}}}\seq\choice{\sinprlbl{\q}{\lab_j}{x_j}\seq\ArbitraryEncoding{\PP_j}}{j \in J}
		\end{array} $\\
		\hline
		2 &
		$ \begin{array}{l}
			\mcbs \blueArrow \scbs\\
			\dmpst \blueArrow \smpst
		\end{array} $ &
		$ \begin{array}{l}
			\ArbitraryEncoding{\left( \choice{\soutprlbl{\q}{\lab_i}{v_i}\seq\PP_i}{i \in I} \right) \OR \left( \choice{\sinprlbl{\q}{\lab_j}{x_j}\seq\PP_j}{j \in J} \right)}\\
			= \begin{cases}
				\begin{array}{l}
					\left( \choice{\soutprlbl{\q}{\lab_i}{v_i}\seq\ArbitraryEncoding{\PP_i}}{i \in I} \right) \OR\\
					\colorbox{red!10}{\ssout{\q}{\lbl{enc_i}}}\seq\left( \left(\choice{\sinprlbl{\q}{\lab_j}{x_j}\seq\ArbitraryEncoding{\PP_j}}{j \in J} \right) \OR \colorbox{yellow!30}{\ssinp{\q}{\lbl{reset}}}\seq \choice{\soutprlbl{\q}{\lab_i}{v_i}\seq\ArbitraryEncoding{\PP_i}}{i \in I} \right)
				\end{array} & \text{if } I \neq \emptyset \neq J \wedge \p < \q\\
				\left( \choice{\sinprlbl{\q}{\lab_j}{x_j}\seq\ArbitraryEncoding{\PP_j}}{j \in J} \right) \OR \colorbox{red!10}{\ssinp{\q}{\lbl{enc_i}}}\seq\choice{\soutprlbl{\q}{\lab_i}{v_i}\seq\ArbitraryEncoding{\PP_i}}{i \in I} & \text{if } I \neq \emptyset \neq J \wedge \p > \q\\
				\choice{\soutprlbl{\q}{\lab_i}{v_i}\seq\ArbitraryEncoding{\PP_i}}{i \in I} & \text{if } J = \emptyset \wedge \p < \q\\
				\colorbox{red!10}{\ssinp{\q}{\lbl{enc_i}}}\seq\choice{\soutprlbl{\q}{\lab_i}{v_i}\seq\ArbitraryEncoding{\PP_i}}{i \in I} & \text{if } J = \emptyset \wedge \p > \q\\
				\colorbox{red!10}{\ssout{\q}{\lbl{enc_i}}}\seq\choice{\sinprlbl{\q}{\lab_j}{x_j}\seq\ArbitraryEncoding{\PP_j}}{j \in J} & \text{if } I = \emptyset \wedge \p < \q\\
				\left( \choice{\sinprlbl{\q}{\lab_j}{x_j}\seq\ArbitraryEncoding{\PP_j}}{j \in J} \right) \OR \colorbox{red!10}{\ssinp{\q}{\lbl{enc_i}}}\seq\colorbox{yellow!30}{\ssout{\q}{\lbl{reset}}}\seq\choice{\sinprlbl{\q}{\lab_j}{x_j}\seq\ArbitraryEncoding{\PP_j}}{j \in J} & \text{if } I = \emptyset \wedge \p > \q
			\end{cases}
		\end{array} $\\
		\hline
		3 &
		$ \begin{array}{l}
			\mcbs \blueArrow \bs\\
			\dmpst \blueArrow \mpst
		\end{array} $ &
		$ \begin{array}{l}
			\ArbitraryEncoding{\left( \choice{\soutprlbl{\q}{\lab_i}{v_i}\seq\PP_i}{i \in I} \right) \OR \left( \choice{\sinprlbl{\q}{\lab_j}{x_j}\seq\PP_j}{j \in J} \right)}\\
			= \begin{cases}
				\begin{array}{l}
					\left( \choice{\colorbox{green!20}{\ssinp{\q}{\lbl{enc_o}}}\seq\soutprlbl{\q}{\lab_i}{v_i}\seq\ArbitraryEncoding{\PP_i}}{i \in I} \right) \OR\\
					\colorbox{green!20}{\ssinp{\q}{\lbl{enc_o}}}\seq\colorbox{red!10}{\ssout{\q}{\lbl{enc_i}}}\seq\left( \left(\choice{\sinprlbl{\q}{\lab_j}{x_j}\seq\ArbitraryEncoding{\PP_j}}{j \in J} \right) \OR \colorbox{yellow!30}{\ssinp{\q}{\lbl{reset}}}\seq \choice{\soutprlbl{\q}{\lab_i}{v_i}\seq\ArbitraryEncoding{\PP_i}}{i \in I} \right)
				\end{array} & \text{if } I \neq \emptyset \neq J \wedge \p < \q\\
				\colorbox{green!20}{\ssout{\q}{\lbl{enc_o}}}\seq\left( \left( \choice{\sinprlbl{\q}{\lab_j}{x_j}\seq\ArbitraryEncoding{\PP_j}}{j \in J} \right) \OR \colorbox{red!10}{\ssinp{\q}{\lbl{enc_i}}}\seq\choice{\soutprlbl{\q}{\lab_i}{v_i}\seq\ArbitraryEncoding{\PP_i}}{i \in I} \right) & \text{if } I \neq \emptyset \neq J \wedge \p > \q\\
				\choice{\colorbox{green!20}{\ssinp{\q}{\lbl{enc_o}}}\seq\soutprlbl{\q}{\lab_i}{v_i}\seq\ArbitraryEncoding{\PP_i}}{i \in I} & \text{if } J = \emptyset \wedge \p < \q\\
				\colorbox{green!20}{\ssout{\q}{\lbl{enc_o}}}\seq\colorbox{red!10}{\ssinp{\q}{\lbl{enc_i}}}\seq\choice{\soutprlbl{\q}{\lab_i}{v_i}\seq\ArbitraryEncoding{\PP_i}}{i \in I} & \text{if } J = \emptyset \wedge \p > \q\\
				\colorbox{green!20}{\ssinp{\q}{\lbl{enc_o}}}\seq\colorbox{red!10}{\ssout{\q}{\lbl{enc_i}}}\seq\choice{\sinprlbl{\q}{\lab_j}{x_j}\seq\ArbitraryEncoding{\PP_j}}{j \in J} & \text{if } I = \emptyset \wedge \p < \q\\
				\colorbox{green!20}{\ssout{\q}{\lbl{enc_o}}}\seq\left( \left( \choice{\sinprlbl{\q}{\lab_j}{x_j}\seq\ArbitraryEncoding{\PP_j}}{j \in J} \right) \OR \colorbox{red!10}{\ssinp{\q}{\lbl{enc_i}}}\seq\colorbox{yellow!30}{\ssout{\q}{\lbl{reset}}}\seq\choice{\sinprlbl{\q}{\lab_j}{x_j}\seq\ArbitraryEncoding{\PP_j}}{j \in J} \right) & \text{if } I = \emptyset \wedge \p > \q
			\end{cases}
		\end{array} $\\
		\hline
		\multirow{2}{*}{4} &
		\multirow{2}{*}{\mpst \redArrow \mcbs} &
		three distributable steps in\\
		&& $ M_{\mpst} = \pa{\role{a}}{\left( \ssout{\role{b}}{l}\seq\success \OR \ssout{\role{b}}{l}\seq\inact \right)} \Par \pa{\role{b}}{\ssinp{\role{a}}{l}\seq\inact} \Par \pa{\role{c}}{\left( \ssout{\role{d}}{l}\seq\success \OR \ssout{\role{d}}{l}\seq\inact \right)} \Par \pa{\role{d}}{\ssinp{\role{c}}{l}\seq\inact} \Par \pa{\role{e}}{\left( \ssout{\role{f}}{l}\seq\success \OR \ssout{\role{f}}{l}\seq\inact \right)} \Par \pa{\role{f}}{\ssinp{\role{e}}{l}\seq\inact} $\\
		\hline
		\multirow{2}{*}{5} &
		\scmpst \redArrow &
		the synchronisation pattern \patternM in\\
		& Bottom Layer &
		$ \PM_{\scmpst} = \pa{\enodei{a}} \left( \ssout{\enodei{b}}{l}. \ssout{\enodei{d}}{l}. \inact \OR \ssout{\enodei{d}}{l'}. \inact \right) \Par \pa{\enodei{b}} \left( \ssinp{\enodei{a}}{l}. \inact \OR \ssinp{\enodei{c}}{l}. \inact \right) \Par \pa{\enodei{c}} \left( \ssout{\enodei{b}}{l}. \inact \OR \ssout{\enodei{d}}{l}. \inact \right) \Par \pa{\enodei{d}} \left( \ssinp{\enodei{c}}{l}. \ssinp{\enodei{a}}{l}. \success \OR \ssinp{\enodei{a}}{l'}. \inact \right) $\\
		\hline
		\multirow{2}{*}{6} &
		\cnv \redArrow &
		the synchronisation pattern \patternM in\\
		& Bottom Layer &
		$ \PM_{\cnv} = \ResCMV{x}{y}{\left( \OutCMV{x}{\true}{\inactCMV} \Par \InpCMV{\linCMV}{y}{z}{\ConditionalCMV{z}{\inact}{\inactCMV}} \Par \OutCMV{x}{\false}{\inactCMV} \Par \InpCMV{\linCMV}{y}{z}{\ConditionalCMV{z}{\inactCMV}{\success}} \right)} $\\
		\hline
		\multirow{2}{*}{7} &
		\cnvms \redArrow &
		the synchronisation pattern \patternM in\\
		& Bottom Layer &
		$ \PM_{\text{\cnvms}} = \ResCMVmix{x}{y}{\left( \ChoiceCMVmix{\linCMV}{x}{\OutCMVmix{l}{\true}{\inactCMVmix}} \Par \ChoiceCMVmix{\linCMV}{y}{\InpCMVmix{l}{z}{\ConditionalCMVmix{z}{\inact}{\inactCMVmix}}} \Par \ChoiceCMVmix{\linCMV}{x}{\OutCMVmix{l}{\false}{\inactCMVmix}} \Par \ChoiceCMVmix{\linCMV}{y}{\InpCMVmix{l}{z}{\ConditionalCMVmix{z}{\inactCMVmix}{\success}}} \right)} $\\
		\hline
		\multirow{4}{*}{8} &
		& the synchronisation pattern \patternStar in\\
		& \msmpst \redArrow &
		$ \PS_{\msmpst} = M_{\enodei{a}} \Par M_{\enodei{b}} \Par M_{\enodei{c}} \Par M_{\enodei{d}} \Par M_{\enodei{e}} \Par M_{\enodei{gc}} \hspace{2em} M_{\enodei{a}} = \pa{\enodei{a}} \left( \ssout{\enodei{e}}{l}. \inact \OR \ssinp{\enodei{b}}{l}. P_{\enodei{a}} \OR \ssinp{\enodei{gc}}{\delete}. \inact \right) $\\
		& Lower Layers &
		$ M_{\enodei{gc}} = \choice{\ssout{\enodei{i}}{\delete} \seq \inact}{\enodei{i} \in \set{\enodei{a}, \enodei{b}, \enodei{c}, \enodei{d},\enodei{e}}} \hspace{2em} \sigma = \left[ \enodei{a} \mapsto \enodei{b}, \enodei{b} \mapsto \enodei{c}, \enodei{c} \mapsto \enodei{d}, \enodei{d} \mapsto \enodei{e}, \enodei{e} \mapsto \enodei{a} \right] $\\
		&& $ M_{\enodei{b}} = M_{\enodei{a}}\sigma \quad M_{\enodei{c}} = M_{\enodei{b}}\sigma \quad M_{\enodei{d}} = M_{\enodei{c}}\sigma \quad M_{\enodei{e}} = M_{\enodei{d}}\sigma \hspace{2em} P_{\enodei{a}}, \ldots, P_{\enodei{e}} \in \left\{ \success, \inact \right\} $\\[1mm]
		\hline
		9 &
		\mcmpst \blueArrow \msmpst &
		$ \begin{array}{l}
\ArbitraryEncoding{\sum_{\q} \left( \left( \choice{\soutprlbl{\q}{\lab_i}{v_i}\seq\PP_i}{i \in I} \right) \OR \left( \choice{\sinprlbl{\q}{\lab_j}{x_j}\seq\PP_j}{j \in J} \right) \right)} 
			= \sum_{\q}
			\begin{cases}
				\text{case distinction of \mcbs \blueArrow \scbs}
			\end{cases}
		\end{array} $\\[1mm]
		\hline
		10 &
		\mcbs \redArrow \lcnvms &
		$ M_{\mcbs} = \pa{\p}{\left( \ssout{\q}{l_1}\seq\inact \OR \colorbox{orange!20}{\ensuremath{\ssinp{\q}{l_2}\seq\success}} \right)} \Par \pa{\q}{\left( \ssinp{\p}{l_1}\seq\inact \OR \colorbox{orange!20}{\ensuremath{\ssinp{\p}{l_3}\seq\success}} \right)} $\\
		\hline
		11 &
		\lcnvms \blueArrow \mcbs &
		$ \begin{array}{l}
\, 			\ArbitraryEncoding{\Gamma \vdash \ResCMVmix{x}{y}{\PP}} = \ArbitraryEncoding{\Gamma' \vdash \PP}_{x, y}\\
			\ArbitraryEncoding{\Gamma \vdash \ChoiceCMVmix{\linCMVmix}{\mathsf{k}}{\left( \PP = \left( \sum_{i \in I} \OutCMVmix{\lab_i}{v_i}{\PP_i} \right) \OR \left( \sum_{j \in J} \InpCMVmix{\lab_j}{z_j}{\PP_j} \right) \right)}}_{x, y}\\
			= \begin{cases}
				\pa{\enodei{k}}{\inact} & \text{if } \fv{\PP} = \Set{x, y} \cup \Set{\mathsf{k}}\\\
				\pa{\enodei{k}}{\left( \sum_{i \in I} \soutprlbl{\enodei{\overline{k}}}{\lab_i \cdot \colorbox{blue!10}{o}}{v_i}\seq\ArbitraryEncoding{\Gamma_i \vdash \PP_i}_{\mathsf{\overline{k}}} \right)} & \text{if } \fv{\PP} \neq \Set{x, y}\text{, } \mathsf{\overline{k}} \in \Set{x, y} \setminus \Set{\mathsf{k}}\text{,}\\
				\hspace{1.5em} \OR \left( \sum_{j \in J} \ssout{\enodei{\overline{k}}}{\lab_i \cdot \colorbox{blue!10}{i}}\seq\sinprlbl{\enodei{\overline{k}}}{\lab_i}{z}\seq\ArbitraryEncoding{\Gamma_j \vdash \PP_j}_{\mathsf{\overline{k}}} \right) & \phantom{if }\text{and } \PP \text{ is typed as internal in } \Gamma\\
				\pa{\enodei{k}}{\left( \sum_{i \in I} \ssinp{\enodei{\overline{k}}}{\lab_i \cdot \colorbox{blue!10}{i}}\seq\soutprlbl{\enodei{\overline{k}}}{\lab_i}{v_i}\seq\ArbitraryEncoding{\Gamma_i \vdash \PP_i}_{\mathsf{\overline{k}}} \right)} & \text{if } \fv{\PP} \neq \Set{x, y}\text{, } \mathsf{\overline{k}} \in \Set{x, y} \setminus \Set{\mathsf{k}}\text{,}\\
				\hspace{1.5em} \OR \left( \sum_{j \in J} \sinprlbl{\enodei{\overline{k}}}{\lab_i \cdot \colorbox{blue!10}{o}}{z}\seq\ArbitraryEncoding{\Gamma_j \vdash \PP_j}_{\mathsf{\overline{k}}} \right) & \phantom{if }\text{and } \PP \text{ is typed as external in } \Gamma
			\end{cases}
		\end{array} $\\
		\hline
	\end{tabular}
	\caption{Keys to Separation and Encodability Results\label{fig:KeysExpressiveness}}
\end{figure*}

The encodings in Figure~\ref{fig:KeysExpressiveness} use a total order $ < $ on participants.
To ensure compositionality and name invariance, $ < $ is not assumed but constructed by the encoding functions from the tree-structure of parallel composition in source terms.

We start with the subcalculi of \mcmpst and go from bottom to top.
The counter in the following subsection headings refers to the number of encodability or separation results, but we are not counting results obtained from calculus inclusion.
Figure~\ref{fig:KeysExpressiveness}($n$) denotes the $n$th row in the figure.

\subsection{Encodability (1--3): Binary Sessions}
\label{subsec:encodability:binary}

In binary sessions, there is no difference in the expressive power between separate or mixed choice.

\begin{restatable}[Binary Sessions]{newthm}{thmbinarySessions}
	\label{thm:binarySessions}
	Let $ \mathcal{L} = \Set{\bs, \scbs, \mcbs} $.
	There is a good encoding from any $ L_1 \in \mathcal{L} $ into any $ L_2 \in \mathcal{L} $.
\end{restatable}

\myparagraph{From \scbs to \bs.}
In contrast to \scbs with separate choice, there are only single outputs in \bs and no choices on outputs.
Therefore the encoding \scbs \blueArrow \bs translates an output-guarded choice to an input-guarded choice with the same input-prefix $ \colorbox{green!20}{\ssinp{\q}{\lbl{enc_o}}} $ and the respective outputs as continuations (see Figure~\ref{fig:KeysExpressiveness}(1)).
As explained in Example~\ref{ex:choiceprocess}(1,2), such choices are typable in \bs because of subtyping.
The encoding of input-guarded choice starts with the matching output $ \colorbox{green!20}{\ssout{\q}{\lbl{enc_o}}} $ followed by the original choice.
Accordingly, a single interaction in the source term between some output-guarded and some input-guarded choice is translated into two steps on the target side.

Deadlock-freedom ensures the existence of the respective communication partner for the first step (with the fresh label $ \colorbox{green!20}{\lbl{enc_o}} $) and safety ensures that for every output that was picked in the first step there will be a matching input in the second step.
The rest of the encoding \scbs \blueArrow \bs is homomorphic.

\myparagraph{From \mcbs to \scbs. }
The encoding \mcbs \blueArrow \scbs has to translate mixed into separate choices (see Figure~\ref{fig:KeysExpressiveness}(2)).
Therefore, it translates a choice from $ \p $ to $ \q $ with $ \p < \q $ into an output-guarded choice and with $ \p > \q $ into an input-guarded choice, where $ \colorbox{red!10}{\ssout{\q}{\lbl{enc_i}}} $ (or $ \colorbox{red!10}{\ssinp{\q}{\lbl{enc_i}}} $) are used to guard the outputs in the input-guarded choice (the inputs in the output-guarded choice).

Accordingly, an interaction of an output in \p and an input in \q with $ \p < \q $ can be emulated by a single step, whereas the case $ \p > \q $ requires two steps: one interaction with the fresh label $ \colorbox{red!10}{\lbl{enc_i}} $ and the interaction of the respective out- and input.  
In the case of $ \p > \q $, deadlock-freedom guarantees the existence of the respective communication partner for the first step, and safety the existence of a matching input for every output in the second step.

In the case $ \p < \q $, additional inputs in $ \p $ without matching outputs in $ \q $ require a special attention.
The type system of the target calculus \scbs forces us to ensure that the output $ \colorbox{red!10}{\ssout{\q}{\lbl{enc_i}}} $, that guards the inputs in the translation of \p, is matched by an input $ \colorbox{red!10}{\ssinp{\p}{\lbl{enc_i}}} $ in the translation of \q (see Corollary~\ref{cor:typesafety}).
We add $ \colorbox{red!10}{\ssinp{\p}{\lbl{enc_i}}}\seq\colorbox{yellow!30}{\ssout{\p}{\lbl{reset}}}\seq\choice{\sinprlbl{\p}{\lab_j}{x_j}\seq\ArbitraryEncoding{\PP_j}}{j \in J} $ to the translation of \q (the last summand in the last line of Figure~\ref{fig:KeysExpressiveness}(2) with \p and \q swapped).
If \p has outputs and inputs but no input has a match in \q, then the emulation of a single source term step takes either one or three steps.

The rest of \mcbs \blueArrow \scbs is homomorphic (except for the construction of $ < $).

\myparagraph{From \mcbs to \bs.} \
The encoding \mcbs \blueArrow \bs combines the ideas of \scbs \blueArrow \bs and \mcbs \blueArrow \scbs (see Figure~\ref{fig:KeysExpressiveness}(3)).

\myparagraph{Inclusion.} \
In the remaining cases of $ L_1, L_2 \in \Set{\bs, \scbs, \mcbs} $ identity is a good encoding, because $ L_1 = L_2 $ or $ L_1 \subset L_2 $ since $ \bs \subset \scbs \subset \mcbs $.

\subsection{Encodability (4--6): Multiparty Sessions with Choices on a Single Participant}
\label{subsec:encodability:multiparty}

In multiparty sessions, in that choice is limited to one participant, mixed choice does not increase the expressive power.

\begin{restatable}[Multiparty Sessions with Choices on a Single Participant]{newthm}{thmMP}
	\label{thm:MP}
	Let $ \mathcal{L} = \Set{\mpst, \smpst, \dmpst} $.
	There is a good encoding from any $ L_1 \in \mathcal{L} $ into any $ L_2 \in \mathcal{L} $.
\end{restatable}

The encoding \smpst \blueArrow \mpst translates choice in the same way as \scbs \blueArrow \bs.
Encoding \dmpst \blueArrow \smpst inherits the encoding of mixed to separate choices from \mcbs \blueArrow \scbs.
\dmpst \blueArrow \mpst combines \smpst \blueArrow \mpst and \dmpst \blueArrow \smpst.
In all three encodings, all remaining operators are translated homomorphically (except for the construction of $ < $).
In the remaining cases of $ L_1, L_2 \in \Set{\mpst, \smpst, \dmpst} $, identity is a good encoding because $ L_1 = L_2 $ or $ L_1 \subset L_2 $ since $ \mpst \subset \smpst \subset \dmpst $.

\subsection{Encodability: Binary into Multiparty with Choice on a Single Participant}
\label{subsec:encodability:BSMS}

We now consider the blue arrow between the grey squares in the bottom layer.
It tells us, that binary sessions with mixed choice can be simulated by the classical multiparty sessions (\mpst), since in both cases mixed choice does not add expressive power.

\begin{corollary}[Binary into Multiparty with Choice on a Single Participant]
	\label{cor:BsIntoMpst}
	There is a good encoding from any $ L_1 \in \Set{\bs, \scbs, \mcbs} $ into any $ L_2 \in \Set{\mpst, \smpst, \dmpst} $.
\end{corollary}

Corollary~\ref{cor:BsIntoMpst} follows from Theorem~\ref{thm:binarySessions} and calculus inclusion.
The encoding \mcbs \blueArrow \bs is also a good encoding from \mcbs into \mpst, since $ \bs \subset \mpst $.
With $ \mpst \subset \smpst $ then this encoding is also a good encoding from \mcbs into \smpst.
Identity is a good encoding from \mcbs into \dmpst, because $ \mcbs \subset \dmpst $.
Similarly, we obtain good encodings for $ L_1 = \scbs $ and $ L_1 = \bs $ from the encoding \scbs \blueArrow \bs and the calculus inclusions $ \bs \subset \mpst \subset \smpst \subset \dmpst $ and $ \scbs \subset \smpst $.

\subsection{Separate (1) Multiparty Sessions with Choice on a Single Participant from Binary Sessions}
\label{subsec:SepMPfromBS}

We now show the first separation result: there exists no good encoding from \mpst into \mcbs.

\begin{restatable}[Separate Multiparty with Choice on a Single Participant from Binary Sessions]{newthm}{thmMpstIntoBs}
	\label{thm:MpstIntoBs}
	There is no good encoding from any $ L_1 \in \Set{\mpst, \smpst, \dmpst} $ into any $ L_2 \in \Set{\bs, \scbs, \mcbs} $.
\end{restatable}

Note that the binary versions \bs, \scbs, and \mcbs include only a single binary session (hence 2 parties).
Such a system cannot perform three distributable steps, nor can it emulate such behaviour while respecting the criterion on the preservation of distributability.
We use $ M_{\mpst} $ from Figure~\ref{fig:KeysExpressiveness}(4) as counterexample.

By $ \mpst \subset \smpst \subset \dmpst $, $ M_{\mpst} $ is also contained in \smpst and \dmpst.
If there would be a good encoding from \mpst into \scbs, then this encoding would also be a good encoding from \mpst into \mcbs, because $ \scbs \subset \mcbs $; \ie such an encoding contradicts \mpst \redArrow \mcbs.
Hence, the remaining separation results in Theorem~\ref{thm:MpstIntoBs} follow from \mpst \redArrow \mcbs and calculus inclusion.

\subsection{Separate (2--13) the Middle from the Bottom}
\label{subsec:sepMiddleFromBottom}

Going further upwards in Figure~\ref{fig:hierarchy}, we observe two dashed lines above \dmpst (see also Figure~\ref{fig:hierarchyPi}).
These two dashed lines divide Figure~\ref{fig:hierarchy} into three layers along the ability to express the synchronisation patterns \patternStar and \patternM from \cite{glabbeekGoltzSchicke12, DBLP:conf/esop/PetersNG13} and thus the different amounts of synchronisation they capture.
Calculi in the bottom layer can express neither \patternM nor \patternStar and are considered as asynchronously distributed calculi.

\myparagraph{The pattern \patternM.} \
A system is an \patternM if it is distributable into two parts that can act independently by performing the distributable steps $ a $ and $ c $; but that may also interact in a step $ b $ that is in conflict to both $ a $ and $ c $.

\begin{definition}[Synchronisation Pattern \patternM]
	\label{def:synchronisationPatternM}
	Let $ \PM $ be a process such that:
	\begin{itemize}[leftmargin=*]
		\item $ \PM $ can perform at least three alternative steps $ a{:}\; \PM \red M_a $, $ b{:}\; \PM \red M_b $, and $ c{:}\; \PM \red M_c $ such that $ M_a $, $ M_b $, and $ M_c $ are pairwise different.
		\item The steps $ a $ and $ c $ are distributable in $ \PM $.
		\item But $ b $ is in conflict with both $ a $ and $ c $.
	\end{itemize}
	In this case, we denote the process $ \PM $ as \patternM.
\end{definition}

In an \patternM, the two parts of the system decide whether they interact or proceed independently, but are able to make this decision consistently without any form of interaction.
This is a minimal form of synchronisation.
The system $ \PM_{\scmpst} $ in Figure~\ref{fig:KeysExpressiveness}(5) is an example of an \patternM in \scmpst, where step $ a $ is an interaction of $ \enodei{a} $ and $ \enodei{b} $, step $ b $ is an interaction of $ \enodei{b} $ and $ \enodei{c} $, and step $ c $ is an interaction of $ \enodei{c} $ and $ \enodei{d} $.

\myparagraph{Mixed Sessions.} \
In the bottom layer we find \lcnv and \lcnvms that we introduce as subcalculi of \cnv and the \emph{mixed sessions} \cnvms introduced in \cite{CASAL202223}.
We briefly recap their main concepts (see \cite{PetersYoshida24tecRep} for more).
The syntax of \cnvms is given as\\[1mm]
\centerline{
	$ P \bnfis \ChoiceCMVmix{q}{x}{\sum_{i \in I} {M_i}} \bnfbar P \mid P \bnfbar \ResCMVmix{x}{y}{P} \bnfbar \ConditionalCMVmix{v}{P}{P} \bnfbar \inactCMVmix $
}\\[1mm]
where $ M \bnfis \BranchCMVmix{\lab}{*}{v}{P} $, $ * \bnfis {!} \bnfbar ? $, and $ q \bnfis \linCMVmix \bnfbar \unCMVmix $ denote \emph{linear} and \emph{unrestricted} choices.
The double restriction $ \ResCMVmix{x}{y}{P} $ introduces two matching \emph{session endpoints} $ x $ and $ y $.
Interaction is limited to two matching endpoints\\[1mm]
\centerline{
	$\begin{array}{l}
		\ResCMVmix{y}{z}{{\left( \ChoiceCMVmix{\linCMVmix}{y}{{\left( \OutCMVmix{\lab}{v}{P} + M \right)}} \mid \ChoiceCMVmix{\linCMVmix}{z}{{\left( \InpCMVmix{\lab}{x}{Q} + N \right)}} \mid R \right)}} \\ \hspace*{5cm}\stepCMVmix 
		\ResCMVmix{y}{z}{{\left( P \mid Q\Set{\Subst{v}{x}} \mid R \right)}}
	\end{array} $
}\\[1mm]
where linear choices are consumed, whereas unrestricted choices are persistent.
In \cnv, choice is replaced by output $ \OutCMV{y}{v}{P} $, input $ \InpCMV{q}{y}{x}{P} $, selection $ \SelCMV{x}{\lab}{P} $, and branching $ \BranCMV{x}{\Set{\BranchCMV{\lab_i}{P_i}}_{i \in I}} $.

\lcnvms and \lcnv result from restricting \cnvms and \cnv to a single sessions, \ie at most one restriction, forbidding delegation, \ie only values can be transmitted, allowing only linear choices that are typed as linear.

\myparagraph{The Bottom Layer.} \
\lcnv, \lcnvms, \bs, \scbs, \mcbs, \mpst, \smpst, and \dmpst are placed in the bottom, because they do not contain the synchronisation pattern \patternM.

\begin{restatable}[No \patternM]{newlemma}{lemNoM}
	\label{lem:noM}
	There are no \patternM in \lcnv, \lcnvms, \bs, \scbs, \mcbs, \mpst, \smpst, and \dmpst.
\end{restatable}

We show that these languages have no distributable steps $ a $ and $ c $ that are both in conflict to a step $ b $.

\myparagraph{Separation.} \
Accordingly, the ability to combine different communication partners in choice is the key to the \patternM in $ \PM_{\scmpst} $, because the typing discipline of \mcmpst and its variants forbids different participants with the same name.

\begin{restatable}[Separate Middle from Bottom]{newthm}{thmSeparateSCMP}
	\label{thm:separateSCMP}
	There is no good encoding from any $ L_1 \in \Set{\cnv, \text{\cnvms}, \scmpst} $ into any calculus $ L_2 \in \Set{\lcnv, \text{\lcnvms}, \bs, \scbs, \mcbs, \mpst, \smpst, \dmpst} $.
\end{restatable}

In the proof we show that the \patternM is preserved by the criteria of a good encoding, \ie to emulate the behaviour of an \patternM the target calculus needs an \patternM.
Then $ \PM_{\scmpst} $ in Figure~\ref{fig:KeysExpressiveness}(5) is used as counterexample if $ L_1 = \scmpst $, $ \PM_{\cnv} $ in Figure~\ref{fig:KeysExpressiveness}(6) if $ L_1 = \cnv $, and $ \PM_{\text{\cnvms}} $ in Figure~\ref{fig:KeysExpressiveness}(7) if $ L_1 = \text{\cnvms} $.

\subsection{Separate (14--18) the Top Layer from the Rest}
\label{subsec:sepTop}

The \patternStar can only be expressed by synchronously distributed calculi in the top layer.
It describes the amount of synchronisation that is \eg necessary to solve leader election in symmetric networks and captures the power of synchronisation of $ \pi $ with mixed choice.

\begin{definition}[Synchronisation Pattern \patternStar]
	\label{def:synchronisationPatternGreatM}
	Let $ \PS $ be a process such that:
	\begin{itemize}[leftmargin=*]
		\item $ \PS $ can perform at least five alternative steps $ i : \PS \red M_i $ for $ i \in \left\{ a, b, c, d, e \right\} $ such that the $ M_i $ are pairwise different;
		\item the steps $ a $, $ b $, $ c $, $ d $, and $ e $ form a circle such that $ a $ is in conflict with $ b $, $ b $ is in conflict with $ c $, $ c $ is in conflict with $ d $, $ d $ is in conflict with $ e $, and $ e $ is in conflict with $ a $; and 
		\item every pair of steps in $ \left\{ a, b, c, d, e \right\} $ that is not in conflict due to the previous condition is distributable in $ \PS $.
	\end{itemize}
	In this case, we denote the process $ \PS $ as \patternStar.
\end{definition}

An example of a \patternStar in \msmpst is $ \PS_{\msmpst} $ in Figure~\ref{fig:KeysExpressiveness}(8).
Since $ \msmpst \subset \mcmpst $, $ \PS_{\msmpst} $ is also a \patternStar in \mcmpst.
All other calculi in Figure~\ref{fig:hierarchy} do not contain \patternStar{s}.

\begin{restatable}[No \patternStar]{newlemma}{lemNoStar}
	\label{lem:noStar}
	There are no \patternStar in \cnv, \cnvms, \scmpst, \lcnv, \lcnvms, \bs, \scbs, \mcbs, \mpst, \smpst, and \dmpst.
\end{restatable}

\noindent
\begin{minipage}{\columnwidth}
	\begin{wrapfigure}{R}{0.25\columnwidth}
		\centering
		\vspace{-1em}
		\hspace{-2em}
		\scalebox{0.7}{
		\begin{tikzpicture}[]
			\foreach \x/\xlabel/\xtext/\ytext in {1/e/$ C_5 $/$ b $,2/d/$ C_4 $/$ a $,3/c/$ C_3 $/$ e $,4/b/$ C_2 $/$ d $,5/a/$ C_1 $/$ c $}
	        {
	            \path (360*\x/5+125:0.8) node (\xlabel) {\xtext};
	            \path (360*\x/5-55:1.75) node (p\x) {\ytext};
	        }
	        \draw[-latex] (p2) -- (a);
	        \draw[-latex] (p2) -- (b);
	        \draw[-latex] (p1) -- (b);
	        \draw[-latex] (p1) -- (c);
	        \draw[-latex] (p5) -- (c);
	        \draw[-latex] (p5) -- (d);
	        \draw[-latex] (p4) -- (d);
	        \draw[-latex] (p4) -- (e);
	        \draw[-latex] (p3) -- (e);
	        \draw[-latex] (p3) -- (a);
		\end{tikzpicture}
		}
	\end{wrapfigure}

	$ \quad $ Assume by contradiction, that \scmpst contains a \patternStar.
Since a step reducing a conditional cannot be in conflict with any other step, the five steps of the \patternStar in \scmpst are interactions that reduce an output-guarded and an input-guarded choice, respectively.
	Then in the steps $ a, b, c, d, e $ five choices $ C_1, \ldots, C_5 $ are reduced as depicted on the right, where \eg the step $ a $ reduces the choices $ C_1 $ and $ C_2 $.
\end{minipage}

\vspace{0.25em}
Since \scmpst does not allow for mixed choices, each of the five choices $ C_1, \ldots, C_5 $ contains either only outputs or only inputs.
W.o.g.\ assume that $ C_1 $ contains only outputs and $ C_2 $ only inputs.
Then the choice $ C_3 $ needs to be on outputs again, because step $ b $ reduces $ C_2 $ (with only inputs) and $ C_3 $.
Then $ C_4 $ is on inputs and $ C_5 $ is on outputs.
But then step $ e $ reduces two choices $ C_1 $ and $ C_5 $ that are both on outputs.
Since the reduction semantics of \scmpst does not allow such a step, this is a contradiction.

The proofs for the absence of \patternStar{s} in \cnv and \cnvms are similar and were already discussed in \cite{PY2022}.
For the remaining cases, Lemma~\ref{lem:noStar} follows from calculus inclusion.

\myparagraph{Separation.} \
We observe, that is indeed mixed choice (in contrast to only separate choice) that is the key to the \patternStar $ \PS_{\msmpst} $ in \msmpst.

\begin{restatable}[Separate the Top Layer]{newthm}{thmSeparateMCMP}
	\label{thm:separateMCMP}
	There is no good encoding from any $ L_1 \in \Set{\msmpst, \mcmpst} $ into any $ L_2 \in \Set{\cnv, \text{\cnvms}} \cup \Set{\scmpst, \lcnv, \text{\lcnvms}, \bs, \scbs, \mcbs, \mpst, \smpst, \dmpst} $.
\end{restatable}

In the proof we show again that the \patternStar is preserved by the criteria of a good encoding, \ie to emulate the behaviour of a \patternStar the target calculus needs a \patternStar.
Then $ \PS_{\msmpst} $ in Figure~\ref{fig:KeysExpressiveness}(8) is used as counterexample if $ L_1 = \msmpst $.
The other case, \ie $ L_1 = \mcmpst $, follows then from $ \msmpst \subset \mcmpst $.

\subsection{Encodability (7): Mixed Choice into Separate Choice per Participant}
\label{subsec:encodability:MCMP}

The smallest calculus in Figure~\ref{fig:hierarchy} that contains a \patternStar is \msmpst, \ie multiparty sessions with mixed choice that allow only separate choice per participant.
In \S~\ref{subsec:sepMiddleFromBottom} we learnt that the key to the synchronisation pattern \patternM is the ability to combine different communication partners in choice.
Here, we learn that also further up in the hierarchy, the important feature is the combination of different communication partners in a choice.
Whether or not the choice limits the summands of the same participant in a mixed choice to either outputs or inputs does not change the expressive power.

\begin{restatable}[Mixed Choice into Separate Choice per Participant]{newthm}{thmMCMP}
	\label{thm:MCMP}
	There is a good encoding from \mcmpst into \msmpst and vice versa.
\end{restatable}

The encoding \mcmpst \blueArrow \msmpst translates mixed choices into choices that are separate per participant.
Therefore, we apply for each participant the idea of \mcbs \blueArrow \scbs and then combine the resulting choices for each participant in a choice (see Figure~\ref{fig:KeysExpressiveness}(9)).

\subsection{Separate (19) \mcbs from \lcnvms}
\label{subsec:sepMCBSfromLCNVms}

Since all variants of \mcmpst describe only a single session without any form of shared names, comparing directly with \cnvms would cause negative results that are completely independent of the respective choice constructs.
Because of that, we introduced \lcnv and \lcnvms with $ \lcnv \subset \cnv $ and $ \text{\lcnvms} \subset \text{\cnvms} $ in Figure~\ref{fig:hierarchy}.
We observe, that \lcnvms cannot emulate all behaviours of \mcbs.

\begin{restatable}[Separate \mcbs from \lcnvms]{newthm}{thmSeparateMcbsFromLcnvms}
	\label{thm:separateMCBS}
	There is no good encoding from \mcbs into \lcnvms.
\end{restatable}

The counterexample used for this separation result, namely $ M_{\mcbs} $ in Figure~\ref{fig:KeysExpressiveness}(10), utilises the different typing mechanisms.
The type system of \cnvms requires duality of the types of the interacting choices, where subtyping increases flexibility but allows only choices typed as external to have additional summands, \ie summands that are not matched by the communication partner such as $ \colorbox{orange!20}{\ensuremath{\ssinp{\p}{l_2}\seq\success}} $ and $ \colorbox{orange!20}{\ensuremath{\ssinp{\p}{l_3}\seq\success}} $ in $ M_{\mcbs} $.
In \mcbs, on the other hand side, additional inputs are allowed in \emph{both} choices.

A translation of $ M_{\mcbs} $ would need to turn one of the two choices into choice that is typed as external without additional summands.
This prevents a good encoding from \mcbs into \lcnvms.

\subsection{Encodability (8): From \lcnvms into \mcbs}
\label{subsec:encodability:LcnvmsIntoMcbs}

In the opposite direction there is an encoding from \lcnvms into \mcbs.

\begin{restatable}[From \lcnvms into \mcbs]{newthm}{thmLcnvmsIntoMcbs}
	\label{thm:LcnvmsIntoMcbs}
	There is a good encoding from \lcnvms into \mcbs.
\end{restatable}

The subtyping in \lcnvms---in contrast to \mcbs---does not only allow for additional inputs but also additional outputs.
Hence, we use the type information in $ \Gamma $, whether the choice we are translating is typed as internal or external.
We translate all summands of a choice typed as external to outputs and all summands of a choice typed as internal to inputs.
As done in \cite{CASAL202223}, we extend the labels by $ \colorbox{blue!10}{o} $ or $ \colorbox{blue!10}{i} $ to ensure that the original matches are respected.

\lcnvms does not guarantee deadlock-freedom.
Fortunately, the limitation to a single session ensures that the only typed and deadlocked cases are systems in \lcnvms which compose both communication partners sequentially.
Accordingly, we translate choices that have both session endpoints as free variables to $ \inact $ in the first case of the translation of choice in Figure~\ref{fig:KeysExpressiveness}(11).
The translations of the continuations $ \ArbitraryEncoding{\PP_i}_{\mathsf{\overline{k}}} $ and $ \ArbitraryEncoding{\PP_j}_{\mathsf{\overline{k}}} $ are similar to the encoding of choice $ \ArbitraryEncoding{\ldots}_{x, y} $ but without the first case and without $ \pa{\enodei{k}}{\,} $.

\subsection{Separation (20) via Leader Election}
\label{subsec:leaderElection}

The synchronisation patterns distribute the hierarchy in Figure~\ref{fig:hierarchyPi} and \ref{fig:hierarchy} into asynchronously and synchronously distributed calculi.
To further support the practical implications of our analysis, we are studying the problem of leader election in symmetric networks.
Also the landmark result in \citep{Palamidessi03} to separate $ \pi $ from $ \pi_{\mathsf{a}} $ uses this problem as distinguishing feature.

A network $ M = M_1 \Par \ldots \Par M_n $ is \emph{symmetric} iff $ M_{\sigma(i)} = M_i\sigma $ for each $ i \in \left\{1, \ldots, n\right\} $ and for all permutations $ \sigma $ on participants.
$ M $ is an \emph{electoral system} if in every maximal execution exactly one leader is announced.
The system $ \mathsf{Election} $ in Example~\ref{ex:leaderElection-process} presents a leader election protocol in a symmetric network with five participants in \msmpst.
Since calculi in the middle or bottom layer do not contain symmetric electoral systems, we can use this problem to separate \msmpst from \scmpst.

\begin{restatable}[Separation via Leader Election]{newthm}{thmseparation}
	\label{thm:separationViaLE}
	There is no good and barb sensitive encoding from \msmpst into \scmpst.
\end{restatable}

We prove first that \scmpst cannot solve leader election in symmetric networks.
To elect a leader, the initially symmetric network has to break its symmetry such that only one leader is elected.
Without mixed choice, any attempt to do so can be counteracted by steps of the symmetric parts of the network that restore the original symmetry.
This leads to an infinite sequence of steps in that no unique leader is elected.
Then, we use $ \mathsf{Election} $ as counterexample to show that there is no good and barb sensitiveness encoding.
Therefore, we show that the combination of operational correspondence, divergence reflection, and barb sensitiveness ensures that the translation of a symmetric electoral system is again a symmetric electoral system.

Note that we use here barb sensitiveness, \ie a source network and its translation may reach the same barbs, as additional encodability criterion, where \emph{barbs} \cite{milnerSangiorgi92} are the standard observables used in $ \pi $. 
Barb sensitiveness ensures that the announcement of the leader is respected by the encoding function, \ie that the translation of an electoral system again announces exactly one leader.

\subsection{Summary}

The considered binary variants of \mcmpst have the same expressive power.
The encodings introduce additional steps on the fresh labels $ \colorbox{green!20}{\lbl{enc_o}} $, $ \colorbox{red!10}{\lbl{enc_i}} $, and $ \colorbox{yellow!30}{\lbl{reset}} $.
Therefore, safety and deadlock-freedom of the source help us to ensure operational correspondence, \ie that the target does not introduce new behaviours by introducing reductions that are stuck.
Similarly, the variants of \mpst with choice on a single participant can be encoded into each other.

There is an increase in the expressive power for separate and mixed choice but \emph{only if we combine different communication partners in a choice}, \ie with \scmpst for separate choice and \mcmpst for mixed choice; 
however, no difference between \msmpst and \mcmpst.

\NY{
Notice that each calculus plays a distinct role, i.e.~it is novel
from the viewpoint of types and/or expressiveness. 
\begin{enumerate}
\item \mcmpst, \msmpst and \scmpst are the first calculi
ensuring safety and deadlock-freedom for which an increasing
expressive power of choice was shown; 

\item \msmpst is a realistic subform of general mixed choice in \mcmpst.
A server may wait to receive requests from clients but also may send
status information \eg for load balancing;

\item 
\scmpst is the only calculus in the middle layer, that guarantees
deadlock-freedom; and 

\item
\dmpst has never appeared in the literature,
but it can be viewed one step extension from \mpst with mixed choice 
construct, and 
\mcbs is a binary version of \dmpst;
\item
  \scbs and \bs are a binary version of \smpst and \mpst respectively,
  and \bs can express \mcbs.    
\end{enumerate}
}
Finally, we observe that \mcbs is \emph{strictly more expressive} than \lcnvms.
The way to type choices in \lcnvms is strictly less expressive than typing for choice in \mcbs (see \S~\ref{subsec:sepMCBSfromLCNVms} and \S~\ref{subsec:encodability:LcnvmsIntoMcbs}).

\section{Related and Future Work}
\label{sec:related}
\myparagraph{Mixed Choice in the $\pi$-Calculus. \ }
Mixed choice has been proposed as a fundamental
process calculi construct 
and extensively studied in the
context of the $\pi$-calculi. Choice $P+Q$ first appeared in the 
original $\pi$-calculus syntax 
\cite{short:MilnerR:calmp1} where $P$ and $Q$ can
contain any form of processes (such as parallel compositions).
Later Milner proposed \emph{guarded mixed choice} 
\cite{milner1991polyadic} 
where each process in the choice branch is either an input or an
output. 
\citet{Palamidessi03} has shown  
that mixed guarded choice cannot be encoded into separate  
guarded choice through a symmetric and divergence preserving encoding. 
Example~\ref{ex:leaderElection-process} 
is motivated by leader election  
used to show this separation result in \cite{Palamidessi03}. 
See Figure~\ref{fig:hierarchyPi} in \S~\ref{sec:intro} for a relationship 
with other calculi.  
Our guarded mixed choice
differs from those in the standard 
$\pi$-calculus, where 
it can freely use distinct channels 
(like $P=a\outsymbol v.Q_1 + b\inpsymbol (x).Q_2+c\outsymbol w.Q_3$);   
in \mcmpst, we use \emph{distinct participants}
to express mixed choice. For instance, we can mimic above $P$ 
as $\role{q}\outsymbol v.Q_1 + 
\role{r}\inpsymbol (x).Q_2+
\p\outsymbol w.Q_3$ (with $\role{q}$,
$\role{r}$, $\role{s}$ pairwise distinct).    

In \cite{DBLP:conf/esop/Glabbeek22}, Glabbeek encodes a variant of the $\pi$-calculus with implicit matching into a variant of CCS, where the result of
a synchronisation of two actions is itself an action subject to relabelling
or restriction.
A comparison between this variant of CCS and 
\mcmpst based on the results in 
\cite{DBLP:conf/esop/Glabbeek22} is 
an interesting future study; maybe using 
a syntactic similarity between CCS and local types.

\myparagraph{Binary Mixed Sessions. \ }
\citet{CASAL202223} have proposed binary mixed sessions and 
provided an encoding from \cnvms into a traditional variant of binary sessions with selection and  branching (\cnv) \cite{VASCONCELOS201252}, and proved the encoding criteria except for operational soundness, which was left open. 
This problem was solved by \citet{PY2022}.
Further \citet{CASAL202223} provided a good encoding from \cnv into \cnvms.
Hence, \cnvms and \cnv have the same expressive power, \ie mixed choice does not increase the expressive power in binary sessions in \cite{VASCONCELOS201252}. Additionally, \citet{PY2022} have proved that 
\cnvms cannot emulate the calculi 
in the top layer of Figure~\ref{fig:hierarchyPi} (hence 
staying in the middle), 
proving leader election in symmetric networks cannot be solved in \cnvms.  
Here we prove that \mcmpst and its weaker form 
($\msmpst$) belong to the top.

\citet{less_is_more} present
a multiparty session type system which 
does not require global type correspondence; rather it
identifies and computes the {\em desired} properties against
a set of local types. 
We apply this general approach 
to build a typing system of mixed choice, but limit our attention to 
a single multiparty session, extending from 
\mpst in \cite{DBLP:journals/jlp/GhilezanJPSY19}. 
This allowed us to focus on which mixed choice construct can strictly raise
the expressive power. Our future work is to analyse 
how much expressive power is added to multiparty session types by inclusions of session delegation from \cite{less_is_more} and 
shared names from \cite{HYC2016,HYC08,CDPY2015}.
A related research question is 
a more detailed comparisons of \scmpst with $\pi_s$ and of \mcmpst with $\pi$ 
in Figure~\ref{fig:hierarchyPi}.
Remember that \scmpst and \mcmpst are typed calculi 
that ensure \eg deadlock-freedom, but the $ \pi$-calculus does not.
Since there are deadlock-free processes in $\pi$ 
which are untypable by \mcmpst, considering of the minimal set of operators
that, if added to \mcmpst, close a gap to $\pi$ is 
an interesting question. 

\myparagraph{Global Types with Flexible Choices. \ }
\NY{In the context of multiparty session types, the 
research so far focuses to explore the \emph{top-down} 
approach extend \emph{global types} to more 
flexible choice. In the top-down approach \cite{HYC2016,YG2020,CDPY2015,YH2024}, 
a global type is projected into a set of local types; and  
if each participant typed by each local type, 
participants in a session automatically satisfies safety and deadlock-freedom 
(\emph{correctness by construction}).} 
 
\citet{DBLP:journals/corr/abs-1203-0780} 
present a \emph{semantic} procedure to check well-formedness of 
global types with parallel composition and mixed choice, 
which is undecidable due to 
infinite FIFO buffered semantics. 
They also propose
a decidable algorithm for projecting a limited class of 
global types with their extension. 
{\citet{JY-ESOP20}} extend global
types with a mixed choice operator,
an existential quantification over roles, 
and unbounded interleaving of subprotocols.
It presents a bisimulation technique 
for developing a correspondence between global
types and local interactions. 
\citet{HJ-TACAS20} propose a 
\emph{runtime verification framework} based on 
the domain-specific language (Disclojure) to 
verify programs against multiparty session types with mixed choice.  
The work concentrates on tool implementation, hence no theorem for
correctness is provided. 
\citet{DBLP:conf/concur/MajumdarMSZ21} present
a generalised decidable projection procedure for multiparty
session types with infinite FIFO buffered semantics,  
which extends the original syntax of global types 
to one sender with multiple receivers. 
They use a message causality
analysis based on message sequence chart techniques to check
the projectability of global types. 
This approach is further extended in \cite{Li2023} to enable a sound and 
complete projection from a global type to 
deadlock-free communicating automata 
\cite{BZ1983}. 
\citet{jongmans_et_al:LIPIcs.ECOOP.2023.42} propose 
a \emph{synthetic} typing system which directly uses 
an operational semantics of \emph{implicit} local types 
in a typing judgement of synchronous multiparty processes. 
An implicit local type 
has no explicit syntax but 
represents abstract behaviour of a global type wrt 
each participant. Their approach allows 
flexible type syntax including mixed choice in global types, 
but requires stronger conditions 
for realisable global types 
which are similar with those in 
\cite{DBLP:journals/corr/abs-1203-0780}. 
\KPCom{Flexible choice for choreographies is studied 
in \cite{cruz2018communications}. 
In \cite{cruz2018communications} selection/branching (separate choice) is combined with multicoms/multisels.
Multicoms/multisels group multiple actions, but as concurrent and not as choice actions (all actions can happen, not just one). }

None of the above work with flexible choice 
\cite{DBLP:journals/corr/abs-1203-0780,DBLP:conf/concur/MajumdarMSZ21,jongmans_et_al:LIPIcs.ECOOP.2023.42,Li2023} 
has studied expressiveness of processes typed by their systems. 
It is an interesting future work to compare their expressive powers 
extending the work by \citet{BeauxisPV08} which studies encodability and 
separation results for the (untyped) $\pi$-calculi with mixed choice, 
stack, bags and FIFO queues. 

\myparagraph{Applications. \ }
The integration of model checking in a type system of the
$\pi$-calculus is piloted by \citet{POPL02ModelChecking}, where the
tool checks LTL formulae against behavioural types.  Those ideas are
applied to MPST in~\cite{less_is_more}, which is extended to a
crash-stop failure model in \cite{barwell2022crashstop}.  As
programming language applications, \citet{DBLP:conf/icse/LangeNTY18,DBLP:conf/popl/LangeNTY17}
and \citet{GY2020} extract behavioural types from Go source code, and
\citet{PLDI19Effpi} design Scala library for communication programs
with behavioural dependent types.  These works use the mCLR2 to
validate safety and deadlock-free properties through type-level
behaviours. Notably, 
\cite{DBLP:conf/icse/LangeNTY18,DBLP:conf/popl/LangeNTY17,GY2020} use internal
and external choices which consist of input, output and
$\tau$-prefixes
to model \texttt{select} construct in Go. 
Extension of the model-checking tool in \cite{less_is_more} to mixed choice 
and studying expressiveness of choice with $\tau$-action 
are interesting topics for future research.

\paragraph*{\bf Acknowledgements. } 
Work partially supported by: EPSRC\\ EP/T006544/2, EP/K011715/1,
EP/K034413/1, EP/L00058X/1,\\ EP/N027833/2,  EP/N028201/1, EP/T014709/2,
EP/V000462/1, \\EP/X015955/1, NCSS/EPSRC VeTSS, and Horizon EU
TaRDIS\\ 101093006. We thank Dimitrios Kouzapas for his collaboration
on an early version of a related topic.



\bibliographystyle{ACM-Reference-Format}
\bibliography{session2}

\ifnotsplit{
\pagebreak


\appendix
\setlength{\proofrightwidth}{0.3\linewidth}
\section{Syntax of Mixed Choice Multiparty Session Types}

\subsection{Omitted Definitions}
\label{app:types_ommited}

\begin{newdefinition}[Guarded recursive variables]
	\label{app:def:guarded}
	We define the \guarded{\local} predicate
	for local types:
	\[
	\begin{array}{|c|}
		\hline
		\multicolumn{1}{|l|}{\guards{\tvar{t}}{\local}}
		\\
		\hline
		\guards{\tvar{t}}{\tinact}
		\qquad
		\guards{\tvar{t}}{\Tor{i\in I}{\tinoutlbli{p}{\lab}{\UType} \locali{i}}}
		\\[3mm]
		\tree{
			\tvar{t} \not= \tvar{t'}
		}{
			\guards{\tvar{t}}{\tvar{t'}}
		}
		\qquad
		\tree {
			\guards{\tvar{t}}{\local}
			\quad
			\tvar{t} \not= \tvar{t'}
		}{
			\guards{\tvar{t}}{\trec{t'}{\local}}
		}{}	
		\\
		\hline	
	\end{array}
\]

\vspace{4mm}

\[
	\begin{array}{|c|}
		\hline
		\multicolumn{1}{|l|}{\guarded{\local}}
		\\
		\hline
		\guarded{\tinact}
		\qquad
		\qquad
		\tree {
			\forall i \in I, \guarded{\locali{i}}
		}{
			\guarded{\Tor{i \in I}{\tinoutlbli{p}{\lab}{\UType} \locali{i}}}
		}{}
		\\[4mm]
		\guarded{\tvar{t}}
		\qquad
		\tree{
			\guards{\tvar{t}}{\local}
			\quad
			\guarded{\local}
		}{
			\guarded{\trec{t}{\local}}
		}{}
		\\
		\hline
	\end{array}
	\]
\end{newdefinition}

\begin{newdefinition}[Set of participants]
	\label{app:def:parts_function}
	The set of participants of 
	a local type, $\parts{\local}$ are defined as:
	\[
	\begin{array}{|rcl|}
		\hline
		\multicolumn{3}{|l|}{\partsf: \Local \to \powerset{\Part}}
		\\
		\hline
		\parts{\tinact}                    & = & \es
		\\[1mm]
		\parts{\tvar{t}}                   & = & \es
		\\[1mm]
		\parts{\trec{t}{\local}}           & = & \parts{\local}
		\\[1mm]
		\parts{\Tor{i \in I}{\tinoutlbli{p}{\lab}{\UType} \locali{i}}} &=& \set{\rolei{p}{i}}_{i \in I} \cup \bigcup_{i \in I} \parts{\locali{i}}	
		\\
		\hline
	\end{array}
	\]
\end{newdefinition}

\begin{newdefinition}[Free type variables]
	\label{app:def:ftv_function}
	The set of free variables in a local type, $\ftv{\local}$, is defined as:
	\[
		\begin{array}{|rcl|}
			\hline
\multicolumn{3}{| l|}{\ftvsf: \Local \to \powerset{\set{\tvar{t} \setbar \tvar{t} \text{ is a type variable}}}}
			\\
			\hline
                        \ftv{\tinact}  & = & \es
			\\[1mm]
                        \ftv{\tvar{t}} & = & \set{\tvar{t}}
			\\[1mm]
			\ftv{\trec{t}{\local}}           & = & \ftv{\local}\backslash\set{\tvar{t}}
			\\[1mm]
			\ftv{\Tor{i \in I}{\tinoutlbli{p}{\lab}{\UType}\locali{i}}} &=& \bigcup_{i \in I} \ftv{\locali{i}}
			\\
			\hline
		\end{array}
	\]
\end{newdefinition}

\section{Proof for Proposition~\ref{prop:subtyping}}
\label{app:subt_properties}


\begin{newlemma}
	\label{lem:wellformedness}
	If $\wfjudgement{\Tor{i \in I{\cup} J}{\locali{i}}}$ then 
	$\wfjudgement{\Tor{i \in I}{\locali{i}}}$
\end{newlemma}
\begin{proof}
	By the definition of well-formedness. 
\end{proof}

\propmergingandsubtyping*


	
	



\begin{proof}
	\hfill
	\begin{enumerate}
		\item[(a)] $\locali{1}\subt \locali{2}$ is 
			a preorder if it is reflexive and 
			transitive.  

			Reflexivity is immediate by induction for all
                        cases other than
                        for 
			the recursive type which requires both the \SRcR and \SRcL rules. 
			\[
				\ttree {
					\ttree{
						\locald\subst{\trec{t}{\locald}}{\tvar{t}} 
						\subt 
						\locald\subst{\trec{t}{\locald}}{\tvar{t}}
					}{
						\locald\subst{\trec{t}{\locald}}{\tvar{t}} 
						\subt 
						\trec{t}{\locald}
					}{\SRcR}	
				}{
					\trec{t}{\locald} \subt \trec{t}{\locald}
				}{\SRcL}
			\]

Transitivity, $\locali{1} \subt \locali{2} \wedge 
							\locali{2} \subt \locali{3}$ then 
							$\locali{1} \subt \locali{3}$, is given by induction. 

Assume $\locali{1} \subt \locali{2}$ is derived by rule (a) and 
and $\locali{2} \subt \locali{3}$ is derived by rule (b). 

\begin{itemize}
\item If (a)=(b), it is straightforward. 
\item Assume (a)=\SChoice and (b)=\SSel. 

$\forall k {\in} K, \locali{k} \subt \localid{k}$ by \SChoice 
and $\localid{i} \subt \localidd{i}$ by \SSel for some $i\in K$.  

By IH, $\localid{i} \subt \localidd{i}$. 

Hence $\forall k {\in} K\setminus \{i\}, 
\locali{k} \subt \localid{k}=\localidd{k}$, which implies 
$\forall k {\in} K, \locali{k} \subt \localidd{k}$.   
By \SChoice, we have: 
$\Tor{k\in K}{\locali{k}}
\subt 
\Tor{k\in K}{ \localidd{k}}$, as required. 
\end{itemize}

			\item[(b)] Reflexivity and transitivity of 
			$\LLi{1} \subt \LLi{2}$ are 
			proven by reflexivity and of transitivity of local types. 
 	\end{enumerate}

The prove that $\LLi{1} \subt \LLi{2}$ is reflexive and transitive, is straightforward by the definition of $\subt$ and
		Part 1.
\begin{itemize}
\item Reflexivity 

It holds that
		\[
			\set{\p_i: \locali{i}}_{i \in I} \subt \set{\p_i: \locali{i}}_{i \in I}
		\]
		since $\forall i \in I, \locali{i} \subt \locali{i}$ (reflexivity of \subt for local types).

\item Transitivity. 

Assume
		\begin{eqnarray*}
			\set{\p_i: \locali{i}}_{i \in I} &\subt& \set{\p_j: \localid{j}}_{j \in J}
			\\
			\set{\p_j: \localid{j}}_{j \in J} &\subt& \set{\p_k: \localidd{k}}_{k \in K}
		\end{eqnarray*}
where (1) for all $i\in I \cap J$, 
	$\locali{i} {\subt} \localid{i}$
        and for all  
$i\in I\setminus J$ and $j\in J\setminus I$, 
$\locali{i}=\localid{j}=\tinact$. Similarly, 
(2) for all $j\in J\cap K$, 
	$\localid{j} {\subt} \localidd{j}$
        and for all  
$j\in J\setminus K$ and $k\in K\setminus J$, 
$\localid{j}=\localidd{k}=\tinact$.

Hence if $i\in I\cap J$ and $i\not\in K$ (i.e.,  $i\not\in I\cap J\cap K$), 
we have $\locali{i}=\localidd{i}=\tinact$.
From the reflexitivity of \subt for local types, we have
		\[
			\forall i \in I\cap J\cap K, \locali{i} \subt \localidd{i}
		\]
and it is then straightforward to imply
		\begin{eqnarray*}
			\set{\p_i: \locali{i}}_{i \in I} &\subt&  \set{\p_k: \localidd{k}}_{k \in K}
		\end{eqnarray*}
from the definition of $\subt$ for local contexts.

\end{itemize}
\end{proof}

\section{Type-level properties of mixed choice multiparty session types}
\subsection{Proof for Subtyping Properties}
\label{app:lem:subtypingproperties}

\lemmasubtypingproperties*

\begin{proof}[Proof for Lemma~\ref{lem:subtypingproperties}]
	\begin{enumerate}
		\item {\bf (a)}

We show that $\safe{\LL, \p:\local}$ and 
$\LLd, \p:\locald \subt \LL,\p:\local$ imply $\safe{\LLd,\p:\locald}$ 
with $\LLd\subt \LL$ and $\locald\subt \local$. 

The proof is done by induction on the definition of $\subt$. 
We proceed by the case analysis on $\locald\subt \local$. \\

{\bf Case} \SChoice:

Straightforward
by induction on $\localid{k} \subt \locali{k}$ with $k\in K$. \\

{\bf Case} \SSel:  

Assume 
that $\safe{\LL, \p: \Tor{i \in I \cup J}{\toutlbl{\q}{\lab_i}{\UTypei{i}} \locali{i}}}$ with 

$\LLd, \p: \Tor{i \in I}{\toutlbl{\q}{\lab_i}{\UTypei{i}} \localid{i}}
\subt \LL, \p: \Tor{i \in I \cup J}{\toutlbl{\q}{\lab_i}{\UTypei{i}}
  \locali{i}}$. There are two cases. 

{\bf (1)} $\neg\exists 
\LL\by{\actinplbl{\q}{\p}{\lab'}{\UType'}}$. 

Then it is obvious that 
$\safe{\LLd, \p: \Tor{i \in I}{\toutlbl{\q}{\lab_i}{\UTypei{i}} \localid{i}}}$.

{\bf (2)} $\LL\by{\actinplbl{\q}{\p}{\lab'}{\UType'}}$ 

Then 
$\safe{\LL, 
\p: \Tor{i \in I \cup J}{\toutlbl{\q}{\lab_i}{\UTypei{i}} \locali{i}}}$ and

$\p:\Tor{i \in I \cup J}{\toutlbl{\q}{\lab_i}{\UTypei{i}}\locali{i}} \by{\actoutlbl{\p}{\q}{\lab_i}{\UTypei{i}}}$, and 

$\LL\by{\actinplbl{\q}{\p}{\lab'}{\UType'}}$ imply 

$\LL,\p: \Tor{i \in I \cup J}{\toutlbl{\q}{\lab_i}{\UTypei{i}} \locali{i}}\by{\actlbl{\p}{\q}{\lab_i}{\UTypei{i}}}$. 

Hence we can write 
$\LL=\q:\Tor{i \in I \cup J \cup H}
{\tinplbl{\p}{\lab_i}{\UTypei{i}}\localidd{i}}\tor \locald,\LLdd$ for some $H$, $\locald$ 
and $\LLdd$, and

$\LLd=\q:\Tor{i \in I \cup J \cup H\cup K}{\tinplbl{\p}{\lab_i}{\UTypei{i}}\localiddd{i}}\tor \localdd,\LLddd$ for some $K$, $\localdd$, 
  and $\LLddd$ with $\LLddd\subt \LLdd$. 

This implies 
  $\LLd, \p: \Tor{i \in I}{\toutlbl{\q}{\lab_i}{\UTypei{i}}
   \locali{i}}\by{\actlbl{\p}{\q}{\lab_i}{\UTypei{i}}}$, as
  required.

{\bf (b)} By (a) above, $\safe{\De}$. 

Then by the definition of safety, $\safe{\LLd}$ and . 
$\safe{\LLdd}$. We show $\LLd\subt \LLdd$ by induction of $\by{}$. 

Assume $\LL=\LLi{0},\p:\locali{0},\q:\localid{0}$
and 
$\LLd=\LLi{1},\p:\locali{1},\q:\localid{1}$. 

By $\LL  \subt \LLd$, we have 
$\locali{0}\subt \locali{1}$ and 
$\localid{0}\subt \localid{1}$. It is sufficient to prove:

$\p:\locali{0},\q:\localid{0}\by{} \p:\locali{3},\q:\localid{3}$  
and 
$\p:\locali{1},\q:\localid{1}\by{} \p:\locali{4},\q:\localid{4}$  
then 
$\locali{3}\subt \locali{4}$ and 
$\localid{3}\subt \localid{4}$.

By assumption, we can write 

$\locali{0}=\p: 
\Tor{i \in I}{\toutlbl{\q}{\lab_i}{\UTypei{i}} \locali{i}}+\locali{5}$ 

$\localid{0}=\q: \Tor{i \in I\cup I_1}{\tinplbl{\p}{\lab_i}{\UTypei{i}} \localid{i}}+\locali{6}$ 

$\locali{3}=\p: \Tor{i \in I\setminus {k}}{\toutlbl{\q}{\lab_i}{\UTypei{i}} \locali{i}}+\locali{5}+\locali{k}$; and

$\localid{3}=\q: \Tor{i \in I\setminus {k}\cup I_1}{\toutlbl{\p}{\lab_i}{\UTypeid{i}} \localid{i}}+\locali{6}+\localid{k}$. 

Then we can write: 

$\locali{1}=\p: \Tor{i \in I}{\toutlbl{\q}{\lab_i}{\UTypei{i}} \locali{i1}}
+\locali{7}$ 

$\localid{1}=\q: \Tor{i \in I\cup I_1}{\tinplbl{\p}{\lab_i}{\UTypei{i}} \localid{i1}}+\locali{8}$ 

$\locali{4}=\p: \Tor{i \in I\setminus {k}}{\toutlbl{\q}{\lab_i}{\UTypei{i}} \locali{i1}}+\locali{7}+\locali{k1}$; and

$\localid{4}=\q: \Tor{i \in I\setminus {k}\cup I_1}{\toutlbl{\p}{\lab_i}{\UTypeid{i}} \localid{i1}}+\locali{8}+\localid{k1}$. 

with $\locali{5}\subt \locali{7}$ and 
$\locali{6}\subt \locali{8}$ and  
$\locali{k}\subt \locali{k1}$ and 
$\localid{k}\subt \localid{k1}$. 

Hence  $\LLdd\subt \LLddd$, as required.

\item	We consider the last applied rule is 
\SSel. Other rules are similar.
  Assume that
\begin{itemize}
\item	$\safe{\LL, \p: \Tor{i \in I \cup J}{\toutlbl{\q}{\lab_i}{\UTypei{i}} \locali{i}}}$ 
\item 
					$\df{\LL, \p: \Tor{i \in I \cup J}{\toutlbl{\q}{\lab_i}{\UTypei{i}} \locali{i}}}$ and 
\item 
				$\LLd, \p: \Tor{i \in I}{\toutlbl{\q}{\lab_i}{\UTypei{i}} \localid{i}} \subt \LL, \p: \Tor{i \in I \cup J}{\toutlbl{\q}{\lab_i}{\UTypei{i}} \locali{i}}$.
\end{itemize}

				We check
$\LLd, \p: \Tor{i \in I}{\toutlbl{\q}{\lab_i}{\UTypei{i}} \localid{i}}$ is deadlock-free. Note that 
$\LLd, \p: \Tor{i \in I}{\toutlbl{\q}{\lab_i}{\UTypei{i}} \localid{i}}$ is safe by (1) above. 

				By the assumption (safety and deadlock-freedom), we can always match the transitions of
				participant \p in the transitions of local context

				$$
					\LL, \p: \Tor{i \in I}{\toutlbl{\q}{\lab_i}{\UTypei{i}} \locali{i}}
				$$

				\noindent
				i.e., from the fact that
				\df{\LL, \p: \Tor{i \in I \cup J}{\toutlbl{\q}{\lab_i}{\UTypei{i}} \locali{i}}},
				there exist \LLdd such that
				\[
					\LL, \p: \Tor{i \in I \cup J}{\toutlbl{\q}{\lab_i}{\UTypei{i}} \locali{i}}
					\by{}^*
					\LLdd, \p: \Tor{i \in I \cup J}{\toutlbl{\q}{\lab_i}{\UTypei{i}} \locali{i}}
					\by{\actlbl{\rr}{\role{r'}}{\lab}{\UType}}
				\]
				with $\p = \rr$ or $\q = \role{r'}$; 
$\lab=\labi{i}$ and $\UType=\UTypei{i}$ for some $i\in I$. 
Hence

$\role{q}:\locald\in \LLdd$ such that 
$\role{q}:\locald\by{\tinplbl{\p}{\lab_i}{\UTypei{i}}}$. 
Hence $\role{q}:
\Tor{i \in I\cup K}{\tinplbl{\p}{\lab_i}{\UTypei{i}} \localid{i}} \in \LLdd$.  

\smallskip 

				Thus there exists $\LLddd$ where we have:

				$$
					\LLd, \p: \Tor{i \in I}{\toutlbl{\q}{\lab_i}{\UTypei{i}} \localid{i}}
					\by{}^*
					\LLddd, \p: \Tor{i \in I}{\toutlbl{\q}{\lab_i}{\UTypei{i}} \localid{i}}
				$$

				\noindent
and \safe{\LLdd, \p: \Tor{i \in I \cup J}{\toutlbl{\q}{\lab_i}{\UTypei{i}} \localid{i}}}.  
By (1-b) above, we have
\[  
 \safe{\LLddd, \p: \Tor{i \in I}{\toutlbl{\q}{\lab_i}{\UTypei{i}} \localid{i}}}\] 
By $\LLddd\leq \LLdd$, 
we have 
$\role{q}:\Tor{i \in I\cup K\cup H}{\tinplbl{\p}{\lab_i}{\UTypei{i}} \localid{i}} \in \LLddd$.  

This concludes $\role{q}:\locald\in \LLddd$ and 

\[\role{q}:\localdd\by{\tinplbl{\p}{\lab_i}{\UTypei{i}}}\]

Hence we have:
				\[
	\LLd, \p: \Tor{i \in I}{\toutlbl{\q}{\lab_i}{\UTypei{i}} \localid{i}}
	\by{\actlbl{\p}{\q}{\lab_i}{\UTypei{i}}}
				\]
Hence $\df{\LLd, \p: \Tor{i \in I}{\toutlbl{\q}{\lab_i}{\UTypei{i}} \localid{i}}}$. 



\item 
Notice that the safety and deadlock free properties are defined 
following \cite{less_is_more}. 
Hence we can follow the same proof method in \cite[Appendix K]{SY2019Technical}; we first construct the 
\emph{behavioural set}, $\beh{(\De)}$, following \cite[Definition K.1]{SY2019Technical}, which is the set of all reductions of $\De$, extended to all unfoldings of all their elements (defined by using the least fixed point). We then prove $\De$ is safe  iff $\beh{(\De)}$ is safe. 
Observe that for any $\De$, the transitive closure of the relation $\red$ 
defined in Figure~\ref{fig:type_semantics} and Definition \ref{def:config} 
induces a finite-state transition system. 
By this lemma, it is straightforward to produce an algorithm for safety and deadlock-freedom checking that inspects all (finite) elements of the behavioural set $\beh{(\De)}$ to verify whether they satisfy the conditions defined in Definitions \ref{def:session_safety} and \ref{def:dfree_live}. See \cite[Appendix K]{SY2019Technical} for the details. 
\end{enumerate}
\end{proof}

\section{Proofs of Section~\ref{subsec:mcmp:properties}}
\label{app:soundness}
\subsection{Proof for Subject Reduction Theorem}
\label{app:preservation}
We first state some auxiliary lemmas that 
are used to prove Subject Reduction Theorem. 

\begin{newlemma}[Typing Inversion]
	\label{app:lem:typing_inversion}

	\begin{enumerate}
		\item	Assume $\Ga \types \PP \as \local$. Then
				\begin{enumerate}
					\item	$\PP = \inact$ implies $\local = \tinact$.
			%
			
					\item	$\PP = \soutprlbl{\p}{\lab}{\val}. \PP'$, where 
							\begin{itemize}
								\item	$\toutlbl{\q}{\lab}{\UType} \locald \subt \local$;
								\item	$\Ga \types v \as \UType$; and
								\item	$\Ga \types \PP' \as \locald$.
							\end{itemize}
			
					\item	$\PP = \sinprlbl{\p}{\lab}{x}. \PP'$ implies
							\begin{itemize}
								\item	$\tinplbl{\p}{\lab}{\UType} \locald \subt \local$;
										and 
								\item	$\Ga, x: \UType \types \PP' \as \local$.
							\end{itemize}
			
			%
							
			
					\item	$\PP = \choice{\pprefix_i. \PP_i}{i \in I}$ implies
							\begin{itemize}
								\item 	$\choice{\pprefix_i. \PP_i}{i \in I} = \choice{\choice{\pprefix_j. \PP_k}{k\in K_j}}{j \in J}$;
								\item	$\Tor{j \in J}{\locali{j}} \subt \local$
								\item	for all $j \in J$ it holds that for all $k\in K_j$:\\
										\indent i) $\sendp{\pprefix_j} = \p$ for some \p; and\\ 
										\indent ii) $\Ga \types \pprefix_j \seq{\PP_k} \as \locali{j}$ 
							\end{itemize}
			
					\item	$\PP = \pvar{X}$ implies
							\begin{itemize}
								\item	$\locald \subt \local$; and
								\item	$\Ga = \Ga, \pvar{X}: \locald$
							\end{itemize}
			
					\item	$\PP = \rec{X}{\PP'}$ implies
							\begin{itemize}
								\item	$\locald \subt \local$; and
								\item	$\Ga, \pvar{X}: \locald \types \PP' \as \locald$.
							\end{itemize}
			
			%
					\item	$\PP = \cond{v}{\PP_1}{\PP_2}$ implies
							\begin{itemize}
								\item	$\localid{1} \subt \local$;
								\item	$\localid{2} \subt \local$;
								\item	$\Gamma \types v \as \bool$; and
								\item	$\Ga\types \PP_1 \as \localid{1}$, $\Ga\types \PP_2 \as \localid{2}$.
							\end{itemize}
				\end{enumerate}
		\item	Assume $\Ga \types \M \as \Delta$.
Then $\M = \Pi_{i \in I} \pa{\pii{i}}{\PP_i}$ implies
						\begin{itemize}
							\item	for all $i \in I$, $\Ga \types \PP_i \as \locali{i}$;
							\item	$\De = \set{\pii{i}: \locali{i}}$; and
							\item 	$\safe{\De}$
						\end{itemize}
	\end{enumerate}
\end{newlemma}

\begin{proof}
	Straightforward by the rules in Figure~\ref{fig:typing}.
	Each case syntactically matches with exactly one typing judgement in Figure~\ref{fig:typing}. The conclusion is then
	implied by using typing rule \TSubs. Specifically:
	\begin{itemize}
		\item	case 1a by rule \TInact;
		\item	case 1b by rule \TSend;
		\item	case 1c by rule \TRcv;
		\item 	case 1d by rule \TSum;
		\item	case 1e by rule \TVar;
		\item	case 1f by rule \TRec;
		\item	case 1g by rule \TCond;
		\item	case 2 by rule \TSessPlus.
	\end{itemize}
\end{proof}


\begin{newlemma}[Substitution]
	\label{app:lem:subst}
	\begin{enumerate}
		\item	If $\Ga \types v \as U$ 
		  and $\Ga, x: U \types \PP \as \local$, then 
                   $\Ga \types \PP\subst{\val}{\xx} \as \local$. 
		\item   If $\Gamma, \pvar{X} : \locald \types \PP \as \local$ 
			and $\Gamma \types Q \as \locald$ 
			then $\Ga \types \PP \subst{Q}{\pvar{X}} \as \local$
	\end{enumerate}
\end{newlemma}

\begin{proof}
	Proof by induction on the syntax of $\PP$. 
	\begin{enumerate}[leftmargin=*]
		\item We consider some characteristic cases. All other 
		cases follow the same logic. 
		\begin{itemize}[leftmargin=*]
			\item \caseof{$P = \soutlbl{\p}{\lab}{\xx}{} \PP'$}

			\pfitem[1]{\Ga, \xx:U \types \soutlbl{\p}{\lab}{\xx}{} \PP' \as \toutlbl{\p}{\lab}{\UType} \locald}
			\pfitem[2]{\toutlbl{\p}{\lab}{\UType} \locald \subt \local}
			\pfitem[3]{\Ga, \xx:U \types \PP' \as \locald}
			\proofstep[4]{Induction hypothesis on \proofref{3}}{\Ga\types \PP'\subst{\val}{\xx} \as \locald}
			\proofstep[5]{\TSend on $v\as U$ and \proofref{4}}{\Ga\types \soutlbl{\p}{\lab}{\val} \PP' \subst{\val}{\xx} \as \toutlbl{\p}{\lab}{\UType} \locald}
			\proofstep[6]{\TSubs on \proofref{2} and \proofref{5} }{\Ga \types \soutlbl{\p}{\lab}{\val} \PP' \subst{\val}{\xx} \as \local}

			\item \caseof{$P = \choice{\pprefix_i
                            \seq{\PP_i}}{i \in I}$} 
By Lemma~\ref{app:lem:typing_inversion}(d), 
we have:\\ 
	$\PP = \choice{\pprefix_i. \PP_i}{i \in I}$ implies
							\begin{itemize}[leftmargin=*]
								\item 	$\choice{\pprefix_i. \PP_i}{i \in I} = \choice{\choice{\pprefix_j. \PP_k}{k\in K_j}}{j \in J}$;
								\item	$\Tor{j \in J}{\locali{j}} \subt \local$
								\item	for all $j \in J$ it holds that for all $k\in K_j$:\\
										\indent i) $\sendp{\pprefix_j} = \p$ for some \p; and\\ 
										\indent ii) $\Ga \types \pprefix_j \seq{\PP_k} \as \locali{j}$ 
																\end{itemize}
                         
					                \pfitem[1]{\Ga,\xx:U \types
         \choice{\choice{\pprefix_j. \PP_k}{k\in K_j}}{j \in J} \as
         \local}
                                                        
					\proofstep[2]{Lemma~\ref{app:lem:typing_inversion} case \TSum}{\Tor{j \in J}{\locali{j}} \subt \local}
					\pfitem[3]{\forall j \in J,~\Ga,\xx:U \types \pprefix_j \seq{\PP_k} \as \locali{j}}
		\proofstep[4]{Induction hypothesis on
                  \proofref{3}}{\Ga \types
                  (\pprefix_j \seq{\PP_k}) \subst{v}{\xx} \as \locali{j}}
					\proofstep[5]{\TSum on
                                          \proofref{4}}{\Ga \types
                                          \choice{(\pprefix_j\seq{\PP_k}) \subst{v}{\xx}}{k \in K_j}
                                          \as \locali{j}}
					\proofstep[6]{\TSubs on
                                          \proofref{5} and
                                          \proofref{2}}{\Ga \types
                                          \choice{\choice{(\pprefix_j. \PP_k)\subst{v}{\xx}}{k\in K_j}}{j \in J} \as \local}

		\end{itemize}
\item Straightforward. 
	\end{enumerate}
\end{proof}


\lemsubjectcongruence*
\label{app:subjectcongruence}

\begin{proof}
	The proof proceeds by a case analysis on the definition of \scong
	and makes use of the Typing Inversion Lemma (Lemma~\ref{app:lem:typing_inversion}). 
	Consider 
	\begin{align}
		\M = \pa{\p}{} \PP \Par \M'
		\label{app:proof:subject_congruence:eq1}
	\end{align}
	with
	$\Ga \types \pa{\p} \PP \Par \M \as \De$ and
	moreover $\PP \scong_\alpha Q$.
	Applying the Inversion Lemma (Lemma~\ref{app:lem:typing_inversion}, case \TSessPlus)
	on result~(\ref{app:proof:subject_congruence:eq1}), 
	there exists \De such that:
	\begin{align}
		&\Ga \types \PP \as \local \subt \locald
		\label{app:proof:subject_congruence:eq3}
		\\
		& \Ga \types \PP_i \as \locali{i} \subt \localid{i}
		\label{app:proof:subject_congruence:eq4}
		\\
                & \De = \set{\pii{i}{:}\localid{i}}\cup \{\p{:}\locald\}
		\label{app:proof:subject_congruence:eq5}
	\end{align}
	By the induction hypothesis,
	we obtain
	\[
		\Ga \types Q \as \local
	\]
	Applying typing rule \TSubs to $\local$, 
	we have:
	\[
		\Ga \types Q \as \locald
	\]
	Applying typing rule \TSessPlus to the last results and
	results~ in 
        \ref{app:proof:subject_congruence:eq4}
        and~\ref{app:proof:subject_congruence:eq5}, we have: 
	\[
		\Ga \types \pa{\p}{} Q \Par \M' \as \De 
	\]
	as required.
	
	The second case to consider is 
	\begin{align}
		\M = \pa{\p}{\inact} \Par \M'
		\label{app:proof:subject_congruence:eq11}
	\end{align}
	with
	$\Ga \types \pa{\p}{\inact} \Par \M' \as \De$.
	Applying the Inversion Lemma (Lemma~\ref{app:lem:typing_inversion}, case \TSessPlus)
	on result~(\ref{app:proof:subject_congruence:eq11}), implies that: 
	\begin{align}
		&\Ga \types \inact \as \local = \tinact \quad
 \text{ (Lemma~\ref{app:lem:typing_inversion}, case \TInact)}
		\label{app:proof:subject_congruence:eq7}
		\\
		& \Ga \types \PP_i \as \locali{i} \subt
\localid{i}
		\label{app:proof:subject_congruence:eq8}
		\\
		& \De = \set{\pii{i}{:}\localid{i}}\cup \{\p{:}\tinact\}
		\label{app:proof:subject_congruence:eq9}
	\end{align}
	From result~(\ref{app:proof:subject_congruence:eq8})
	we apply typing rule \TSubs to obtain 
	\[
		\forall i \in I, \Ga \types \PP_i \as \localid{i}
	\]
	By applying typing rule \TSessPlus to the last result to derive that
	\[
		\Ga \types \M' \as \De
	\]
by the fact 
$\safe{\LL,\AT{\p}{\tinact}}$ iff 
$\safe{\LL}$. 
\end{proof}

We now prove Theorem~\ref{th:subject_reduction}.

\thmsubjectreduction*


We give the proof below. 
\begin{proof}
	The proof is done by induction on the reduction relation.
	We provide with the most interesting cases.
	\begin{itemize}
		\item	Case \RChoice
				\[
					\pa{\p}{(\sinprlbl{\role{q}}{\lab}{x} \seq\PP \OR R_1)}
					\Par 
					\pa{\q}{(\soutprlbl{\role{p}}{\lab}{v} \seq Q \OR R_2)} \Par \M 
					\red 
					\pa{\p}{\PP\subst{v}{x}} \Par \pa{\q}{Q} \Par \M 
				\]
				with
				\begin{align*}
					\types \pa{\p}{(\sinprlbl{\role{q}}{\lab}{x} \seq\PP \OR R_1)}
					\Par
					\pa{\q}{(\soutprlbl{\role{p}}{\lab}{v} \seq Q \OR R_2)} \Par \M  \as \De
					\label{app:app:proof:preservation:choice:eq1}
				\end{align*}
				Applying the Inversion Lemma
                                (Lemma~\ref{app:lem:typing_inversion})
                                (cases \TSend, \TRcv, \TSum and
                                \TSessPlus), we have: 
				\begin{align*}
					& \xx:U\types \PP \as \locali{1}
					\\
					& \types Q \as \locali{2}
                                        					\\
					& \types v \as U
					                                        \\
                                                                                					& \types M \as \set{\pii{i}: \locali{i}}_{i \in I}
					\\
					& \p: \tinplbl{\q}{\lab}{U} \locali{1} \tor \localid{1}, \q: \toutlbl{\p}{\lab}{U} \locali{2} \tor \localid{2}, \set{\pii{i}: \locali{i}}_{i \in I} \subt \De 
					\\
					& \safe{\De}
				\end{align*}
				First we have:
				\[
					\De \red \Ded
				\]
				and
\[
\p: \locali{1}, \q: \locali{2}, \set{\pii{i}: \locali{i}}_{i \in I} \subt \Ded
\]
				with 
\[ 
\safe{\p: \locali{1}, \q: \locali{2}, \set{\pii{i}: \locali{i}}_{i \in I}}
\]
				from Lemma~\ref{lem:subtypingproperties}.
				We apply the Substitution Lemma (Lemma~\ref{app:lem:subst}) to obtain:
				$\types \PP\subst{v}{x} \as \locali{1}$. The last result and the fact that $\safe{\p: \locali{1}, \q: \locali{2}, \set{\pii{i}: \locali{i}}_{i \in I}}$ enable typing
				rule \TSessPlus to obtain:
				\[
					\types \pa{\p}{\PP\subst{v}{x}} \Par \pa{\q}{Q} \Par \M \as \p: \locali{1}, \q: \locali{2}, \set{\pii{i}: \locali{i}}_{i \in I}
				\]
				as required.

		\item	Case \RCong: by Lemma~\ref{app:lem:subject_congruence}.
\item Other rules are identical with \cite{DBLP:journals/jlp/GhilezanJPSY19}.
\end{itemize}
\end{proof}

\onecolumn

\section{Encodings and Distributability}
\label{app:criteria}

We introduce a few concepts that we use in the following.

A term is denoted as \emph{output}, if its outermost operator is an output prefix.
Similarly, an \emph{input} is a term, whose outermost operator in an input prefix.
Terms, \ie networks or processes, that appear as subterm of a term with some action prefix (output or input) or conditional as outermost operator are called \emph{guarded}, because the guarded subterm can not be executed before the guarding action or conditional has been reduced.
Terms are unguarded if they are not a subterm of an action prefix or conditional.

To reason about environments of terms, we use functions on process terms called contexts. More precisely, a \emph{context} $ \Context{}{}{\hole_1, \ldots, \hole_n} $ with $ n $ holes is a function from $ n $ terms into one term, \ie given $ M_1, \ldots, M_n $, the term $ \Context{}{}{M_1, \ldots, M_n} $ is the result of inserting $ M_1, \ldots, M_n $ in the corresponding order into the $ n $ holes of~$ \context $.
We consider contexts from processes into a process and contexts from processes and sessions into a session.

Two terms of a calculus are usually compared using some kind of a behavioural simulation relation.
The most commonly known behavioural simulation relation is bisimulation.
A relation $ \mathcal{R} $ is a bisimulation if any two related terms mutually simulate their respective sequences of steps, such that the derivatives are again related.

\begin{definition}[Bisimulation]
	\label{def:bisimulation}
	$ \mathcal{R} $ is a (weak reduction, barbed) bisimulation if for each $ {\left( M_1, M_2 \right)} \in \mathcal{R} $:
	\begin{itemize}
		\item $ M_1 \red^\ast M_1' $ implies that there exists some $ M_2' $ such that $ M_2 \red^\ast M_2' $ and $ {\left( M_1', M_2' \right)} \in \mathcal{R} $
		\item $ M_2 \red^\ast M_2' $ implies that there exists some $ M_1' $ such that $ M_1 \red^\ast M_1' $ and $ {\left( M_1', M_2' \right)} \in \mathcal{R} $
	\end{itemize}
	Two terms are bisimilar if there exists a bisimulation that relates them.
	Let $ \approx $ denote bisimilarity.
\end{definition}

\subsection{Encodability Criteria}

Assume two process calculi, denoted as \emph{source} ($ \indexSource $) and \emph{target} calculus ($ \indexTarget $).
An \emph{encoding} from the source into the target calculus is a function $ \ArbitraryEncoding{\cdot} $ from the terms of the source calculus into the terms of the target calculus.
Encodings often translate single source term steps into a sequence or pomset of target term steps.
We call such a sequence or pomset an \emph{emulation} of the corresponding source term step.

To analyse the quality of encodings and to rule out trivial or meaningless encodings, they are augmented with a set of quality criteria.
One such set of criteria that is well suited for separation as well as encodability results between process calculi was proposed in \cite{DBLP:journals/iandc/Gorla10}.
From this set we inherit the criteria:
\begin{description}
	\item[Compositionality:] For every operator $ \mathbf{op} $ with arity $ n $ of the source calculus and for every subset of participants $ \mathbb{P} $, there exists a context $ \Context{\mathbb{P}}{\mathbf{op}}{\hole_1, \ldots, \hole_n} $ such that, for all source terms $ M_1, \ldots, M_n $ with $ \parts{M_1} \cup \ldots \cup \parts{M_n} = {\mathbb{P}} $, $ \ArbitraryEncoding{{\mathbf{op}}\left( M_1, \ldots, M_n \right)} = \Context{\mathbb{P}}{\mathbf{op}}{\ArbitraryEncoding{M_1}, \ldots, \ArbitraryEncoding{M_n}} $.
	\item[Name Invariance:] For every source term $ M $ and every substitution $ \sigma $ on participants and variables such that $ M\sigma $ is defined, it holds that
		\begin{align*}
			\ArbitraryEncoding{M\sigma}
			\begin{cases}
				= \ArbitraryEncoding{M}\sigma' & \text{if } \sigma \text{ is injective}\\
				\asymp \ArbitraryEncoding{M}\sigma' & \text{otherwise}
			\end{cases}
		\end{align*}
		where $ \sigma' $ is such that $ \varphi(\sigma(a)) = \sigma'{\left( \varphi(a) \right)} $ for all participants or variables $ a $.
	\item[Operational Correspondence:] Every computation of a source term $ M $ can be emulated by its translation, \ie $ M \red^\ast M' $ implies $ \ArbitraryEncoding{M} \red^\ast \asymp \ArbitraryEncoding{M'} $ (completeness), and every computation of a target term corresponds to some computation of the corresponding source term, \ie $ \ArbitraryEncoding{M} \red^\ast N $ implies $ M \red^\ast M' $ and $ N \red^\ast \asymp \ArbitraryEncoding{M'} $ (soundness).
	\item[Divergence Reflection:] For every source term $ M $, $ \ArbitraryEncoding{M} \red^\omega $ implies $ M \red^\omega $.
	\item[Success Sensitiveness:] For every source term $ M $, $ M \reachSuccess $ iff $ \ArbitraryEncoding{M} \reachSuccess $.
\end{description}

Compositionality requires that every operator is translated by a context that is in \citep{DBLP:journals/iandc/Gorla10} allowed to be parametrised by the set of free names.
Name invariance ensures that encodings are independent of specific names in the source.
Therefore, it considers the translations of names in the source.
In \mcmpst (and its variants) names are either names of participants or names of variables.
For simplicity, let in compositionality names be the set of participants only.
This decision is not crucial, \ie all proofs remain valid if compositionality also considers free variables as long as we can clearly distinguish between the sets of participants and variables.
We naturally  extend the function $ \parts{\cdot} $ to terms, \ie $ \parts{M} $ denotes the set of participants in $ M $.
Let $ \varphi $ be a \emph{renaming policy}, \ie a mapping from a (source term) name to a vector of (target term) names that can be used by encodings to split names and to reserve special names, such that no two different names are translated into overlapping vectors of names and reserved names are not confused with translated source term names.
By translating a single participant to a vector of participants, an encoding may \eg translate an interaction between two participants by a protocol that involves more than two participants.

A behavioural equivalence $ \asymp $ on the target calculus is assumed for the definition of operational correspondence and name invariance.
Its purpose is to describe the abstract behaviour of a target process, where abstract refers to the behaviour of the source term.
By \cite{DBLP:journals/iandc/Gorla10} the equivalence $ \asymp $ is often defined in the form of a barbed equivalence (as described \eg in \cite{milnerSangiorgi92}) or can be derived directly from the reduction semantics and is often a congruence, at least with respect to parallel composition.
We require only that $ \asymp $ is a weak reduction bisimulation, \ie for all target terms $ M_1, M_2 $ such that $ M_1 \asymp M_2 $, for all $ M_1 \red^\ast M_1' $ there exists a $ M_2' $ such that $ M_2 \red^\ast M_2' $ and $ M_1' \asymp M_2' $.
In the concrete encodings presented below, $ \asymp $ is instantiated with the bisimulation $ \approx $ of the respective target calculus, where Definition~\ref{def:bisimulation} is extended by the condition:
\begin{itemize}
	\item $ M_1\reachSuccess $ iff $ M_2\reachSuccess $
\end{itemize}

Operational correspondence ensures that the encoding preserves and reflects the behaviour of the considered calculi.
Since it does consider not-labelled but only reduction steps, it has to be combined with the testing of some barbs.
Following \cite{DBLP:journals/iandc/Gorla10}, we choose the weakest form of barb that can be used for this purpose, namely a test for success $ \success $.
Accordingly, we assume that $ \success $ is a constant process without free variables in \mcmpst and all calculi considered in this paper.
Let $ \success $ be typed in the same way as $ \inact $, \ie by the rules
\begin{align*}
	\Ga \types \success \as \tinact
	\hspace{3em}
	\dfrac{\UnT{\Gamma}}{\Gamma \vdash \success}
\end{align*}
in all variants of \mcmpst and all variants of \cnvms (defined in \S~\ref{app:mixedSessions}), respectively.
We choose may-testing to instantiate the test for success in success sensitiveness, \ie $ P\hasSuccess $ if $ P $ contains a top-level unguarded occurrence of $ \success $, $ M\hasSuccess $ if the session $ M $ contains a process $ P $ such that $ P\hasSuccess $, and $ M\reachSuccess $ if there is some $ M' $ such that $ M \red^\ast M' $ and $ M'\hasSuccess $.
As usual $ M\not\reachSuccess $ denotes $ \neg \left( M\reachSuccess \right) $.
However, as we claim, this choice is not crucial.

Note that success sensitiveness only links the behaviours of source terms and their literal translations, but not of their derivatives. To do so, Gorla relates success sensitiveness and operational correspondence by requiring that the equivalence on the target calculus respects success, \ie $ M_1 \reachSuccess $ and $ M_2 \not\reachSuccess $ implies $ M_1 \not\asymp M_2 $.

To reason about leader election in Section~\ref{app:separationLeaderElection}, we introduce the criterion barb sensitiveness that is slightly stronger than success sensitiveness.
\begin{description}
	\item[Barb Sensitiveness] For every source term $ M $ and every barb $ y $, $ M\WeakBarb{y} $ iff $ \ArbitraryEncoding{M}\WeakBarb{y} $.
\end{description}
A participant $ \pa{\p}{P} $ has an output barb $ \actoutlbl{\p}{\q}{\lab}{v} $, denoted as $ P\HasBarb{\actoutlbl{\p}{\q}{\lab}{v}} $, if it has an unguarded output $ \soutprlbl{\q}{\lab}{v}.P' $.
Similarly, $ M\HasBarb{\actoutlbl{\p}{\q}{\lab}{v}} $ if $ M \equiv \pa{\p}{\soutprlbl{\q}{\lab}{v}.P'} \Par M' $.
A participant $ \pa{\p}{P} $ has an input barb $ \actinp{\p}{\q}{\lab} $, denoted as $ P\HasBarb{\actinp{\p}{\q}{\lab}} $, if it has an unguarded input $ \sinprlbl{\q}{\lab}{x}.P' $.
Similarly, $ M\HasBarb{\actinp{\p}{\q}{\lab}} $ if $ M \equiv \pa{\p}{\sinprlbl{\q}{\lab}{x}.P'} \Par M' $.
Moreover, $ M\WeakBarb{y} $ if there is some $ M' $ such that $ M \red^\ast M' $ and $ M'\HasBarb{y} $.
Similarly to success sensitiveness, we require that $ \asymp $ respects barbs if the encoding is barb sensitive, \ie $ M_1\WeakBarb{y} $ and $ M_2\not\WeakBarb{y} $ imply $ M_1 \not\asymp M_2 $.
In this case, we additionally extend Definition~\ref{def:bisimulation} by the condition:
\begin{itemize}
	\item $ M_1\WeakBarb{y} $ iff $ M_2\WeakBarb{y} $ for all barbs $ y $
\end{itemize}

Since we consider typed-calculi here, one might assume that we need an additional criterion on types.
However, it suffices instead to consider only the respective typed fragments of the terms in the source and target calculus, \ie we consider only typed and deadlock-free terms in \mcmpst, \msmpst, \scmpst, \dmpst, \smpst, \mpst, \mcbs, \scbs, and \bs and only typed terms in \cnvms, \cnv, \lcnvms, and \lcnv.
We do not forbid to use type informations in encodings as done in the encoding from \cnvms into \cnv presented in \citep{CASAL202223}.

\subsection{Distributability and Distributability Preservation}

Palamidessi in \citep{Palamidessi03} requires as additional criterion that the parallel operator is translated homomorphically.
As explained in \citep{Palamidessi03} this criterion was used to ensure that encodings preserve the degree of distribution in terms.
Indeed, \citep{PetersN12} presents an encoding of the $ \pi $-calculus with mixed choice into the asynchronous $ \pi $-calculus without choice that respects all of the above criteria.
Requiring that the degree of distribution is preserved is essential for the separation result in \citep{Palamidessi03} or more general to reason about the expressive power of mixed choice.
Unfortunately, as explained in \citep{PetersN12, peters12} the homomorphic translation of the parallel operator is rather strict and rules out encodings that intuitively do preserve the degree of distribution.
Because of that, \citep{PetersN12, peters12, DBLP:conf/esop/PetersNG13} propose an alternative criterion for the preservation of the degree of distribution that we will use here to strengthen our separation results.

Intuitively, a distribution of a process means the extraction (or separation) of its (sequential) components and their association to different locations.
In order to formalise the identification of sequential components and following \citep{peters12, DBLP:conf/esop/PetersNG13}, we assume for each process calculus a so-called \emph{labelling} on the \emph{capabilities} of processes.
As capabilities we denote the parts of a term that are removed in reduction steps.
The capabilities of the \mcmpst and \scmpst are the action prefixes, where the capability of a choice is the conjunction of the prefixes of all its branches---considered as single capability, and the conditionals.

The labelling has to ensure that
\begin{enumerate}
	\item each capability has a label,
	\item no label occurs more than once in a labelled term,
	\item a label disappears only when the corresponding capability is reduced in a reduction step, and (4), once it has disappeared, it will not appear in the execution any more.
\end{enumerate}
The last three conditions are called unicity, disappearance, and persistence in \citep{cacciagranoCorradiniPalamidessi09} which defines a labelling method to establish such a labelling for processes of the $ \pi $-calculus.
Note that such a labelling can be derived from the syntax tree of processes possibly augmented with some informations about the history of the process, as it is done in \citep{cacciagranoCorradiniPalamidessi09}.
However, we assume that, once the labelling of a term is fixed, the labels are preserved by the rules of structural congruence as well as by the reduction semantics.
Because of recursion, new subterms with fresh labels for their capabilities may arise from applications of structural congruence.
Since we need the labels only to distinguish syntactically similar components of a term, and to track them alongside reductions, we do not restrict the domain of the labels nor the method used to obtain them as long as the resulting labelling satisfies the above properties for all terms and all their derivatives in the respective calculus.
In order not to clutter the development with the details of labelling, we prefer to argue at the corresponding informal level.
More precisely, we assume that \textbf{all} processes in the following are implicitly labelled.
Remember that we need these labels only to distinguish between syntactical equivalent capabilities.

Since all calculi considered in this paper are based on the $ \pi $-calculus, we can rely on the intuition that the parallel operator splits locations.
Accordingly, a system $ M $ is distributable into $ M_1, \ldots, M_n $ if and only if we have $ M \equiv M_1 \Par \ldots \Par M_n $, where $ \equiv $ does not unfold any recursions.
This also reflects our intuition that a session $ \pa{\p}{\PP} $ implements a single participant, \ie a single location.
Since we require that structural congruence preserves the labels of capabilities and forbid to unfold recursion, $ M $ and $ M_1, \ldots, M_n $ contain the same capabilities and there are no two occurrences of the same capability in $ M_1, \ldots, M_n $, \ie no label occurs twice.
If $ M $ is distributable into $ M_1, \ldots, M_n $ then we also say that $ M_1, \ldots, M_n $ are distributable within $ M $.

Preservation of distributability means that the target term is at least as distributable as the source term.

\begin{definition}[Preservation of Distributability, \citep{DBLP:conf/esop/PetersNG13}]
	\label{def:distributabilityPreservation}
	An encoding $ \arbitraryEncoding $ \emph{preserves distributability} if for every $ M $ in the source calculus and for all terms $ M_1, \ldots, M_n $ that are distributable within $ M $ there are some $ N_1, \ldots, N_n $ that are distributable within $ \ArbitraryEncoding{M} $ such that $ N_i \asymp \ArbitraryEncoding{M_i} $ for all $ 1 \leq i \leq n $.
\end{definition}

In essence, this requirement is a distributability-enhanced adaptation of operational completeness.
It respects both the intuition on distribution as separation on different locations---an encoded source term is at least as distributable as the source term itself---as well as the intuition on distribution as independence of processes and their executions---implemented by $ N_i \asymp \ArbitraryEncoding{M_i} $.

The preservation of distributability completes our set of criteria for encodings.

\begin{definition}[Good Encoding]
	\label{def:goodEncoding}
	We consider an encoding $ \arbitraryEncoding $ to be \emph{good} if it
	(1) is compositional,
	(2) name invariant,
	(3) satisfies operational correspondence,
	(4) reflects divergence,
	(5) is success sensitive, and
	(6) preserves distributability.
	Moreover we require that the equivalence $ \asymp $ is a success respecting (weak) reduction bisimulation.
\end{definition}

Note that success sensitiveness is slightly weaker than barb sensitiveness.
Hence, it results in stronger separation results.
Accordingly, we use barb sensitiveness only for one separation result (to separate \mcmpst from \scmpst via leader election in Section~\ref{app:separationLeaderElection}) and present for this result an alternative proof (in Section~\ref{app:separationStar}) with success sensitiveness instead of barb sensitiveness.
The hierarchy in Figure~\ref{fig:hierarchy} is based on the above notion of a good encoding.

We inherit some of the machinery introduced in \citep{DBLP:conf/esop/PetersNG13} to work with distributability.
We write $ M \reachSuccessFin $ if $ M $ reaches success in all finite maximal executions, \ie for convergent $ M $ this means that $ M $ has to eventually reach success whatever happens.
A success respecting reduction bisimulation also respects the ability to reach success in all finite maximal executions.

\begin{lemma}[\citep{DBLP:conf/esop/PetersNG13}]
	For all success respecting reduction bisimulations $ \approx $ and all convergent terms $ M_1, M_2 $ such that $ M_1 \approx M_2 $, it holds that $ M_1 \reachSuccessFin $ iff $ M_2 \reachSuccessFin $.
\end{lemma}

Then success sensitiveness preserves the ability to reach success in all finite maximal executions.

\begin{lemma}[\citep{DBLP:conf/esop/PetersNG13}]
	For all operationally sound, divergence reflecting, and success-sensitive encodings $ \arbitraryEncoding $ and for all convergent source terms $ M $, if $ M \reachSuccessFin $ then $ \ArbitraryEncoding{M} \reachSuccessFin $.
\end{lemma}

As explained in \citep{DBLP:conf/esop/PetersNG13} there are some calculi such as the join-calculus, for that the parallel operator does not sufficiently reflects distribution.
Because of that, \citep{DBLP:conf/esop/PetersNG13} distinguishes between parallel and distributable processes or steps.
Since all considered calculi in this paper are based on the $ \pi $-calculus, in that the parallel operator sufficiently reflects distribution, we do not need to distinguish between parallel and distributable processes or steps.
If a single process can perform two different steps, then we call these steps alternative to each other, where we identify steps modulo structural congruence.
Since every step in the three considered calculus reduces exactly one or exactly two capabilities, two steps of one process are different if they do not reduce the same set of capabilities.
Two alternative steps are in \emph{conflict}, if performing one step disables the other step, \ie if both reduce the same capability.
Otherwise they are \emph{distributable}.
Remember that two capabilities are the same only if they have the same label.
Then, two steps in the considered calculi are in conflict if they reduce the same choice.
Note that reducing the same choice does not necessarily mean to reduce the same summand in this choice.
Distributable steps are independent, \ie confluent.

We lift the definition of conflict and distributable steps to executions, \ie sequences of steps.

\begin{definition}[Distributable Executions]
	\label{def:distributableSequences}
	Let $ M $ be a term, and let $ A $ and $ B $ denote two executions of $ M $.
	$ A $ and $ B $ are in \emph{conflict}, if a step of $ A $ and a step of $ B $ are in conflict, else $ A $ and $ B $ are \emph{distributable}.
\end{definition}

As shown in \citep{peters12}, two executions of a term $ M $ are distributable iff $ M $ is distributable into two subterms such that each performs one of these executions.

\begin{lemma}[Distributable Executions, \citep{DBLP:conf/esop/PetersNG13}]
	\label{lem:distributabilityReductionsVsProcesses}
	Let $ M $ be a term and $ N_1, \ldots, N_n $ a set of executions of $ M $.
	The executions $ N_1, \ldots, N_n $ are pairwise distributable within $ M $ iff $ M $ is distributable into $ M_1, \ldots, M_n $ such that, for all $ 1 \leq i \leq n $, $ N_i $ is an execution of $ M_i $, \ie during $ N_i $ only capabilities of $ M_i $ are reduced or removed.
\end{lemma}

Because of that, an operationally complete encoding is distributability-preserving only if it preserves the distributability of sequences of source term steps.

\begin{lemma}[Distributability-Preservation, \citep{DBLP:conf/esop/PetersNG13}]
	\label{lem:distributabilityPreservation}
	An operationally complete encoding $ \arbitraryEncoding $ that preserves distributability also preserves distributability of executions, \ie for all source terms $ M $ and all sets of pairwise distributable executions of $ M $, there exists an emulation of each execution in this set such that all these emulations are pairwise distributable in $ \ArbitraryEncoding{M} $.
\end{lemma}

We also lift the definition of conflict and distributable steps to steps on types.
\section{Mixed Sessions}
\label{app:mixedSessions}

We briefly recap the main concepts of \cnvms as introduced in \citep{CASAL202223} and its sub-calculi considered in this paper.
Readers familiar with \cnvms can safely skip this section.

In \citep{CASAL202223} \emph{expressions}, ranged over by $ e, e', \ldots $, are constructed from variables, unit, and standard boolean operators.
Instead we use variables, numbers, and booleans but no boolean operators as defined in \S~\ref{sec:calculus}.
As we claim, this decision does not influence the presented results but slightly simplifies their presentation.

\emph{Mixed sessions} are a variant of binary session types introduced by Casal, Mordido, and Vasconcelos in \citep{CASAL202223} with a choice-construct that combines prefixes for sending and receiving.
We denote this calculus as \cnvms.

A central idea of \cnvms (and the calculus \cnv it is based on) is that channels are separated in two \emph{channel endpoints} and that interaction is by two processes acting on the respective different ends of such a channel.

\begin{definition}[Mixed Sessions]
	The set of untyped processes $ \procCMVmixUntyped $ of \cnvms is given as:
	\begin{align*}
		P & \bnfis \ChoiceCMVmix{q}{y}{\sum_{i \in I}M_i} \bnfbar P \mid P \bnfbar \ResCMVmix{y}{z}{P} \bnfbar \ConditionalCMVmix{v}{P}{P} \bnfbar \inactCMVmix & \text{Processes}\\
		M & \bnfis \BranchCMVmix{\lab}{*}{v}{P} & \text{Branches}\\
		* & \bnfis ! \bnfbar ? & \text{Polarities}\\
		q & \bnfis \linCMVmix \bnfbar \unCMVmix & \text{Qualifiers}
	\end{align*}
\end{definition}

A choice $ \ChoiceCMVmix{q}{y}{\sum_{i \in I}M_i} $ is declared as either linear ($ \linCMVmix $) or unrestricted ($ \unCMVmix $) by the qualifier $ q $.
It proceeds on a single channel endpoint $ y $.
For every $ i $ in the index set $ I $ it offers a branch $ M_i $.
A branch $ \BranchCMVmix{\lab}{*}{v}{P} $ specifies a label $ \lab $, a polarity $ * $ ($ ! $ for sending or $ ? $ for receiving), a name $ v $ (a value in output actions or a variable for input actions), and a continuation $ P $.
We abbreviate the empty sum, \ie $ \ChoiceCMVmix{q}{y}{\sum_{i \in I}M_i} $ for $ I = \emptyset $, by $ \inactCMVmix $.
Moreover, we often write $ q \; y \left( M_1 + \ldots + M_n \right) $ for a choice $ \ChoiceCMVmix{q}{y}{\sum_{i \in \Set{ 1, \ldots, n }}M_i} $.
Restriction $ \ResCMVmix{y}{z}{P} $ binds the two channel endpoints $ y $ and $ z $ of a single channel to $ P $.
The remaining operators introduce parallel composition $ P \mid P $, conditionals $ \ConditionalCMVmix{v}{P}{P} $, and inaction $ \inactCMVmix $.
We sometimes abbreviate $ P_1 \mid \ldots \mid P_n $ by $ \prod_{i \in \Set{1, \ldots, n}} P_i $.

The variable $ x $ is bound in $ P $ by input branches $ \BranchCMVmix{\lab}{?}{x}{P} $ and the two endpoints of a channel $ x, y $ are bound in $ P $ by restriction $ \ResCMVmix{x}{y}{P} $.
All other variables or channel-endpoints are free.
Let $ \fv{P} $ denote the set of free variables and free channel-endpoints in $ P $.

\begin{figure*}[t]
\centering 
\[
\begin{array}{c}
		\ruleRIfTCMVmix \quad \ConditionalCMVmix{\true}{P}{Q} \stepCMVmix P
		\hspace{2em}
		\ruleRIfFCMVmix \quad \ConditionalCMVmix{\false}{P}{Q} \stepCMVmix Q
		\vspace{0.5em}\\
		\ruleRLinLinCMVmix \; \begin{array}{l}
				\ResCMVmix{y}{z}{{\left( \ChoiceCMVmix{\linCMVmix}{y}{{\left( \OutCMVmix{\lab}{v}{P} + M \right)}} \mid \ChoiceCMVmix{\linCMVmix}{z}{{\left( \InpCMVmix{\lab}{x}{Q} + N \right)}} \mid R \right)}} \stepCMVmix 
				\ResCMVmix{y}{z}{{\left( P \mid Q\Set{\Subst{v}{x}} \mid R \right)}}
			\end{array}
		\vspace{0.5em}\\
		\ruleRLinUnCMVmix \; \begin{array}{l}
				\ResCMVmix{y}{z}{{\left( \ChoiceCMVmix{\linCMVmix}{y}{{\left( \OutCMVmix{\lab}{v}{P} + M \right)}} \mid \ChoiceCMVmix{\unCMVmix}{z}{{\left( \InpCMVmix{\lab}{x}{Q} + N \right)}} \mid R \right)}} \stepCMVmix
				\ResCMVmix{y}{z}{{\left( P \mid Q\Set{\Subst{v}{x}} \mid \ChoiceCMVmix{\unCMVmix}{z}{{\left( \InpCMVmix{\lab}{x}{Q} + N \right)}} \mid R \right)}}
			\end{array}
		\vspace{0.5em}\\
		\ruleRUnLinCMVmix \; \begin{array}{l}
				\ResCMVmix{y}{z}{{\left( \ChoiceCMVmix{\unCMVmix}{y}{{\left( \OutCMVmix{\lab}{v}{P} + M \right)}} \mid \ChoiceCMVmix{\linCMVmix}{z}{{\left( \InpCMVmix{\lab}{x}{Q} + N \right)}} \mid R \right)}} \stepCMVmix
				\ResCMVmix{y}{z}{{\left( P \mid Q\Set{\Subst{v}{x}} \mid \ChoiceCMVmix{\unCMVmix}{y}{{\left( \OutCMVmix{\lab}{v}{P} + M \right)}} \mid R \right)}}
			\end{array}
		\vspace{0.5em}\\
		\ruleRUnUnCMVmix \; \begin{array}{l}
				\ResCMVmix{y}{z}{{\left( \ChoiceCMVmix{\unCMVmix}{y}{{\left( \OutCMVmix{\lab}{v}{P} + M \right)}} \mid \ChoiceCMVmix{\unCMVmix}{z}{{\left( \InpCMVmix{\lab}{x}{Q} + N \right)}} \mid R \right)}} \stepCMVmix
				\ResCMVmix{y}{z}{{\left( P \mid Q\Set{\Subst{v}{x}} \mid \ChoiceCMVmix{\unCMVmix}{y}{{\left( \OutCMVmix{\lab}{v}{P} + M \right)}} \mid \ChoiceCMVmix{\unCMVmix}{z}{{\left( \InpCMVmix{\lab}{x}{Q} + N \right)}} \mid R \right)}}
			\end{array}
		\vspace{0.5em}\\
		\ruleRParCMVmix \; \dfrac{P \stepCMVmix P'}{P \mid Q \stepCMVmix P' \mid Q}
		\hspace{2em}
		\ruleRResCMVmix \; \dfrac{P \stepCMVmix P'}{\ResCMVmix{y}{z}{P} \stepCMVmix \ResCMVmix{y}{z}{P'}}
		\hspace{2em}
		\ruleRStructCMVmix \; \dfrac{P \equiv Q \quad Q \stepCMVmix Q' \quad Q' \equiv P'}{P \stepCMVmix P'}
	\end{array}
\]
	where structural congruence $ \scCMVmix $ is the least congruence that contains $ \alpha $-conversion and satisfies:
\[
\begin{array}{c}
		P \mid Q \scCMVmix Q \mid P
		\hspace{1.5em}
		{\left( P \mid Q \right)} \mid R \scCMVmix P \mid {\left( Q \mid R \right)}
		\hspace{1.5em}
		P \mid \inactCMVmix \scCMVmix P
		\hspace{1.5em}
		\ResCMVmix{y}{z}{\inactCMVmix} \scCMVmix \inactCMVmix
		\hspace{1.5em}
		\ResCMVmix{y}{z}{P} \scCMVmix \ResCMVmix{z}{y}{P}
		\vspace{0.5em}\\
		P \mid \ResCMVmix{y}{z}{Q} \scCMVmix \ResCMVmix{y}{z}{{\left( P \mid Q \right)}} \quad \text{if } y, z \notin \fv{P}
		\hspace{2em}
		\ResCMVmix{w}{x}{\ResCMVmix{y}{z}{P}} \scCMVmix \ResCMVmix{y}{z}{\ResCMVmix{w}{x}{P}}
	\end{array}
\]
	\caption{Reduction Rules ($ \stepCMVmix $) of \cnvms.}
	\label{fig:semanticsCMVmix}
\end{figure*}

The semantics of \cnvms is given by the rules in Figure~\ref{fig:semanticsCMVmix}.
The commutativity and associativity of summands within choices follows from choices being defined via a set of summands.

To obtain from $ \procCMVmixUntyped $ the set $ \procCMVmix $ of typed processes of \cnvms, a type system is introduced in \citep{CASAL202223}.
The syntax of types is given as:
\begin{align*}
	T & \bnfis \ChoiceTCMVmix{q}{\#}{\Set{B_i}_{i \in I}} \bnfbar \finCMVmix \bnfbar \nat \bnfbar \bool \bnfbar \RecCMVmix{t}{T} \bnfbar t & \text{Types}\\
	B & \bnfis \BranchCMVmix{\lab}{*}{T}{T} & \text{Branches}\\
	\# & \bnfis \oplus \bnfbar \& & \text{Views}\\
	\Gamma & \bnfis \cdot \bnfbar \Gamma, \At{x}{T} & \text{Contexts}
\end{align*}

A type of the form $ \ChoiceTCMVmix{q}{\#}{\Set{B_i}_{i \in I}} $ denotes a channel endpoint, where the view $ \# $ is either $ \oplus $ for internal choice or $ \& $ for external choice.
We often call it a choice type.
In a branch $ \BranchCMVmix{\lab}{*}{T_1}{T_2} $ the type $ T_1 $ specifies the communicated value whereas $ T_2 $ is the type of the continuation.
Besides channel endpoints there are types for inaction, the base types for unit and boolean, and types for recursion.

Following \citep{CASAL202223}, we assume that the index sets $ I $ in types are not empty, that the label-polarity-pairs $ \lab * $ are pairwise distinct in the branches of a choice type, and recursive types are contractive, \ie contain no subterm of the form $ \RecCMVmix{t_1}{\ldots\RecCMVmix{t_n}{t_1}} $ with $ n \geq 1 $.
A type variable $ t $ is bound in $ T $ by $ \RecCMVmix{t}{T} $.
All other type variables are free.

Type equivalence $ \simeq $ is coinductively defined by the rules:
\begin{displaymath}\begin{array}{c}
	\finCMVmix \simeq \finCMVmix
	\hspace{2em}
	\nat \simeq \nat
	\hspace{2em}
	\bool \simeq \bool
	\vspace{0.5em}\\
	\dfrac{T_i \simeq T_i' \quad U_i \simeq U_i' \quad {\left( \forall i \in I \right)}}{\ChoiceTCMVmix{q}{\#}{\Set{\BranchCMVmix{\lab}{*_i}{T_i}{U_i}}_{i \in I}} \simeq \ChoiceTCMVmix{q}{\#}{\Set{\BranchCMVmix{\lab}{*_i}{T_i'}{U_i'}}_{i \in I}}}
	\hspace{2em}
	\dfrac{T\Set{\Subst{\RecCMVmix{t}{T}}{t}} \simeq U}{\RecCMVmix{t}{T} \simeq U}
	\hspace{2em}
	\dfrac{T \simeq U\Set{\Subst{\RecCMVmix{t}{U}}{t}}}{T \simeq \RecCMVmix{t}{U}}
\end{array}\end{displaymath}

Two types are dual to each other if they describe well-coordinated behaviour of the two endpoints of a channel.
In particular, input is dual to output and internal choice is dual to external choice.
The operator $ \Dual{\cdot}{\cdot} $ for type duality is defined coinductively by the rules:
\begin{displaymath}\begin{array}{c}
	\Dual{!}{?}
	\hspace{2em}
	\Dual{?}{!}
	\hspace{2em}
	\Dual{{\oplus}}{\&}
	\hspace{2em}
	\Dual{\&}{\oplus}
	\hspace{2em}
	\Dual{\finCMVmix}{\finCMVmix}
	\vspace{0.5em}\\
	\dfrac{\Dual{\#}{\flat} \quad \Dual{{*_i}}{\bullet_i} \quad T_i \simeq T_i' \quad \Dual{U_i}{U_i'} \quad {\left( \forall i \in I \right)}}{\Dual{\ChoiceTCMVmix{q}{\#}{\Set{\BranchCMVmix{\lab}{*_i}{T_i}{U_i}}_{i \in I}}}{\ChoiceTCMVmix{q}{\flat}{\Set{\BranchCMVmix{\lab}{\bullet_i}{T_i'}{U_i'}}_{i \in I}}}}
	\hspace{2em}
	\dfrac{\Dual{T\Set{\Subst{\RecCMVmix{t}{T}}{t}}}{U}}{\Dual{\RecCMVmix{t}{T}}{U}}
	\hspace{2em}
	\dfrac{\Dual{T}{U\Set{\Subst{\RecCMVmix{t}{U}}{t}}}}{\Dual{T}{\RecCMVmix{t}{U}}}
\end{array}\end{displaymath}

Subtyping introduces more flexibility to the usage of types.
In external choices subtyping allows additional branches in the supertype; for internal choice we have the opposite.
The operator $ \Subtype{T_1}{T_2} $ ($ T_1 $ is a subtype of $ T_2 $) is defined coinductively by the rules:
\begin{displaymath}\begin{array}{c}
	\dfrac{\Subtype{T_2}{T_1} \quad \Subtype{U_1}{U_2}}{\Subtype{\OutCMVmix{\lab}{T_1}{U_1}}{\OutCMVmix{\lab}{T_2}{U_2}}}
	\hspace{2em}
	\dfrac{\Subtype{T_1}{T_2} \quad \Subtype{U_1}{U_2}}{\Subtype{\InpCMVmix{\lab}{T_1}{U_1}}{\InpCMVmix{\lab}{T_2}{U_2}}}
	\vspace{0.5em}\\
	\Subtype{\finCMVmix}{\finCMVmix}
	\hspace{2em}
	\Subtype{\nat}{\nat}
	\hspace{2em}
	\Subtype{\bool}{\bool}
	\vspace{0.5em}\\
	\dfrac{J \subseteq I \quad \Subtype{B_j}{C_j} \quad {\left( \forall j \in J \right)}}{\Subtype{\IntCMVmix{q}{\Set{B_i}_{i \in I}}}{\IntCMVmix{q}{\Set{C_j}_{j \in J}}}}
	\hspace{2em}
	\dfrac{I \subseteq J \quad \Subtype{B_i}{C_i} \quad {\left( \forall i \in I \right)}}{\Subtype{\ExtCMVmix{q}{\Set{B_i}_{i \in I}}}{\ExtCMVmix{q}{\Set{C_j}_{j \in J}}}}
	\vspace{0.5em}\\
	\dfrac{\Subtype{T\Set{\Subst{\RecCMVmix{t}{T}}{t}}}{U}}{\Subtype{\RecCMVmix{t}{T}}{U}}
	\hspace{2em}
	\dfrac{\Subtype{T}{U\Set{\Subst{\RecCMVmix{t}{U}}{t}}}}{\Subtype{T}{\RecCMVmix{t}{U}}}
\end{array}\end{displaymath}

The predicate $ \UnT{\cdot} $ that is defined by the rules
\begin{displaymath}\begin{array}{c}
	\UnT{\finCMVmix}
	\hspace{2em}
	\UnT{\nat}
	\hspace{2em}
	\UnT{\bool}
	\hspace{2em}
	\UnT{\ChoiceTCMVmix{\unCMVmix}{\#}{\Set{B_i}_{i \in I}}}
	\hspace{2em}
	\dfrac{\UnT{T}}{\UnT{\RecCMVmix{t}{T}}}
\end{array}\end{displaymath}
identifies unrestricted types, \ie types without an unguarded linear choice type.

Typing contexts $ \Gamma $ collect assignments $ \At{x}{T} $ of names to their types.
We extend the predicate $ \UnT{\cdot} $ to a typing context $ \Gamma $, by requiring that for $ \UnT{\Gamma} $ all types in $ \Gamma $ are unrestricted.
In contrast, all typing contexts are linear, denoted as $ \Gamma \, \linCMVmix $.
The operation $ \cdot \circ \cdot $ allows to split a typing context into two typing contexts provided that all assignments with linear types are on distinct names.
Assignments with unrestricted types can be shared by the two parts.
\begin{displaymath}\begin{array}{c}
	\cdot = \left( \cdot \circ \cdot \right)
	\hspace{2em}
	\dfrac{\Gamma_1 \circ \Gamma_2 = \Gamma \quad \UnT{T}}{\Gamma, \At{x}{T} = {\left( \Gamma_1, \At{x}{T} \right)} \circ {\left( \Gamma_2, \At{x}{T} \right)}}
	\vspace{0.5em}\\
	\dfrac{\Gamma_1 \circ \Gamma_2 = \Gamma}{\Gamma, \At{x}{\linCMVmix \, p} = {\left( \Gamma_1, \At{x}{\linCMVmix \, p} \right)} \circ \Gamma_2}
	\hspace{2em}
	\dfrac{\Gamma_1 \circ \Gamma_2 = \Gamma}{\Gamma, \At{x}{\linCMVmix \, p} = \Gamma_1 \circ {\left( \Gamma_2, \At{x}{\linCMVmix \, p} \right)}}
\end{array}\end{displaymath}

The operation $ \cdot + \cdot $ adds a new assignment to a typing context, while ensuring that in a typing context all assignments are on pairwise distinct names and an assignment can be added to a typing context twice only if its type is unrestricted.
\begin{displaymath}\begin{array}{c}
	\dfrac{\At{x}{U} \notin \Gamma}{\Gamma + \At{x}{T} = \Gamma, \At{x}{T}}
	\hspace{2em}
	\dfrac{\UnT{T}}{{\left( \Gamma, \At{x}{T} \right)} + \At{x}{T} = \Gamma, \At{x}{T}}
\end{array}\end{displaymath}

\begin{figure}[tp]
	\begin{displaymath}\begin{array}{c}
		\ruleTUnitCMVmix \; \dfrac{\UnT{\Gamma}}{\Gamma \vdash \At{n}{\nat}} 		\hspace{2em}
		\ruleTTrueCMVmix \; \dfrac{\UnT{\Gamma}}{\Gamma \vdash \At{\true}{\bool}}
		\hspace{2em}
		\ruleTFalseCMVmix \; \dfrac{\UnT{\Gamma}}{\Gamma \vdash \At{\false}{\bool}}
		\vspace{0.5em}\\
		\ruleTVarCMVmix \; \dfrac{\UnT{\Gamma_1, \Gamma_2}}{\Gamma_1, \At{x}{T}, \Gamma_2 \vdash \At{x}{T}}
		\hspace{2em}
		\ruleTSubCMVmix \; \dfrac{\Gamma \vdash \At{v}{T} \quad \Subtype{T}{U}}{\Gamma \vdash \At{v}{U}}
		\vspace{0.5em}\\
		\ruleTOutCMVmix \; \dfrac{\Gamma_1 \vdash \At{v}{T} \quad \Gamma_2 \vdash P}{\Gamma_1 \circ \Gamma_2 \vdash \At{\OutCMVmix{\lab}{v}{P}}{\OutCMVmix{\lab}{T}{U}}}
		\hspace{2em}
		\ruleTInCMVmix \; \dfrac{\Gamma, \At{x}{T} \vdash P}{\Gamma \vdash \At{\InpCMVmix{\lab}{x}{P}}{\InpCMVmix{\lab}{T}{U}}}
		\vspace{0.5em}\\
		\ruleTInactCMVmix \; \dfrac{\UnT{\Gamma}}{\Gamma \vdash \inactCMVmix}
		\hspace{2em}
		\ruleTParCMVmix \; \dfrac{\Gamma_1 \vdash P_1 \quad \Gamma_2 \vdash P_2}{\Gamma_1 \circ \Gamma_2 \vdash P_1 \mid P_2}
		\vspace{0.5em}\\
		\ruleTIfCMVmix \; \dfrac{\Gamma_1 \vdash \At{v}{\bool} \quad \Gamma_2 \vdash P \quad \Gamma_2 \vdash Q}{\Gamma_1 \circ \Gamma_2 \vdash \ConditionalCMVmix{v}{P}{Q}}
		\hspace{1em}
		\ruleTResCMVmix \; \dfrac{\Gamma, \At{x}{T}, \At{y}{U} \vdash P \quad \Dual{T}{U}}{\Gamma \vdash \ResCMVmix{x}{y}{P}}
		\vspace{0.5em}\\
		\ruleTChoiceCMVmix \; \dfrac{\begin{array}{c} {\left( \Gamma_1 \circ \Gamma_2 \right)} \, q_1 \quad \Gamma_1 \vdash \At{x}{\ChoiceTCMVmix{q_2}{\#}{\Set{\BranchCMVmix{\lab}{*_i}{T_i}{U_i}}_{i \in I}}} \quad \Set{\lab *_j}_{j \in J} = \Set{\lab *_i}_{i \in I}\\ \Gamma_2 + \At{x}{U_j} \vdash \At{\BranchCMVmix{\lab}{*_j}{v_j}{P_j}}{\BranchCMVmix{\lab}{*_j}{T_j}{U_j}} \quad {\left( \forall j \in J \right)} \end{array}}{\Gamma_1 \circ \Gamma_2 \vdash \ChoiceCMVmix{q_1}{x}{\sum_{j \in J} \BranchCMVmix{\lab}{*_j}{v_j}{P_j}}}
	\end{array}\end{displaymath}
	where $ n \in \Set{1, 2, \ldots} $.
	\caption{Typing Rules of \cnvms.}
	\label{fig:typingRulesCMVmix}
\end{figure}

A process $ P $ is typed if there is some typing context $ \Gamma $ such that the \emph{type judgement} $ \Gamma \vdash P $ can be derived from the typing rules in Figure~\ref{fig:typingRulesCMVmix}.

Choices are checked with Rule~\ruleTChoiceCMVmix.
It requires that the typing environment is unrestricted ($ \unCMVmix $) if and only if the analysed choice is qualified as $ \unCMVmix $; else both need to be linear ($ \linCMVmix $).
Then the typing context needs to assign an external or internal choice type to the channel endpoint of this choice, where the qualifier in the type is $ \linCMVmix $ if the choice is qualified as $ \linCMVmix $.

The calculus \cnvms is composed of typed processes and the semantics in Figure~\ref{fig:semanticsCMVmix}, where $ \procCMVmix $ is the typed fragment of $ \procCMVmixUntyped $.

The calculus \cnv is the fragment of \cnvms with a standard branching construct instead of mixed choice (compare to \citep{CASAL202223}).

\begin{definition}[\cnv]
	The set of untyped processes $ \procCMVUntyped $ replaces the choice construct of $ \procCMVmixUntyped $ by the following four constructs
	\begin{align*}
		\OutCMV{y}{v}{P} \bnfbar \InpCMV{q}{y}{x}{P} \bnfbar \SelCMV{x}{\lab}{P} \bnfbar \BranCMV{x}{\Set{\BranchCMV{\lab_i}{P_i}}_{i \in I}}
	\end{align*}
	and keeps the constructs for parallel composition, restriction, conditionals, and inaction.
\end{definition}

The output action is implemented by $ \OutCMV{y}{v}{P} $ and $ \InpCMV{q}{y}{x}{P} $ implements an input action.
Selection $ \SelCMV{x}{\lab}{P} $ allows to select the branch with label $ \lab $ from a branching $ \BranCMV{x}{\Set{\BranchCMV{\lab_i}{P_i}}_{i \in I}} $ provided that $ \lab \in \Set{ \lab_i }_{i \in I} $.

\begin{figure}[tp]
	\begin{displaymath}\begin{array}{c}
		\ruleRLinComCMV \; \ResCMV{x}{y}{{\left( \OutCMV{x}{v}{P} \mid \InpCMV{\linCMV}{y}{z}{Q} \mid R \right)}} \stepCMV \ResCMV{x}{y}{{\left( P \mid Q\Set{\Subst{v}{z}} \mid R \right)}}
		\vspace{0.5em}\\
		\ruleRUnComCMV \; \ResCMV{x}{y}{{\left( \OutCMV{x}{v}{P} \mid \InpCMV{\unCMV}{y}{z}{Q} \mid R \right)}} \stepCMV \ResCMV{x}{y}{{\left( P \mid Q\Set{\Subst{v}{z}} \mid \InpCMV{\unCMV}{y}{z}{Q} \mid R \right)}}
		\vspace{0.5em}\\
		\ruleRCaseCMV \; \dfrac{j \in I}{\ResCMV{x}{y}{{\left( \SelCMV{x}{\lab_j}{P} \mid \BranCMV{y}{\Set{\BranchCMV{\lab_i}{Q_i}}_{i \in I}} \mid R \right)}} \stepCMV \ResCMV{x}{y}{{\left( P \mid Q_j \mid R \right)}}}
	\end{array}\end{displaymath}
	and structural congruence $ \equiv $ as well as the Rules \ruleRIfTCMVmix, \ruleRIfFCMVmix, \ruleRParCMVmix, \ruleRResCMVmix, and \ruleRStructCMVmix from Figure~\ref{fig:semanticsCMVmix}.
	\caption{Reduction Rules ($ \stepCMV $) of \cnv.}
	\label{fig:semanticsCMV}
\end{figure}

The reduction semantics of \cnv is given in Figure~\ref{fig:semanticsCMV}.

The set of types of \cnv replaces the choice construct in the definition of types of \cnvms by the following two constructs
\begin{align*}
	\ComTCMV{q}{{*}}{T}{T} \bnfbar \ChoiceTCMV{q}{\#}{\Set{\BranchCMV{\lab_i}{T_i}}_{i \in I}}
\end{align*}
and keeps the constructs for inaction, base types, and recursion.

We adapt the typing rule that allows to compare types for choices to the simpler rule
\begin{displaymath}\begin{array}{c}
	\dfrac{T_i \simeq T_i'}{\ChoiceTCMV{q}{\#}{\Set{\BranchCMV{\lab_i}{T_i}}_{i \in I}} \simeq \ChoiceTCMV{q}{\#}{\Set{\BranchCMV{\lab_i}{T_i'}}_{i \in I}}}
\end{array}\end{displaymath}
and keep the remaining rules for inaction, base types, and recursion as well as the rules for type equivalence.

The following two rules replace the rule for choice in the definition of duality.
\begin{displaymath}\begin{array}{c}
	\dfrac{\Dual{\bullet}{{*}} \quad T_1 \simeq T_2 \quad \Dual{U_1}{U_2}}{\Dual{\ComTCMV{q}{\bullet}{T_1}{U_1}}{\ComTCMV{q}{{*}}{T_2}{U_2}}}
	\hspace{2em}
	\dfrac{\Dual{\#}{\flat} \quad \Dual{T_i}{U_i} \quad {\left( \forall i \in I \right)}}{\Dual{\ChoiceTCMV{q}{\#}{\Set{\BranchCMV{\lab_i}{T_i}}_{i \in I}}}{\ChoiceTCMV{q}{\flat}{\Set{\BranchCMV{\lab_i}{U_i}}_{i \in I}}}}
\end{array}\end{displaymath}
We keep the rules for polarities, views, inaction, and recursion.

The following four rules replace the subtyping rules for inputs, outputs, and choice.
\begin{displaymath}\begin{array}{c}
	\dfrac{\Subtype{U}{T} \quad \Subtype{T'}{U'}}{\Subtype{\ComTCMV{q}{!}{T}{T'}}{\ComTCMV{q}{!}{U}{U'}}}
	\hspace{2em}
	\dfrac{\Subtype{T}{U} \quad \Subtype{T'}{U'}}{\Subtype{\ComTCMV{q}{?}{T}{T'}}{\ComTCMV{q}{?}{U}{U'}}}
	\vspace{0.5em}\\
	\dfrac{J \subseteq I \quad \Subtype{T_j}{U_j} \quad {\left( \forall j \in J \right)}}{\Subtype{\ChoiceTCMV{q}{\oplus}{\Set{\BranchCMV{\lab_i}{T_i}}_{i \in I}}}{\ChoiceTCMV{q}{\oplus}{\Set{\BranchCMV{\lab_j}{U_j}}_{j \in J}}}}
	\hspace{2em}
	\dfrac{I \subseteq J \quad \Subtype{T_i}{U_i} \quad {\left( \forall i \in I \right)}}{\Subtype{\ChoiceTCMV{q}{\&}{\Set{\BranchCMV{\lab_i}{T_i}}_{i \in I}}}{\ChoiceTCMV{q}{\&}{\Set{\BranchCMV{\lab_j}{U_j}}_{j \in J}}}}
\end{array}\end{displaymath}
We keep the subtyping rules for inaction, base types, and recursion.

The following two rules replace the rule for choice in the definition of the predicate $ \UnT{\cdot} $.
\begin{displaymath}\begin{array}{c}
	\UnT{\ComTCMV{\unCMV}{{*}}{T}{U}}
	\hspace{2em}
	\UnT{\ChoiceTCMV{\unCMV}{\#}{\Set{\BranchCMV{\lab_i}{T_i}}_{i \in I}}}
\end{array}\end{displaymath}
We keep the rules for inaction, base types, and recursion.

\begin{figure}[tp]
	\begin{displaymath}\begin{array}{c}
		\ruleTOutCMV \; \dfrac{\Gamma_1 \vdash \At{x}{\ComTCMV{q}{!}{T}{U}} \quad \Gamma_2 + \At{x}{U} = \Gamma_3 \circ \Gamma_4 \quad \Gamma_3 \vdash \At{v}{T} \quad \Gamma_4 \vdash P}{\Gamma_1 \circ \Gamma_2 \vdash \OutCMV{x}{v}{P}}
		\vspace{0.5em}\\
		\ruleTInCMV \; \dfrac{{\left( \Gamma_1 \circ \Gamma_2 \right)} \, q_1 \quad \Gamma_1 \vdash \At{x}{\ComTCMV{q_2}{?}{T}{U}} \quad {\left( \Gamma_2 + \At{x}{U} \right)}, \At{y}{T} \vdash P}{\Gamma_1 \circ \Gamma_2 \vdash \InpCMV{q_1}{x}{y}{P}}
		\vspace{0.5em}\\
		\ruleTBranchCMV \; \dfrac{\Gamma_1 \vdash \At{x}{\ChoiceTCMV{q}{\&}{\Set{\BranchCMV{\lab_i}{T_i}}_{i \in I}}} \quad \Gamma_2 + \At{x}{T_i} \vdash P_i \quad {\left( \forall i \in I \right)}}{\Gamma_1 \circ \Gamma_2 \vdash \BranCMV{x}{\Set{\BranchCMV{\lab_i}{P_i}}_{i \in I}}}
\quad 
		\ruleTSelCMV \; \dfrac{\Gamma_1 \vdash \At{x}{\ChoiceTCMV{q}{\oplus}{\Set{\BranchCMV{\lab}{T}}}} \quad \Gamma_2 + \At{x}{T} \vdash P}{\Gamma_1 \circ \Gamma_2 \vdash \SelCMV{x}{\lab}{P}}
	\end{array}\end{displaymath}
	and \ruleTUnitCMVmix, \ruleTTrueCMVmix, \ruleTFalseCMVmix, \ruleTVarCMVmix, \ruleTIfCMVmix, \ruleTSubCMVmix, \ruleTInactCMVmix, \ruleTParCMVmix, and \ruleTResCMVmix.
	\caption{Typing Rules of \cnv.}
	\label{fig:typingRulesCMV}
\end{figure}

The typing rules of \cnv are depicted in Figure~\ref{fig:typingRulesCMV}.
In the \cnv-calculus we consider only typed terms.
The linear fragments \lcnvms and \lcnv are obtained from \cnvms and \cnv by forbidding $ \unCMV $ in types.
Since choices typed as $ \linCMV $ have to be $ \linCMV $-choices, there are also no $ \unCMV $-choices in \lcnvms and \lcnv.
We also restricting \lcnvms and \lcnv to a single sessions, \ie at most one restriction, and forbid delegation, \ie only values can be transmitted.
\section{Encodability and Separation Results}
\label{app:encodabilitySeparation}

We prove the encodability and separation results depicted in Figure~\ref{fig:hierarchy}.
The green arrows \greenArrow for calculus inclusion do not need proving.
If $ L_1 $ is a subcalculus of $ L_2 $ then clearly the identity function is a good encoding from $ L_1 $ into $ L_2 $.

To separate the three layers we show, which synchronisation patterns can be described by the respective calculi in \S~\ref{app:synchronisationPatterns} and then use such a synchronisation pattern as counterexample to derive the necessary separation results in \S~\ref{app:separationM} and \S~\ref{app:separationStar}.
Thereby, the general definition of the synchronisation patterns allow us to prove separation between several calculi uniformly.

The remaining separation results for the bottom layer are presented in \S~\ref{app:separateMultipartyFromBinary} and \S~\ref{app:separationMcbsFromLcnvms}.
The subsections \S~\ref{app:encodingScbsIntoBs}--\S~\ref{app:LcnvmsIntoMcbs} present encodability results.

%
%
In Section~\ref{app:separationLeaderElection} we show that the leader election problem can be used to separate \msmpst from \scmpst and thereby reprove this result already presented in Section~\ref{app:separationStar} by using instead the synchronisation pattern \patternStar.
Of course it would be sufficient to provide one proof for this result.
The proof in Section~\ref{app:separationLeaderElection} uses with leader election in symmetric networks a practical problem that can be solved in \msmpst but not in \scmpst.
Thereby, we show that the difference in the expressive power is of practical relevance.
We need, however, the slightly stricter criterion barb sensitiveness.
In Section~\ref{app:separationStar} we require only the weaker success sensitiveness, and thus produce a stronger separation result, but we are using a more abstract feature to distinguish the calculi.
Since we choose the weaker criterion of success sensitiveness for our notion of good encoding, we need the proof via synchronisation patterns to build the hierarchy in Figure~\ref{fig:hierarchy}.

\subsection{Synchronisation Patterns}
\label{app:synchronisationPatterns}

%
%
%
%
%

In \citep{DBLP:conf/esop/PetersNG13} the technique used in \citep{Palamidessi03} and its relation to synchronisation are analysed.
Two synchronisation patterns, the pattern \patternM and the pattern \patternStar, are identified that describe two different levels of synchronisation and allow to more clearly separate calculi along their ability to express synchronisation.
These patterns are called \patternM and \patternStar, because their respective representations as a Petri net (see \S~\ref{sec:intro}) have these shapes.
The pattern \patternStar captures the power of synchronisation of the $ \pi $-calculus.
In particular it captures what is necessary to solve the leader election problem.


The pattern \patternM captures a very weak form of synchronisation, not enough to solve leader election but enough to make a fully distributed implementation of calculi with this pattern difficult (see also \citep{petersNestmannIC20}).
This pattern was originally identified in \citep{Glabbeek2008} when studying the relevance of synchrony and distribution on Petri nets.
As shown in \citep{peters12, DBLP:conf/esop/PetersNG13}, the ability to express these different amounts of synchronisation in the $ \pi $-calculus lies in its different forms of choices: to express the pattern \patternStar the $ \pi $-calculus needs mixed choice, whereas the separate choice allows to express the pattern \patternM.

\subsubsection{The Pattern \patternM}

We inherit the definition of the synchronisation pattern \patternM in Definition~\ref{def:synchronisationPatternM} from \citep{DBLP:conf/esop/PetersNG13}, but we do not distinguish between local and non-local \patternM since in the $ \pi $-calculus---and its variants such as \scmpst---there is no difference between parallel and distributable steps.

In \scmpst we can find \eg the following \patternM given in Figure~\ref{fig:KeysExpressiveness}(5).

\begin{newexample}[The \patternM in \scmpst]
	\label{ex:MinScmpst}
	Consider the session
	\begin{align*}
		\PM_{\scmpst} &= \pa{\enodei{a}} \left( \ssout{\enodei{b}}{l}. \ssout{\enodei{d}}{l}. \inact \OR \ssout{\enodei{d}}{l'}. \inact \right) \Par \pa{\enodei{b}} \left( \ssinp{\enodei{a}}{l}. \inact \OR \ssinp{\enodei{c}}{l}. \inact \right) \Par \pa{\enodei{c}} \left( \ssout{\enodei{b}}{l}. \inact \OR \ssout{\enodei{d}}{l}. \inact \right) \Par \pa{\enodei{d}} \left( \ssinp{\enodei{c}}{l}. \ssinp{\enodei{a}}{l}. \success \OR \ssinp{\enodei{a}}{l'}. \inact \right)
	\end{align*}
	with the steps
	\begin{description}
		\item[Step $ a $:] $ \PM_{\scmpst} \red \PM_{\scmpst, a} = \pa{\enodei{a}} \ssout{\enodei{d}}{l}. \inact \Par \pa{\enodei{b}} \inact \Par \pa{\enodei{c}} \left( \ssout{\enodei{b}}{l}. \inact \OR \ssout{\enodei{d}}{l}. \inact \right) \Par \pa{\enodei{d}} \left( \ssinp{\enodei{c}}{l}. \ssinp{\enodei{a}}{l}. \success \OR \ssinp{\enodei{a}}{l'}. \inact \right) $
		\item[Step $ b $:] $ \PM_{\scmpst} \red \PM_{\scmpst, b} = \pa{\enodei{a}} \left( \ssout{\enodei{b}}{l}. \ssout{\enodei{d}}{l}. \inact \OR \ssout{\enodei{d}}{l'}. \inact \right) \Par \pa{\enodei{b}} \inact \Par \pa{\enodei{c}} \inact \Par \pa{\enodei{d}} \left( \ssinp{\enodei{c}}{l}. \ssinp{\enodei{a}}{l}. \success \OR \ssinp{\enodei{a}}{l'}. \inact \right) $
		\item[Step $ c $:] $ \PM_{\scmpst} \red \PM_{\scmpst, c} = \pa{\enodei{a}} \left( \ssout{\enodei{b}}{l}. \ssout{\enodei{d}}{l}. \inact \OR \ssout{\enodei{d}}{l'}. \inact \right) \Par \pa{\enodei{b}} \left( \ssinp{\enodei{a}}{l}. \inact \OR \ssinp{\enodei{c}}{l}. \inact \right) \Par \pa{\enodei{c}} \inact \Par \pa{\enodei{d}} \ssinp{\enodei{a}}{l}. \success $
	\end{description}
	where $ a $ and $ c $ are distributable but step $ b $ is in conflict with $ a $ and $ c $.
	Note that $ \PM_{\scmpst} $ reaches success if and only if it performs both of the distributable steps $ a $ and $ c $.
	Then $ \PM_{\scmpst, a} \reachSuccessFin $, $ \PM_{\scmpst, b} \not\reachSuccess $, and $ \PM_{\scmpst, c} \reachSuccessFin $.
	\qed
\end{newexample}

By calculus inclusion then also the calculi \msmpst and \mcmpst contain \patternM.

\begin{lemma}
	The session $ \PM_{\scmpst} $ in Example~\ref{ex:MinScmpst} is typed and deadlock-free.
\end{lemma}

\begin{proof}
	In Example~\ref{ex:MinScmpst} we omit the objects in communication, because they are not relevant for the protocol.
	Assume that in all communications $ \true $ is transmitted and the receiver stores the received value in the variable $ \xx $.
	By Figure~\ref{fig:typing}, we get
	\begin{align*}
		\types \ssout{\enodei{b}}{l}. \ssout{\enodei{d}}{l}. \inact \OR \ssout{\enodei{d}}{l'}. \inact \as T_{\role{a}} &= \tout{\role{b}}{l}\tout{\role{d}}{l}\tinact \; \tor \; \tout{\role{d}}{l'}\tinact\\
		\types \ssinp{\enodei{a}}{l}. \inact \OR \ssinp{\enodei{c}}{l}. \inact \as T_{\role{b}} &= \tinp{\role{a}}{l}\tinact \; \tor \; \tinp{\role{c}}{l}\tinact\\
		\types \ssout{\enodei{b}}{l}. \inact \OR \ssout{\enodei{d}}{l}. \inact \as T_{\role{c}} &= \tout{\role{b}}{l}\tinact \; \tor \; \tout{\role{d}}{l}\tinact\\
		\types \ssinp{\enodei{c}}{l}. \ssinp{\enodei{a}}{l}. \success \OR \ssinp{\enodei{a}}{l'}. \inact \as T_{\role{d}} &= \tinp{\role{c}}{l}\tinp{\role{a}}{l}\tinact \; \tor \; \tinp{\role{a}}{l'}\tinact
	\end{align*}
	and the omitted payload type is $ \bool $ in all cases.
	Let $ \LL = \localise{\enodei{a}} T_{\enodei{a}}, \localise{\enodei{b}} T_{\enodei{b}}, \localise{\enodei{c}} T_{\enodei{c}}, \localise{\enodei{d}} T_{\enodei{d}} $.
	Since there is no recursion, $ \PM_{\scmpst} $ has only finitely many maximal executions.
	The reductions of the local types follow closely the reductions of the protocol.
	For instance we have the sequence:
	\begin{align*}
		\LL & \red \localise{\enodei{a}} \tout{\role{d}}{l}\tinact, \localise{\enodei{b}} \tinact, \localise{\enodei{c}} T_{\enodei{c}}, \localise{\enodei{d}} T_{\enodei{d}}
		\red \localise{\enodei{a}} \tout{\role{d}}{l}\tinact, \localise{\enodei{b}} \tinact, \localise{\enodei{c}} \tinact, \localise{\enodei{d}} \tinp{\role{a}}{l}\tinact
		\red \localise{\enodei{a}} \tinact, \localise{\enodei{b}} \tinact, \localise{\enodei{c}} \tinact, \localise{\enodei{d}} \tinact
		\noRed
	\end{align*}
	By checking all sequences of reductions of the local types, we obtain $ \safe{\LL} $.
	By Figure~\ref{fig:typing}, then $ \types \PM_{\scmpst} \as \LL $.
	Similarly by checking all sequences of reductions of local types, we see that $ \LL $ is deadlock-free, \ie $ \df{\LL} $.
	By Theorem~\ref{thm:deadlockfree}, then $ \PM_{\scmpst} $ is deadlock-free.
\end{proof}

An \patternM in \cnvms was given in Example~4.1 in \citep{PY2022}.
To emphasise that the ability to express the pattern \patternM in \cnvms results from channels typed as unrestricted and not choice, we give with $ \PM_{\text{\cnvms}} $ in Figure~\ref{fig:KeysExpressiveness}(7) another \patternM in \cnvms without choice.

\begin{newexample}[The \patternM in \cnvms]
	\label{ex:MinCnvms}
	Consider the session
	\begin{align*}
		\PM_{\text{\cnvms}} &= \ResCMVmix{x}{y}{\left( \ChoiceCMVmix{\linCMV}{x}{\OutCMVmix{l}{\true}{\inactCMVmix}} \Par \ChoiceCMVmix{\linCMV}{y}{\InpCMVmix{l}{z}{\ConditionalCMVmix{z}{\inact}{\inactCMVmix}}} \Par \ChoiceCMVmix{\linCMV}{x}{\OutCMVmix{l}{\false}{\inactCMVmix}} \Par \ChoiceCMVmix{\linCMV}{y}{\InpCMVmix{l}{z}{\ConditionalCMVmix{z}{\inactCMVmix}{\success}}} \right)}
	\end{align*}
	with the steps
	\begin{description}
		\item[Step $ a $:] $ \PM_{\text{\cnvms}} \red \PM_{\text{\cnvms}, a} = \ResCMVmix{x}{y}{\left( \inactCMVmix \Par \ConditionalCMVmix{\true}{\inact}{\inactCMVmix} \Par \ChoiceCMVmix{\linCMV}{x}{\OutCMVmix{l}{\false}{\inactCMVmix}} \Par \ChoiceCMVmix{\linCMV}{y}{\InpCMVmix{l}{z}{\ConditionalCMVmix{z}{\inactCMVmix}{\success}}} \right)} $
		\item[Step $ b $:] $ \PM_{\text{\cnvms}} \red \PM_{\text{\cnvms}, b} = \ResCMVmix{x}{y}{\left( \ChoiceCMVmix{\linCMV}{x}{\OutCMVmix{l}{\true}{\inactCMVmix}} \Par \ConditionalCMVmix{\false}{\inact}{\inactCMVmix} \Par \inactCMVmix \Par \ChoiceCMVmix{\linCMV}{y}{\InpCMVmix{l}{z}{\ConditionalCMVmix{z}{\inactCMVmix}{\success}}} \right)} $
		\item[Step $ c $:] $ \PM_{\text{\cnvms}} \red \PM_{\text{\cnvms}, c} = \ResCMVmix{x}{y}{\left( \ChoiceCMVmix{\linCMV}{x}{\OutCMVmix{l}{\true}{\inactCMVmix}} \Par \ChoiceCMVmix{\linCMV}{y}{\InpCMVmix{l}{z}{\ConditionalCMVmix{z}{\inact}{\inactCMVmix}}} \Par \inactCMVmix \Par \ConditionalCMVmix{\false}{\inactCMVmix}{\success} \right)} $
	\end{description}
	where $ a $ and $ c $ are distributable but step $ b $ is in conflict with $ a $ and $ c $.
	Moreover, $ \PM_{\text{\cnvms}, a} \reachSuccessFin $, $ \PM_{\text{\cnvms}, b} \not\reachSuccess $, and $ \PM_{\text{\cnvms}, c} \reachSuccessFin $.
	\qed
\end{newexample}

The system $ \PM_{\cnv} $ in Figure~\ref{fig:KeysExpressiveness}(6) is $ \PM_{\text{\cnvms}} $ adapted to \cnv.

\begin{newexample}[The \patternM in \cnv]
	\label{ex:MinCnv}
	Consider the session
	\begin{align*}
		\PM_{\cnv} &= \ResCMV{x}{y}{\left( \OutCMV{x}{\true}{\inactCMV} \Par \InpCMV{\linCMV}{y}{z}{\ConditionalCMV{z}{\inact}{\inactCMV}} \Par \OutCMV{x}{\false}{\inactCMV} \Par \InpCMV{\linCMV}{y}{z}{\ConditionalCMV{z}{\inactCMV}{\success}} \right)}
	\end{align*}
	with the steps
	\begin{description}
		\item[Step $ a $:] $ \PM_{\cnv} \red \PM_{\cnv, a} = \ResCMV{x}{y}{\left( \inactCMV \Par \ConditionalCMV{\true}{\inact}{\inactCMV} \Par \OutCMV{x}{\false}{\inactCMV} \Par \InpCMV{\linCMV}{y}{z}{\ConditionalCMV{z}{\inactCMV}{\success}} \right)} $
		\item[Step $ b $:] $ \PM_{\cnv} \red \PM_{\cnv, b} = \ResCMV{x}{y}{\left( \OutCMV{x}{\true}{\inactCMV} \Par \ConditionalCMV{\false}{\inact}{\inactCMV} \Par \inactCMV \Par \InpCMV{\linCMV}{y}{z}{\ConditionalCMV{z}{\inactCMV}{\success}} \right)} $
		\item[Step $ c $:] $ \PM_{\cnv} \red \PM_{\cnv, c} = \ResCMV{x}{y}{\left( \OutCMV{x}{\true}{\inactCMV} \Par \InpCMV{\linCMV}{y}{z}{\ConditionalCMV{z}{\inact}{\inactCMV}} \Par \inactCMV \Par \ConditionalCMV{\false}{\inactCMV}{\success} \right)} $
	\end{description}
	where $ a $ and $ c $ are distributable but step $ b $ is in conflict with $ a $ and $ c $.
	Moreover, $ \PM_{\cnv, a} \reachSuccessFin $, $ \PM_{\cnv, b} \not\reachSuccess $, and $ \PM_{\cnv, c} \reachSuccessFin $.
	\qed
\end{newexample}

None of the remaining calculi, \ie none of the calculi in the bottom layer of Figure~\ref{fig:hierarchy}, has an \patternM.
We start with \mcbs.

\begin{newlemma}[No \patternM in \mcbs]
	\label{lem:noMinMcbs}
	There are no \patternM in \mcbs.
\end{newlemma}

\begin{proof}
	Assume the contrary, \ie assume that there is a term $ \PM_{\mcbs} $ in the \mcbs-calculus that is an \patternM.
	Then $ \PM_{\mcbs} $ can perform at least three alternative reduction steps $ a, b, c $ such that $ b $ is in conflict to $ a $ and $ c $ but $ a $ and $ c $ are distributable.
	By Figure~\ref{fig:reduction}, step $ a $ reduces a conditional or performs a communication by reducing two choices.
	Since a conditional cannot be in conflict with any other step, none of the steps in $ \left\{ a, b, c \right\} $ reduces a conditional.
	Then all steps in $ \left\{ a, b, c \right\} $ are communication steps that reduce an output and an input that both are part of choices (with at least one summand).
	Let $ \p $ and $ \q $ be the participants, whose choices are reduced in $ a $.
	Because of the conflict between $ a $ and $ b $, these two steps reduce the same choice.
	Without loss of generality, assume that $ a $ and $ b $ compete for the choice of participant $ \q $.
	Since \mcbs is limited to binary sessions and since single participants cannot contain parallel composition, then also step $ b $ reduces the same choices of $ \p $ and $ \q $ that are reduced in step $ a $ (picking different summands in at least one of the choices).
	Similarly, because of the conflict between $ b $ and $ c $, then also step $ c $ has to reduce the same choices of $ \p $ and $ \q $ (again picking different summands).
	But then also the steps $ a $ and $ c $ are in conflict and cannot be distributable.
	This is a contradiction.
	We conclude that there are no \patternM in the \mcbs-calculus.
\end{proof}

The proof of the Lemma~\ref{lem:noMinMcbs} relies on the following observation.
Since sessions in \mcbs are binary, two steps can only be in conflict if they reduce exactly the same choice.
Since step $ b $ is in conflict with $ a $ and $ c $ then all three steps reduce the same choices and thus $ a $ and $ c $ cannot be distributable.
By calculus inclusion, then there are no \patternM in \scbs or \bs.
Next, we analyse $ \dmpst $.

\begin{newlemma}[No \patternM in \dmpst]
	\label{lem:noMinDmpst}
	There are no \patternM in \dmpst.
\end{newlemma}

\begin{proof}
	Assume the contrary, \ie assume that there is a term $ \PM_{\dmpst} $ in the \dmpst-calculus that is an \patternM.
	Then $ \PM_{\dmpst} $ can perform at least three alternative reduction steps $ a, b, c $ such that $ b $ is in conflict to $ a $ and $ c $ but $ a $ and $ c $ are distributable.
	By Figure~\ref{fig:reduction}, step $ a $ reduces a conditional or performs a communication by reducing two choices.
	Since a conditional cannot be in conflict with any other step, none of the steps in $ \left\{ a, b, c \right\} $ reduces a conditional.
	Then all steps in $ \left\{ a, b, c \right\} $ are communication steps that reduce an output and an input that both are part of choices (with at least one summand).
	Let $ \p $ and $ \q $ be the participants, whose choices are reduced in $ a $.
	Because of the conflict between $ a $ and $ b $, these two steps reduce the same choice.
	Without loss of generality, assume that $ a $ and $ b $ compete for the choice of participant $ \q $.
	Since choice in \dmpst is limited to a single participant and since single participants cannot contain parallel composition, then also step $ b $ reduces the same choices of $ \p $ and $ \q $ that are reduced in step $ a $ (picking different summands in at least one of the choices).
	Similarly, because of the conflict between $ b $ and $ c $, then also step $ c $ has to reduce the same choices of $ \p $ and $ \q $ (again picking different summands).
	But then also the steps $ a $ and $ c $ are in conflict and cannot be distributable.
	This is a contradiction.
	We conclude that there are no \patternM in the \dmpst-calculus.
\end{proof}

In contrast to \mcbs, \dmpst has sessions with more than two participants.
But the limitation of choice to a single participant still ensures, that all three steps $ a $, $ b $, and $ c $ reduce the same choices.
Again, $ a $ and $ c $ cannot be distributable.
By calculus inclusion, then there are no \patternM in \smpst or \mpst.

The \patternM in \cnvms and \cnv requires channels typed as unrestricted.
The linear fragments \lcnvms and \lcnv do not contain \patternM.

\begin{newlemma}[No \patternM in \lcnvms]
	\label{lem:noMinLcnvms}
	There are no \patternM in \lcnvms.
\end{newlemma}

\begin{proof}
	Assume the contrary, \ie assume that there is a term $ \PM_{\text{\lcnvms}} $ in the \lcnvms-calculus that is an \patternM.
	Then $ \PM_{\text{\lcnvms}} $ can perform at least three alternative reduction steps $ a, b, c $ such that $ b $ is in conflict to $ a $ and $ c $ but $ a $ and $ c $ are distributable.
	By the reduction semantics in Figure~\ref{fig:semanticsCMVmix} (see \citep{CASAL202223}), step $ a $ reduces a conditional or performs a communication by reducing two choices.
	Since a conditional cannot be in conflict with any other step, none of the steps in $ \left\{ a, b, c \right\} $ reduces a conditional.
	Then all steps in $ \left\{ a, b, c \right\} $ are communication steps that reduce an output and an input that both are part of choices (with at least one summand) of some session endpoints $ x $ and $ y $.
	Because of the conflict between $ a $ and $ b $, these two steps reduce the same choice.
	Without loss of generality, assume that $ a $ and $ b $ compete for the choice of session endpoint $ x $.
	Since \lcnvms is limited to the linear fragment of \cnvms there is at most one unguarded choice of $ x $ and $ y $, respectively.
	Then also step $ b $ reduces the same choices of $ x $ and $ y $ that are reduced in step $ a $ (picking different summands in at least one of the choices).
	Similarly, because of the conflict between $ b $ and $ c $, then also step $ c $ has to reduce the same choices of $ x $ and $ y $ (again picking different summands).
	But then also the steps $ a $ and $ c $ are in conflict and cannot be distributable.
	This is a contradiction.
	We conclude that there are no \patternM in the \lcnvms-calculus.
\end{proof}

\begin{newlemma}[No \patternM in \lcnv]
	\label{lem:noMinLcnv}
	There are no \patternM in \lcnv.
\end{newlemma}

\begin{proof}
	Assume the contrary, \ie assume that there is a term $ \PM_{\lcnv} $ in the \lcnv-calculus that is an \patternM.
	Then $ \PM_{\lcnv} $ can perform at least three alternative reduction steps $ a, b, c $ such that $ b $ is in conflict to $ a $ and $ c $ but $ a $ and $ c $ are distributable.
	By the reduction semantics in Figure~\ref{fig:semanticsCMV} (see \citep{CASAL202223}), step $ a $ reduces a conditional, performs a communication by reducing an output and an input, or reduces a selection and a branching construct.
	Since a conditional cannot be in conflict with any other step, none of the steps in $ \left\{ a, b, c \right\} $ reduces a conditional.
	Then, because of the conflicts, all steps in $ \left\{ a, b, c \right\} $ are either communication steps that reduce an output and an input or all steps resolve a branching.
	\begin{description}
		\item[Communication:] Let $ x $ and $ y $ be the session endpoints, whose output and input are reduced in $ a $.
			Because of the conflict between $ a $ and $ b $, these two steps reduce the same capability.
			Without loss of generality, assume that $ a $ and $ b $ compete for the output of session endpoint $ x $.
			Since \lcnv is limited to the linear fragment of \cnv, there is at most one unguarded output or input per session endpoint but not both.
			Then also step $ b $ reduces the same output of $ x $ and input of $ y $ that are reduced in step $ a $.
			But then $ a $ and $ b $ are the same step and cannot be in conflict.
			This is a contradiction.
		\item[Branching:] Let $ x $ and $ y $ be the session endpoints, whose selection and branching constructs are reduced in $ a $.
			Because of the conflict between $ a $ and $ b $, these two steps reduce the same capability.
			Without loss of generality, assume that $ a $ and $ b $ compete for the selection of session endpoint $ x $.
			Since \lcnv is limited to the linear fragment of \cnv, there is at most one unguarded selection or branching per session endpoint but not both.
			Then also step $ b $ reduces the same selection of $ x $ and branching of $ y $ that are reduced in step $ a $ (picking a different case of the branching construct).
			Similarly, because of the conflict between $ b $ and $ c $, then also step $ c $ has to reduce the same selection of $ x $ and branching of $ y $ (again picking a different case in branching).
			But then also the steps $ a $ and $ c $ are in conflict and cannot be distributable.
			This is a contradiction.
	\end{description}
	We conclude that there are no \patternM in the \lcnv-calculus.
\end{proof}

The differences in the proofs of Lemma~\ref{lem:noMinMcbs} and the Lemmata~\ref{lem:noMinLcnvms} and \ref{lem:noMinLcnv} are due to the syntactical differences of the calculi.
In \mcbs participants cannot contain parallel compositions, which ensures that there is at most one unguarded choice of each participant, \ie at most one choice that can be currently used for a reduction step.
In \cnv and \cnvms several capabilities of the same session endpoint can be combined in parallel.
But in the linear fragments of these calculi, where all choices are typed as linear, there can again be at most one unguarded capability per session endpoint.

We summarise and show that there are no \patternM in any of the calculi in the bottom layer of Figure~\ref{fig:hierarchy}.

\lemNoM*

\begin{proof}
	By Lemma~\ref{lem:noMinLcnv}, there is no \patternM in \lcnv.
	By Lemma~\ref{lem:noMinLcnvms}, there is no \patternM in \lcnvms.
	By Lemma~\ref{lem:noMinMcbs}, there is no \patternM in \mcbs.
	Since $ \bs \subset \scbs \subset \mcbs $, then there are no \patternM in \bs and \scbs.
	By Lemma~\ref{lem:noMinDmpst}, there is no \patternM in \dmpst.
	Since $ \mpst \subset \smpst \subset \dmpst $, then there are no \patternM in \mpst and \smpst.
\end{proof}

In \S~\ref{app:separationM} we use the \patternM to separate calculi from the two upper layers from the bottom layer.

\subsubsection{The Pattern \patternStar}

Intuitively, a \patternStar consists of three \patternM that are overlapping to form a cycle.
We also inherit the definition of the synchronisation pattern \patternStar (Definition~\ref{def:synchronisationPatternGreatM}) from \citep{DBLP:conf/esop/PetersNG13}.

The process $ \mathsf{Election} $ in Example~\ref{ex:leaderElection-process} is a \patternStar.
However, as counterexample to separate \scmpst from the calculi of the lower layers, we use a simplified version of $ \mathsf{Election} $, \ie its first stage, and some notion of reaching success.
The session $ \PS_{\msmpst} $ is given in Figure~\ref{fig:KeysExpressiveness}(8).

\begin{newexample}[A \patternStar in \msmpst]
	\label{ex:StarInMsmpst}
	Consider the protocol
	\begin{align*}
		\PS_{\msmpst} &= M_{\enodei{a}} \Par M_{\enodei{b}} \Par M_{\enodei{c}} \Par M_{\enodei{d}} \Par M_{\enodei{e}} \Par M_{\enodei{gc}}\\
		M_{\enodei{a}} &= \pa{\enodei{a}} \left( \ssout{\enodei{e}}{l}. \inact \OR \ssinp{\enodei{b}}{l}. P_{\enodei{a}} \OR \ssinp{\enodei{gc}}{\delete}. \inact \right)
		\hspace{2em}
		M_{\enodei{b}} = \pa{\enodei{b}} \left( \ssout{\enodei{a}}{l}. \inact \OR \ssinp{\enodei{c}}{l}. P_{\enodei{b}} \OR \ssinp{\enodei{gc}}{\delete}. \inact \right)
		\\
		M_{\enodei{c}} &= \pa{\enodei{c}} \left( \ssout{\enodei{b}}{l}. \inact \OR \ssinp{\enodei{d}}{l}. P_{\enodei{c}} \OR \ssinp{\enodei{gc}}{\delete}. \inact \right)
		\hspace{2em}
		M_{\enodei{d}} = \pa{\enodei{d}} \left( \ssout{\enodei{c}}{l}. \inact \OR \ssinp{\enodei{e}}{l}. P_{\enodei{d}} \OR \ssinp{\enodei{gc}}{\delete}. \inact \right)
		\\
		M_{\enodei{e}} &= \pa{\enodei{e}} \left( \ssout{\enodei{d}}{l}. \inact \OR \ssinp{\enodei{a}}{l}. P_{\enodei{e}} \OR \ssinp{\enodei{gc}}{\delete}. \inact \right)
		\hspace{2em}
		M_{\enodei{gc}} = \choice{\ssout{\enodei{i}}{\delete} \seq \inact}{\enodei{i} \in \set{\enodei{a}, \enodei{b}, \enodei{c}, \enodei{d},\enodei{e}}}
	\end{align*}
	where $ P_{\enodei{a}}, \ldots, P_{\enodei{e}} \in \left\{ \success, \inact \right\} $.
	Each of the continuations $ P_{\enodei{a}}, \ldots, P_{\enodei{e}} $ (and thereby possibly $ \success $) is unguarded by exactly one of the following five steps.
	\begin{description}
		\item[Step $ a $:] $ \PS_{\msmpst} \red \PS_{\msmpst, a} = \pa{\enodei{a}} P_{\enodei{a}} \Par \pa{\enodei{b}} \inact \Par M_{\enodei{c}} \Par M_{\enodei{d}} \Par M_{\enodei{e}} \Par M_{\enodei{gc}} $
		\item[Step $ b $:] $ \PS_{\msmpst} \red \PS_{\msmpst, b} = M_{\enodei{a}} \Par \pa{\enodei{b}} P_{\enodei{b}} \Par \pa{\enodei{c}} \inact \Par M_{\enodei{d}} \Par M_{\enodei{e}} \Par M_{\enodei{gc}} $
		\item[Step $ c $:] $ \PS_{\msmpst} \red \PS_{\msmpst, c} = M_{\enodei{a}} \Par M_{\enodei{b}} \Par \pa{\enodei{c}} P_{\enodei{c}} \Par \pa{\enodei{d}} \inact \Par M_{\enodei{e}} \Par M_{\enodei{gc}} $
		\item[Step $ d $:] $ \PS_{\msmpst} \red \PS_{\msmpst, d} = M_{\enodei{a}} \Par M_{\enodei{b}} \Par M_{\enodei{c}} \Par \pa{\enodei{d}} P_{\enodei{d}} \Par \pa{\enodei{e}} \inact \Par M_{\enodei{gc}} $
		\item[Step $ e $:] $ \PS_{\msmpst} \red \PS_{\msmpst, e} = \pa{\enodei{a}} \inact \Par M_{\enodei{b}} \Par M_{\enodei{c}} \Par M_{\enodei{d}} \Par \pa{\enodei{e}} P_{\enodei{e}} \Par M_{\enodei{gc}} $
	\end{description}
	Since the steps $ a, \ldots, e $ form conflicts and are distributable steps as described in Definition~\ref{def:synchronisationPatternGreatM}, the process $ \PS_{\msmpst} $ is a \patternStar.
\end{newexample}

By calculus inclusion, then also \mcmpst has a \patternStar.

\begin{lemma}
	The session $ \PS_{\msmpst} $ in Example~\ref{ex:StarInMsmpst} is typed and deadlock-free.
\end{lemma}

\begin{proof}
	In Example~\ref{ex:StarInMsmpst} we omit the objects in communication, because they are not relevant for the protocol.
	Assume that in all communications $ \true $ is transmitted and the receiver stores the received value in the variable $ \xx $.
	By Figure~\ref{fig:typing}, we get
	\begin{align*}
		\types \ssout{\enodei{e}}{l}. \inact \OR \ssinp{\enodei{b}}{l}. P_{\enodei{a}} \OR \ssinp{\enodei{gc}}{\delete}. \inact \as T_{\role{a}} &= \tout{\role{e}}{l}\tinact \; \tor \; \tinp{\role{b}}{l}\tinact \; \tor \; \tinp{\role{gc}}{\delete}\tinact\\
		\types \ssout{\enodei{a}}{l}. \inact \OR \ssinp{\enodei{c}}{l}. P_{\enodei{b}} \OR \ssinp{\enodei{gc}}{\delete}. \inact \as T_{\role{b}} &= \tout{\role{a}}{l}\tinact \; \tor \; \tinp{\role{c}}{l}\tinact \; \tor \; \tinp{\role{gc}}{\delete}\tinact\\
		\types \ssout{\enodei{b}}{l}. \inact \OR \ssinp{\enodei{d}}{l}. P_{\enodei{c}} \OR \ssinp{\enodei{gc}}{\delete}. \inact \as T_{\role{c}} &= \tout{\role{b}}{l}\tinact \; \tor \; \tinp{\role{d}}{l}\tinact \; \tor \; \tinp{\role{gc}}{\delete}\tinact\\
		\types \ssout{\enodei{c}}{l}. \inact \OR \ssinp{\enodei{e}}{l}. P_{\enodei{d}} \OR \ssinp{\enodei{gc}}{\delete}. \inact \as T_{\role{d}} &= \tout{\role{c}}{l}\tinact \; \tor \; \tinp{\role{e}}{l}\tinact \; \tor \; \tinp{\role{gc}}{\delete}\tinact\\
		\types \ssout{\enodei{d}}{l}. \inact \OR \ssinp{\enodei{a}}{l}. P_{\enodei{e}} \OR \ssinp{\enodei{gc}}{\delete}. \inact \as T_{\role{e}} &= \tout{\role{d}}{l}\tinact \; \tor \; \tinp{\role{a}}{l}\tinact \; \tor \; \tinp{\role{gc}}{\delete}\tinact\\
		types \choice{\ssout{\enodei{i}}{\delete} \seq \inact}{\enodei{i} \in \set{\enodei{a}, \enodei{b}, \enodei{c}, \enodei{d},\enodei{e}}} \as T_{\role{gc}} &= \Tor{\enodei{i} \in \set{\enodei{a}, \enodei{b}, \enodei{c}, \enodei{d},\enodei{e}}}{\tout{\role{i}}{\delete}\tinact}
	\end{align*}
	and the omitted payload type is $ \bool $ in all cases.
	Let $ \LL = \localise{\enodei{a}} T_{\enodei{a}}, \localise{\enodei{b}} T_{\enodei{b}}, \localise{\enodei{c}} T_{\enodei{c}}, \localise{\enodei{d}} T_{\enodei{d}}, \localise{\enodei{e}} T_{\enodei{e}}, \localise{\enodei{gc}} T_{\enodei{gc}} $.
	Since there is no recursion, $ \PS_{\msmpst} $ has only finitely many maximal executions.
	The reductions of the local types follow closely the reductions of the protocol.
	For instance we have the sequence:
	\begin{align*}
		\LL & \red \localise{\enodei{a}} \tinact, \localise{\enodei{b}} \tinact, \localise{\enodei{c}} T_{\enodei{c}}, \localise{\enodei{d}} T_{\enodei{d}}, \localise{\enodei{e}} T_{\enodei{e}}, \localise{\enodei{gc}} T_{\enodei{gc}}\\
		& \red \localise{\enodei{a}} \tinact, \localise{\enodei{b}} \tinact, \localise{\enodei{c}} \tinact, \localise{\enodei{d}} \tinact, \localise{\enodei{e}} T_{\enodei{e}}, \localise{\enodei{gc}} T_{\enodei{gc}}\\
		& \red \localise{\enodei{a}} \tinact, \localise{\enodei{b}} \tinact, \localise{\enodei{c}} \tinact, \localise{\enodei{d}} \tinact, \localise{\enodei{e}} \tinact, \localise{\enodei{gc}} \tinact
		\noRed
	\end{align*}
	By checking all sequences of reductions of the local types, we obtain $ \safe{\LL} $.
	By Figure~\ref{fig:typing}, then $ \types \PS_{\msmpst} \as \LL $.
	Similarly by checking all sequences of reductions of local types, we see that $ \LL $ is deadlock-free, \ie $ \df{\LL} $.
	By Theorem~\ref{thm:deadlockfree}, then $ \PS_{\msmpst} $ is deadlock-free.
\end{proof}

Since \mcbs does not contain \patternM, it cannot contain \patternStar.

\begin{newlemma}[No \patternStar in \mcbs]
	\label{lem:noStarInMcbs}
	There are no \patternStar in the \mcbs-calculus.
\end{newlemma}

\begin{proof}
	By Lemma~\ref{lem:noMinMcbs}, there are no \patternM in the \mcbs-calculus, \ie there are no terms with at least three steps $ a, b, c $ such that $ b $ is in conflict with $ a $ and $ c $ but $ a $ and $ c $ are distributable.
	By Definition~\ref{def:synchronisationPatternGreatM}, then the \mcbs-calculus cannot contain a \patternStar.
\end{proof}

By calculus inclusion then there is no \patternStar in \scbs or \bs.
Similarly, there are no \patternStar in \dmpst.

\begin{newlemma}[No \patternStar in \dmpst]
	\label{lem:noStarInDmpst}
	There are no \patternStar in the \dmpst-calculus.
\end{newlemma}

\begin{proof}
	By Lemma~\ref{lem:noMinDmpst}, there are no \patternM in the \dmpst-calculus, \ie there are no terms with at least three steps $ a, b, c $ such that $ b $ is in conflict with $ a $ and $ c $ but $ a $ and $ c $ are distributable.
	By Definition~\ref{def:synchronisationPatternGreatM}, then the \dmpst-calculus cannot contain a \patternStar.
\end{proof}

By calculus inclusion then there is no \patternStar in \smpst or \mpst.
To prove that there are no \patternStar in \scmpst is more difficult.

\begin{newlemma}[No \patternStar in \scmpst]
	\label{lem:noStarInScmpst}
	There are no \patternStar in the \scmpst-calculus.
\end{newlemma}

\begin{proof}
	Assume the contrary, \ie assume that there is a term $ \PS_{\scmpst} $ in the \scmpst-calculus that is a \patternStar.
	Then $ \PS_{\scmpst} $ can perform at least five alternative reduction steps $ a, b, c, d, e $ such that neighbouring steps in the sequence $ a, b, c, d, e, a $ are pairwise in conflict and non-neighbouring steps are distributable.
	By Figure~\ref{fig:reduction}, step $ a $ reduces a conditional or performs a communication by reducing two choices.
	Since steps reducing a conditional cannot be in conflict with any other step, none of the steps in $ \left\{a, b, c, d, e\right\} $ reduces a conditional.
	Then all steps in $ \left\{a, b, c, d, e\right\} $ are communication steps that reduce an output and an input that both are part of choices (with at least one summand).
	Because of the conflict between $ a $ and $ b $, these two steps reduce the same choice but this choice is not reduced in $ c $, because $ a $ and $ c $ are distributable.

	\noindent
	\begin{minipage}{\textwidth}
		\begin{wrapfigure}{R}{0.25\textwidth}
			\centering
			\vspace{-1em}
			\scalebox{0.8}{
			\begin{tikzpicture}[]
				\foreach \x/\xlabel/\xtext/\ytext in {1/e/$ C_5 $/$ b $,2/d/$ C_4 $/$ a $,3/c/$ C_3 $/$ e $,4/b/$ C_2 $/$ d $,5/a/$ C_1 $/$ c $}
		        {
		            \path (360*\x/5+125:0.8) node (\xlabel) {\xtext};
		            \path (360*\x/5-55:1.75) node (p\x) {\ytext};
		        }
		        \draw[-latex] (p2) -- (a);
		        \draw[-latex] (p2) -- (b);
		        \draw[-latex] (p1) -- (b);
		        \draw[-latex] (p1) -- (c);
		        \draw[-latex] (p5) -- (c);
		        \draw[-latex] (p5) -- (d);
		        \draw[-latex] (p4) -- (d);
		        \draw[-latex] (p4) -- (e);
		        \draw[-latex] (p3) -- (e);
		        \draw[-latex] (p3) -- (a);
			\end{tikzpicture}
			}
		\end{wrapfigure}

		$ \quad $ By repeating this argument, we conclude that in the steps $ a, b, c, d, e $ five choices $ C_1, \ldots, C_5 $ are reduced as depicted on the right, where \eg the step $ a $ reduces the choices $ C_1 $ and $ C_2 $.
		Since \scmpst does not allow for mixed choices, each of the five choices $ C_1, \ldots, C_5 $ contains either only outputs or only inputs.
		To enable communication, a choice with outputs has to interact with a choice on inputs.
		Without loss of generality, assume that $ C_1 $ contains only outputs and $ C_2 $ only inputs.
		Then the choice $ C_3 $ needs to be on outputs again, because step $ b $ reduces $ C_2 $ (with only inputs) and $ C_3 $.
		By repeating this argument, then $ C_4 $ is on inputs and $ C_5 $ is on outputs.
		But then step $ e $ reduces two choices $ C_1 $ and $ C_5 $ that are both on outputs.
		Since the reduction semantics of \scmpst does not allow such a step, this is a contradiction.
	\end{minipage}

	We conclude that there are no \patternStar in the \scmpst-calculus.
\end{proof}

A \patternStar is a circle of steps of odd degree, where neighbouring steps are in conflict.
More precisely, the star with five points in \patternStar is the smallest odd cycle of steps, where neighbouring steps are in conflict and that contains non-neighbouring distributable steps.
The proof shows that the limitation of choice to a single kind of action prefixes and the requirement of the semantics that outputs have to interact with inputs forbids for \patternStar-structures in \scmpst.

That there are no \patternStar in \cnvms was shown in \citep{PY2022}.
This proof is very similar to the proof of Lemma~\ref{lem:noStarInScmpst}.
By calculus inclusion there are no \patternStar in \lcnvms.
Again a very similar proof can be used to show that there are also no \patternStar in \cnv.

\begin{newlemma}[No \patternStar in \cnv]
	\label{lem:noStarInCnv}
	There are no \patternStar in the \cnv-calculus.
\end{newlemma}

\begin{proof}
	Assume the contrary, \ie assume that there is a term $ \PS_{\cnv} $ in the \mpst-calculus that is a \patternStar.
	Then $ \PS_{\cnv} $ can perform at least five alternative reduction steps $ a, b, c, d, e $ such that neighbouring steps in the sequence $ a, b, c, d, e, a $ are pairwise in conflict and non-neighbouring steps are distributable.
	By the reduction semantics of \cnv given in \citep{CASAL202223}, step $ a $ reduces a conditional, performs a communication by reducing an output and an input, or reduces a select and a branching.
	Since steps reducing a conditional cannot be in conflict with any other step, none of the steps in $ \left\{a, b, c, d, e\right\} $ reduces a conditional.
	Since communication steps cannot be in conflict with steps that resolve branching, either all steps in $ \left\{a, b, c, d, e\right\} $ are communication steps or all steps in $ \left\{a, b, c, d, e\right\} $ resolve branching.
	\begin{description}
		\item[Communication:] Assume that all steps in $ \left\{a, b, c, d, e\right\} $ are communication steps that reduce an output and an input.
			Because of the conflict between $ a $ and $ b $, these two steps reduce the same capability but this capability is not reduced in $ c $, because $ a $ and $ c $ are distributable.

\smallskip
			\begin{minipage}{\textwidth-1.9em}
				\begin{wrapfigure}{R}{0.25\textwidth}
					\centering
					\vspace{-1em}
					\scalebox{0.8}{
					\begin{tikzpicture}[]
						\foreach \x/\xlabel/\xtext/\ytext in {1/e/$ C_5 $/$ b $,2/d/$ C_4 $/$ a $,3/c/$ C_3 $/$ e $,4/b/$ C_2 $/$ d $,5/a/$ C_1 $/$ c $}
				        {
				            \path (360*\x/5+125:0.8) node (\xlabel) {\xtext};
				            \path (360*\x/5-55:1.75) node (p\x) {\ytext};
				        }
				        \draw[-latex] (p2) -- (a);
				        \draw[-latex] (p2) -- (b);
				        \draw[-latex] (p1) -- (b);
				        \draw[-latex] (p1) -- (c);
				        \draw[-latex] (p5) -- (c);
				        \draw[-latex] (p5) -- (d);
				        \draw[-latex] (p4) -- (d);
				        \draw[-latex] (p4) -- (e);
				        \draw[-latex] (p3) -- (e);
				        \draw[-latex] (p3) -- (a);
					\end{tikzpicture}
					}
				\end{wrapfigure}

				\quad By repeating this argument, we conclude that in the steps $ a, b, c, d, e $ five capabilities $ C_1, \ldots, C_5 $ are reduced as depicted on the right, where \eg the step $ a $ reduces the capabilities $ C_1 $ and $ C_2 $.
				All of these five capabilities are outputs or inputs.
				To enable communication, an output has to interact with an input.
				Without loss of generality, assume that $ C_1 $ is an output and $ C_2 $ is an input.
				Then $ C_3 $ needs to be an output again, because step $ b $ reduces $ C_2 $ (an input) and $ C_3 $.
				By repeating this argument, then $ C_4 $ is an input and $ C_5 $ is an output.
				But then step $ e $ reduces two outputs $ C_1 $ and $ C_5 $.
				Since the reduction semantics in \citep{CASAL202223} of \cnv does not allow such a step, this is a contradiction.
			\end{minipage}
\smallskip

		\item[Branching:] The case of branching is very similar to the case before, where selection replaces the output and the branching replaces the input.
			Again it is not possible to close the cycle, since selection has to interact with branching.
	\end{description}
	We conclude that there are no \patternStar in the \cnv-calculus.
\end{proof}

By calculus inclusion then there are no \patternStar in \lcnv.
We summarise and show that there are no \patternStar in any of the calculi in the bottom or middle layer of Figure~\ref{fig:hierarchy}.

\lemNoStar*

\begin{proof}
	By Lemma~4.4 in \citep{PY2022}, there are no \patternStar in \cnvms.
	Since $ \text{\lcnvms} \subset \text{\cnvms} $, then there are no \patternStar in \lcnvms.
	By Lemma~\ref{lem:noStarInCnv}, there are no \patternStar in \cnv.
	Since $ \lcnv \subset \cnv $, then there are no \patternStar in \lcnv.
	By Lemma~\ref{lem:noStarInScmpst}, there are no \patternStar in \scmpst.
	By Lemma~\ref{lem:noStarInMcbs}, there are no \patternStar in \mcbs.
	Since $ \bs \subset \scbs \subset \mcbs $, then there are no \patternStar in \bs and \scbs.
	By Lemma~\ref{lem:noStarInDmpst}, there are no \patternStar in \dmpst.
	Since $ \mpst \subset \smpst \subset \dmpst $, then there are no \patternStar in \mpst and \smpst.
\end{proof}

\subsection{Separation via the Synchronisation Pattern M}
\label{app:separationM}

In \S~\ref{app:synchronisationPatterns} we give an \patternM in \scmpst (see Example~\ref{ex:MinScmpst}), whereas \eg \dmpst does not contain \patternM{s} (see Lemma~\ref{lem:noM}).
Accordingly, we use $ \PM_{\scmpst} $ of Example~\ref{ex:MinScmpst} as counterexample to prove that there is no good encoding from \scmpst into \dmpst.
Synchronisation pattern allow to capture important features of a calculus in an abstract way.
Thereby, they allow to generalise separation results with these synchronisation pattern as counterexample.
We will use this to not only separate \scmpst from \dmpst but generalise, to separate more calculi of the middle from the bottom layer of Figure~\ref{fig:hierarchy}.
We start by proving that the conflict in the source term $ \PM_{\scmpst} $ of Example~\ref{ex:MinScmpst} has to be split up in the corresponding emulations.

\begin{lemma}
	\label{lem:translateConflictsM}
	Any good encoding $ \arbitraryEncoding $ from the \scmpst-calculus into a target calculus from the set $ \left\{ \text{\dmpst, \mcbs, \lcnv, \lcnvms} \right\} $ has to split up the conflict in $ \PM_{\scmpst} $ given in Example~\ref{ex:MinScmpst} of $ b $ with $ a $ and $ c $ such that there exists a maximal execution of $ \ArbitraryEncoding{\PM_{\scmpst}} $ in which $ a $ is emulated but not $ c $ as well as a maximal execution in which $ c $ is emulated but not $ a $.
\end{lemma}

\begin{proof}
	By operational completeness, all three steps of the term $ \PM_{\scmpst} $ have to be emulated in $ \ArbitraryEncoding{\PM_{\scmpst}} $, \ie there exist some $ N_a, N_b, N_c $ in the considered target calculus such that $ \ArbitraryEncoding{\PM_{\scmpst}} \red^\ast N_x \asymp \ArbitraryEncoding{\PM_{\scmpst, x}} $ for all $ x \in \left\{ a, b, c \right\} $.
	Because $ \PM_{\scmpst} $ is convergent and because of divergence reflection, $ \ArbitraryEncoding{\PM_{\scmpst}} $ is convergent.
	By success sensitiveness and since $ \asymp $ respects success, $ N_a \reachSuccessFin $, $ N_b \not\reachSuccess $, $ N_c \reachSuccessFin $ and thus $ N_a \not\asymp N_b \not\asymp N_c $.
	We conclude that for all $ N_a, N_b, N_c $ such that $ N_a \asymp \ArbitraryEncoding{\PM_{\scmpst, a}} $, $ N_b \asymp \ArbitraryEncoding{\PM_{\scmpst, b}} $, and $ N_c \asymp \ArbitraryEncoding{\PM_{\scmpst, c}} $ and for all sequences $ A: \ArbitraryEncoding{\PM_{\scmpst}} \red^\ast N_a $, $ B: \ArbitraryEncoding{\PM_{\scmpst}} \red^\ast N_b $, and $ C: \ArbitraryEncoding{\PM_{\scmpst}} \red^\ast N_c $, there is a conflict between a step of $ A $ and a step of $ B $, and there is a conflict between a step of $ B $ and a step of $ C $.
	Note that, since $ N_b \not\reachSuccess $ but $ N_a \reachSuccessFin $ and $ N_c \reachSuccessFin $, the conflict between $ a $ and $ b $ (or $ b $ and $ c $) has to be translated into a conflict of $ A $ and $ B $ (or $ B $ and $ C $).
	It is not possible, that the emulation of $ b $ disables all ways to reach success after $ N_a $ or $ N_c $ is reached.

	Because $ \arbitraryEncoding $ preserves distributability and because of Lemma~\ref{lem:distributabilityPreservation}, the distributable steps $ a $ and $ c $ of $ \PM_{\scmpst} $ have to be translated into distributable executions, \ie there is at least one $ A $ and one $ C $ such that these two executions are distributable.
	By Lemma~\ref{lem:distributabilityReductionsVsProcesses}, this implies that $ \ArbitraryEncoding{\PM_{\scmpst}} $ is distributable into $ N_1, N_2 $ such that $ A $ is an execution of $ N_1 $ and $ C $ is an execution of $ N_2 $.
	If a single step of $ B $ is in conflict with $ A $ as well as $ C $ then this step is an interaction between $ N_1 $ and $ N_2 $.
	Here we have to consider the different target calculi separately.
	\begin{description}
		\item[\dmpst:] Then both conflicts are ruled out by a communication step reducing two choices, \ie $ A $ and $ B $ compete for one of these choices and $ B $ and $ C $ compete for the respective other choice.
			Without loss of generality, assume that this step reduces a choice of participant $ \p $ in $ A $ and a choice of participant $ \q $ is reduced in $ C $ (compare to Lemma~\ref{lem:noMinDmpst}).
			Since choice in \dmpst is limited to a single participant and since single participants cannot contain parallel composition, then also the step in $ A $ and the step in $ C $ reduce the same choices of $ \p $ and $ \q $.
			But then the steps of $ A $ and $ C $ are in conflict and cannot be distributable, \ie $ A $ and $ C $ are not distributable.
			This is a contradiction.
		\item[\mcbs:] Then both conflicts are ruled out by a communication step reducing two choices, \ie $ A $ and $ B $ compete for one of these choices and $ B $ and $ C $ compete for the respective other choice.
			Without loss of generality, assume that this step reduces a choice of participant $ \p $ in $ A $ and a choice of participant $ \q $ is reduced in $ C $ (compare to Lemma~\ref{lem:noMinMcbs}).
			Since \mcbs is limited to binary sessions and since single participants cannot contain parallel composition, then also the step in $ A $ and the step in $ C $ reduce the same choices of $ \p $ and $ \q $.
			But then the steps of $ A $ and $ C $ are in conflict and cannot be distributable, \ie $ A $ and $ C $ are not distributable.
			This is a contradiction.
		\item[\lcnvms:] Then both conflicts are ruled out by a communication step reducing two choices, \ie $ A $ and $ B $ compete for one of these choices and $ B $ and $ C $ compete for the respective other choice.
			Without loss of generality, assume that this step reduces a choice of session endpoint $ x $ in $ A $ and a choice of session-endpoint $ y $ is reduced in $ C $ (compare to Lemma~\ref{lem:noMinLcnvms}).
			Since in \lcnvms there is at most one unguarded capability per session endpoint, then also the step in $ A $ and the step in $ C $ reduce the same choices of $ x $ and $ y $.
			But then the steps of $ A $ and $ C $ are in conflict and cannot be distributable, \ie $ A $ and $ C $ are not distributable.
			This is a contradiction.
		\item[\lcnv:] Then both conflicts are ruled out by reducing a selection and a branching construct, \ie $ A $ and $ B $ compete for one of these capabilities and $ B $ and $ C $ compete for the respective other capability.
			Without loss of generality, assume that this step reduces a selection of session endpoint $ x $ in $ A $ and a branching of session-endpoint $ y $ is reduced in $ C $ (compare to Lemma~\ref{lem:noMinLcnv}).
			Since in \lcnv there is at most one unguarded capability per session endpoint, then also the step in $ A $ and the step in $ C $ reduce the same capabilities of $ x $ and $ y $.
			But then the steps of $ A $ and $ C $ are in conflict and cannot be distributable, \ie $ A $ and $ C $ are not distributable.
			This is a contradiction.
	\end{description}
	Since we reach a contradiction in all cases, the conflict of $ B $ with $ A $ and $ C $ cannot be ruled out in a single step.
	Moreover, the reduction steps of $ A $ that lead to the conflicting step with $ B $ and the reduction steps of $ C $ that lead to the conflicting step with $ B $ are distributable, because $ A $ and $ C $ are distributable.
	We conclude that there is at least one emulation of $ b $, \ie one execution $ B: \ArbitraryEncoding{\PM_{\scmpst}} \red^\ast N_b \asymp \ArbitraryEncoding{\PM_{\scmpst, b}} $, starting with two distributable executions such that one is (in its last step) in conflict with the emulation of $ a $ in $ A: \ArbitraryEncoding{\PM_{\scmpst}} \red^\ast N_a \asymp \ArbitraryEncoding{\PM_{\scmpst, a}} $ and the other is in conflict with the emulation of $ c $ in $ C: \ArbitraryEncoding{\PM_{\scmpst}} \red^\ast N_c \asymp \ArbitraryEncoding{\PM_{\scmpst, c}} $.
	In particular this means that also the two steps of $ B $ that are in conflict with a step of $ A $ and a step of $ C $ are distributable.

	Hence, there is no possibility to ensure that these conflicts are decided consistently, \ie there is a maximal execution of $ \ArbitraryEncoding{\PM_{\scmpst}} $ that emulates $ a $ but not $ c $ as well as a maximal execution of $ \ArbitraryEncoding{\PM_{\scmpst}} $ that emulates $ c $ but not $ a $.
\end{proof}

We observe that the only information about $ \PM_{\scmpst} $ used in the proof of Lemma~\ref{lem:translateConflictsM} is that it is an \patternM, \ie about its conflicting and distributable steps, and that after step $ a $ and after step $ c $ success is reached in all maximal executions, whereas after step $ b $ success can not be reached.
$ \PM_{\cnv} $ in Example~\ref{ex:MinCnv} has the same properties.
Accordingly, we can repeat the proof of Lemma~\ref{lem:translateConflictsM} for $ \PM_{\cnv} $ instead of $ \PM_{\scmpst} $.

\begin{corollary}
	\label{cor:translateConflictsMcnv}
	Any good encoding $ \arbitraryEncoding $ from the \cnv-calculus into a target calculus from the set $ \left\{ \text{\dmpst, \mcbs, \lcnv, \lcnvms} \right\} $ has to split up the conflict in $ \PM_{\cnv} $ given in Example~\ref{ex:MinCnv} of $ b $ with $ a $ and $ c $ such that there exists a maximal execution of $ \ArbitraryEncoding{\PM_{\cnv}} $ in which $ a $ is emulated but not $ c $ as well as a maximal execution in which $ c $ is emulated but not $ a $.
\end{corollary}

Similarly, we can repeat the proof of Lemma~\ref{lem:translateConflictsM} for $ \PM_{\text{\cnvms}} $ instead of $ \PM_{\scmpst} $.

\begin{corollary}
	\label{cor:translateConflictsMcnvms}
	Any good encoding $ \arbitraryEncoding $ from the \cnvms-calculus into a target calculus from the set $ \left\{ \text{\dmpst, \mcbs, \lcnv, \lcnvms} \right\} $ has to split up the conflict in $ \PM_{\text{\cnvms}} $ given in Example~\ref{ex:MinCnvms} of $ b $ with $ a $ and $ c $ such that there exists a maximal execution of $ \ArbitraryEncoding{\PM_{\text{\cnvms}}} $ in which $ a $ is emulated but not $ c $ as well as a maximal execution in which $ c $ is emulated but not $ a $.
\end{corollary}

With Lemma~\ref{lem:translateConflictsM} we prove that there is no good encoding from \scmpst into \dmpst.
Again we immediately generalise to more target calculi.

\begin{lemma}[Separation Via \patternM, \scmpst]
	\label{lem:separateViaM}
	There is no good encoding from the \scmpst-calculus into a target calculus $ L $, where $ L $ is from the set $ \left\{ \text{\dmpst, \mcbs, \lcnv, \lcnvms} \right\} $.
\end{lemma}

\begin{proof}
	Assume the opposite, \ie assume that there is a good encoding $ \arbitraryEncoding $ from the \scmpst-calculus into a target calculus from the set $ \left\{ \text{\dmpst, \mcbs, \lcnv, \lcnvms} \right\} $, and, thus, also of $ \PM_{\scmpst} $ given by Example~\ref{ex:MinScmpst}.
	By Lemma~\ref{lem:translateConflictsM} there exists a maximal execution in $ \ArbitraryEncoding{\PM_{\scmpst}} $ in which $ a $ but not $ c $ is emulated or vice versa.
	Since $ \PM_{\scmpst, a} \reachSuccessFin $ and $ \PM_{\scmpst, c} \reachSuccessFin $ and because of success sensitiveness, the corresponding emulation leads to success.
	So there is an execution such that the emulation of $ a $ leads to success without the emulation of $ c $ or vice versa.
	Let us assume that $ a $ but not $ c $ is emulated.
	The other case is similar.

	By compositionality, the translation of $ \pa{\enodei{a}} \ssout{\enodei{b}}{l}. \ssout{\enodei{d}}{l}. \inact \Par \pa{\enodei{b}} \left( \ssinp{\enodei{a}}{l}. \inact \OR \ssinp{\enodei{c}}{l}. \inact \right) $ is independent of the translation of $ \pa{\enodei{c}} \left( \ssout{\enodei{b}}{l}. \inact \OR \ssout{\enodei{d}}{l}. \inact \right) \Par \pa{\enodei{d}} \ssinp{\enodei{c}}{l}. \ssinp{\enodei{a}}{l}. \success $.
	Let $ \PM_{\scmpst, \text{no } \success} = \pa{\enodei{a}} \ssout{\enodei{b}}{l}. \ssout{\enodei{d}}{l}. \inact \Par \pa{\enodei{b}} \left( \ssinp{\enodei{a}}{l}. \inact \OR \ssinp{\enodei{c}}{l}. \inact \right) \Par \pa{\enodei{c}} \left( \ssout{\enodei{b}}{l}. \inact \OR \ssout{\enodei{d}}{l}. \inact \right) \Par \pa{\enodei{d}} \ssinp{\enodei{c}}{l}. \ssinp{\enodei{a}}{l}. \inact $, \ie $ \PM_{\scmpst} $ and $ \PM_{\scmpst, \text{no } \success} $ differ only in their respective last symbol that is $ \success $ in the former and $ \inact $ in the latter case.
	Since $ \success $ and $ \inact $ have no free names, compositionality ensures that the translation of $ \ArbitraryEncoding{\PM_{\scmpst, \text{no } \success}} $ and $ \ArbitraryEncoding{\PM_{\scmpst}} $ differ only in the encoding of $ \success $.
	We conclude that, since $ \ArbitraryEncoding{\PM_{\scmpst}} $ reaches some $ N \reachSuccessFin $ without the emulation of $ c $, then $ \ArbitraryEncoding{\PM_{\scmpst, \text{no } \success}} $ reaches at least some state $ N' $ such that $ N' \reachSuccess $.
	Hence, $ \ArbitraryEncoding{\PM_{\scmpst, \text{no } \success}} \reachSuccess $ but $ \PM_{\scmpst, \text{no } \success} \not\reachSuccess $, which contradicts success sensitiveness.
	We conclude that there cannot be such an encoding.
\end{proof}

By calculus inclusion, then there is also no good encoding from \scmpst into \bs, \scbs, \mpst, or \smpst.

With Corollary~\ref{cor:translateConflictsMcnv} we prove that there is no good encoding from \cnv into \eg \dmpst, by adapting the proof of Lemma~\ref{lem:separateViaM} to the counterexample $ \PM_{\cnv} $.

\begin{lemma}[Separation Via \patternM, \cnv]
	\label{lem:separateViaMcmv}
	There is no good encoding from the \cnv-calculus into a target calculus $ L $, where $ L $ is from the set $ \left\{ \text{\dmpst, \mcbs, \lcnv, \lcnvms} \right\} $.
\end{lemma}

\begin{proof}
	Assume the opposite, \ie assume that there is a good encoding $ \arbitraryEncoding $ from the \cnv-calculus into a target calculus from the set $ \left\{ \text{\dmpst, \mcbs, \lcnv, \lcnvms} \right\} $, and, thus, also of $ \PM_{\cnv} $ given by Example~\ref{ex:MinCnv}.
	By Lemma~\ref{cor:translateConflictsMcnv} there exists a maximal execution in $ \ArbitraryEncoding{\PM_{\cnv}} $ in which $ a $ but not $ c $ is emulated or vice versa.
	Since $ \PM_{\cnv, a} \reachSuccessFin $ and $ \PM_{\cnv, c} \reachSuccessFin $ and because of success sensitiveness, the corresponding emulation leads to success.
	So there is an execution such that the emulation of $ a $ leads to success without the emulation of $ c $ or vice versa.
	Let us assume that $ a $ but not $ c $ is emulated.
	The other case is similar.

	The translation of $ \OutCMV{x}{\true}{\inactCMV} \Par \InpCMV{\linCMV}{y}{z}{\ConditionalCMV{z}{\inact}{\inactCMV}} $ is independent of the translation of $ \OutCMV{x}{\false}{\inactCMV} \Par \InpCMV{\linCMV}{y}{z}{\ConditionalCMV{z}{\inactCMV}{\success}} $, because of compositionality.
	Let $ \PM_{\cnv, \text{no } \success} = \ResCMV{x}{y}{\left( \OutCMV{x}{\true}{\inactCMV} \Par \InpCMV{\linCMV}{y}{z}{\ConditionalCMV{z}{\inact}{\inactCMV}} \Par \OutCMV{x}{\false}{\inactCMV} \Par \InpCMV{\linCMV}{y}{z}{\ConditionalCMV{z}{\inactCMV}{\inact}} \right)} $, \ie $ \PM_{\cnv} $ and $ \PM_{\cnv, \text{no } \success} $ differ only in their respective last symbol that is $ \success $ in the former and $ \inact $ in the latter case.
	Since $ \success $ and $ \inact $ have no free names, compositionality ensures that the translation of $ \ArbitraryEncoding{\PM_{\cnv, \text{no } \success}} $ and $ \ArbitraryEncoding{\PM_{\cnv}} $ differ only in the encoding of $ \success $.
	We conclude that, since $ \ArbitraryEncoding{\PM_{\cnv}} $ reaches some $ N \reachSuccessFin $ without the emulation of $ c $, then $ \ArbitraryEncoding{\PM_{\cnv, \text{no } \success}} $ reaches at least some state $ N' $ such that $ N' \reachSuccess $.
	Hence, $ \ArbitraryEncoding{\PM_{\cnv, \text{no } \success}} \reachSuccess $ but $ \PM_{\cnv, \text{no } \success} \not\reachSuccess $, which contradicts success sensitiveness.
	We conclude that there cannot be such an encoding.
\end{proof}

By calculus inclusion, then there is also no good encoding from \cnv into \bs, \scbs, \mpst, or \smpst.

With Corollary~\ref{cor:translateConflictsMcnvms} we prove that there is no good encoding from \cnv into \eg \dmpst, by adapting the proof of Lemma~\ref{lem:separateViaM} to the counterexample $ \PM_{\text{\cnvms}} $.

\begin{lemma}[Separation Via \patternM, \cnvms]
	\label{lem:separateViaMcmvms}
	There is no good encoding from the \cnvms-calculus into a target calculus $ L $, where $ L $ is from the set $ \left\{ \text{\dmpst, \mcbs, \lcnv, \lcnvms} \right\} $.
\end{lemma}

\begin{proof}
	Assume the opposite, \ie assume that there is a good encoding $ \arbitraryEncoding $ from the \cnvms-calculus into a target calculus from the set $ \left\{ \text{\dmpst, \mcbs, \lcnv, \lcnvms} \right\} $, and, thus, also of $ \PM_{\text{\cnvms}} $ given by Example~\ref{ex:MinCnvms}.
	By Lemma~\ref{cor:translateConflictsMcnvms} there exists a maximal execution in $ \ArbitraryEncoding{\PM_{\text{\cnvms}}} $ in which $ a $ but not $ c $ is emulated or vice versa.
	Since $ \PM_{\text{\cnvms}, a} \reachSuccessFin $ and $ \PM_{\text{\cnvms}, c} \reachSuccessFin $ and because of success sensitiveness, the corresponding emulation leads to success.
	So there is an execution such that the emulation of $ a $ leads to success without the emulation of $ c $ or vice versa.
	Let us assume that $ a $ but not $ c $ is emulated.
	The other case is similar.

	The translation of $ \ChoiceCMVmix{\linCMV}{x}{\OutCMVmix{l}{\true}{\inactCMVmix}} \Par \ChoiceCMVmix{\linCMV}{y}{\InpCMVmix{l}{z}{\ConditionalCMVmix{z}{\inact}{\inactCMVmix}}} $ is independent of the translation of $ \ChoiceCMVmix{\linCMV}{x}{\OutCMVmix{l}{\false}{\inactCMVmix}} \Par \ChoiceCMVmix{\linCMV}{y}{\InpCMVmix{l}{z}{\ConditionalCMVmix{z}{\inactCMVmix}{\success}}} $, because of compositionality.
	Let $ \PM_{\text{\cnvms}, \text{no } \success} = \ResCMVmix{x}{y}{\left( \ChoiceCMVmix{\linCMV}{x}{\OutCMVmix{l}{\true}{\inactCMVmix}} \Par \ChoiceCMVmix{\linCMV}{y}{\InpCMVmix{l}{z}{\ConditionalCMVmix{z}{\inact}{\inactCMVmix}}} \Par \ChoiceCMVmix{\linCMV}{x}{\OutCMVmix{l}{\false}{\inactCMVmix}} \Par \ChoiceCMVmix{\linCMV}{y}{\InpCMVmix{l}{z}{\ConditionalCMVmix{z}{\inactCMVmix}{\inact}}} \right)} $, \ie $ \PM_{\text{\cnvms}} $ and $ \PM_{\text{\cnvms}, \text{no } \success} $ differ only in their respective last symbol that is $ \success $ in the former and $ \inact $ in the latter case.
	Since $ \success $ and $ \inact $ have no free names, compositionality ensures that the translation of $ \ArbitraryEncoding{\PM_{\text{\cnvms}, \text{no } \success}} $ and $ \ArbitraryEncoding{\PM_{\text{\cnvms}}} $ differ only in the encoding of $ \success $.
	We conclude that, since $ \ArbitraryEncoding{\PM_{\text{\cnvms}}} $ reaches some $ N \reachSuccessFin $ without the emulation of $ c $, then $ \ArbitraryEncoding{\PM_{\text{\cnvms}, \text{no } \success}} $ reaches at least some state $ N' $ such that $ N' \reachSuccess $.
	Hence, $ \ArbitraryEncoding{\PM_{\text{\cnvms}, \text{no } \success}} \reachSuccess $ but $ \PM_{\text{\cnvms}, \text{no } \success} \not\reachSuccess $, which contradicts success sensitiveness.
	We conclude that there cannot be such an encoding.
\end{proof}

By calculus inclusion, then there is also no good encoding from \cnvms into \bs, \scbs, \mpst, or \smpst.

We summarise and show that there are no good encodings from a calculus of the middle layer into a calculus of the bottom layer.

\thmSeparateSCMP*

\begin{proof}
	By Lemma~\ref{lem:separateViaM}, there is no good encoding from \scmpst into any $ L_2 \in \Set{\lcnv, \text{\lcnvms}, \mcbs, \dmpst} $.
	Since $ \bs \subset \scbs \subset \mcbs $ and $ \mpst \subset \smpst \subset \dmpst $, then there is no good encoding from \scmpst into any $ L_2 \in \Set{\bs, \scbs, \mpst, \smpst} $.
	\begin{quote}
		Consider \eg $ L_2 = \bs $.
		Assume by contradiction, that there is a good encoding from \scmpst into \bs.
		Since $ \bs \subset \mcbs $, then for every source term $ M $ in \scmpst the translation $ \ArbitraryEncoding{M} $ is not only in \bs but also in \mcbs.
		Then this encoding is also a good encoding from \scmpst into \mcbs.
		This contradicts Lemma~\ref{lem:separateViaM}.
		The argumentation for \scbs, \mpst, and \smpst is similar.
	\end{quote}
	By Lemma~\ref{lem:separateViaMcmv}, there is no good encoding from \cnv into any $ L_2 \in \Set{\lcnv, \text{\lcnvms}, \mcbs, \dmpst} $.
	Since $ \bs \subset \scbs \subset \mcbs $ and $ \mpst \subset \smpst \subset \dmpst $, then there is no good encoding from \cnv into any $ L_2 \in \Set{\bs, \scbs, \mpst, \smpst} $.
	By Lemma~\ref{lem:separateViaMcmvms}, there is no good encoding from \cnvms into any $ L_2 \in \Set{\lcnv, \text{\lcnvms}, \mcbs, \dmpst} $.
	Since $ \bs \subset \scbs \subset \mcbs $ and $ \mpst \subset \smpst \subset \dmpst $, then there is no good encoding from \cnvms into any $ L_2 \in \Set{\bs, \scbs, \mpst, \smpst} $.
\end{proof}

\subsection{Separation via the Synchronisation Pattern Star}
\label{app:separationStar}

In \S~\ref{app:synchronisationPatterns} we present a \patternStar in \msmpst (see Example~\ref{ex:StarInMsmpst}), whereas \eg \scmpst does not contain any \patternStar (see Lemma~\ref{lem:noStarInScmpst}).
Accordingly, we use $ \PS_{\msmpst} $ of Example~\ref{ex:StarInMsmpst} as counterexample to prove that there is no good encoding from \msmpst into \scmpst.
Again we generalise this result to show more separation results between the top layer and the lower layers of Figure~\ref{fig:hierarchy}.
We show first that the conflicts in the source term $ \PS_{\msmpst} $ of Example~\ref{ex:StarInMsmpst} have to be translated into conflicts of the corresponding emulations.

\begin{lemma}
	\label{lem:translateConflictsStar}
	Any good encoding $ \arbitraryEncoding $ from the \msmpst-calculus into some arbitrary target calculus has to translate the conflicts in $ \PS_{\msmpst} $ given in Example~\ref{ex:StarInMsmpst} into conflicts of the corresponding emulations.
\end{lemma}

\begin{proof}
	By operational completeness, all five steps of $ \PS_{\msmpst} $ have to be emulated in $ \ArbitraryEncoding{\PS_{\msmpst}} $, \ie there exist some $ N_a, N_b, N_c, N_d, N_e $ in the target calculus such that $ \ArbitraryEncoding{\PS_{\msmpst}} \red^\ast N_x \asymp \ArbitraryEncoding{\PS_{\msmpst, x}} $ for all $ x \in \left\{ a, b, c, d, e \right\} $.
	Because $ \PS_{\msmpst} $ is convergent and because of divergence reflection, $ \ArbitraryEncoding{\PS_{\msmpst}} $ is convergent.
	Because $ \arbitraryEncoding $ preserves distributability, for each pair of steps $ x $ and $ y $ that are distributable in $ \PS_{\msmpst} $, the emulations $ X: \ArbitraryEncoding{\PS_{\msmpst}} \red^\ast N_x $ and $ Y: \ArbitraryEncoding{\PS_{\msmpst}} \red^\ast N_y $ such that $ N_x \asymp \ArbitraryEncoding{\PS_{\msmpst, x}} $ and $ N_y \asymp \ArbitraryEncoding{\PS_{\msmpst, y}} $ are distributable.
	Note that $ X $ and $ Y $ refer to the upper case variants of $ x $ and $ y $, respectively.

	In Example~\ref{ex:StarInMsmpst} we do not fix $ M_{\enodei{a}}', \ldots, M_{\enodei{e}}' $ in $ \PS_{\msmpst} $.
	Instead we consider all variants of $ \PS_{\msmpst} $, where $ M_{\enodei{a}}', \ldots, M_{\enodei{e}}' \in \left\{\success, \inact\right\} $, \ie each of these terms is either chosen to be $ \inact $ or as an unguarded occurrence of $ \success $.
	Because of compositionality, the encodings of these variants of $ \PS_{\msmpst} $ differ only by the encodings of $ M_{\enodei{a}}', \ldots, M_{\enodei{e}}' $.
	The remaining operators and, hence, the remaining term has to be translated in exactly the same way.

	Consider each triple of steps $ x, y, z \in \left\{ a, b, c, d, e \right\} $ in $ \PS_{\msmpst} $ such that $ y $ is in conflict with $ x $ and $ z $ but $ x $ and $ z $ are parallel.
	Then, choose $ M_{\enodei{x}}' = \success = M_{\enodei{z}}' $ and $ M_{\enodei{i}}' = \inact $ for all $ x \neq i \neq z $.
	Since $ M_{\enodei{x}}'\reachSuccessFin $, $ M_{\enodei{y}}'\not\reachSuccess $, and $ M_{\enodei{z}}'\reachSuccessFin $, also $ \PS_{\msmpst, x}\reachSuccessFin $, $ \PS_{\msmpst, y}\not\reachSuccess $, and $ \PS_{\msmpst, z}\reachSuccessFin $.
	Since $ \arbitraryEncoding $ as well as $ \asymp $ respect success, $ N_x\reachSuccessFin $, $ N_y\not\reachSuccess $, and $ N_z\reachSuccessFin $, and thus $ N_x \not\asymp N_y \not\asymp N_z $.
	We conclude that, for all $ N_x, N_y, N_z $ such that $ N_x \asymp \ArbitraryEncoding{\PS_{\msmpst, x}} $, $ N_y \asymp \ArbitraryEncoding{\PS_{\msmpst, y}} $, and $ N_z \asymp \ArbitraryEncoding{\PS_{\msmpst, z}} $ and for all sequences $ X : \ArbitraryEncoding{\PS_{\msmpst}} \red^\ast N_x $, $ Y : \ArbitraryEncoding{\PS_{\msmpst}} \red^\ast N_y $, and $ Z : \ArbitraryEncoding{\PS_{\msmpst}} \red^\ast N_z $, there is a conflict between a step of $ X $ and a step of $ Y $, and there is a conflict between a step of $ Y $ and a step of $ Z $.
\end{proof}

Then we show that each good encoding of the counterexample $ \PS_{\msmpst} $ has to distribute one of its conflicts.

\begin{lemma}
	\label{lem:distributeMixedChoice}
	Any good encoding $ \arbitraryEncoding $ from the \msmpst-calculus into a target calculus from the set $ \left\{ \text{\scmpst, \dmpst, \mcbs, \cnv, \cnvms} \right\} $ has to split up at least one of the conflicts in $ \PS_{\msmpst} $ given by Example~\ref{ex:StarInMsmpst} such that there exists a maximal execution in $ \ArbitraryEncoding{\PS_{\msmpst}} $ that emulates only one source term step.
\end{lemma}

\begin{proof}
	By operational completeness, all five steps of $ \PS_{\msmpst} $ have to be emulated in $ \ArbitraryEncoding{\PS_{\msmpst}} $, \ie there exist some $ N_a, N_b, N_c, N_d, N_e $ in the considered target calculus such that $ X: \ArbitraryEncoding{\PS_{\msmpst}} \red^\ast N_x \asymp \ArbitraryEncoding{\PS_{\msmpst, x}} $ for all $ x \in \left\{ a, b, c, d, e \right\} $, where $ X $ is the upper case variant of $ x $.
	Because $ \PS_{\msmpst} $ is convergent and because of divergence reflection, $ \ArbitraryEncoding{\PS_{\msmpst}} $ is convergent.
	By Lemma~\ref{lem:translateConflictsStar}, for all $ N_a, N_b, N_c, N_d, N_e $ and all $ x \in \left\{ a, b, c, d, e \right\} $ such that $ N_x \asymp \ArbitraryEncoding{\PS_{\msmpst, x}} $, there is a conflict between a step of the following pairs of emulations: $ A $ and $ B $, $ B $ and $ C $, $ C $ and $ D $, $ D $ and $ E $, and $ E $ and $ A $.

	Since $ \arbitraryEncoding $ preserves distributability and by Lemma~\ref{lem:distributabilityPreservation}, each pair of distributable steps in $ \PS_{\msmpst} $ has to be translated into emulations that are distributable within $ \ArbitraryEncoding{\PS_{\msmpst}} $.
	Let $ X, Y, Z \in \left\{ A, B, C, D, E \right\} $ be such that $ X $ and $ Z $ are distributable within $ \ArbitraryEncoding{\PS_{\msmpst}} $ but $ Y $ is in conflict with $ X $ as well as $ Z $.
	By Lemma~\ref{lem:distributabilityReductionsVsProcesses}, this implies that $ \ArbitraryEncoding{\PS_{\msmpst}} $ is distributable into $ N_1, N_2 $ such that $ X $ is an execution of $ N_1 $ and $ Z $ is an execution of $ N_2 $.
	Since $ Y $ is in conflict with $ X $ and $ Z $ and because all three emulations are executions of $ \ArbitraryEncoding{\PS_{\msmpst}} $, there is one step of $ Y $ that is in conflict with one step of $ X $ and there is one (possibly the same) step of $ Y $ that is in conflict with one step of $ Z $.
	Moreover, since $ X $ and $ Z $ are distributable, if a single step of $ Y $ is in conflict with $ X $ as well as $ Z $ then this step is an interaction between $ N_1 $ and $ N_2 $.

	Assume that for all such combinations $ X $, $ Y $, and $ Z $, the conflicts between $ Y $ and $ X $ or $ Z $ are ruled out by a single step of $ Y $.
	Here we have to consider the different target calculi separately.
	\begin{description}
		\item[\scmpst:] Then both conflicts are ruled out by a communication step between some choice of $ X $ and some choice of $ Z $.
			Then this step reduces one choice in one of the executions $ X $ and $ Z $ and a complementary choice in the respective other execution, \ie $ X $ and $ Y $ compete for one choice and $ Y $ and $ Z $ compete for the respective other choice, where one of these choices is on outputs and the other on inputs (compare to Lemma~\ref{lem:noStarInScmpst}).
			Without loss of generality let us assume that $ A $ and $ B $ compete for a choice on outputs and, thus, $ B $ and $ C $ compete for a choice on inputs, $ C $ and $ D $ compete for a choice on outputs, $ D $ and $ E $ compete for a choice on inputs, $ E $ and $ A $ compete for a choice on outputs, and $ A $ and $ B $ compete for a choice on inputs.
			This is a contradiction, because $ A $ and $ B $ cannot compete for both: a choice on outputs and a choice on inputs in a single step.
		\item[\dmpst:] Remember that $ X $ and $ Z $ are distributable.
			Hence, if a single step of $ Y $ rules out the conflict with $ X $ and $ Z $, then there are three alternative steps, one of each sequence $ X, Y, Z $, such that the step of $ Y $ is in conflict with the step of $ X $ and $ Z $ but the steps of $ X $ and $ Z $ are distributable.
			This is not possible in the \dmpst-calculus (compare to Lemma~\ref{lem:noMinDmpst}).
			Again, we reach a contradiction.
		\item[\mcbs:] Remember that $ X $ and $ Z $ are distributable.
			Hence, if a single step of $ Y $ rules out the conflict with $ X $ and $ Z $, then there are three alternative steps, one of each sequence $ X, Y, Z $, such that the step of $ Y $ is in conflict with the step of $ X $ and $ Z $ but the steps of $ X $ and $ Z $ are distributable.
			This is not possible in the \mcbs-calculus (compare to Lemma~\ref{lem:noMinMcbs}).
			Again, we reach a contradiction.
		\item[\cnvms:] Then both conflicts are ruled out by a communication step between some choice of $ X $ and some choice of $ Z $.
			Then this step reduces one choice on some session endpoint $ x $ in one of the executions $ X $ and $ Z $ and a choice on the respective other session endpoint $ y $ in the respective other execution, \ie $ X $ and $ Y $ compete for one choice on $ x $ and $ Y $ and $ Z $ compete a choice on $ y $.
			Without loss of generality let us assume that $ A $ and $ B $ compete for a choice on $ x $ and, thus, $ B $ and $ C $ compete for a choice on $ y $, $ C $ and $ D $ compete for a choice on $ x $, $ D $ and $ E $ compete for a choice on $ y $, $ E $ and $ A $ compete for a choice on $ x $, and $ A $ and $ B $ compete for a choice on $ y $.
			This is a contradiction, because $ A $ and $ B $ cannot compete for both: a choice on $ x $ and a choice on $ y $ in a single step.
		\item[\cnv:] Then both conflicts are ruled out either by a communication step between some capability of $ X $ and some capability of $ Z $.
			In \cnv either the capabilities are outputs and inputs or selection and branching (compare to Lemma~\ref{lem:noStarInCnv}).
			We consider the case of outputs and inputs.
			The case of selection and branching is similar with selection instead of outputs and branching instead of inputs.
			Then this step reduces one output or input in one of the executions $ X $ and $ Z $ and the respective other kind of capability on the respective other execution, \ie $ X $ and $ Y $ compete for one output or input and $ Y $ and $ Z $ compete for the respective other kind of capability.
			Without loss of generality let us assume that $ A $ and $ B $ compete for an output and, thus, $ B $ and $ C $ compete for an input, $ C $ and $ D $ compete for an output, $ D $ and $ E $ compete for an input, $ E $ and $ A $ compete for an output, and $ A $ and $ B $ compete for an input.
			This is a contradiction, because $ A $ and $ B $ cannot compete for both: an output and an input in a single step.
	\end{description}
	We reach a contradiction in each case.

	We conclude that there is at least one triple of emulations $ X $, $ Y $, and $ Z $ such that the conflict of $ Y $ with $ X $ and with $ Z $ results from two different steps in $ Y $.
	Because $ X $ and $ Z $ are distributable, the reduction steps of $ X $ that lead to the conflicting step with $ Y $ and the reduction steps of $ Z $ that lead to the conflicting step with $ Y $ are distributable.
	We conclude that there is at least one emulation of $ y $, \ie one execution $ Y: \ArbitraryEncoding{\PS_{\msmpst}} \red^\ast N_y \asymp \ArbitraryEncoding{\PS_{\msmpst, y}} $, starting with two distributable executions such that one is (in its last step) in conflict with the emulation of $ x $ in $ X: \ArbitraryEncoding{\PS_{\msmpst}} \red^\ast N_x \asymp \ArbitraryEncoding{\PS_{\msmpst, x}} $ and the other one is in conflict with the emulation of $ z $ in $ Z: \ArbitraryEncoding{\PS_{\msmpst}} \red^\ast N_z \asymp \ArbitraryEncoding{\PS_{\msmpst, z}} $.
	In particular this means that also the two steps of $ Y $ that are in conflict with a step in $ X $ and a step in $ Z $ are distributable.
	Hence, it is impossible to ensure that these two conflicts are decided consistently, \ie there is a maximal execution of $ \ArbitraryEncoding{\PS_{\msmpst}} $ that emulates $ X $ but neither $ Y $ nor $ Z $.

	In the set $ \left\{ A, B, C, D, E \right\} $ there are---apart from $ X $, $ Y $, and $ Z $---two remaining executions.
	One of them, say $ X' $, is in conflict with $ X $ and the other one, say $ Z' $, is in conflict with $ Z $.
	Since $ X $ is emulated successfully, $ X' $ cannot be emulated.
	Moreover, note that $ Y $ and $ Z' $ are distributable.
	Thus, also $ Z' $ and the partial execution of $ Y $ that leads to the conflict with $ Z $ are distributable.
	Moreover, also the step of $ Y $ that already rules out $ Z $ cannot be in conflict with a step of $ Z' $.
	Thus, although the successful completion of $ Z $ is already ruled out by the conflict with $ Y $, there is some step of $ Z $ left, that is in conflict with one step in $ Z' $.
	Hence, the conflict between $ Z $ and $ Z' $ cannot be ruled out by the partial execution described so far that leads to the emulation of $ X $ but forbids to complete the emulations of $ X' $, $ Y $, and $ Z $.
	Thus, it cannot be avoided that $ Z $ wins this conflict, \ie that also $ Z' $ cannot be completed.
	We conclude that there is a maximal execution of $ \ArbitraryEncoding{\PS_{\msmpst}} $ such that only one of the five source term steps of $ \PS_{\msmpst} $ is emulated.
\end{proof}

Since each maximal execution of $ \PS_{\msmpst} $ given by Example~\ref{ex:StarInMsmpst} consists of exactly two distributable steps, Lemma~\ref{lem:distributeMixedChoice} violates the requirements on a good encoding.

\begin{lemma}[Separation Via \patternStar]
	\label{lem:separateViaStar}
	There is no good encoding from the \msmpst-calculus into a target calculus $ L $, where $ L $ is from the set $ \left\{ \text{\scmpst, \dmpst, \mcbs, \cnv, \cnvms} \right\} $.
\end{lemma}

\begin{proof}
	Assume the opposite, \ie assume that there is a good encoding $ \arbitraryEncoding $ from the \msmpst-calculus into a target calculus from the set $ \left\{ \text{\scmpst, \mcbs, \cnv, \cnvms} \right\} $, and, thus, also of $ \PS_{\msmpst} $ given by Example~\ref{ex:StarInMsmpst}.
	By Lemma~\ref{lem:distributeMixedChoice} there exists a maximal execution in $ \ArbitraryEncoding{\PS_{\msmpst}} $ in which only one source term step is emulated.
	Let us denote this step by $ x \in \left\{ a, b, c, d, e \right\} $, \ie there is a maximal execution $ X: \ArbitraryEncoding{\PS_{\msmpst}} \red^\ast N_x \red^\ast \ldots $ with $ N_x \asymp \ArbitraryEncoding{\PS_{\msmpst, x}} $ in that only step $ x $ is emulated.
	Choose $ M_{\enodei{x}}' = \inact $ and $ M_{\enodei{i}}' = \success $ for all $ i \neq x $.
	Then $ M_{\enodei{x}}'\not\reachSuccess $ but, since $ \PS_{\msmpst} $ reaches success in all maximal executions, $ \PS_{\msmpst, x}\reachSuccessFin $.
	Because $ \arbitraryEncoding $ is operationally corresponding and respects success and because no other source term step is emulated, $ M_{\enodei{x}}' = \inact $ implies $ N_x\not\reachSuccess $.
	Moreover, since the encoding respects success, $ \PS_{\msmpst, x}\reachSuccessFin $ implies $ \ArbitraryEncoding{\PS_{\msmpst, x}}\reachSuccessFin $.
	Since $ N_x \asymp \ArbitraryEncoding{\PS_{\msmpst, x}} $, $ N_x\not\reachSuccess $ and $ \ArbitraryEncoding{\PS_{\msmpst, x}}\reachSuccessFin $ violate the assumption that $ \asymp $ respects success.
	We conclude that there cannot be such an encoding.
\end{proof}

By calculus inclusion, then there is also no good encoding from \mcmpst into \scmpst, \dmpst, \mcbs, \cnv, or \cnvms, since $ \PS_{\msmpst} $ is also a term of the \mcmpst-calculus.
We summarise and prove that there is no good encoding from any calculus of the top layer into any calculus of the lower layers of Figure~\ref{fig:hierarchy}.

\thmSeparateMCMP*

\begin{proof}
	By Lemma~\ref{lem:separateViaStar} there is no good encoding from the \msmpst-calculus into any $ L_2 \in \left\{ \text{\scmpst, \dmpst, \mcbs, \cnv, \cnvms} \right\} $.
	Since $ \msmpst \subset \mcmpst $, then there is no good encoding from the \mcmpst-calculus into any $ L_2 \in \left\{ \text{\scmpst, \dmpst, \mcbs, \cnv, \cnvms} \right\} $.
	Since $ \mpst \subset \smpst \subset \dmpst $, then there is no good encoding from any $ L_1 \in \Set{ \msmpst, \mcmpst } $ into any $ L_2 \in \Set{ \mpst, \smpst } $.
	Since $ \bs \subset \scbs \subset \mcbs $, then there is no good encoding from any $ L_1 \in \Set{ \msmpst, \mcmpst } $ into any $ L_2 \in \Set{ \bs, \scbs } $.
	Since $ \lcnv \subset \cnv $, then there is no good encoding from any $ L_1 \in \Set{ \msmpst, \mcmpst } $ into \lcnv.
	Since $ \text{\lcnvms} \subset \text{\cnvms} $, then there is no good encoding from any $ L_1 \in \Set{ \msmpst, \mcmpst } $ into \lcnvms.
\end{proof}

\subsection{Separating a Multiparty Session from a Binary Session}
\label{app:separateMultipartyFromBinary}

To separate a multiparty session with choice on a single participant (\mpst, \smpst, and \dmpst) from a binary session (\bs, \scbs, and \mcbs), we use $ M_{\mpst} = \pa{\role{a}}{\left( \ssout{\role{b}}{l}\seq\success \OR \ssout{\role{b}}{l}\seq\inact \right)} \Par \pa{\role{b}}{\ssinp{\role{a}}{l}\seq\inact} \Par \pa{\role{c}}{\left( \ssout{\role{d}}{l}\seq\success \OR \ssout{\role{d}}{l}\seq\inact \right)} \Par \pa{\role{d}}{\ssinp{\role{c}}{l}\seq\inact} \Par \pa{\role{e}}{\left( \ssout{\role{f}}{l}\seq\success \OR \ssout{\role{f}}{l}\seq\inact \right)} \Par \pa{\role{f}}{\ssinp{\role{e}}{l}\seq\inact} $ in Figure~\ref{fig:KeysExpressiveness}(4) as counterexample.

\begin{lemma}
	The session $ M_{\mpst} $ in Figure~\ref{fig:KeysExpressiveness}(4) is typed and deadlock-free.
\end{lemma}

\begin{proof}
	In $ M_{\mpst} $ in Figure~\ref{fig:KeysExpressiveness}(4) we omit the objects in communication, because they are not relevant for the protocol.
	Assume that in all communications $ \true $ is transmitted and the receiver stores the received value in the variable $ \xx $.
	By Figure~\ref{fig:typing}, we get
	\begin{align*}
		\types \ssout{\role{b}}{l}\seq\success \OR \ssout{\role{b}}{l}\seq\inact \as T_{\role{a}} &= \tout{\role{b}}{l}\tinact\\
		\types \ssinp{\role{a}}{l}\seq\inact \as T_{\role{b}} &= \tinp{\role{a}}{l}\tinact\\
		\types \ssout{\role{d}}{l}\seq\success \OR \ssout{\role{d}}{l}\seq\inact \as T_{\role{c}} &= \tout{\role{d}}{l}\tinact\\
		\types \ssinp{\role{c}}{l}\seq\inact \as T_{\role{d}} &= \tinp{\role{c}}{l}\tinact\\
		\types \ssout{\role{f}}{l}\seq\success \OR \ssout{\role{f}}{l}\seq\inact \as T_{\role{e}} &= \tout{\role{f}}{l}\tinact\\
		\types \ssinp{\role{e}}{l}\seq\inact \as T_{\role{f}} &= \tinp{\role{e}}{l}\tinact
	\end{align*}
	and the omitted payload type is $ \bool $ in all cases.
	Let $ \LL = \localise{\enodei{a}} T_{\enodei{a}}, \localise{\enodei{b}} T_{\enodei{b}}, \localise{\enodei{c}} T_{\enodei{c}}, \localise{\enodei{d}} T_{\enodei{d}}, \localise{\enodei{e}} T_{\enodei{e}}, \localise{\enodei{f}} T_{\enodei{f}} $.
	Since there is no recursion, $ M_{\mpst} $ has only finitely many maximal executions.
	The reductions of the local types follow closely the reductions of the protocol.
	For instance we have the sequence:
	\begin{align*}
		\LL & \red \localise{\enodei{a}} \tinact, \localise{\enodei{b}} \tinact, \localise{\enodei{c}} T_{\enodei{c}}, \localise{\enodei{d}} T_{\enodei{d}}, \localise{\enodei{e}} T_{\enodei{e}}, \localise{\enodei{f}} T_{\enodei{f}}\\
		& \red \localise{\enodei{a}} \tinact, \localise{\enodei{b}} \tinact, \localise{\enodei{c}} \tinact, \localise{\enodei{d}} \tinact, \localise{\enodei{e}} T_{\enodei{e}}, \localise{\enodei{f}} T_{\enodei{f}}\\
		& \red \localise{\enodei{a}} \tinact, \localise{\enodei{b}} \tinact, \localise{\enodei{c}} \tinact, \localise{\enodei{d}} \tinact, \localise{\enodei{e}} \tinact, \localise{\enodei{f}} \tinact
		\noRed
	\end{align*}
	By checking all sequences of reductions of the local types, we obtain $ \safe{\LL} $.
	By Figure~\ref{fig:typing}, then $ \types M_{\mpst} \as \LL $.
	Similarly by checking all sequences of reductions of local types, we see that $ \LL $ is deadlock-free, \ie $ \df{\LL} $.
	By Theorem~\ref{thm:deadlockfree}, then $ M_{\mpst} $ is deadlock-free.
\end{proof}

We prove first that there is no good encoding from \mpst---the smallest of the three multiparty calculi---into \mcbs---the largest of the three binary calculi.

\begin{lemma}[Separate \mpst from \mcbs]
	\label{lem:sepMpstFromMcbs}
	There is no good encoding from the \mpst-calculus into the \mcbs-calculus.
\end{lemma}

\begin{proof}
	Assume the opposite, \ie assume that there is a good encoding $ \arbitraryEncoding $ from \mpst into \mcbs and, thus, also of $ M_{\mpst} $ in Figure~\ref{fig:KeysExpressiveness}(4).
	$ M_{\mpst} $ can perform six alternative steps, \eg an interaction of $ \role{a} $ and $ \role{b} $ may unguard $ \success $ or $ \inact $.
	Two alternative steps on different participants are distributable.
	By operational completeness, all six steps of $ M_{\mpst} $ have to be emulated in $ \ArbitraryEncoding{M_{\mpst}} $.
	Since $ \arbitraryEncoding $ preserves distributability and by Lemma~\ref{lem:distributabilityPreservation}, every triple of distributable steps in $ M_{\mpst} $ has to be translated into three emulations that are pairwise distributable within $ \ArbitraryEncoding{M_{\mpst}} $.
	Then there are $ N_1, N_2, N_3 $ such that the emulations $ \ArbitraryEncoding{M_{\mpst}} \red^\ast N_1 $, $ \ArbitraryEncoding{M_{\mpst}} \red^\ast N_2 $, and $ \ArbitraryEncoding{M_{\mpst}} \red^\ast N_3 $ are pairwise distributable, where $ \ArbitraryEncoding{M_{\mpst}} \red^\ast N_1 $ is the emulation of a step on $ \role{a} $ and $ \role{b} $, $ \ArbitraryEncoding{M_{\mpst}} \red^\ast N_2 $ is the emulation of a step on $ \role{c} $ and $ \role{d} $, and $ \ArbitraryEncoding{M_{\mpst}} \red^\ast N_3 $ is the emulation of a step on $ \role{e} $ and $ \role{f} $.
	Then $ \ArbitraryEncoding{M_{\mpst}}\reachSuccess $, $ \ArbitraryEncoding{M_{\mpst}}\not\reachSuccessFin $, either $ N_1\hasSuccess $ or $ N_1\not\reachSuccess $, either $ N_2\hasSuccess $ or $ N_2\not\reachSuccess $, and either $ N_3\hasSuccess $ or $ N_3\not\reachSuccess $.
	By Lemma~\ref{lem:distributabilityReductionsVsProcesses}, then $ \ArbitraryEncoding{M_{\mpst}} $ is distributable into some $ N_{\role{ab}}, N_{\role{cd}}, N_{\role{ef}} $, \ie $ \ArbitraryEncoding{M_{\mpst}} \equiv N_{\role{ab}} \Par N_{\role{cd}} \Par N_{\role{ef}} $, such that $ N_{\role{ab}} \red^\ast N_{\role{ab}}' $ is the emulation of a step on $ \role{a} $ and $ \role{b} $, $ N_{\role{cd}} \red^\ast N_{\role{cd}}' $ is the emulation of a step on $ \role{c} $ and $ \role{d} $, and $ N_{\role{ef}} \red^\ast N_{\role{ef}}' $ is the emulation of a step on $ \role{e} $ and $ \role{f} $.
	Since $ \ArbitraryEncoding{M_{\mpst}}\reachSuccess $, $ \ArbitraryEncoding{M_{\mpst}}\not\reachSuccessFin $ and either $ N_1\hasSuccess $ or $ N_1\not\reachSuccess $, then $ N_{\role{ab}}\reachSuccess $, $ N_{\role{ab}}\not\reachSuccessFin $.
	Hence, the sequence $ N_{\role{ab}} \red^\ast N_{\role{ab}}' $ contains at least one interaction step, \ie a step reducing two choices.
	Similarly, $ N_{\role{cd}} \red^\ast N_{\role{cd}}' $ and $ N_{\role{ef}} \red^\ast N_{\role{ef}}' $ contain at least one step.
	But there is no such session $ \ArbitraryEncoding{M_{\mpst}} \equiv N_{\role{ab}} \Par N_{\role{cd}} \Par N_{\role{ef}} $ with three pairwise distributable steps in the target calculus \mcbs.
	This is a contradiction.
	We conclude that there cannot be such an encoding.
\end{proof}

The criteria of a good encoding force us to translate $ M_{\mpst} $ into a target session that has again three pairwise distributable steps.
But there is no such session in \mcbs, since a binary session in \mcbs cannot contain more than two interacting participants.
The remaining separations in Theorem~\ref{thm:MpstIntoBs} follow from calculus inclusion.

\thmMpstIntoBs*

\begin{proof}
	By Lemma~\ref{lem:sepMpstFromMcbs}, there is no good encoding from \mpst into \mcbs.
	Since $ \mpst \subset \smpst \subset \dmpst $, then there is no good encoding from $ L_1 \in \Set{\mpst, \smpst, \dmpst} $ into \mcbs.
	Since $ \bs \subset \scbs \subset \mcbs $, then there is no good encoding from $ L_1 \in \Set{\mpst, \smpst, \dmpst} $ into $ L_2 \in \Set{\bs, \scbs, \mcbs} $.
\end{proof}

\subsection{Separating \mcbs from \lcnvms}
\label{app:separationMcbsFromLcnvms}

To show that there is no good encoding from \mcbs into \lcnvms, we use $ M_{\mcbs} = \pa{\p}{\left( \ssout{\q}{l_1}\seq\inact \OR \colorbox{orange!20}{\ensuremath{\ssinp{\q}{l_2}\seq\success}} \right)} \Par \pa{\q}{\left( \ssinp{\p}{l_1}\seq\inact \OR \colorbox{orange!20}{\ensuremath{\ssinp{\p}{l_3}\seq\success}} \right)} $ in Figure~\ref{fig:KeysExpressiveness}(10) as counterexample.

\begin{lemma}
	The session $ M_{\mcbs} $ in Figure~\ref{fig:KeysExpressiveness}(10) is typed and deadlock-free.
\end{lemma}

\begin{proof}
	In $ M_{\mcbs} $ in Figure~\ref{fig:KeysExpressiveness}(10) we omit the objects in communication, because they are not relevant for the protocol.
	Assume that in all communications $ \true $ is transmitted and the receiver stores the received value in the variable $ \xx $.
	By Figure~\ref{fig:typing}, we get
	\begin{align*}
		\types \ssout{\q}{l_1}\seq\inact \OR \colorbox{orange!20}{\ensuremath{\ssinp{\q}{l_2}\seq\success}} \as T_{\p} &= \tout{\q}{l_1}\tinact \; \tor \; \tinp{\q}{l_2}\tinact\\
		\types \ssinp{\p}{l_1}\seq\inact \OR \colorbox{orange!20}{\ensuremath{\ssinp{\p}{l_3}\seq\success}} \as T_{\q} &= \tinp{\p}{l_1}\tinact \; \tor \; \tinp{\p}{l_3}\tinact
	\end{align*}
	and the omitted payload type is $ \bool $ in all cases.
	Let $ \LL = \localise{\p} T_{\p}, \localise{\q} T_{\q} $.
	$ M_{\mcbs} $ has only a single maximal execution.
	The reductions of the local types follow closely this one execution of the protocol:
	\begin{align*}
		\LL &
		\red \localise{\p} \tinact, \localise{\q} \tinact
		\noRed
	\end{align*}
	By checking this sequence of reductions of local types, we obtain $ \safe{\LL} $.
	By Figure~\ref{fig:typing}, then $ \types M_{\mcbs} \as \LL $.
	Similarly by checking this sequence of reductions of local types, we see that $ \LL $ is deadlock-free, \ie $ \df{\LL} $.
	By Theorem~\ref{thm:deadlockfree}, then $ M_{\mcbs} $ is deadlock-free.
\end{proof}

We show that the criteria of a good encoding, do not allow to simply drop the additional summands $ \colorbox{orange!20}{\ensuremath{\ssinp{\q}{l_2}\seq\success}} $ and $ \colorbox{orange!20}{\ensuremath{\ssinp{\p}{l_3}\seq\success}} $ but that the type system of the target calculus \lcnvms also does not allow to keep them.

\thmSeparateMcbsFromLcnvms*

\begin{proof}
	Assume the opposite, \ie assume that there is a good encoding $ \arbitraryEncoding $ from \mcbs into \lcnvms and, thus, also of $ M_{\mcbs} $ in Figure~\ref{fig:KeysExpressiveness}(10).
	By compositionality, the translation of $ \pa{\p}{\left( \ssout{\q}{l_1}\seq\inact \OR \colorbox{orange!20}{\ensuremath{\ssinp{\q}{l_2}\seq\success}} \right)} $ is independent of the translation of $ \pa{\q}{\left( \ssinp{\p}{l_1}\seq\inact \OR \colorbox{orange!20}{\ensuremath{\ssinp{\p}{l_3}\seq\success}} \right)} $.
	Then the translation of $ M_{\mcbs} $ differs from the translation of $ M_{\mcbs, \diamond \success} = \pa{\p}{\left( \ssout{\q}{l_1}\seq\inact \OR \colorbox{orange!20}{\ensuremath{\ssinp{\q}{l_2}\seq\success}} \right)} \Par \pa{\q}{\left( \ssinp{\p}{l_1}\seq\inact \OR \ssout{\p}{l_2}\seq\inact \right)} $ only by the translation of the respective right hand sides of the parallel operator.
	Note that $ M_{\mcbs, \diamond \success} $ can unguard $ \success $ by an interaction with the formerly additional summand $ \colorbox{orange!20}{\ensuremath{\ssinp{\q}{l_2}\seq\success}} $ but $ M_{\mcbs}\not\reachSuccess $.
	Also note that $ M_{\mcbs, \diamond \success} $ may reach success, \ie $ M_{\mcbs, \diamond \success}\reachSuccess $, but not must reach success, \ie $ M_{\mcbs, \diamond \success}\not\reachSuccessFin $.
	Such behaviour cannot result from reducing a conditional, but requires an interaction between the two session endpoints in the translation.
	Moreover, the translation of $ \pa{\q}{\left( \ssinp{\p}{l_1}\seq\inact \OR \ssout{\p}{l_2}\seq\inact \right)} $ cannot introduce a way to unguard success, because that would lead to success in the translation of $ M_{\mcbs, \text{no } \success} = \pa{\p}{\left( \ssout{\q}{l_1}\seq\inact \OR \ssinp{\q}{l_2}\seq\inact \right)} \Par \pa{\q}{\left( \ssinp{\p}{l_1}\seq\inact \OR \ssout{\p}{l_2}\seq\inact \right)} $.
	Hence, the translation of $ \pa{\p}{\left( \ssout{\q}{l_1}\seq\inact \OR \colorbox{orange!20}{\ensuremath{\ssinp{\q}{l_2}\seq\success}} \right)} $ has to lead (possibly after some initial sequence of steps that may or may not involve interaction with the translation of $ \q $) to a choice that offers the translation of the additional summand $ \colorbox{orange!20}{\ensuremath{\ssinp{\q}{l_2}\seq\success}} $ as option.
	Since $ M_{\mcbs}\not\reachSuccess $ and \eg $ M_{\mcbs}'\not\reachSuccess $ with $ M_{\mcbs}' = \pa{\p}{\left( \ssout{\q}{l_1}\seq\inact \OR \colorbox{orange!20}{\ensuremath{\ssinp{\q}{l_2}\seq\success}} \right)} \Par \pa{\q}{\left( \ssinp{\p}{l_1}\seq\inact \right)} $ but $ M_{\mcbs, \diamond \success}\reachSuccess $ and $ M_{\mcbs}''\reachSuccess $ with $ M_{\mcbs}'' = \pa{\p}{\left( \ssout{\q}{l_1}\seq\inact \OR \colorbox{orange!20}{\ensuremath{\ssinp{\q}{l_2}\seq\success}} \right)} \Par \pa{\q}{\left( \ssinp{\p}{l_1}\seq\success \right)} $, taking or not taking this option cannot be decided non-deterministically.

	By repeating this line of argument, then $ \ArbitraryEncoding{M_{\mcbs}} \red^\ast N $ such that $ N $ has an unguarded choice for each session endpoint, where the choice that origins from the left hand side of the encoding has an additional, \ie unmatched, option resulting from the translation of $ \colorbox{orange!20}{\ensuremath{\ssinp{\q}{l_2}\seq\success}} $ and, similarly, the choice that origins from the right hand side of the encoding has an additional, \ie unmatched, option resulting from the translation of $ \colorbox{orange!20}{\ensuremath{\ssinp{\p}{l_3}\seq\success}} $.
	Such a target term cannot be typed in the target calculus, \ie $ \ArbitraryEncoding{M_{\mcbs}} $ is not typed in \lcnvms.
	Since we consider only encodings between the respective typed fragments of the calculi, this contradicts our assumption that $ \arbitraryEncoding $ is a good encoding from \mcbs into \lcnvms.
	We conclude that there cannot be such an encoding.
\end{proof}

\subsection{Encoding \scbs into \bs}
\label{app:encodingScbsIntoBs}

We show that there is a good encoding \scbs \blueArrow \bs, \ie a good encoding from \scbs into \bs.
Therefore, we fix for this subsection the function $ \arbitraryEncoding $ to be the encoding given in Figure~\ref{fig:encodingScbsIntoBs} and show that $ \arbitraryEncoding $ is a good encoding from \scbs into \bs.
Moreover, we strengthen the result by instantiating $ \asymp $ with $ = $ (instead of the weaker $ \approx $).
The most interesting part of $ \arbitraryEncoding $ is the translation of choice that was already given in Figure~\ref{fig:KeysExpressiveness}(1).
The remaining operators of \scbs are translated homomorphically by $ \arbitraryEncoding $.

\begin{figure}
	\begin{align*}
		\ArbitraryEncoding{M_1 \Par M_2} &= \ArbitraryEncoding{M_1} \Par \ArbitraryEncoding{M_2}\\
		\ArbitraryEncoding{\pa{\p}{\PP}} &= \pa{\p}{\ArbitraryEncoding{\PP}}\\
		\ArbitraryEncoding{\inact} &= \inact\\
		\ArbitraryEncoding{\success} &= \success\\
		\ArbitraryEncoding{\pvar{X}} &= \pvar{X}\\
		\ArbitraryEncoding{\rec{X} \PP} &= \rec{X} \ArbitraryEncoding{\PP}\\
		\ArbitraryEncoding{\choice{\soutprlbl{\q}{\lab_i}{v_i}\seq\PP_i}{i \in I}} &= \choice{\colorbox{green!20}{\ssinp{\q}{\lbl{enc_o}}}\seq\soutprlbl{\q}{\lab_i}{v_i}\seq\ArbitraryEncoding{\PP_i}}{i \in I}\\
		\ArbitraryEncoding{\choice{\sinprlbl{\p}{\lab_j}{x_j}\seq\PP_j}{j \in J}} &= \colorbox{green!20}{\ssout{\q}{\lbl{enc_o}}}\seq\choice{\sinprlbl{\q}{\lab_j}{x_j}\seq\ArbitraryEncoding{\PP_j}}{j \in J}\\
		\ArbitraryEncoding{\cond{v}{\PP}{Q}} &= \cond{v}{\ArbitraryEncoding{\PP}}{\ArbitraryEncoding{Q}}
	\end{align*}
	where $ \lbl{enc_o} $ is a label that does not occur in any source term.
	\caption{An Encoding from \scbs into \bs\label{fig:encodingScbsIntoBs}}
\end{figure}

In contrast to \scbs with separate choice, there are only single outputs in \bs and no choices on outputs.
Therefore, $ \arbitraryEncoding $ translates an output-guarded choice to an input-guarded choice with the same input-prefix $ \colorbox{green!20}{\ssinp{\q}{\lbl{enc_o}}} $ and the respective outputs as continuations.
As explained in Example~\ref{ex:choiceprocess}(1,2), such choices are typable in \bs, because of subtyping.
The encoding of input-guarded choice starts with the matching output $ \colorbox{green!20}{\ssout{\q}{\lbl{enc_o}}} $ followed by the original choice.

Accordingly, a single interaction in the source term between some output-guarded and some input-guarded choice is translated into two steps on the target side.
The first step with the fresh label $ \colorbox{green!20}{\lbl{enc_o}} $ allows the encoding of the output-guarded choice to non-deterministically pick an output.
The second step then is the interaction of this output with one of the inputs from the input-guarded choice.

The encoding induces a translation on types.
Let $ \LL $ not contain any mixed choices on types.
Then $ \TypeEncoding{\LL} $ is inductively defined as:
\begin{displaymath}
	\begin{array}{rclcrclcrclcrcl}
		\TypeEncoding{\emptyset} &=& \emptyset
		&& \TypeEncoding{\tinact} &=& \tinact
		&& \TypeEncoding{\tvar{t}} &=& \tvar{t}
		&& \TypeEncoding{\Tor{i \in I}{\toutlbl{\q}{\lab_i}{\UType_i}\local_i}} &=& \tinp{\q}{\lbl{enc_o}}\Tor{i \in I}{\toutlbl{\q}{\lab_i}{\UType_i}\TypeEncoding{\local_i}}\\
		\TypeEncoding{\LL', \p{:}\local} &=& \TypeEncoding{\LL'}, \p{:}\TypeEncoding{\local}
		&&&&
		&& \TypeEncoding{\trec{t} \local} &=& \trec{t} \TypeEncoding{\local}
		&& \TypeEncoding{\Tor{j \in J}{\tinplbl{\q}{\lab_j}{\UType_j}\local_j}} &=& \tout{\q}{\lbl{enc_o}}\Tor{j \in J}{\tinplbl{\q}{\lab_j}{\UType_j}\TypeEncoding{\local_j}}
	\end{array}
\end{displaymath}

We show that the translation of types satisfies a variant of operational correspondence for types.

\begin{lemma}
	\label{lem:ocTypesSCBSintoBS}
	Let $ \LL $ not contain any mixed choices on types and assume $ \safe{\LL} $ and $ \df{\LL} $.
	Then:
	\begin{description}
		\item[Completeness:] If $ \LL \by{\lambda} \LL' $, then $ \TypeEncoding{\LL} \red\by{\lambda} \TypeEncoding{\LL'} $.
		\item[Soundness:] If $ \TypeEncoding{\LL} \red^\ast \LL_{\mathsf{enc}} $, then there is some $ \LL' $ such that $ \LL \red^\ast \LL' $ and either $ \LL_{\mathsf{enc}} = \TypeEncoding{\LL'} $ or $ \LL_{\mathsf{enc}} \red \TypeEncoding{\LL'} $.
	\end{description}
\end{lemma}

\begin{proof}
	\begin{description}
		\item[Completeness:] Assume $ \LL \by{\actlbl{\p}{\q}{\lab}{\UType}} \LL' $.
			By Figure~\ref{fig:type_semantics}, then $ \LL = \LL_{\text{no }\p\q}, \p{:}T_{\p}, \q{:}T_{\q} $, $ \lab_n = \lab = \lab_m' $, $ \UTypei{n} = \UType = \UTypei{m}' $, and $ \LL' = \LL_{\text{no }\p\q}, \p{:}T_{\p, n}, \q{:}T_{\q, m} $ for some $ n \in I $ and some $ m \in J $, where $ T_{\p} $ is either $ \Tor{i \in I}{\toutlbl{\q}{\lab_i}{\UTypei{i}} T_{\p, i}} $ or $ \trec{t} \Tor{i \in I}{\toutlbl{\q}{\lab_i}{\UTypei{i}} T_{\p, i}} $ and $ T_{\q} $ is either $ \Tor{j \in J}{\tinplbl{\p}{\lab_j'}{\UTypei{j}'} T_{\q, j}} $ or $ \trec{t} \Tor{j \in J}{\tinplbl{\p}{\lab_j'}{\UTypei{j}'} T_{\q, j}} $.
			Then $ \TypeEncoding{\LL} = \TypeEncoding{\LL_{\text{no }\p\q}}, \p{:}\TypeEncoding{T_{\p}}, \q{:}\TypeEncoding{T_{\q}} $, where $ \TypeEncoding{T_{\p}} $ is either $ \tinp{\q}{\lbl{enc_o}}\Tor{i \in I}{\toutlbl{\q}{\lab_i}{\UTypei{i}} \TypeEncoding{T_{\p, i}}} $ or $ \trec{t} \tout{\q}{\lbl{enc_o}}\Tor{i \in I}{\toutlbl{\q}{\lab_i}{\UTypei{i}} \TypeEncoding{T_{\p, i}}} $ and $ \TypeEncoding{T_{\q}} $ is either $ \tout{\q}{\lbl{enc_o}}\Tor{j \in J}{\tinplbl{\p}{\lab_j'}{\UTypei{j}'} \TypeEncoding{T_{\q, j}}} $ or $ \trec{t} \tout{\q}{\lbl{enc_o}}\Tor{j \in J}{\tinplbl{\p}{\lab_j'}{\UTypei{j}'} \TypeEncoding{T_{\q, j}}} $.
			Moreover, then $ \TypeEncoding{\LL'} = \TypeEncoding{\LL_{\text{no }\p\q}}, \p{:}\TypeEncoding{T_{\p, n}}, \q{:}\TypeEncoding{T_{\q, m}} $.
			Then $ \TypeEncoding{\LL} \red \by{\actlbl{\p}{\q}{\lab}{\UType}} \TypeEncoding{\LL'} $.
		\item[Soundness:] We consider the first two steps of $ \TypeEncoding{\LL} \red^\ast \LL_{\mathsf{enc}} $.
			\begin{description}
				\item[$ \TypeEncoding{\LL} \red \LL_{\mathsf{enc}} $:] In this case $ \TypeEncoding{\LL} = \LL_{\mathsf{enc}, \text{no }\p\q}, \p{:}T_{\p, \mathsf{enc}}, \q{:}T_{\q, \mathsf{enc}} $ and $ \LL_{\mathsf{enc}} = \LL_{\mathsf{enc}, \text{no }\p\q}, \p{:}\Tor{i \in I}{\toutlbl{\q}{\lab_i}{\UType_i}\TypeEncoding{\local_i}}, \q{:}\Tor{j \in J}{\tinplbl{\p}{\lab_j'}{\UType_j'}\TypeEncoding{\local_j}} $ for some $ \LL_{\mathsf{enc}, \text{no }\p\q} $, where $ T_{\p, \mathsf{enc}} $ is either $ \tinp{\q}{\lbl{enc_o}}\Tor{i \in I}{\toutlbl{\q}{\lab_i}{\UType_i}\TypeEncoding{\local_i}} $ or $ \trec{t} \tinp{\q}{\lbl{enc_o}}\Tor{i \in I}{\toutlbl{\q}{\lab_i}{\UType_i}\TypeEncoding{\local_i}} $ and $ T_{\q, \mathsf{enc}} $ is $ \tout{\p}{\lbl{enc_o}}\Tor{j \in J}{\tinplbl{\p}{\lab_j'}{\UType_j'}\TypeEncoding{\local_j}} $ or $ \trec{t} \tout{\p}{\lbl{enc_o}}\Tor{j \in J}{\tinplbl{\p}{\lab_j'}{\UType_j'}\TypeEncoding{\local_j}} $.
					Then $ \LL = \LL_{\text{no }\p\q}, \p{:}T_{\p}, \q{:}T_{\q} $ such that $ \TypeEncoding{\LL_{\text{no }\p\q}} = \LL_{\mathsf{enc}, \text{no }\p\q} $, $ T_{\p} $ is either $ \Tor{i \in I}{\toutlbl{\q}{\lab_i}{\UType_i}\local_i} $ or $ \trec{t} \Tor{i \in I}{\toutlbl{\q}{\lab_i}{\UType_i}\local_i} $, and $ T_{\q} $ is either $ \Tor{j \in J}{\tinplbl{\p}{\lab_j'}{\UType_j'}\local_j} $ or $ \trec{t} \Tor{j \in J}{\tinplbl{\p}{\lab_j'}{\UType_j'}\local_j} $.
					By Figure~\ref{fig:type_semantics} and since $ \safe{\LL} $ and $ \df{\LL} $, then $ \LL \red \LL' = \LL_{\text{no }\p\q}, \p{:}\local_n, \q{:}\local_m $ for some $ n \in I $ and some $ m \in J $ such that $ \lab_n = \lab_m' $.
					Then $ \TypeEncoding{\LL'} = \LL_{\mathsf{enc}, \text{no }\p\q}, \p{:}\local_n, \q{:}\local_m $.
					Then $ \LL_{\mathsf{enc}} \red \TypeEncoding{\LL'} $.
				\item[$ \TypeEncoding{\LL} \red\red \LL_{\mathsf{enc}} $:] Then $ \TypeEncoding{\LL} = \LL_{\mathsf{enc}, \text{no }\p\q}, \p{:}T_{\p, \mathsf{enc}}, \q{:}T_{\q, \mathsf{enc}} $ and $ \LL_{\mathsf{enc}} = \LL_{\mathsf{enc}, \text{no }\p\q}, \p{:}\TypeEncoding{\local_n}, \q{:}\TypeEncoding{\local_m} $ for some $ \LL_{\mathsf{enc}, \text{no }\p\q} $, some $ n \in I $, and some $ m \in J $, where $ T_{\p, \mathsf{enc}} = \Tor{i \in I}{\toutlbl{\q}{\lab_i}{\UType_i}\TypeEncoding{\local_i}} $ and $ T_{\q, \mathsf{enc}} = \Tor{j \in J}{\tinplbl{\p}{\lab_j'}{\UType_j'}\TypeEncoding{\local_j}} $.
					Then $ \LL = \LL_{\text{no }\p\q}, \p{:}T_{\p}, \q{:}T_{\q} $ such that $ \TypeEncoding{\LL_{\text{no }\p\q}} = \LL_{\mathsf{enc}, \text{no }\p\q} $, $ T_{\p} $ is either $ \Tor{i \in I}{\toutlbl{\q}{\lab_i}{\UType_i}\local_i} $ or $ \trec{t} \Tor{i \in I}{\toutlbl{\q}{\lab_i}{\UType_i}\local_i} $, and $ T_{\q} $ is either $ \Tor{j \in J}{\tinplbl{\p}{\lab_j'}{\UType_j'}\local_j} $ or $ \trec{t} \Tor{j \in J}{\tinplbl{\p}{\lab_j'}{\UType_j'}\local_j} $.
					Then $ \LL \red \LL' = \LL_{\text{no }\p\q}, \p{:}\local_n, \q{:}\local_m $.
					Then $ \TypeEncoding{\LL'} = \LL_{\mathsf{enc}} $.
			\end{description}
			Then soundness follows from an induction on the length of the sequence $ \TypeEncoding{\LL} \red^\ast \LL_{\mathsf{enc}} $ that alternates between the above two cases.
			\qedhere
	\end{description}
\end{proof}

In the formulation of the completeness part, we formalise that the encoding translates every step of the source into two steps in the target, which is true for sessions as well as the reductions of their types.
We also adapted the formulation of soundness to the properties of the encoding function.
In the proof of soundness, we use $ \safe{\LL} $ and $ \df{\LL} $ to ensure that there is a matching input and output type for the second step in the translation of types.
With completeness and soundness of the translation of types we show that the translation on types preserves safety and deadlock-freedom.

\begin{lemma}
	\label{lem:DFofSCBSintoBS}
	Let $ \LL $ not contain any mixed choices on types.
	If $ \safe{\LL} $ and $ \df{\LL} $ then $ \safe{\TypeEncoding{\LL}} $ and $ \df{\TypeEncoding{\LL}} $.
\end{lemma}

\begin{proof}
	\begin{description}
		\item[Safe:] Assume $ \TypeEncoding{\LL} \red^\ast \LL_{\mathsf{enc}} $ and $ \Set{ \p{:}T_{\mathsf{enc}, 1}, \q{:}T_{\mathsf{enc}, 2} } \subseteq \LL_{\mathsf{enc}} $ such that $ \p{:}T_{\mathsf{enc}, 1} \by{\actoutlbl{\p}{\q}{\lab}{\UType}} $ and $ \q{:}T_{\mathsf{enc}, 2} \by{\actinplbl{\q}{\p}{\labd}{\UTyped}} $.
			By soundness in Lemma~\ref{lem:ocTypesSCBSintoBS}, there is some $ \LL' $ such that $ \LL \red^\ast \LL' $ and either $ \LL_{\mathsf{enc}} = \TypeEncoding{\LL'} $ or $ \LL_{\mathsf{enc}} \red \TypeEncoding{\LL'} $.
			\begin{description}
				\item[$ \LL_{\mathsf{enc}} = \TypeEncoding{\LL'} $:] Then $ \lab = \lbl{enc_o} = \lab' $ and there is exactly one output $ \colorbox{green!20}{\ssout{\q}{\lbl{enc_o}}} $ and one input-guarded choice, where all prefixes are $ \colorbox{green!20}{\ssinp{\q}{\lbl{enc_o}}} $.
					Then $ \LL_{\mathsf{enc}} \by{\actlbl{\p}{\q}{\lab}{\UType}} \LL_{\mathsf{enc}}' $ for some $ \LL_{\mathsf{enc}}' $.
				\item[$ \LL_{\mathsf{enc}} \red \TypeEncoding{\LL'} $:] Then there are $ \LL'', T_1, T_2 $ such that $ \LL \red^\ast \LL'' \red \LL' $, $ \LL_{\mathsf{enc}} = \TypeEncoding{\LL''} $, $ T_{\mathsf{enc}, 1} = \TypeEncoding{T_1} $, $ T_{\mathsf{enc}, 2} = \TypeEncoding{T_2} $, $ \Set{ \p{:}T_1, \q{:}T_2 } \subseteq \LL'' $, $ \p{:}T_1 \by{\actoutlbl{\p}{\q}{\lab}{\UType}} $, and $ \q{:}T_2 \by{\actinplbl{\q}{\p}{\labd}{\UTyped}} $.
					Since $ \safe{\LL} $ and $ \LL \red^\ast \LL'' $, $ \safe{\LL''} $.
					Then $ \LL'' \by{\actlbl{\p}{\q}{\lab}{\UType}} \LL''' $ for some $ \LL''' $.
					By completeness in Lemma~\ref{lem:ocTypesSCBSintoBS}, then $ \LL_{\mathsf{enc}} \by{\actlbl{\p}{\q}{\lab}{\UType}} \LL_{\mathsf{enc}}' = \TypeEncoding{\LL'''} $.
			\end{description}
			In both cases, we have $ \LL_{\mathsf{enc}} \by{\actlbl{\p}{\q}{\lab}{\UType}} \LL_{\mathsf{enc}}' $ and thus $ \safe{\TypeEncoding{\LL}} $.
		\item[Deadlock-Freedom:] Assume $ \TypeEncoding{\LL} \red^\ast \LL_{\mathsf{enc}} \noRed $.
			By soundness in Lemma~\ref{lem:ocTypesSCBSintoBS}, then there is some $ \LL' $ such that $ \LL_{\mathsf{enc}} = \TypeEncoding{\LL'} $ and $ \LL \red^\ast \LL' $.
			By completeness in Lemma~\ref{lem:ocTypesSCBSintoBS}, $ \LL_{\mathsf{enc}} \noRed $ implies $ \LL' \noRed $.
			Since $ \df{\LL} $, then all types in $ \LL' $ are $ \tinact $.
			Then $ \LL_{\mathsf{enc}} = \TypeEncoding{\LL'} $ implies that all types in $ \LL_{\mathsf{enc}} $ are $ \tinact $.
			Then $ \df{\TypeEncoding{\LL}} $.
			\qedhere
	\end{description}
\end{proof}

Deadlock-freedom ensures the existence of the respective communication partner for the first step (with the fresh label $ \colorbox{green!20}{\lbl{enc_o}} $).
Because of that, we have to assume deadlock-freedom for $ \LL $ to show deadlock-freedom for $ \TypeEncoding{\LL} $ in the proof above.
With Lemma~\ref{lem:ocTypesSCBSintoBS} and Lemma~\ref{lem:DFofSCBSintoBS}, we show that $ \arbitraryEncoding $ is an encoding from \scbs into \bs, \ie that it translates typed and deadlock-free \scbs-sessions into typed and deadlock-free \bs-sessions.

\begin{lemma}
	\label{lem:typeSCBSintoBS}
	$ \ArbitraryEncoding{M} $ is typed and deadlock-free.
\end{lemma}

\begin{proof}
	Fix an arbitrary typed and deadlock-free $ M $ in \scbs.
	Then there is some $ \types M \as \LL_{\mathsf{nsc}} $ such that $ \safe{\LL_{\mathsf{nsc}}} $ and $ \df{\LL_{\mathsf{nsc}}} $.
	All choices in $ M $ are separated, but, because of subtyping, not all choices of $ \LL_{\mathsf{nsc}} $ have to be separated.
	By \TSum and subsumption \TSubs and because $ M $ has no mixed choices, there is some $ \LL $ with only separated choice such that $ \types M \as \LL $ and $ \LL \subt \LL_{\mathsf{nsc}} $.
	By Lemma~\ref{lem:subtypingproperties}, then $ \safe{\LL} $ and $ \df{\LL} $.
	By induction over the structure of $ P $, we show that $ \Gamma \types P \as T $ implies $ \TypeEncoding{\Gamma} \types \ArbitraryEncoding{P} \as \TypeEncoding{T} $, where $ \TypeEncoding{\Gamma} $ replaces all $ T $ in $ \Gamma $ by $ \TypeEncoding{T} $.
	\begin{description}
		\item[$ P = \inact $:] Then $ \ArbitraryEncoding{P} = \inact $.
			By Lemma~\ref{app:lem:typing_inversion}, then $ T = \tinact = \TypeEncoding{T} $.
			By \TInact, then $ \TypeEncoding{\Gamma} \types \ArbitraryEncoding{P} \as \TypeEncoding{T} $.
		\item[$ P = \success $:] Then $ \ArbitraryEncoding{P} = \success $.
			By Lemma~\ref{app:lem:typing_inversion} and since $ \success $ in the same way as $ \inact $, then $ T = \tinact = \TypeEncoding{T} $.
			By thy typing rule for $ \success $ that is similar to \TInact, then $ \TypeEncoding{\Gamma} \types \ArbitraryEncoding{P} \as \TypeEncoding{T} $.
		\item[$ P = \pvar{X} $:] Then $ \ArbitraryEncoding{P} = \pvar{X} $.
			By Lemma~\ref{app:lem:typing_inversion}, then $ X : T' \in \Gamma $ and $ T' \subt T $.
			Then $ X : \TypeEncoding{T'} \in \TypeEncoding{\Gamma} $.
			By \TVar, then $ \TypeEncoding{\Gamma} \types \ArbitraryEncoding{P} \as \TypeEncoding{T'} $.
			Since $ T' \subt T $ implies $ \TypeEncoding{T'} \subt \TypeEncoding{T} $, then $ \TypeEncoding{\Gamma} \types \ArbitraryEncoding{P} \as \TypeEncoding{T} $.
		\item[$ P = \rec{X} Q $:] Then $ \ArbitraryEncoding{P} = \rec{X} \ArbitraryEncoding{Q} $.
			By Lemma~\ref{app:lem:typing_inversion}, then $ \Gamma, X : T' \types Q \as T' $ and $ T' \subt T $.
			By the induction hypothesis, then $ \TypeEncoding{\Gamma}, X : \TypeEncoding{T'} \types Q \as \TypeEncoding{T'} $.
			By \TReq, then $ \TypeEncoding{\Gamma} \types \ArbitraryEncoding{P} \as \TypeEncoding{T'} $.
			Since $ T' \subt T $ implies $ \TypeEncoding{T'} \subt \TypeEncoding{T} $, then $ \TypeEncoding{\Gamma} \types \ArbitraryEncoding{P} \as \TypeEncoding{T} $.
		\item[$ P = \choice{\choice{\soutprlbl{\q}{\lab_i}{v_i}\seq\PP_j}{i \in I}}{j \in J} $:] Then $ \ArbitraryEncoding{P} = \choice{\choice{\colorbox{green!20}{\ssinp{\q}{\lbl{enc_o}}}\seq\soutprlbl{\q}{\lab_i}{v_i}\seq\ArbitraryEncoding{\PP_j}}{i \in I}}{j \in J} $.
			By Lemma~\ref{app:lem:typing_inversion}, then $ \Gamma \types \soutprlbl{\q}{\lab_i}{v_i}\seq\PP_j \as \toutlbl{\q}{\lab_i}{\UType_i}T_i $ for all $ i \in I $ and all $ j \in J $ and $ \Tor{i \in I}{\toutlbl{\q}{\lab_i}{\UType_i}T_i} \subt T $.
			Then $ \Gamma \types \PP_j \as T_i $.
			By the induction hypothesis, then $ \TypeEncoding{\Gamma} \types \ArbitraryEncoding{\PP_j} \as \TypeEncoding{T_i} $.
			By \TSend and \TSubs, then $ \TypeEncoding{\Gamma} \types \soutprlbl{\q}{\lab_i}{v_i}\seq\ArbitraryEncoding{\PP_j} \as \Tor{i \in I}{\toutlbl{\q}{\lab_i}{\UType_i}\TypeEncoding{T_i}} $.
			By \TSum and \TRcv, then $ \TypeEncoding{\Gamma} \types \ArbitraryEncoding{P} \as \tinp{\q}{\lbl{enc_o}}\Tor{i \in I}{\toutlbl{\q}{\lab_i}{\UType_i}\TypeEncoding{T_i}} $.
			Since $ \Tor{i \in I}{\toutlbl{\q}{\lab_i}{\UType_i}T_i} \subt T $ implies $ \TypeEncoding{\Tor{i \in I}{\toutlbl{\q}{\lab_i}{\UType_i}T_i}} \subt \TypeEncoding{T} $, then $ \tinp{\q}{\lbl{enc_o}}\Tor{i \in I}{\toutlbl{\q}{\lab_i}{\UType_i}\TypeEncoding{T_i}} \subt \TypeEncoding{T} $ and thus $ \TypeEncoding{\Gamma} \types \ArbitraryEncoding{P} \as \TypeEncoding{T} $.
		\item[$ P = \choice{\choice{\sinprlbl{\q}{\lab_i}{x_i}\seq\PP_j}{i \in I}}{j \in J} $:] Then $ \ArbitraryEncoding{P} = \colorbox{green!20}{\ssout{\q}{\lbl{enc_o}}}\seq\choice{\choice{\sinprlbl{\q}{\lab_i}{x_i}\seq\ArbitraryEncoding{\PP_j}}{i \in I}}{j \in J} $.
			By Lemma~\ref{app:lem:typing_inversion}, then $ \Gamma \types \sinprlbl{\q}{\lab_i}{x_i}\seq\PP_j \as \tinplbl{\q}{\lab_i}{\UType_i}T_i $ for all $ i \in I $ and all $ j \in J $ and $ \Tor{i \in I}{\tinplbl{\q}{\lab_i}{\UType_i}T_i} \subt T $.
			Then $ \Gamma, x_i : \UType_i \types \PP_j \as T_i $.
			By the induction hypothesis, then $ \TypeEncoding{\Gamma}, x_i : \UType_i \types \ArbitraryEncoding{\PP_j} \as \TypeEncoding{T_i} $.
			By \TRcv, then $ \TypeEncoding{\Gamma} \types \sinprlbl{\q}{\lab_i}{x_i}\seq\ArbitraryEncoding{\PP_j} \as \tinplbl{\q}{\lab_i}{\UType_i}\TypeEncoding{T_i} $.
			By \TSum and \TSend, then $ \TypeEncoding{\Gamma} \types \ArbitraryEncoding{P} \as \tout{\q}{\lbl{enc_o}}\Tor{i \in I}{\tinplbl{\q}{\lab_i}{\UType_i}\TypeEncoding{T_i}} $.
			Since $ \Tor{i \in I}{\tinplbl{\q}{\lab_i}{\UType_i}T_i} \subt T $ implies $ \TypeEncoding{\Tor{i \in I}{\tinplbl{\q}{\lab_i}{\UType_i}T_i}} \subt \TypeEncoding{T} $, then $ \tout{\q}{\lbl{enc_o}}\Tor{i \in I}{\tinplbl{\q}{\lab_i}{\UType_i}\TypeEncoding{T_i}} \subt \TypeEncoding{T} $ and thus $ \TypeEncoding{\Gamma} \types \ArbitraryEncoding{P} \as \TypeEncoding{T} $.
		\item[$ P = \cond{v}{Q}{R} $:] Then $ \ArbitraryEncoding{P} = \cond{v}{\ArbitraryEncoding{Q}}{\ArbitraryEncoding{R}} $.
			By Lemma~\ref{app:lem:typing_inversion}, $ \Gamma \types v \as \bool $, $ \Gamma \types Q \as T_Q $, $ \Gamma \types R \as T_T $, $ T_Q \subt T $, and $ T_R \subt T $.
			By the induction hypothesis, then $ \TypeEncoding{\Gamma} \types v \as \bool $, $ \TypeEncoding{\Gamma} \types \ArbitraryEncoding{Q} \as \TypeEncoding{T_Q} $, and $ \TypeEncoding{\Gamma} \types \ArbitraryEncoding{R} \as \TypeEncoding{T_R} $.
			Since $ T_Q \subt T $ implies $ \TypeEncoding{T_Q} \subt \TypeEncoding{T} $, then $ \TypeEncoding{\Gamma} \types \ArbitraryEncoding{Q} \as \TypeEncoding{T} $.
			Since $ T_R \subt T $ implies $ \TypeEncoding{T_R} \subt \TypeEncoding{T} $, then $ \TypeEncoding{\Gamma} \types \ArbitraryEncoding{R} \as \TypeEncoding{T} $.
			By \TCond, then $ \TypeEncoding{\Gamma} \types \ArbitraryEncoding{P} \as \TypeEncoding{T} $.
	\end{description}
	Since $ M $ is in \mcbs, $ M = \Pi_{i \in I} \pa{\p_i}{P_i} $ for some $ I, \Set{\p_i}_{i \in I}, \Set{P_i}_{i \in I} $ with $ I \subseteq \Set{1, 2} $.
	By Lemma~\ref{app:lem:typing_inversion}, then $ \types M \as \LL $ implies that $ \LL = \Set{ \p_i{:}T_i }_{i \in I} $ and $ \types P_i \as T_i $ for all $ i \in I $.
	Then $ \types \ArbitraryEncoding{P_i} \as \TypeEncoding{T_i} $ for all $ i \in I $.
	Since $ \LL = \Set{ \p_i{:}T_i }_{i \in I} $, $ \TypeEncoding{\LL} = \Set{ \p_i{:}\TypeEncoding{T_i} }_{i \in I} $.
	By Lemma~\ref{lem:DFofSCBSintoBS}, $ \safe{\LL} $ and $ \df{\LL} $ imply $ \safe{\TypeEncoding{\LL}} $ and $ \df{\TypeEncoding{\LL}} $.
	By \TSessPlus and $ \safe{\TypeEncoding{\LL}} $, then $ \types \ArbitraryEncoding{M} \as \TypeEncoding{\LL} $.
\end{proof}

The proof shows that typed terms are translated into typed terms, where the type of the translation is the translation of the type in the source.
In the translation of an output-guarded choice, we use subtyping to connect $ \ArbitraryEncoding{\choice{\choice{\soutprlbl{\q}{\lab_i}{v_i}\seq\PP_j}{i \in I}}{j \in J}} = \choice{\choice{\colorbox{green!20}{\ssinp{\q}{\lbl{enc_o}}}\seq\soutprlbl{\q}{\lab_i}{v_i}\seq\ArbitraryEncoding{\PP_j}}{i \in I}}{j \in J} $ and its type $ \TypeEncoding{\Tor{i \in I}{\toutlbl{\q}{\lab_i}{\UType_i}T_i}} = \tinp{\q}{\lbl{enc_o}}\Tor{i \in I}{\toutlbl{\q}{\lab_i}{\UType_i}\TypeEncoding{T_i}} $.

To prove that the encoding \mcbs \blueArrow \bs is good, we show first that it respects structural congruence.

\begin{lemma}
	\label{lem:scSCBSintoBS}
	\begin{enumerate}
		\item $ P_1 \equiv P_2 $ iff $ \ArbitraryEncoding{P_1} \equiv \ArbitraryEncoding{P_2} $.
		\item $ M_1 \equiv M_2 $ iff $ \ArbitraryEncoding{M_1} \equiv \ArbitraryEncoding{M_2} $.
		\item If $ \ArbitraryEncoding{M_1} \equiv M $ then there is some $ M_2 $ such that $ M = \ArbitraryEncoding{M_2} $.
	\end{enumerate}
\end{lemma}

\begin{proof}
	The proof is by structural induction on the definition of structural congruence in Figure~\ref{fig:reduction}.
	Since the encoding does not change names and translates inaction, recursion, and parallel composition homomorphically, all cases follow directly from the respective assumptions or the induction hypothesis.
\end{proof}

Compositionality is immediate from Figure~\ref{fig:encodingScbsIntoBs}.
Name invariance can be shown by structural induction.

\begin{lemma}
	\label{lem:niSCBSintoBS}
	The encoding $ \arbitraryEncoding $ is name invariant, \ie $ \ArbitraryEncoding{M \sigma} = \ArbitraryEncoding{M} \sigma $ for all substitutions $ \sigma $ such that $ M\sigma $ is defined.
\end{lemma}

\begin{proof}
	By induction on the structure of $ M $.
	Since, $ \arbitraryEncoding $ translates names of the source to themselves and does neither reserve names nor introduces additional names, all cases follow directly from the respective assumptions or the induction hypothesis.
\end{proof}

For operational completeness, remember that a single interaction in the source is translated into two steps on the target, where the first is on the fresh label $ \colorbox{green!20}{\lbl{enc_o}} $ and the second is the step of the source.

\begin{lemma}
	\label{lem:occSCBSintoBS}
	The encoding $ \arbitraryEncoding $ is operationally complete, \ie $ M \red^\ast M' $ implies $ \ArbitraryEncoding{M} \red^\ast \ArbitraryEncoding{M'} $.
\end{lemma}

\begin{proof}
	Assume $ M \red M' $.
	We proceed with an induction on the derivation of $ M \red M' $ by the rules in Figure~\ref{fig:reduction} and show $ \ArbitraryEncoding{M} \red^\ast \ArbitraryEncoding{M'} $.
	\begin{description}
		\item[\RChoice:] Then $ M = \pa{\p}{\choice{\soutprlbl{\q}{\lab_i}{v_i}\seq\PP_i}{i \in I}} \Par \pa{\q}{\choice{\sinprlbl{\p}{\lab_j'}{x_j}\seq Q_j}{j \in J}} $ and $ M' = \pa{\p}{\PP_n} \Par \pa{\q}{Q_m\subst{v_n}{x_m}} $ for some $ n \in I $ and some $ m \in J $ such that $ \lab_n = \lab_m' $.
			Then $ \ArbitraryEncoding{M} = \pa{\p}{\choice{\colorbox{green!20}{\ssinp{\q}{\lbl{enc_o}}}\seq\soutprlbl{\q}{\lab_i}{v_i}\seq\ArbitraryEncoding{\PP_i}}{i \in I}} \Par \pa{\q}{\colorbox{green!20}{\ssout{\p}{\lbl{enc_o}}}\seq\choice{\sinprlbl{\p}{\lab_j'}{x_j}\seq\ArbitraryEncoding{Q_j}}{j \in J}} $ and $ \ArbitraryEncoding{M'} = \pa{\p}{\ArbitraryEncoding{\PP_n}} \Par \pa{\q}{\ArbitraryEncoding{Q_m\subst{v_n}{x_m}}} $.
			By Lemma~\ref{lem:niSCBSintoBS}, then $ \ArbitraryEncoding{M'} = \pa{\p}{\ArbitraryEncoding{\PP_n}} \Par \pa{\q}{\ArbitraryEncoding{Q_m}\subst{v_n}{x_m}} $.
			By \RChoice, then $ \ArbitraryEncoding{M} \red \pa{\p}{\soutprlbl{\q}{\lab_n}{v_n}\seq\ArbitraryEncoding{\PP_n}} \Par \pa{\q}{\choice{\sinprlbl{\p}{\lab_j'}{x_j}\seq\ArbitraryEncoding{Q_j}}{j \in J}} \red \ArbitraryEncoding{M'} $ and thus $ \ArbitraryEncoding{M} \red^\ast \ArbitraryEncoding{M'} $.
		\item[\RCondT:] Then $ M = \pa{\p}{\cond{\true}{P_1}{P_2}} \Par M_1 $ and $ M' = \pa{\p}{P_1} \Par M_1 $.
			Then $ \ArbitraryEncoding{M} = \pa{\p}{\cond{\true}{\ArbitraryEncoding{P_1}}{\ArbitraryEncoding{P_2}}} \Par \ArbitraryEncoding{M_1} $ and $ \ArbitraryEncoding{M'} = \pa{\p}{\ArbitraryEncoding{P_1}} \Par \ArbitraryEncoding{M_1} $.
			By \RCondT, then $ \ArbitraryEncoding{M} \red \ArbitraryEncoding{M'} $ and thus $ \ArbitraryEncoding{M} \red^\ast \ArbitraryEncoding{M'} $.
		\item[\RCondF:] Then $ M = \pa{\p}{\cond{\false}{P_1}{P_2}} \Par M_1 $ and $ M' = \pa{\p}{P_2} \Par M_1 $.
			Then $ \ArbitraryEncoding{M} = \pa{\p}{\cond{\false}{\ArbitraryEncoding{P_1}}{\ArbitraryEncoding{P_2}}} \Par \ArbitraryEncoding{M_1} $ and $ \ArbitraryEncoding{M'} = \pa{\p}{\ArbitraryEncoding{P_2}} \Par \ArbitraryEncoding{M_1} $.
			By \RCondF, then $ \ArbitraryEncoding{M} \red \ArbitraryEncoding{M'} $ and thus $ \ArbitraryEncoding{M} \red^\ast \ArbitraryEncoding{M'} $.
		\item[\RCong:] In this case $ M \equiv M_1 \red M_2 \equiv M' $.
			By Lemma~\ref{lem:scSCBSintoBS}, then $ \ArbitraryEncoding{M} \equiv \ArbitraryEncoding{M_1} $ and $ \ArbitraryEncoding{M_2} \equiv \ArbitraryEncoding{M'} $.
			By the induction hypothesis, $ M_1 \red M_2 $ implies $ \ArbitraryEncoding{M_1} \red \ArbitraryEncoding{M_2} $.
			By \RCong, then $ \ArbitraryEncoding{M} \red \ArbitraryEncoding{M'} $ and thus $ \ArbitraryEncoding{M} \red^\ast \ArbitraryEncoding{M'} $.
	\end{description}
	By induction on the number of steps in $ M \red^\ast M' $, then $ M \red^\ast M' $ implies $ \ArbitraryEncoding{M} \red^\ast \ArbitraryEncoding{M'} $, \ie $ \arbitraryEncoding $ is operationally complete with $ \asymp $ chosen as $ = $.
\end{proof}

For operational soundness we have to show that every step of the translation is the start or end of an emulation and that every emulation can be completed.
Remember that $ \arbitraryEncoding $ translates interactions into two steps but a step that reduces a conditional is translated to a single step.

\begin{lemma}
	\label{lem:ocsSCBSintoBS}
	The encoding $ \arbitraryEncoding $ is operationally sound, \ie $ \ArbitraryEncoding{M} \red^\ast N $ implies $ M \red^\ast M' $ and $ N \red^\ast \ArbitraryEncoding{M'} $, where $ \ArbitraryEncoding{M} \red^\ast \ArbitraryEncoding{M'} $ contains at most twice as much steps as $ M \red^\ast M' $.
\end{lemma}

\begin{proof}
	We consider the first two steps of $ \ArbitraryEncoding{M} \red^\ast N $.
	\begin{description}
		\item[$ \ArbitraryEncoding{M} \red N $:] We proceed with an induction on the derivation of $ \ArbitraryEncoding{M} \red N $ by the rules in Figure~\ref{fig:reduction} and show $ M \red^\ast M' $ and $ N \red^\ast \ArbitraryEncoding{M'} $, where $ \ArbitraryEncoding{M} \red^\ast \ArbitraryEncoding{M'} $ contains at most twice as much steps as $ M \red^\ast M' $.
			\begin{description}
				\item[\RChoice:] Then $ \ArbitraryEncoding{M} = \pa{\p}{\choice{\colorbox{green!20}{\ssinp{\q}{\lbl{enc_o}}}\seq\soutprlbl{\q}{\lab_i}{v_i}\seq\ArbitraryEncoding{\PP_i}}{i \in I}} \Par \pa{\q}{\colorbox{green!20}{\ssout{\p}{\lbl{enc_o}}}\seq\choice{\sinprlbl{\p}{\lab_j'}{x_j}\seq\ArbitraryEncoding{Q_j}}{j \in J}} $ and $ N = \pa{\q}{\soutprlbl{\q}{\lab_n}{v_n}\seq\ArbitraryEncoding{\PP_n}} \Par \pa{\q}{\choice{\sinprlbl{\p}{\lab_j'}{x_j}\seq\ArbitraryEncoding{Q_j}}{j \in J}} $ for some $ n \in I $.
					Then $ M = \pa{\p}{\choice{\soutprlbl{\q}{\lab_i}{v_i}\seq\PP_i}{i \in I}} \Par \pa{\q}{\choice{\sinprlbl{\p}{\lab_j'}{x_j}\seq Q_j}{j \in J}} $.
					By safety, there is some $ m \in J $ such that $ \lab_n = \lab_m' $.
					By \RChoice, then $ N \red \pa{\p}{\ArbitraryEncoding{\PP_n}} \Par \pa{\q}{\ArbitraryEncoding{Q_m}\subst{v_n}{x_m}} $.
					By Lemma~\ref{lem:niSCBSintoBS}, $ \ArbitraryEncoding{Q_m}\subst{v_n}{x_m} = \ArbitraryEncoding{Q_m\subst{v_n}{x_m}} $.
					Then $ M \red M' $ and $ N \red \ArbitraryEncoding{M'} $, where $ M' = \pa{\p}{\PP_n} \Par \pa{\q}{Q_m\subst{v_n}{x_m}}  $.
					Thus $ M \red^\ast M' $ and $ N \red^\ast \ArbitraryEncoding{M'} $.
				\item[\RCondT:] Then $ \ArbitraryEncoding{M} = \pa{\p}{\cond{\true}{\ArbitraryEncoding{P_1}}{\ArbitraryEncoding{P_2}}} \Par \ArbitraryEncoding{M_1} $ and $ N = \pa{\p}{\ArbitraryEncoding{P_1}} \Par \ArbitraryEncoding{M_1} $.
					Then $ M = \pa{\p}{\cond{\true}{P_1}{P_2}} \Par M_1 $ and $ N = \ArbitraryEncoding{M'} $, where $ M' = \pa{\p}{P_1} \Par M_1 $.
					By \RCondT, then $ M \red M' $ and thus $ M \red^\ast M' $ and $ N \red^\ast \ArbitraryEncoding{M'} $.
				\item[\RCondF:] Then $ \ArbitraryEncoding{M} = \pa{\p}{\cond{\false}{\ArbitraryEncoding{P_1}}{\ArbitraryEncoding{P_2}}} \Par \ArbitraryEncoding{M_1} $ and $ N = \pa{\p}{\ArbitraryEncoding{P_2}} \Par \ArbitraryEncoding{M_1} $.
					Then $ M = \pa{\p}{\cond{\false}{P_1}{P_2}} \Par M_1 $ and $ N = \ArbitraryEncoding{M'} $, where $ M' = \pa{\p}{P_2} \Par M_1 $.
					By \RCondF, then $ M \red M' $ and thus $ M \red^\ast M' $ and $ N \red^\ast \ArbitraryEncoding{M'} $.
				\item[\RCong:] In this case $ \ArbitraryEncoding{M} \equiv N_1 \red N_2 \equiv N $.
					By Lemma~\ref{lem:scSCBSintoBS}, then $ M \equiv M_1 $ for some $ M_1 $ such that $ N_1 = \ArbitraryEncoding{M_1} $.
					By the induction hypothesis, $ \ArbitraryEncoding{M_1} \red N_2 $ implies $ M_1 \red^\ast M' $ and $ N_2 \red^\ast \ArbitraryEncoding{M'} $, where $ \ArbitraryEncoding{M_1} \red^\ast \ArbitraryEncoding{M'} $ contains at most twice as much steps as $ M_1 \red^\ast M' $.
					By \RCong, then $ M \red^\ast M' $ and $ N \red^\ast \ArbitraryEncoding{M'} $.
			\end{description}
		\item[$ \ArbitraryEncoding{M} \red N_1 \red N_2 $:] We proceed with an induction on the derivation of the step in $ N_1 \red N_2 $ by the rules in Figure~\ref{fig:reduction} and show $ M \red^\ast M' $ and $ N_2 \red^\ast \ArbitraryEncoding{M'} $, where $ \ArbitraryEncoding{M} \red^\ast \ArbitraryEncoding{M'} $ contains at most twice as much steps as $ M \red^\ast M' $.
			\begin{description}
				\item[\RChoice:] Then we have $ \ArbitraryEncoding{M} = \pa{\p}{\choice{\colorbox{green!20}{\ssinp{\q}{\lbl{enc_o}}}\seq\soutprlbl{\q}{\lab_i}{v_i}\seq\ArbitraryEncoding{\PP_i}}{i \in I}} \Par \pa{\q}{\colorbox{green!20}{\ssout{\p}{\lbl{enc_o}}}\seq\choice{\sinprlbl{\p}{\lab_j'}{x_j}\seq\ArbitraryEncoding{Q_j}}{j \in J}} $ and we have $ N_2 = \pa{\p}{\ArbitraryEncoding{\PP_n}} \Par \pa{\q}{\ArbitraryEncoding{Q_m}\subst{v_n}{x_m}} $ for some $ n \in I $ and some $ m \in J $ such that $ \lab_n = \lab_m' $.
					Then $ M = \pa{\p}{\choice{\soutprlbl{\q}{\lab_i}{v_i}\seq\PP_i}{i \in I}} \Par \pa{\q}{\choice{\sinprlbl{\p}{\lab_j'}{x_j}\seq Q_j}{j \in J}} $.
					By Lemma~\ref{lem:niSCBSintoBS}, $ \ArbitraryEncoding{Q_m}\subst{v_n}{x_m} = \ArbitraryEncoding{Q_m\subst{v_n}{x_m}} $.
					Then $ N_2 = \ArbitraryEncoding{M'} $, where $ M' = \pa{\p}{\PP_n} \Par \pa{\q}{Q_m\subst{v_n}{x_m}} $.
					By \RChoice, then $ M \red M' $.
					Thus $ M \red^\ast M' $ and $ N_2 \red^\ast \ArbitraryEncoding{M'} $.
				\item[\RCondT:] Then also $ \ArbitraryEncoding{M} \red N_1 $ reduces a conditional, \ie was derived by \RCondT or \RCondF.
					By the cases \RCondT and \RCondF in the above considered case of a single step from $ \ArbitraryEncoding{M} $, then $ M \red M'' $ and $ N_1 = \ArbitraryEncoding{M''} $.
					Then $ \ArbitraryEncoding{M''} = \pa{\p}{\cond{\true}{\ArbitraryEncoding{P_1}}{\ArbitraryEncoding{P_2}}} \Par \ArbitraryEncoding{M_1} $ and $ N_2 = \pa{\p}{\ArbitraryEncoding{P_1}} \Par \ArbitraryEncoding{M_1} $.
					Then $ M'' = \pa{\p}{\cond{\true}{P_1}{P_2}} \Par M_1 $ and $ N_2 = \ArbitraryEncoding{M'} $, where $ M' = \pa{\p}{P_1} \Par M_1 $.
					By \RCondT, then $ M'' \red M' $ and thus $ M \red^\ast M' $ and $ N_2 \red^\ast \ArbitraryEncoding{M'} $.
				\item[\RCondF:] Then also $ \ArbitraryEncoding{M} \red N_1 $ reduces a conditional, \ie was derived by \RCondT or \RCondF.
					By the cases \RCondT and \RCondF in the above considered case of a single step from $ \ArbitraryEncoding{M} $, then $ M \red M'' $ and $ N_1 = \ArbitraryEncoding{M''} $.
					Then $ \ArbitraryEncoding{M''} = \pa{\p}{\cond{\false}{\ArbitraryEncoding{P_1}}{\ArbitraryEncoding{P_2}}} \Par \ArbitraryEncoding{M_1} $ and $ N_2 = \pa{\p}{\ArbitraryEncoding{P_2}} \Par \ArbitraryEncoding{M_1} $.
					Then $ M'' = \pa{\p}{\cond{\false}{P_1}{P_2}} \Par M_1 $ and $ N_2 = \ArbitraryEncoding{M'} $, where $ M' = \pa{\p}{P_2} \Par M_1 $.
					By \RCondF, then $ M'' \red M' $ and thus $ M \red^\ast M' $ and $ N_2 \red^\ast \ArbitraryEncoding{M'} $.
				\item[\RCong:] In this case $ N_1 \equiv N_3 \red N_4 \equiv N_2 $.
					By \RCong. then $ \ArbitraryEncoding{M} \red N_3 $.
					By the induction hypothesis, $ \ArbitraryEncoding{M} \red N_3 \red N_4 $ implies $ M \red^\ast M' $ and $ N_4 \red^\ast \ArbitraryEncoding{M'} $, where $ \ArbitraryEncoding{M} \red^\ast \ArbitraryEncoding{M'} $ contains at most twice as much steps as $ M \red^\ast M' $.
					By \RCong, then $ N_2 \red^\ast \ArbitraryEncoding{M'} $.
			\end{description}
	\end{description}
	Then soundness with $ \asymp $ chosen as $ = $ follows from an induction on the length of the sequence $ \ArbitraryEncoding{M} \red^\ast N $ that alternates the above two cases.
\end{proof}

Safety ensures that for every output that was picked in the first step (with the fresh label $ \colorbox{green!20}{\lbl{enc_o}} $) there will be a matching input in the second step.
Because of that, we need safety in the case of a single step from $ \ArbitraryEncoding{M} $ that is an interaction.

\begin{theorem}[From \scbs into \bs]
	\label{thm:scbsIntoBs}
	$ \arbitraryEncoding $ in Figure~\ref{fig:encodingScbsIntoBs} is a good encoding from \scbs into \bs.
\end{theorem}

\begin{proof}
	By Lemma~\ref{lem:typeSCBSintoBS}, $ \arbitraryEncoding $ is an encoding from \scbs into \bs.
	Compositionality follows from the definition of $ \arbitraryEncoding $ in Figure~\ref{fig:encodingScbsIntoBs}.
	By Lemma~\ref{lem:niSCBSintoBS}, $ \arbitraryEncoding $ is name invariant.
	By the Lemmata~\ref{lem:occSCBSintoBS} and \ref{lem:ocsSCBSintoBS}, $ \arbitraryEncoding $ satisfies operational correspondence.
	Divergence reflection follows from Lemma~\ref{lem:ocsSCBSintoBS}, since it proves a bound on the number of steps to emulate a single source term step.
	By Figure~\ref{fig:encodingScbsIntoBs}, $ M\hasSuccess $ iff $ \ArbitraryEncoding{M}\hasSuccess $.
	By operational correspondence, then $ \arbitraryEncoding $ is success sensitive.
	Because the parallel operator is translated homomorphically, $ \arbitraryEncoding $ preserves distributability.
	We instantiated $ \asymp $ with $ = $, which is an equivalence and a success respecting reduction bisimulation.
	By Definition~\ref{def:goodEncoding}, then $ \arbitraryEncoding $ is a good encoding from \scbs into \bs.
\end{proof}

\subsection{Creating a Total Order on Participants}
\label{app:totalOrder}

Before we look at \mcbs \blueArrow \scbs, we show how the translations of sessions can be used, to compute a total order of the participants in a session.
We will use this order in \mcbs \blueArrow \scbs and other encodings.
The algorithm is given by the partial encoding function in Figure~\ref{fig:totalOrder}.

\begin{figure}
	\begin{align*}
		\ArbitraryEncoding{M_1 \Par M_2} &= \ArbitraryEncoding{M_1}^l_{\Set{ \left( \p, \q \right) \mid \p, \q \in \parts{M_1 \Par M_2} \wedge \p \neq \q }} \Par \ArbitraryEncoding{M_2}^r_{\Set{ \left( \p, \q \right) \mid \p, \q \in \parts{M_1 \Par M_2} \wedge \p \neq \q }}\\
		\ArbitraryEncoding{M_1 \Par M_2}^l_< &= \ArbitraryEncoding{M_1}^l_{\mathsf{lu}(<, \parts{M_1 \Par M_2})} \Par \ArbitraryEncoding{M_2}^r_{\mathsf{lu}(<, \parts{M_1 \Par M_2})}\\
		\ArbitraryEncoding{M_1 \Par M_2}^r_< &= \ArbitraryEncoding{M_1}^l_{\mathsf{ru}(<, \parts{M_1 \Par M_2})} \Par \ArbitraryEncoding{M_2}^r_{\mathsf{ru}(<, \parts{M_1 \Par M_2})}\\
		\ArbitraryEncoding{\pa{\p}{\PP}}^l_< &= \pa{\p}{\ArbitraryEncoding{\PP}_{\mathsf{lu}(<, \parts{M_1 \Par M_2})}}\\
		\ArbitraryEncoding{\pa{\p}{\PP}}^r_< &= \pa{\p}{\ArbitraryEncoding{\PP}_{\mathsf{ru}(<, \parts{M_1 \Par M_2})}}
	\end{align*}
	where $ \mathsf{lu}(<, F) = \Set{ \left( \p, \q \right) \mid \p < \q \wedge \p \in F } \cup \Set{ \left( \p, \q \right) \mid \q \in F \wedge \p \notin \mathsf{dom}(\Set{ \left( \p, \q \right) \mid \p < \q \wedge \p \in F }) } $ and $ \mathsf{ru}(<, F) = \Set{ \left( \p, \q \right) \mid \p < \q \wedge \q \in F } \cup \Set{ \left( \p, \q \right) \mid \p \in F \wedge \q \notin \mathsf{dom}(\Set{ \left( \p, \q \right) \mid \p < \q \wedge \q \in F }) } $.
	\caption{Creating a Total Order on Participants\label{fig:totalOrder}}
\end{figure}

Assuming a total order, would violate name invariance \eg in \mcbs \blueArrow \scbs.
Accordingly, we start without any order in $ \ArbitraryEncoding{M_1 \Par M_2} $.
Then we use the fact, that the contexts in the definition of compositionality can be parametrised by the free names, \ie free participants.
Also not that we cannot use the set of free participants of a subterm, but we can let the encoding function carry the information from the outer context to the translations of the subterm.
Hence, $ \ArbitraryEncoding{M_1 \Par M_2} $ is used only for the outer-most parallel operator.
Inside the tree we are using the functions $ \arbitraryEncoding^x_< $, where $ x \in \Set{l, r} $ indicates a left or right branch and $ < $ is the current relation.

\begin{newexample}
	We start with the $ \forall $-relation in the root and remove invalid pairs from the relation, while moving downwards.
	The union of the leafs gives the total order, where each leaf has the pairs that are relevant for its encoding.
	\begin{align*}
		\ArbitraryEncoding{\left( \pa{\role{a}}{\ldots} \Par \pa{\role{b}}{\ldots} \right) \Par \pa{\role{c}}{\ldots}} &=
		\ArbitraryEncoding{\pa{\role{a}}{\ldots} \Par \pa{\role{b}}{\ldots}}^l_{\Set{\role{a} < \role{b}, \role{a} < \role{c}, \role{b} < \role{a}, \role{b} < \role{c}, \role{c} < \role{a}, \role{c} < \role{b}}} \Par \ArbitraryEncoding{\pa{\role{c}}{\ldots}}^r_{\Set{\role{a} < \role{b}, \role{a} < \role{c}, \role{b} < \role{a}, \role{b} < \role{c}, \role{c} < \role{a}, \role{c} < \role{b}}}\\
		&= \ArbitraryEncoding{\pa{\role{a}}{\ldots}}^l_{\Set{\role{a} < \role{b}, \role{a} < \role{c}, \role{b} < \role{a}, \role{b} < \role{c}}} \Par \ArbitraryEncoding{\pa{\role{b}}{\ldots}}^r_{\Set{\role{a} < \role{b}, \role{a} < \role{c}, \role{b} < \role{a}, \role{b} < \role{c}}} \Par \pa{\role{c}}{\ArbitraryEncoding{\ldots}^r_{\Set{\role{a} < \role{c}, \role{b} < \role{c}}}}\\
		&= \pa{\role{a}}{\ArbitraryEncoding{\ldots}_{\Set{\role{a} < \role{b}, \role{a} < \role{c}}}} \Par \pa{\role{b}}{\ArbitraryEncoding{\ldots}_{\Set{\role{a} < \role{b}, \role{b} < \role{c}}}} \Par \pa{\role{c}}{\ArbitraryEncoding{\ldots}_{\Set{\role{a} < \role{c}, \role{b} < \role{c}}}}
	\end{align*}
\end{newexample}

A drawback of this order is, that structural congruent terms can be translated into structurally very different terms, due to the rule for commutativity of the parallel operator.
More precisely, $ M_1 \equiv M_2 $ does not imply $ \ArbitraryEncoding{M_1} \equiv \ArbitraryEncoding{M_2} $.
Therefore, we use $ M_1 \equiv M_2 $ implies $ \ArbitraryEncoding{M_1} \approx \ArbitraryEncoding{M_2} $ instead with encodings that rely on this order.

Also note that applying structural congruence after the translation in the derivation of steps does not cause any problems.

We do not display the parameters $ x $ and $ < $ of $ \arbitraryEncoding^x_< $ outside of this subsection with the exception of the definition of encoding functions.

\subsection{Encoding \mcbs into \scbs}
\label{app:encodingMcbsIntoScbs}

We show that there is a good encoding \mcbs \blueArrow \scbs, \ie a good encoding from \mcbs into \scbs.
Therefore, we fix for this subsection the function $ \arbitraryEncoding $ to be the encoding given in Figure~\ref{fig:encodingMcbsIntoScbs} and show that $ \arbitraryEncoding $ is a good encoding from \mcbs into \scbs.
The most interesting part of $ \arbitraryEncoding $ is the translation of choice that was already given in Figure~\ref{fig:KeysExpressiveness}(2).
The remaining operators of \mcbs are translated homomorphically by $ \arbitraryEncoding $ except for the construction of $ < $ in Figure~\ref{fig:totalOrder}.

\begin{figure}
	\begin{align*}
		\ArbitraryEncoding{\inact}_{<} &= \inact\\
		\ArbitraryEncoding{\success}_{<} &= \success\\
		\ArbitraryEncoding{\pvar{X}}_{<} &= \pvar{X}\\
		\ArbitraryEncoding{\rec{X} \PP}_{<} &= \rec{X} \ArbitraryEncoding{\PP}_{<}\\
		\ArbitraryEncoding{\left( \choice{\soutprlbl{\q}{\lab_i}{v_i}\seq\PP_i}{i \in I} \right) \OR \left( \choice{\sinprlbl{\q}{\lab_j'}{x_j}\seq Q_j}{j \in J} \right)}_{<}
		&= \begin{cases}
			\begin{array}{l}
				\left( \choice{\soutprlbl{\q}{\lab_i}{v_i}\seq\ArbitraryEncoding{\PP_i}_{<}}{i \in I} \right) \OR\\
				\colorbox{red!10}{\ssout{\q}{\lbl{enc_i}}}\seq\left( \left(\choice{\sinprlbl{\q}{\lab_j'}{x_j}\seq\ArbitraryEncoding{Q_j}_{<}}{j \in J} \right) \OR \colorbox{yellow!30}{\ssinp{\q}{\lbl{reset}}}\seq \choice{\soutprlbl{\q}{\lab_i}{v_i}\seq\ArbitraryEncoding{\PP_i}}{i \in I} \right)
			\end{array} & \text{if } I \neq \emptyset \neq J \wedge \p < \q\\
			\left( \choice{\sinprlbl{\q}{\lab_j'}{x_j}\seq\ArbitraryEncoding{Q_j}_{<}}{j \in J} \right) \OR \colorbox{red!10}{\ssinp{\q}{\lbl{enc_i}}}\seq\choice{\soutprlbl{\q}{\lab_i}{v_i}\seq\ArbitraryEncoding{\PP_i}_{<}}{i \in I} & \text{if } I \neq \emptyset \neq J \wedge \p > \q\\
			\choice{\soutprlbl{\q}{\lab_i}{v_i}\seq\ArbitraryEncoding{\PP_i}_{<}}{i \in I} & \text{if } J = \emptyset \wedge \p < \q\\
			\colorbox{red!10}{\ssinp{\q}{\lbl{enc_i}}}\seq\choice{\soutprlbl{\q}{\lab_i}{v_i}\seq\ArbitraryEncoding{\PP_i}_{<}}{i \in I} & \text{if } J = \emptyset \wedge \p > \q\\
			\colorbox{red!10}{\ssout{\q}{\lbl{enc_i}}}\seq\choice{\sinprlbl{\q}{\lab_j'}{x_j}\seq\ArbitraryEncoding{Q_j}_{<}}{j \in J} & \text{if } I = \emptyset \wedge \p < \q\\
			\left( \choice{\sinprlbl{\q}{\lab_j'}{x_j}\seq\ArbitraryEncoding{Q_j}_{<}}{j \in J} \right) \OR \colorbox{red!10}{\ssinp{\q}{\lbl{enc_i}}}\seq\colorbox{yellow!30}{\ssout{\q}{\lbl{reset}}}\seq\choice{\sinprlbl{\q}{\lab_j'}{x_j}\seq\ArbitraryEncoding{Q_j}_{<}}{j \in J} & \text{if } I = \emptyset \wedge \p > \q
		\end{cases}\\
		\ArbitraryEncoding{\cond{v}{\PP}{Q}}_{<} &= \cond{v}{\ArbitraryEncoding{\PP}_{<}}{\ArbitraryEncoding{Q}_{<}}
	\end{align*}
	where $ \lbl{enc_i} $ and $ \lbl{reset} $ are labels that do not occur in any source term.
	\caption{Encoding from \mcbs into \scbs\label{fig:encodingMcbsIntoScbs}}
\end{figure}

In contrast to \scbs with only separate choice, \mcbs has mixed choices.
To split mixed choice, $ \arbitraryEncoding $ translates a choice from $ \p $ to $ \q $ with $ \p < \q $ into an output-guarded choice and with $ \p > \q $ into an input-guarded choice, where $ \colorbox{red!10}{\ssout{\q}{\lbl{enc_i}}} $ (or $ \colorbox{red!10}{\ssinp{\q}{\lbl{enc_i}}} $) are used to guard the outputs in the input-guarded choice (the inputs in the output-guarded choice).

An interaction of an output in \p and an input in \q with $ \p < \q $ can be emulated by a single step, whereas the case $ \p > \q $ requires two steps: one interaction with the fresh label $ \colorbox{red!10}{\lbl{enc_i}} $ and the interaction of the respective out- and input.
In the case $ \p < \q $, additional inputs in $ \p $ without matching outputs in $ \q $ require special attention.
The type system of the target calculus \scbs forces us to ensure that the output $ \colorbox{red!10}{\ssout{\q}{\lbl{enc_i}}} $, that guards the inputs in the translation of \p, is matched by an input $ \colorbox{red!10}{\ssinp{\p}{\lbl{enc_i}}} $ in the translation of \q (see Corollary~\ref{cor:typesafety}).
We add $ \colorbox{red!10}{\ssinp{\p}{\lbl{enc_i}}}\seq\colorbox{yellow!30}{\ssout{\p}{\lbl{reset}}}\seq\choice{\sinprlbl{\p}{\lab_j}{x_j}\seq\ArbitraryEncoding{\PP_j}}{j \in J} $ to the translation of \q (the last summand in the last line in the translation of choice in Figure~\ref{fig:encodingMcbsIntoScbs} with \p and \q swapped).
If \p has outputs and inputs but no input has a match in \q, then the emulation of a single source term step takes either one or three steps.

Outputs of \p can directly interact with inputs of \q emulating in one step.
Alternatively, there can be a step with label $ \colorbox{red!10}{\lbl{enc_i}} $ discarding temporarily the outputs of \p and the inputs of \q, followed by a step with label $ \colorbox{yellow!30}{\lbl{reset}} $, restoring the outputs and inputs but not allowing for another step with label $ \colorbox{red!10}{\lbl{enc_i}} $, and ultimately the interaction of an output of \p and an input of \q.

The encoding induces a translation on types.
$ \TypeEncoding{\LL} $ is inductively defined as:
\begin{align*}
	\TypeEncoding{\emptyset} &= \emptyset\\
	\TypeEncoding{\LL', \p{:}\local} &= \TypeEncoding{\LL'}, \p{:}\TypeEncoding{\local}\\
	\TypeEncoding{\tinact} &= \tinact\\
	\TypeEncoding{\tvar{t}} &= \tvar{t}\\
	\TypeEncoding{\trec{t} \local} &= \trec{t} \TypeEncoding{\local}\\
	\TypeEncoding{\Tor{i \in I}{\toutlbl{\q}{\lab_i}{\UType_i}\local_i} \; \tor \; \Tor{j \in J}{\tinplbl{\q}{\lab_j'}{\UType_j'}\local_j}} &=
		\begin{cases}
			\begin{array}{l}
				\left( \Tor{i \in I}{\toutlbl{\q}{\lab_i}{\UType_i}\seq\TypeEncoding{T_i}} \right) \; \tor\\
				\tout{\q}{\lbl{enc_i}}\seq\left( \left(\Tor{j \in J}{\tinplbl{\q}{\lab_j'}{\UType_j'}\seq\TypeEncoding{T_j'}} \right) \; \tor \; \tinp{\q}{\lbl{reset}}\seq \Tor{i \in I}{\toutlbl{\q}{\lab_i}{\UType_i}\seq\TypeEncoding{T_i}} \right)
			\end{array} & \text{if } I \neq \emptyset \neq J \wedge \p < \q\\
			\left( \choice{\tinplbl{\q}{\lab_j'}{\UType_j'}\seq\TypeEncoding{T_j'}}{j \in J} \right) \; \tor \; \tinp{\q}{\lbl{enc_i}}\seq\Tor{i \in I}{\toutlbl{\q}{\lab_i}{\UType_i}\seq\TypeEncoding{T_i}} & \text{if } I \neq \emptyset \neq J \wedge \p > \q\\
			\Tor{i \in I}{\toutlbl{\q}{\lab_i}{\UType_i}\seq\TypeEncoding{T_i}} & \text{if } J = \emptyset \wedge \p < \q\\
			\tinp{\q}{\lbl{enc_i}}\seq\Tor{i \in I}{\toutlbl{\q}{\lab_i}{\UType_i}\seq\TypeEncoding{T_i}} & \text{if } J = \emptyset \wedge \p > \q\\
			\tout{\q}{\lbl{enc_i}}\seq\Tor{j \in J}{\tinplbl{\q}{\lab_j'}{\UType_j'}\seq\TypeEncoding{T_j'}} & \text{if } I = \emptyset \wedge \p < \q\\
			\left( \Tor{j \in J}{\tinplbl{\q}{\lab_j'}{\UType_j'}\seq\TypeEncoding{T_j'}} \right) \; \tor \; \tinp{\q}{\lbl{enc_i}}\seq\tout{\q}{\lbl{reset}}\seq\Tor{j \in J}{\tinplbl{\q}{\lab_j'}{\UType_j'}\seq\TypeEncoding{T_j'}} & \text{if } I = \emptyset \wedge \p > \q
		\end{cases}
\end{align*}

We show that the translation of types satisfies a variant of operational correspondence for types.

\begin{lemma}
	\label{lem:ocTypesMCBSintoSCBS}
	Assume $ \safe{\LL} $ and $ \df{\LL} $.
	Then:
	\begin{description}
		\item[Completeness:] If $ \LL \by{\lambda} \LL' $, then $ \TypeEncoding{\LL} \red^\ast\by{\lambda} \TypeEncoding{\LL'} $.
		\item[Soundness:] If $ \TypeEncoding{\LL} \red^\ast \LL_{\mathsf{enc}} $, then there is some $ \LL' $ such that $ \LL \red^\ast \LL' $ and $ \LL_{\mathsf{enc}} \red^\ast \TypeEncoding{\LL'} $.
	\end{description}
\end{lemma}

\begin{proof}
	The proof is similar to the proof of Lemma~\ref{lem:ocTypesSCBSintoBS}.
	We revisit the completeness case and adapt it to the present setting, to illustrate the differences.
	\begin{description}
		\item[Completeness:] Assume $ \LL \by{\actlbl{\p}{\q}{\lab}{\UType}} \LL' $.
			By Figure~\ref{fig:type_semantics}, then $ \LL = \LL_{\text{no }\p\q}, \p{:}T_{\p}, \q{:}T_{\q} $ and $ \LL' = \LL_{\text{no }\p\q}, \p{:}T_{\p, n}, \q{:}T_{\q, m} $, where $ T_{\p} $ is either $ \toutlbl{\q}{\lab}{\UType} T_{\p}' \; \tor \; T_{\Sigma, \p} $ or $ \trec{t} \toutlbl{\q}{\lab}{\UType} T_{\p}' \; \tor \; T_{\Sigma, \p} $ and $ T_{\q} $ is either $ \tinplbl{\p}{\lab}{\UType} T_{\q}' \; \tor \; T_{\Sigma, \q} $ or $ \trec{t} \tinplbl{\p}{\lab}{\UType} T_{\q}' \; \tor \; T_{\Sigma, \q} $.
			Then $ \TypeEncoding{\LL} = \TypeEncoding{\LL_{\text{no }\p\q}}, \p{:}\TypeEncoding{T_{\p}}, \q{:}\TypeEncoding{T_{\q}} $ and $ \TypeEncoding{\LL'} = \TypeEncoding{\LL_{\text{no }\p\q}}, \p{:}\TypeEncoding{T_{\p}'}, \q{:}\TypeEncoding{T_{\q}'} $.
			\begin{description}
				\item[$ \p < \q $:] Then $ \TypeEncoding{T_{\p}} $ is either $ \toutlbl{\q}{\lab}{\UType} \TypeEncoding{T_{\p}'} \; \tor \; T_{\mathsf{enc}, \Sigma, \p} $ or $ \trec{t} \left( \toutlbl{\q}{\lab}{\UType} \TypeEncoding{T_{\p}'} \; \tor \; T_{\mathsf{enc}, \Sigma, \p} \right) $ and $ \TypeEncoding{T_{\q}} $ is either $ \tinplbl{\p}{\lab}{\UType} \TypeEncoding{T_{\q}'} \; \tor \; T_{\mathsf{enc}, \Sigma, \q} $ or we have $ \trec{t} \left( \tinplbl{\p}{\lab}{\UType} \TypeEncoding{T_{\q}'} \; \tor \; T_{\mathsf{enc}, \Sigma, \q} \right) $.
					
					Then $ \TypeEncoding{\LL} \by{\actlbl{\p}{\q}{\lab}{\UType}} \TypeEncoding{\LL'} $ and thus $ \TypeEncoding{\LL} \red^\ast \by{\actlbl{\p}{\q}{\lab}{\UType}} \TypeEncoding{\LL'} $.
				\item[$ \p > \q $:] Then $ \TypeEncoding{T_{\p}} $ is either $ \tinp{\q}{\lbl{enc_i}}\toutlbl{\q}{\lab}{\UType} \TypeEncoding{T_{\p}'} \; \tor \; T_{\mathsf{enc}, \Sigma, \p} $ or $ \trec{t} \left( \tinp{\q}{\lbl{enc_i}}\toutlbl{\q}{\lab}{\UType} \TypeEncoding{T_{\p}'} \; \tor \; T_{\mathsf{enc}, \Sigma, \p} \right) $ and $ \TypeEncoding{T_{\q}} $ is $ \tout{\p}{\lbl{enc_i}}\tinplbl{\p}{\lab}{\UType} \TypeEncoding{T_{\q}'} \; \tor \; T_{\mathsf{enc}, \Sigma, \q} $ or $ \trec{t} \left( \tout{\p}{\lbl{enc_i}}\tinplbl{\p}{\lab}{\UType} \TypeEncoding{T_{\q}'} \; \tor \; T_{\mathsf{enc}, \Sigma, \q} \right) $.
					Then $ \TypeEncoding{\LL} \red \by{\actlbl{\p}{\q}{\lab}{\UType}} \TypeEncoding{\LL'} $ and thus $ \TypeEncoding{\LL} \red^\ast \by{\actlbl{\p}{\q}{\lab}{\UType}} \TypeEncoding{\LL'} $.
					\qedhere
			\end{description}
	\end{description}
\end{proof}

With completeness and soundness of the translation of types we show that the translation on types preserves safety and deadlock-freedom.

\begin{lemma}
\label{lem:DFofMCBSintoSCBS}
If $ \safe{\LL} $ and $ \df{\LL} $ then $ \safe{\TypeEncoding{\LL}} $ and $ \df{\TypeEncoding{\LL}} $.
\end{lemma}

\begin{proof}
	Similar to the proof of Lemma~\ref{lem:DFofSCBSintoBS}.
\end{proof}

With Lemma~\ref{lem:ocTypesMCBSintoSCBS} and Lemma~\ref{lem:DFofMCBSintoSCBS}, we show that $ \arbitraryEncoding $ is an encoding from \mcbs into \scbs, \ie that it translates typed and deadlock-free \mcbs-sessions into typed and deadlock-free \scbs-sessions.

\begin{lemma}
\label{lem:typeMCBSintoSCBS}
$ \ArbitraryEncoding{M} $ is typed and deadlock-free.
\end{lemma}

\begin{proof}
	Similar to the proof of Lemma~\ref{lem:typeSCBSintoBS}.
\end{proof}


To prove that the encoding \mcbs \blueArrow \scbs is good, we show first that it respects structural congruence up-to $ \approx $.
Let $ \equiv_{-c} $ be structural congruence without the rule $ M \Par M' \equiv M' \Par M $.

\begin{lemma}
	\label{lem:scMCBSintoSCBS}
	\begin{enumerate}
		\item $ P_1 \equiv P_2 $ iff $ \ArbitraryEncoding{P_1} \equiv \ArbitraryEncoding{P_2} $.
		\item $ M_1 \equiv_{-c} M_2 $ iff $ \ArbitraryEncoding{M_1} \equiv_{-c} \ArbitraryEncoding{M_2} $.
		\item $ M_1 \equiv M_2 $ iff $ \ArbitraryEncoding{M_1} \approx \ArbitraryEncoding{M_2} $.
		\item If $ \ArbitraryEncoding{M_1} \equiv M $ then there is some $ M_2 $ such that $ M = \ArbitraryEncoding{M_2} $.
	\end{enumerate}
\end{lemma}

\begin{proof}
	We prove the properties (1), (2), and (4) by structural induction on the definition of structural congruence in Figure~\ref{fig:reduction}.
	Since the encoding does not change names and translates inaction, recursion, and parallel composition homomorphically, all cases follow directly from the respective assumptions or the induction hypothesis.
	The only interesting case is commutativity of the parallel operator in property (3).
	By reordering the participants, the total order generated by the partial translation in Figure~\ref{fig:totalOrder} changes.
	The encoding function in Figure~\ref{fig:encodingMcbsIntoScbs} ensures, that if a participant receives dual treatment due to reordering then so does its communication partner in the respective translations of choice.
	Since in all cases, either both summands remain unguarded or both are guarded by dual prefixes with the label $ \colorbox{red!10}{\lbl{enc_i}} $, $ M_1 \equiv M_2 $ still implies $ \ArbitraryEncoding{M_1} \approx \ArbitraryEncoding{M_2} $.
\end{proof}

Compositionality is again immediate from the definition of the encoding function in the Figures~\ref{fig:encodingMcbsIntoScbs} and \ref{fig:totalOrder}.
Name invariance can be shown by structural induction.

\begin{lemma}
	\label{lem:niMCBSintoSCBS}
	The encoding $ \arbitraryEncoding $ is name invariant, \ie $ \ArbitraryEncoding{M \sigma} = \ArbitraryEncoding{M} \sigma $ for all substitutions $ \sigma $ such that $ M\sigma $ is defined.
\end{lemma}

\begin{proof}
	By induction on the structure of $ M $.
	Since, $ \arbitraryEncoding $ translates names of the source to themselves and does neither reserve names nor introduces additional names, all case follow directly from the respective assumptions or the induction hypothesis.
\end{proof}

The encoding \mcbs \blueArrow \scbs translates a source term step into 1-3 steps of the target.

\begin{lemma}
	\label{lem:occMCBSintoSCBS}
	The encoding $ \arbitraryEncoding $ is operationally complete, \ie $ M \red^\ast M' $ implies $ \ArbitraryEncoding{M} \red^\ast \approx \ArbitraryEncoding{M'} $.
\end{lemma}

\begin{proof}
	Assume $ M \red M' $.
	We proceed with an induction on the derivation of $ M \red M' $ by the rules in Figure~\ref{fig:reduction} and show $ \ArbitraryEncoding{M} \red^\ast \approx \ArbitraryEncoding{M'} $.
	\begin{description}
		\item[\RChoice:] Then $ M = \pa{\p}{(\sinprlbl{\role{q}}{\lab}{x} \seq\PP + R_1)} \Par \pa{\q}{(\soutprlbl{\role{p}}{\lab}{v} \seq Q + R_2)} $ and $ M' = \pa{\p}{\PP\subst{v}{x}} \Par \pa{\q}{Q} $.
			Then $ \p < \q $ from the structure of $ M $.
			Then $ \ArbitraryEncoding{M} = \pa{\p}{\left( \colorbox{red!10}{\ssout{\q}{\lbl{enc_i}}}\seq\left( \sinprlbl{\role{q}}{\lab}{x} \seq\PP + R_3 \right) + R_4 \right)} \Par \pa{\q}{\left( \colorbox{red!10}{\ssout{\p}{\lbl{enc_i}}}\seq\left( \soutprlbl{\p}{\lab}{v}\seq\ArbitraryEncoding{Q} + R_5 \right) + R_6 \right)} $ and $ \ArbitraryEncoding{M'} = \pa{\p}{\ArbitraryEncoding{\PP\subst{v}{x}}} \Par \pa{\q}{\ArbitraryEncoding{Q}} $.
			By Lemma~\ref{lem:niMCBSintoSCBS}, then $ \ArbitraryEncoding{M'} = \pa{\p}{\ArbitraryEncoding{\PP}\subst{v}{x}} \Par \pa{\q}{\ArbitraryEncoding{Q}} $.
			By \RChoice, then $ \ArbitraryEncoding{M} \red \pa{\p}{ \left( \sinprlbl{\role{q}}{\lab}{x} \seq\ArbitraryEncoding{\PP} + R_3 \right)} \Par \pa{\q}{\left( \soutprlbl{\p}{\lab}{v}\seq\ArbitraryEncoding{Q} + R_5 \right)} \red \ArbitraryEncoding{M'} $ and thus $ \ArbitraryEncoding{M} \red^\ast \approx \ArbitraryEncoding{M'} $.
		\item[\RCondT:] Then $ M = \pa{\p}{\cond{\true}{P_1}{P_2}} \Par M_1 $ and $ M' = \pa{\p}{P_1} \Par M_1 $.
			Then $ \ArbitraryEncoding{M} = \pa{\p}{\cond{\true}{\ArbitraryEncoding{P_1}}{\ArbitraryEncoding{P_2}}} \Par \ArbitraryEncoding{M_1} $ and $ \ArbitraryEncoding{M'} = \pa{\p}{\ArbitraryEncoding{P_1}} \Par \ArbitraryEncoding{M_1} $.
			By \RCondT, then $ \ArbitraryEncoding{M} \red \ArbitraryEncoding{M'} $ and thus $ \ArbitraryEncoding{M} \red^\ast \approx \ArbitraryEncoding{M'} $.
		\item[\RCondF:] Then $ M = \pa{\p}{\cond{\false}{P_1}{P_2}} \Par M_1 $ and $ M' = \pa{\p}{P_2} \Par M_1 $.
			Then $ \ArbitraryEncoding{M} = \pa{\p}{\cond{\false}{\ArbitraryEncoding{P_1}}{\ArbitraryEncoding{P_2}}} \Par \ArbitraryEncoding{M_1} $ and $ \ArbitraryEncoding{M'} = \pa{\p}{\ArbitraryEncoding{P_2}} \Par \ArbitraryEncoding{M_1} $.
			By \RCondF, then $ \ArbitraryEncoding{M} \red \ArbitraryEncoding{M'} $ and thus $ \ArbitraryEncoding{M} \red^\ast \approx \ArbitraryEncoding{M'} $.
		\item[\RCong:] In this case $ M \equiv M_1 \red M_2 \equiv M' $.
			By Lemma~\ref{lem:scMCBSintoSCBS}, then $ \ArbitraryEncoding{M} \approx \ArbitraryEncoding{M_1} $ and $ \ArbitraryEncoding{M_2} \approx \ArbitraryEncoding{M'} $.
			By the induction hypothesis, $ M_1 \red M_2 $ implies $ \ArbitraryEncoding{M_1} \red^\ast \approx \ArbitraryEncoding{M_2} $.
			Then $ \ArbitraryEncoding{M} \red^\ast \approx \ArbitraryEncoding{M'} $.
	\end{description}
	By induction on the number of steps in $ M \red^\ast M' $, then $ M \red^\ast M' $ implies $ \ArbitraryEncoding{M} \red^\ast \approx \ArbitraryEncoding{M'} $.
\end{proof}

For operational soundness we have to show that every step of the translation belongs to an emulation and that every emulation can be completed.

\begin{lemma}
	\label{lem:ocsMCBSintoSCBS}
	The encoding $ \arbitraryEncoding $ is operationally sound, \ie $ \ArbitraryEncoding{M} \red^\ast N $ implies $ M \red^\ast M' $ and $ N \red^\ast \approx \ArbitraryEncoding{M'} $, where $ \ArbitraryEncoding{M} \red^\ast \approx \ArbitraryEncoding{M'} $ contains at most three times as much steps as $ M \red^\ast M' $.
\end{lemma}

\begin{proof}
	The proof is by induction on the number of steps in $ \ArbitraryEncoding{M} \red^\ast N $.
	\begin{description}
		\item[Base Case ($ \ArbitraryEncoding{M} = N $):] Then $ M \red^\ast M $ and $ N \red^\ast \approx \ArbitraryEncoding{M} $.
		\item[Induction Step ($ \ArbitraryEncoding{M} \red^\ast N_1 \red N $):] By the induction hypothesis, $ M \red^\ast M_1 $ and $ N_1 \red^\ast \approx \ArbitraryEncoding{M_1} $, where $ \ArbitraryEncoding{M} \red^\ast \approx \ArbitraryEncoding{M_1} $ contains at most three times as much steps as $ M \red^\ast M_1 $.
			Then (1)~$ N_1 \red N $ is contained in the sequence $ N_1 \red^\ast \approx \ArbitraryEncoding{M_1} $, (2)~$ N_1 \red N $ and any step in this sequence are distributable, or (3)~$ N_1 \red N $ is in conflict with one of the steps in this sequence.
			\begin{enumerate}
				\item Assume that $ N_1 \red N $ is contained in $ N_1 \red^\ast \approx \ArbitraryEncoding{M_1} $, \ie a step of the sequence $ N_1 \red^\ast \approx \ArbitraryEncoding{M_1} $ and $ N_1 \red N $ reduce exactly the same capabilities in the same way (picking the same summands).
					Since distributable steps are confluent, then we can reorder the steps in the sequence $ N_1 \red^\ast \approx \ArbitraryEncoding{M_1} $ such that $ N_1 \red N \red^\ast \approx \ArbitraryEncoding{M_1} $.
					Then $ M \red^\ast M_1 = M' $ and $ N \red^\ast \approx \ArbitraryEncoding{M'} $, where $ \ArbitraryEncoding{M} \red^\ast \approx \ArbitraryEncoding{M'} $ contains at most three times as much steps as $ M \red^\ast M' $.
				\item Assume that $ N_1 \red N $ and any step in $ N_1 \red^\ast \approx \ArbitraryEncoding{M_1} $ are distributable.
					We proceed with an induction on the derivation of $ N_1 \red N $ by the rules in Figure~\ref{fig:reduction} to show that $ M_1 \red M' $ and $ N \red^\ast \approx \ArbitraryEncoding{M'} $, where $ \ArbitraryEncoding{M_1} \red^\ast \approx \ArbitraryEncoding{M'} $ contains at most three times as much steps as $ M \red^\ast M_1 $.
					\begin{description}
						\item[\RChoice:] In binary sessions an interaction is not distributable with any other step, \ie in this case the sequence $ N_1 \red^\ast \approx \ArbitraryEncoding{M_1} $ is empty and $ N_1 \approx \ArbitraryEncoding{M_1} $.
							There are steps on the labels $ \lbl{enc_i} $ and $ \lbl{reset} $ as well as steps on labels of the source term.
							\begin{description}
								\item[$ \lbl{enc_i} $:] Then $ N_1 = \pa{\p}{N_P} \Par \pa{\q}{N_Q} $, where either we have (a)~$ N_P = \colorbox{red!10}{\ssinp{\q}{\lbl{enc_i}}}\seq\left( \soutprlbl{\q}{\lab}{v}\seq\ArbitraryEncoding{\PP} + N_1 \right) + N_2 $ with $ \p > \q $ or we have (b)~$ N_P = \colorbox{red!10}{\ssinp{\q}{\lbl{enc_i}}}\seq\colorbox{yellow!30}{\ssinp{\q}{\lbl{reset}}}\seq\left( \soutprlbl{\q}{\lab}{v}\seq\ArbitraryEncoding{\PP} + N_1 \right) + N_2 $ with $ \p < \q $.
									\begin{enumerate}
										\item Since the source was typed, safety ensures that there is a matching input for $ \soutprlbl{\q}{\lab}{v} $ in the source.
											Then $ N_Q = \colorbox{red!10}{\ssout{\q}{\lbl{enc_i}}}\seq\left( \sinprlbl{\q}{\lab}{x}\seq\ArbitraryEncoding{Q} + N_3 \right) + N_4 $ and $ N = \pa{\p}{\soutprlbl{\q}{\lab}{v}\seq\ArbitraryEncoding{\PP} + N_1} \Par \pa{\q}{\sinprlbl{\p}{\lab}{x}\seq\ArbitraryEncoding{Q} + N_3} $.
										\item Then again safety in the source ensures that there is a matching input in $ N_Q $ for the output $ \soutprlbl{\q}{\lab}{v} $ in $ N_P $, \ie $ N_Q = \colorbox{red!10}{\ssinp{\q}{\lbl{enc_i}}}\seq\colorbox{yellow!30}{\ssout{\q}{\lbl{reset}}}\seq\left( \sinprlbl{\q}{\lab}{x}\seq\ArbitraryEncoding{Q} + N_3 \right) + N_4 $ and $ N = \pa{\p}{\colorbox{yellow!30}{\ssinp{\q}{\lbl{reset}}}\seq\left( \soutprlbl{\q}{\lab}{v}\seq\ArbitraryEncoding{\PP} + N_1 \right)} \Par \pa{\q}{\colorbox{yellow!30}{\ssout{\p}{\lbl{reset}}}\seq\left( \sinprlbl{\p}{\lab}{x}\seq\ArbitraryEncoding{Q} + N_3 \right)} $.
									\end{enumerate}
									In both cases, then $ \ArbitraryEncoding{M_1} \approx N_1 \red N \red^\ast \ArbitraryEncoding{M'} $ by 2-3 applications of \RChoice (in (a) $ N_1 \red N \red $ or (b) $ N_1 \red N \red \red $), where $ M' = \pa{\p}{\ArbitraryEncoding{\PP}} \Par \pa{\q}{\ArbitraryEncoding{Q}\subst{v}{x}} $.
									By Lemma~\ref{lem:niMCBSintoSCBS}, $ \ArbitraryEncoding{Q}\subst{v}{x} = 	\ArbitraryEncoding{Q\subst{v}{x}} $.
									Then $ M_1 \red M' $.
								\item[$ \lbl{reset} $:] Similar to the second step of case (b) above.
								\item[other labels:] In this case $ N_1 = \pa{\p}{\left( \soutprlbl{\q}{\lab}{v}\seq\ArbitraryEncoding{\PP} + N_1 \right)} \Par \pa{\q}{\sinprlbl{\q}{\lab}{x}\seq\ArbitraryEncoding{Q} + N_2} $ with $ \p < \q $.
									Then $ \ArbitraryEncoding{M_1} \approx N_1 \red N \approx \ArbitraryEncoding{M'} $, where $ M' = \pa{\p}{\ArbitraryEncoding{\PP}} \Par \pa{\q}{\ArbitraryEncoding{Q}\subst{v}{x}} $.
									By Lemma~\ref{lem:niMCBSintoSCBS}, $ \ArbitraryEncoding{Q}\subst{v}{x} = 	\ArbitraryEncoding{Q\subst{v}{x}} $.
									Then $ M_1 \red M' $.
							\end{description}
						\item[\RCondT:] Then $ N_1 = \pa{\p}{\cond{\true}{\ArbitraryEncoding{P_1}}{\ArbitraryEncoding{P_2}}} \Par N' $ and $ N = \pa{\p}{\ArbitraryEncoding{P_1}} \Par N' $.
							Then $ M_1 = \pa{\p}{\cond{\false}{P_1}{P_2}} \Par M_1' $, $ M' = \pa{\p}{P_1} \Par M_1' $, and $ N \red^\ast \ArbitraryEncoding{M'} $.
							By \RCondT, then $ M_1 \red M' $.
						\item[\RCondF:] Then $ N_1 = \pa{\p}{\cond{\false}{\ArbitraryEncoding{P_1}}{\ArbitraryEncoding{P_2}}} \Par N' $ and $ N = \pa{\p}{\ArbitraryEncoding{P_2}} \Par N' $.
							Then $ M_1 = \pa{\p}{\cond{\false}{P_1}{P_2}} \Par M_1' $, $ M' = \pa{\p}{P_2} \Par M_1' $, and $ N \red^\ast \ArbitraryEncoding{M'} $.
							By \RCondF, then $ M_1 \red M' $.
						\item[\RCong:] In this case $ N_1 \equiv N_2 \red N_3 \equiv N $.
							By \RCong, then $ N_1 \red N_3 $.
							By the induction hypothesis, $ M_1 \red M' $ and $ N_3 \red^\ast \approx \ArbitraryEncoding{M'} $, where $ \ArbitraryEncoding{M_1} \red^\ast \ArbitraryEncoding{M'} $ contains at most three times as much steps as $ M_1 \red^\ast M' $.
							Then $ M \red^\ast M_1 \red M' $ and thus $ M \red^\ast M' $.
							By \RCong, $ N \red^\ast \approx \ArbitraryEncoding{M'} $.
					\end{description}
				\item Assume that $ N_1 \red N $ is in conflict with one of the steps in $ N_1 \red^\ast N_1' \approx \ArbitraryEncoding{M_1} $.
					Then $ N_1 \red N $ is an interaction.
					Then $ N_1 \red N $ does not reduce the label $ \lbl{reset} $, since steps on $ \lbl{reset} $ cannot be in conflict with other steps.
					If $ N_1 \red N $ does not reduce the label $ \lbl{enc_i} $, it emulates a source term step by a single step.
					Then $ N_1 \red N \red^\ast \approx \ArbitraryEncoding{M'} $ is obtained by removing in $ N_1 \red^\ast N_1' $ the conflicting step and all steps enabled by it.
					$ M' $ is then the result of the source steps emulated in $ N_1 \red N \red^\ast \approx \ArbitraryEncoding{M'} $, \ie $ M \red^\ast M' $.

					If $ N_1 \red N $ reduces the label $ \lbl{enc_i} $, then this step starts an emulation that has to be grouped with the second or second and third step necessary to complete this emulation as described in the cases (a) and (b) above.
					Accordingly, after removing the conflicting step and the steps it enabled from $ N_1 \red^\ast N_1' $ the steps that are necessary to complete the emulation started by $ N_1 \red N $ are added.
					Then again $ N_1 \red N \red^\ast \approx \ArbitraryEncoding{M'} $, \ie $ M \red^\ast M' $.
					\qedhere
			\end{enumerate}
	\end{description}
\end{proof}

In the case of $ \p > \q $ (see case~(b) in the proof above), deadlock-freedom guarantees the existence of the respective communication partner for the first step, and safety the existence of a matching input for every output in the second step.

\begin{theorem}[From \mcbs into \scbs]
	\label{thm:mcbsIntoScbs}
	$ \arbitraryEncoding $ in Figure~\ref{fig:encodingMcbsIntoScbs} is a good encoding from \mcbs into \scbs.
\end{theorem}

\begin{proof}
	By Lemma~\ref{lem:typeMCBSintoSCBS}, $ \arbitraryEncoding $ is an encoding from \mcbs into \scbs.
	Compositionality follows from the definition of $ \arbitraryEncoding $ in the Figures~\ref{fig:encodingMcbsIntoScbs} and \ref{fig:totalOrder}.
	By Lemma~\ref{lem:niMCBSintoSCBS}, $ \arbitraryEncoding $ is name invariant.
	By the Lemmata~\ref{lem:occMCBSintoSCBS} and \ref{lem:ocsMCBSintoSCBS}, $ \arbitraryEncoding $ satisfies operational correspondence.
	Divergence reflection follows from Lemma~\ref{lem:ocsMCBSintoSCBS}, since it proves a bound on the number of steps to emulate a single source term step.
	By Figure~\ref{fig:encodingMcbsIntoScbs}, $ M\hasSuccess $ iff $ \ArbitraryEncoding{M}\hasSuccess $.
	By operational correspondence, then $ \arbitraryEncoding $ is success sensitive.
	Because the parallel operator is translated homomorphically, $ \arbitraryEncoding $ preserves distributability.
	We instantiated $ \asymp $ with $ \approx $, which is an equivalence and a success respecting reduction bisimulation.
	By Definition~\ref{def:goodEncoding}, then $ \arbitraryEncoding $ is a good encoding from \mcbs into \scbs.
\end{proof}

\subsection{Encoding \mcbs into \bs}
\label{app:encodingMcbsIntoBs}

We show that there is a good encoding \mcbs \blueArrow \bs, \ie a good encoding from \mcbs into \bs.
Therefore, we fix for this subsection the function $ \arbitraryEncoding $ to be the encoding given in Figure~\ref{fig:encodingMcbsIntoBs} and show that $ \arbitraryEncoding $ is a good encoding from \mcbs into \bs.
The most interesting part of $ \arbitraryEncoding $ is the translation of choice that was already given in Figure~\ref{fig:KeysExpressiveness}(3).
The remaining operators of \mcbs are translated homomorphically by $ \arbitraryEncoding $ except for the construction of $ < $ in Figure~\ref{fig:totalOrder}.

\begin{figure}
	\begin{align*}
		\ArbitraryEncoding{\inact}_{<} &= \inact\\
		\ArbitraryEncoding{\success}_{<} &= \success\\
		\ArbitraryEncoding{\pvar{X}}_{<} &= \pvar{X}\\
		\ArbitraryEncoding{\rec{X} \PP}_{<} &= \rec{X} \ArbitraryEncoding{\PP}_{<}\\
		\ArbitraryEncoding{\begin{array}{l}
				\left( \choice{\soutprlbl{\q}{\lab_i}{v_i}\seq\PP_i}{i \in I} \right) \OR\\
				\left( \choice{\sinprlbl{\q}{\lab_j}{x_j}\seq\PP_j}{j \in J} \right)
			\end{array}}_{<}
		&= \begin{cases}
			\begin{array}{l}
				\left( \choice{\colorbox{green!20}{\ssinp{\q}{\lbl{enc_o}}}\seq\soutprlbl{\q}{\lab_i}{v_i}\seq\ArbitraryEncoding{\PP_i}_{<}}{i \in I} \right) \OR\\
				\colorbox{green!20}{\ssinp{\q}{\lbl{enc_o}}}\seq\colorbox{red!10}{\ssout{\q}{\lbl{enc_i}}}\seq\left( \left(\choice{\sinprlbl{\q}{\lab_j}{x_j}\seq\ArbitraryEncoding{\PP_j}_{<}}{j \in J} \right) \OR \colorbox{yellow!30}{\ssinp{\q}{\lbl{reset}}}\seq \choice{\soutprlbl{\q}{\lab_i}{v_i}\seq\ArbitraryEncoding{\PP_i}_{<}}{i \in I} \right)
			\end{array} & \text{if } I \neq \emptyset \neq J \wedge \p < \q\\
			\colorbox{green!20}{\ssout{\q}{\lbl{enc_o}}}\seq\left( \left( \choice{\sinprlbl{\q}{\lab_j}{x_j}\seq\ArbitraryEncoding{\PP_j}_{<}}{j \in J} \right) \OR \colorbox{red!10}{\ssinp{\q}{\lbl{enc_i}}}\seq\choice{\soutprlbl{\q}{\lab_i}{v_i}\seq\ArbitraryEncoding{\PP_i}_{<}}{i \in I} \right) & \text{if } I \neq \emptyset \neq J \wedge \p > \q\\
			\choice{\colorbox{green!20}{\ssinp{\q}{\lbl{enc_o}}}\seq\soutprlbl{\q}{\lab_i}{v_i}\seq\ArbitraryEncoding{\PP_i}_{<}}{i \in I} & \text{if } J = \emptyset \wedge \p < \q\\
			\colorbox{green!20}{\ssout{\q}{\lbl{enc_o}}}\seq\colorbox{red!10}{\ssinp{\q}{\lbl{enc_i}}}\seq\choice{\soutprlbl{\q}{\lab_i}{v_i}\seq\ArbitraryEncoding{\PP_i}_{<}}{i \in I} & \text{if } J = \emptyset \wedge \p > \q\\
			\colorbox{green!20}{\ssinp{\q}{\lbl{enc_o}}}\seq\colorbox{red!10}{\ssout{\q}{\lbl{enc_i}}}\seq\choice{\sinprlbl{\q}{\lab_j}{x_j}\seq\ArbitraryEncoding{\PP_j}_{<}}{j \in J} & \text{if } I = \emptyset \wedge \p < \q\\
			\colorbox{green!20}{\ssout{\q}{\lbl{enc_o}}}\seq\left( \left( \choice{\sinprlbl{\q}{\lab_j}{x_j}\seq\ArbitraryEncoding{\PP_j}_{<}}{j \in J} \right) \OR \colorbox{red!10}{\ssinp{\q}{\lbl{enc_i}}}\seq\colorbox{yellow!30}{\ssout{\q}{\lbl{reset}}}\seq\choice{\sinprlbl{\q}{\lab_j}{x_j}\seq\ArbitraryEncoding{\PP_j}_{<}}{j \in J} \right) & \text{if } I = \emptyset \wedge \p > \q
		\end{cases}\\
		\ArbitraryEncoding{\cond{v}{\PP}{Q}}_{<} &= \cond{v}{\ArbitraryEncoding{\PP}_{<}}{\ArbitraryEncoding{Q}_{<}}
	\end{align*}
	for all $ \mathsf{x} \in \left\{ \mathsf{o}, \mathsf{i} \right\} $, where $ \lbl{enc_i} $, $ \lbl{enc_o} $, and $ \lbl{reset} $ are labels that do not occur in any source term.
	\caption{Encoding from \mcbs into \bs\label{fig:encodingMcbsIntoBs}}
\end{figure}

The encoding \mcbs \blueArrow \bs translates typed and deadlock-free sessions from \mcbs into typed and deadlock-free sessions of \bs.

\begin{lemma}
	\label{lem:typeMCBSintoBS}
	$ \ArbitraryEncoding{M} $ is typed and deadlock-free.
\end{lemma}

\begin{proof}
	By combining the Lemmata~\ref{lem:typeSCBSintoBS} and \ref{lem:typeMCBSintoSCBS}.
\end{proof}

To prove that the encoding \mcbs \blueArrow \bs is good, we show again that it respects structural congruence up-to $ \approx $.
Let $ \equiv_{-c} $ be structural congruence without the rule $ M \Par M' \equiv M' \Par M $.

\begin{lemma}
	\label{lem:scMCBSintoBS}
	\begin{enumerate}
		\item $ P_1 \equiv P_2 $ iff $ \ArbitraryEncoding{P_1} \equiv \ArbitraryEncoding{P_2} $.
		\item $ M_1 \equiv_{-c} M_2 $ iff $ \ArbitraryEncoding{M_1} \equiv_{-c} \ArbitraryEncoding{M_2} $.
		\item $ M_1 \equiv M_2 $ iff $ \ArbitraryEncoding{M_1} \approx \ArbitraryEncoding{M_2} $.
		\item If $ \ArbitraryEncoding{M_1} \equiv M $ then there is some $ M_2 $ such that $ M = \ArbitraryEncoding{M_2} $.
	\end{enumerate}
\end{lemma}

\begin{proof}
	We prove the properties (1), (2), and (4) by structural induction on the definition of structural congruence in Figure~\ref{fig:reduction}.
	Since the encoding does not change names and translates inaction, recursion, and parallel composition homomorphically, all cases follow directly from the respective assumptions or the induction hypothesis.
	The only interesting case is commutativity of the parallel operator in property (3).
	By reordering the participants, the total order generated by the partial translation in Figure~\ref{fig:totalOrder} changes.
	By combining the ideas of translating choice from \scbs \blueArrow \bs and \mcbs \blueArrow, the encoding function in Figure~\ref{fig:encodingMcbsIntoBs} ensures, that if a participant receives dual treatment due to reordering then so does its communication partner in the respective translations of choice.
	As in \scbs \blueArrow \bs all summands in all choices are guard by a prefix with the label $ \colorbox{green!20}{\lbl{enc_o}} $, where for every matching pairs of choices the two parties of the session are treated dually.
	After this first prefix, following the ideas of \mcbs \blueArrow \scbs one or two more prefixes with the label $ \colorbox{red!10}{\lbl{enc_i}} $ an $ \colorbox{yellow!30}{\lbl{reset}} $ may be added, but if so again in a dual way to both participants.
	Hence, $ M_1 \equiv M_2 $ still implies $ \ArbitraryEncoding{M_1} \approx \ArbitraryEncoding{M_2} $.
\end{proof}

Compositionality is again immediate from the definition of the encoding function in the Figures~\ref{fig:encodingMcbsIntoBs} and \ref{fig:totalOrder}.
Name invariance can be shown by structural induction.

\begin{lemma}
	\label{lem:niMCBSintoBS}
	The encoding $ \arbitraryEncoding $ is name invariant, \ie $ \ArbitraryEncoding{M \sigma} = \ArbitraryEncoding{M} \sigma $ for all substitutions $ \sigma $ such that $ M\sigma $ is defined.
\end{lemma}

\begin{proof}
	By induction on the structure of $ M $.
	Since, $ \arbitraryEncoding $ translates names of the source to themselves and does neither reserve names nor introduces additional names, all case follow directly from the respective assumptions or the induction hypothesis.
\end{proof}

By combining the ideas of \scbs \blueArrow \bs and \mcbs \blueArrow \scbs, the encoding \mcbs \blueArrow \bs translates a source term step into 2-4 steps of the target.

\begin{lemma}
	\label{lem:occMCBSintoBS}
	The encoding $ \arbitraryEncoding $ is operationally complete, \ie $ M \red^\ast M' $ implies $ \ArbitraryEncoding{M} \red^\ast \approx \ArbitraryEncoding{M'} $.
\end{lemma}

\begin{proof}
	Assume $ M \red M' $.
	We proceed with an induction on the derivation of $ M \red M' $ by the rules in Figure~\ref{fig:reduction} and show $ \ArbitraryEncoding{M} \red^\ast \approx \ArbitraryEncoding{M'} $.
	\begin{description}
		\item[\RChoice:] Then $ M = \pa{\p}{(\sinprlbl{\role{q}}{\lab}{x} \seq\PP + R_1)} \Par \pa{\q}{(\soutprlbl{\role{p}}{\lab}{v} \seq Q + R_2)} $ and $ M' = \pa{\p}{\PP\subst{v}{x}} \Par \pa{\q}{Q} $.
		Then $ \p < \q $ from the structure of $ M $.
		Then $ \ArbitraryEncoding{M} = \pa{\p}{\left( \colorbox{green!20}{\ssinp{\q}{\lbl{enc_o}}}\seq\colorbox{red!10}{\ssout{\q}{\lbl{enc_i}}}\seq\left( \sinprlbl{\role{q}}{\lab}{x} \seq\PP + R_3 \right) + R_4 \right)} \Par \pa{\q}{\left( \colorbox{green!20}{\ssout{\p}{\lbl{enc_o}}}\seq\colorbox{red!10}{\ssout{\p}{\lbl{enc_i}}}\seq\left( \soutprlbl{\p}{\lab}{v}\seq\ArbitraryEncoding{Q} + R_5 \right) + R_6 \right)} $ and $ \ArbitraryEncoding{M'} = \pa{\p}{\ArbitraryEncoding{\PP\subst{v}{x}}} \Par \pa{\q}{\ArbitraryEncoding{Q}} $.
		By Lemma~\ref{lem:niMCBSintoBS}, then $ \ArbitraryEncoding{M'} = \pa{\p}{\ArbitraryEncoding{\PP}\subst{v}{x}} \Par \pa{\q}{\ArbitraryEncoding{Q}} $.
		By \RChoice, then $ \ArbitraryEncoding{M} \red\red \pa{\p}{ \left( \sinprlbl{\role{q}}{\lab}{x} \seq\ArbitraryEncoding{\PP} + R_3 \right)} \Par \pa{\q}{\left( \soutprlbl{\p}{\lab}{v}\seq\ArbitraryEncoding{Q} + R_5 \right)} \red \ArbitraryEncoding{M'} $ and thus $ \ArbitraryEncoding{M} \red^\ast \approx \ArbitraryEncoding{M'} $.
		\item[\RCondT:] Then $ M = \pa{\p}{\cond{\true}{P_1}{P_2}} \Par M_1 $ and $ M' = \pa{\p}{P_1} \Par M_1 $.
		Then $ \ArbitraryEncoding{M} = \pa{\p}{\cond{\true}{\ArbitraryEncoding{P_1}}{\ArbitraryEncoding{P_2}}} \Par \ArbitraryEncoding{M_1} $ and $ \ArbitraryEncoding{M'} = \pa{\p}{\ArbitraryEncoding{P_1}} \Par \ArbitraryEncoding{M_1} $.
		By \RCondT, then $ \ArbitraryEncoding{M} \red \ArbitraryEncoding{M'} $ and thus $ \ArbitraryEncoding{M} \red^\ast \approx \ArbitraryEncoding{M'} $.
		\item[\RCondF:] Then $ M = \pa{\p}{\cond{\false}{P_1}{P_2}} \Par M_1 $ and $ M' = \pa{\p}{P_2} \Par M_1 $.
		Then $ \ArbitraryEncoding{M} = \pa{\p}{\cond{\false}{\ArbitraryEncoding{P_1}}{\ArbitraryEncoding{P_2}}} \Par \ArbitraryEncoding{M_1} $ and $ \ArbitraryEncoding{M'} = \pa{\p}{\ArbitraryEncoding{P_2}} \Par \ArbitraryEncoding{M_1} $.
		By \RCondF, then $ \ArbitraryEncoding{M} \red \ArbitraryEncoding{M'} $ and thus $ \ArbitraryEncoding{M} \red^\ast \approx \ArbitraryEncoding{M'} $.
		\item[\RCong:] In this case $ M \equiv M_1 \red M_2 \equiv M' $.
		By Lemma~\ref{lem:scMCBSintoBS}, then $ \ArbitraryEncoding{M} \approx \ArbitraryEncoding{M_1} $ and $ \ArbitraryEncoding{M_2} \approx \ArbitraryEncoding{M'} $.
		By the induction hypothesis, $ M_1 \red M_2 $ implies $ \ArbitraryEncoding{M_1} \red^\ast \approx \ArbitraryEncoding{M_2} $.
		Then $ \ArbitraryEncoding{M} \red^\ast \approx \ArbitraryEncoding{M'} $.
	\end{description}
	By induction on the number of steps in $ M \red^\ast M' $, then $ M \red^\ast M' $ implies $ \ArbitraryEncoding{M} \red^\ast \approx \ArbitraryEncoding{M'} $.
\end{proof}

Similarly, operational soundness is a consequence of the soundness of the two encodings in \S~\ref{app:encodingScbsIntoBs} and \ref{app:encodingMcbsIntoScbs}.

\begin{lemma}
\label{lem:ocsMCBSintoBS}
The encoding $ \arbitraryEncoding $ is operationally sound, \ie $ \ArbitraryEncoding{M} \red^\ast N $ implies $ M \red^\ast M' $ and $ N \red^\ast \approx \ArbitraryEncoding{M'} $, where $ \ArbitraryEncoding{M} \red^\ast \approx \ArbitraryEncoding{M'} $ contains at most four times as much steps as $ M \red^\ast M' $.
\end{lemma}

\begin{proof}
	By combining the proofs of the Lemmata~\ref{lem:ocsSCBSintoBS} and \ref{lem:ocsMCBSintoSCBS}.
\end{proof}


We combine the above results, to show that \mcbs \blueArrow \bs is a good encoding.

\begin{theorem}[From \mcbs into \bs]
\label{thm:mcbsIntoBs}
$ \arbitraryEncoding $ in Figure~\ref{fig:encodingMcbsIntoBs} is a good encoding from \mcbs into \bs.
\end{theorem}

\begin{proof}
By Lemma~\ref{lem:typeMCBSintoBS}, $ \arbitraryEncoding $ is an encoding from \mcbs into \bs.
Compositionality follows from the definition of $ \arbitraryEncoding $ in the Figures~\ref{fig:encodingMcbsIntoBs} and \ref{fig:totalOrder}.
By Lemma~\ref{lem:niMCBSintoBS}, $ \arbitraryEncoding $ is name invariant.
By the Lemmata~\ref{lem:occMCBSintoBS} and \ref{lem:ocsMCBSintoBS}, $ \arbitraryEncoding $ satisfies operational correspondence.
Divergence reflection follows from Lemma~\ref{lem:ocsMCBSintoBS}, since it proves a bound on the number of steps to emulate a single source term step.
By Figure~\ref{fig:encodingMcbsIntoBs}, $ M\hasSuccess $ iff $ \ArbitraryEncoding{M}\hasSuccess $.
By operational correspondence, then $ \arbitraryEncoding $ is success sensitive.
Because the parallel operator is translated homomorphically, $ \arbitraryEncoding $ preserves distributability.
We instantiated $ \asymp $ with $ \approx $, which is an equivalence and a success respecting reduction bisimulation.
By Definition~\ref{def:goodEncoding}, then $ \arbitraryEncoding $ is a good encoding from \mcbs into \bs.
\end{proof}

We combine the three encodings \scbs \blueArrow \bs, \mcbs \blueArrow \scbs, and \mcbs \blueArrow \bs and add calculi inclusions, to prove the mutual encodability results depicted by the \GreyBubble{grey square} containing \bs, \scbs, and \mcbs in the bottom layer of Figure~\ref{fig:hierarchy}.

\thmbinarySessions*

\begin{proof}
	By Theorem~\ref{thm:scbsIntoBs}, there is a good encoding from \scbs into \bs.
	By Theorem~\ref{thm:mcbsIntoScbs}, there is a good encoding from \mcbs into \scbs.
	By Theorem~\ref{thm:mcbsIntoBs}, there is a good encoding from \mcbs into \bs.
	Since $ \bs \subset \scbs \subset \mcbs $, the remaining encodings follow from calculi inclusions.
\end{proof}

\subsection{Encoding \smpst into \mpst}
\label{app:encodingSmpstIntoMpst}

The encoding \smpst \blueArrow \mpst encodes choice in the same way as \scbs \blueArrow \bs in \S~\ref{app:encodingScbsIntoBs}.
The source and the target language of \scbs \blueArrow \bs are binary sessions, whereas \smpst and \mpst are multiparty sessions.
We fix for this subsection the function $ \arbitraryEncoding $ to be the encoding given in Figure~\ref{fig:encodingSmpstIntoMpst} and show that $ \arbitraryEncoding $ is a good encoding from \smpst into \mpst.
Note the depicted encoding function in Figure~\ref{fig:encodingSmpstIntoMpst} is the same as in Figure~\ref{fig:encodingScbsIntoBs}.
Again we strengthen the result by instantiating $ \asymp $ with $ = $ (instead of the weaker $ \approx $).

\begin{figure}
	\begin{align*}
		\ArbitraryEncoding{M_1 \Par M_2} &= \ArbitraryEncoding{M_1} \Par \ArbitraryEncoding{M_2}\\
		\ArbitraryEncoding{\pa{\p}{\PP}} &= \pa{\p}{\ArbitraryEncoding{\PP}}\\
		\ArbitraryEncoding{\inact} &= \inact\\
		\ArbitraryEncoding{\success} &= \success\\
		\ArbitraryEncoding{\pvar{X}} &= \pvar{X}\\
		\ArbitraryEncoding{\rec{X} \PP} &= \rec{X} \ArbitraryEncoding{\PP}\\
		\ArbitraryEncoding{\choice{\soutprlbl{\q}{\lab_i}{v_i}\seq\PP_i}{i \in I}} &= \choice{\colorbox{green!20}{\ssinp{\q}{\lbl{enc_o}}}\seq\soutprlbl{\q}{\lab_i}{v_i}\seq\ArbitraryEncoding{\PP_i}}{i \in I}\\
		\ArbitraryEncoding{\choice{\sinprlbl{\p}{\lab_j}{x_j}\seq\PP_j}{j \in J}} &= \colorbox{green!20}{\ssout{\q}{\lbl{enc_o}}}\seq\choice{\sinprlbl{\q}{\lab_j}{x_j}\seq\ArbitraryEncoding{\PP_j}}{j \in J}\\
		\ArbitraryEncoding{\cond{v}{\PP}{Q}} &= \cond{v}{\ArbitraryEncoding{\PP}}{\ArbitraryEncoding{Q}}
	\end{align*}
	where $ \lbl{enc_o} $ is a label that does not occur in any source term.
	\caption{An Encoding from \smpst into \mpst\label{fig:encodingSmpstIntoMpst}}
\end{figure}

Also the translation on types is the same as in \S~\ref{app:encodingScbsIntoBs}.
Let $ \LL $ not contain any mixed choices on types.
Then $ \TypeEncoding{\LL} $ is inductively defined as:
\begin{displaymath}
	\begin{array}{rclcrclcrclcrcl}
		\TypeEncoding{\emptyset} &=& \emptyset
		&& \TypeEncoding{\tinact} &=& \tinact
		&& \TypeEncoding{\tvar{t}} &=& \tvar{t}
		&& \TypeEncoding{\Tor{i \in I}{\toutlbl{\q}{\lab_i}{\UType_i}\local_i}} &=& \tinp{\q}{\lbl{enc_o}}\Tor{i \in I}{\toutlbl{\q}{\lab_i}{\UType_i}\TypeEncoding{\local_i}}\\
		\TypeEncoding{\LL', \p{:}\local} &=& \TypeEncoding{\LL'}, \p{:}\TypeEncoding{\local}
		&&&&
		&& \TypeEncoding{\trec{t} \local} &=& \trec{t} \TypeEncoding{\local}
		&& \TypeEncoding{\Tor{j \in J}{\tinplbl{\q}{\lab_j}{\UType_j}\local_j}} &=& \tout{\q}{\lbl{enc_o}}\Tor{j \in J}{\tinplbl{\q}{\lab_j}{\UType_j}\TypeEncoding{\local_j}}
	\end{array}
\end{displaymath}

We show that the translation of types satisfies a variant of operational correspondence for types.
In contrast to \scbs \blueArrow \bs, emulations of different steps in the source can interleave in the target.
Because of that, we have to relax the formulation of soundness.

\begin{lemma}
	\label{lem:ocTypesSMPintoMP}
	Let $ \LL $ not contain any mixed choices on types and assume $ \safe{\LL} $ and $ \df{\LL} $.
	Then:
	\begin{description}
		\item[Completeness:] If $ \LL \by{\lambda} \LL' $, then $ \TypeEncoding{\LL} \red\by{\lambda} \TypeEncoding{\LL'} $.
		\item[Soundness:] If $ \TypeEncoding{\LL} \red^\ast \LL_{\mathsf{enc}} $, then there is some $ \LL' $ such that $ \LL \red^\ast \LL' $ and $ \LL_{\mathsf{enc}} \red^\ast \TypeEncoding{\LL'} $, where $ \LL_{\mathsf{enc}} \red^\ast \TypeEncoding{\LL'} $ does not reduce the label $ \lbl{enc_o} $.
	\end{description}
\end{lemma}

\begin{proof}
	Similar to the proof of Lemma~\ref{lem:ocTypesSCBSintoBS}.
\end{proof}

With completeness and soundness of the translation of types we show that the translation on types preserves safety and deadlock-freedom.

\begin{lemma}
	\label{lem:DFofSMPintoMP}
	Let $ \LL $ not contain any mixed choices on types.
	If $ \safe{\LL} $ and $ \df{\LL} $ then $ \safe{\TypeEncoding{\LL}} $ and $ \df{\TypeEncoding{\LL}} $.
\end{lemma}

\begin{proof}
	Similar to the proof of Lemma ~\ref{lem:DFofSCBSintoBS}.
\end{proof}

Again deadlock-freedom ensures the existence of the respective communication partner for the first step (with the fresh label $ \colorbox{green!20}{\lbl{enc_o}} $).
With Lemma~\ref{lem:ocTypesSMPintoMP} and Lemma~\ref{lem:DFofSMPintoMP}, we show that $ \arbitraryEncoding $ is an encoding from \smpst into \mpst, \ie that it translates typed and deadlock-free \smpst-sessions into typed and deadlock-free \mpst-sessions.

\begin{lemma}
	\label{lem:typeSMPintoMP}
	$ \ArbitraryEncoding{M} $ is typed and deadlock-free.
\end{lemma}

\begin{proof}
	Similar to the proof of Lemma~\ref{lem:typeSCBSintoBS}.
\end{proof}
%

To prove that the encoding \smpst \blueArrow \mpst is good, we show first that it respects structural congruence.

\begin{lemma}
	\label{lem:scSMPintoMP}
	\begin{enumerate}
		\item $ P_1 \equiv P_2 $ iff $ \ArbitraryEncoding{P_1} \equiv \ArbitraryEncoding{P_2} $.
		\item $ M_1 \equiv M_2 $ iff $ \ArbitraryEncoding{M_1} \equiv \ArbitraryEncoding{M_2} $.
		\item If $ \ArbitraryEncoding{M_1} \equiv M $ then there is some $ M_2 $ such that $ M = \ArbitraryEncoding{M_2} $.
	\end{enumerate}
\end{lemma}

\begin{proof}
	The same as the proof of Lemma~\ref{lem:scSCBSintoBS}.
\end{proof}

Compositionality is immediate from Figure~\ref{fig:encodingSmpstIntoMpst}.
Name invariance can be shown by structural induction.

\begin{lemma}
	\label{lem:niSMPintoMPT}
	The encoding $ \arbitraryEncoding $ is name invariant, \ie $ \ArbitraryEncoding{M \sigma} = \ArbitraryEncoding{M} \sigma $ for all substitutions $ \sigma $ such that $ M\sigma $ is defined.
\end{lemma}

\begin{proof}
	The proof is the same as for Lemma~\ref{lem:niSCBSintoBS}.
\end{proof}

For operational completeness, remember that a single interaction in the source is translated into two steps on the target, where the first is on the fresh label $ \colorbox{green!20}{\lbl{enc_o}} $ and the second is the step of the source.
In contrast to Lemma~\ref{lem:occSCBSintoBS}, we only have to mind parallel contexts.

\begin{lemma}
	\label{lem:occSMPintoMP}
	The encoding $ \arbitraryEncoding $ is operationally complete, \ie $ M \red^\ast M' $ implies $ \ArbitraryEncoding{M} \red^\ast \ArbitraryEncoding{M'} $.
\end{lemma}

\begin{proof}
	Assume $ M \red M' $.
	We proceed with an induction on the derivation of $ M \red M' $ by the rules in Figure~\ref{fig:reduction} and show $ \ArbitraryEncoding{M} \red^\ast \ArbitraryEncoding{M'} $.
	\begin{description}
		\item[\RChoice:] Then $ M = \pa{\p}{\choice{\soutprlbl{\q}{\lab_i}{v_i}\seq\PP_i}{i \in I}} \Par \pa{\q}{\choice{\sinprlbl{\p}{\lab_j'}{x_j}\seq Q_j}{j \in J}} \Par M_1 $ and $ M' = \pa{\p}{\PP_n} \Par \pa{\q}{Q_m\subst{v_n}{x_m}} \Par M_1 $ for some $ n \in I $ and some $ m \in J $ such that $ \lab_n = \lab_m' $.
			Then $ \ArbitraryEncoding{M} = \pa{\p}{\choice{\colorbox{green!20}{\ssinp{\q}{\lbl{enc_o}}}\seq\soutprlbl{\q}{\lab_i}{v_i}\seq\ArbitraryEncoding{\PP_i}}{i \in I}} \Par \pa{\q}{\colorbox{green!20}{\ssout{\p}{\lbl{enc_o}}}\seq\choice{\sinprlbl{\p}{\lab_j'}{x_j}\seq\ArbitraryEncoding{Q_j}}{j \in J}} \Par \ArbitraryEncoding{M_1} $ and $ \ArbitraryEncoding{M'} = \pa{\p}{\ArbitraryEncoding{\PP_n}} \Par \pa{\q}{\ArbitraryEncoding{Q_m\subst{v_n}{x_m}}} \Par \ArbitraryEncoding{M_1} $.
			By Lemma~\ref{lem:niSCBSintoBS}, then $ \ArbitraryEncoding{M'} = \pa{\p}{\ArbitraryEncoding{\PP_n}} \Par \pa{\q}{\ArbitraryEncoding{Q_m}\subst{v_n}{x_m}} \Par \ArbitraryEncoding{M_1} $.
			By \RChoice, then $ \ArbitraryEncoding{M} \red \pa{\p}{\soutprlbl{\q}{\lab_n}{v_n}\seq\ArbitraryEncoding{\PP_n}} \Par \pa{\q}{\choice{\sinprlbl{\p}{\lab_j'}{x_j}\seq\ArbitraryEncoding{Q_j}}{j \in J}} \Par \ArbitraryEncoding{M_1} \red \ArbitraryEncoding{M'} $ and thus $ \ArbitraryEncoding{M} \red^\ast \ArbitraryEncoding{M'} $.
		\item[\RCondT:] The same as case \RCondT in the proof of Lemma~\ref{lem:occSCBSintoBS}.
		\item[\RCondF:] The same as case \RCondF in the proof of Lemma~\ref{lem:occSCBSintoBS}.
		\item[\RCong:] In this case $ M \equiv M_1 \red M_2 \equiv M' $.
			By Lemma~\ref{lem:scSMPintoMP}, then $ \ArbitraryEncoding{M} \equiv \ArbitraryEncoding{M_1} $ and $ \ArbitraryEncoding{M_2} \equiv \ArbitraryEncoding{M'} $.
			By the induction hypothesis, $ M_1 \red M_2 $ implies $ \ArbitraryEncoding{M_1} \red \ArbitraryEncoding{M_2} $.
			By \RCong, then $ \ArbitraryEncoding{M} \red \ArbitraryEncoding{M'} $ and thus $ \ArbitraryEncoding{M} \red^\ast \ArbitraryEncoding{M'} $.
	\end{description}
	By induction on the number of steps in $ M \red^\ast M' $, then $ M \red^\ast M' $ implies $ \ArbitraryEncoding{M} \red^\ast \ArbitraryEncoding{M'} $, \ie $ \arbitraryEncoding $ is operationally complete with $ \asymp $ chosen as $ = $.
\end{proof}

For operational soundness we have to show that every step of the translation belongs to an emulation and that every emulation can be completed.
Remember that $ \arbitraryEncoding $ translates interactions into two steps but a step that reduces a conditional is translated to a single step.
Proving this is more difficult than Lemma~\ref{lem:ocsSCBSintoBS}, because we have to consider the interleaving of different emulations.

\begin{lemma}
	\label{lem:ocsSMPintoMP}
	The encoding $ \arbitraryEncoding $ is operationally sound, \ie $ \ArbitraryEncoding{M} \red^\ast N $ implies $ M \red^\ast M' $ and $ N \red^\ast \ArbitraryEncoding{M'} $, where $ \ArbitraryEncoding{M} \red^\ast \ArbitraryEncoding{M'} $ contains at most twice as much steps as $ M \red^\ast M' $ and where $ N \red^\ast \ArbitraryEncoding{M'} $ contains no step on label $ \lbl{enc_o} $.
\end{lemma}

\begin{proof}
	The proof is by induction on the number of steps in $ \ArbitraryEncoding{M} \red^\ast N $.
	\begin{description}
		\item[Base Case ($ \ArbitraryEncoding{M} = N $):] Then $ M \red^\ast M $ and $ N \red^\ast \ArbitraryEncoding{M} $.
		\item[Induction Step ($ \ArbitraryEncoding{M} \red^\ast N_1 \red N $):] By the induction hypothesis, $ M \red^\ast M_1 $ and $ N_1 \red^\ast \ArbitraryEncoding{M_1} $, where $ \ArbitraryEncoding{M} \red^\ast \ArbitraryEncoding{M_1} $ contains at most twice as much steps as $ M \red^\ast M_1 $ and where $ N_1 \red^\ast \ArbitraryEncoding{M_1} $ contains no step on label $ \lbl{enc_o} $.
			Since all emulations of source steps start in \smpst \blueArrow \mpst with a step on label $ \lbl{enc_o} $, none of the steps in $ N_1 \red^\ast \ArbitraryEncoding{M_1} $ starts an emulation.
			Then $ N_1 \red^\ast \ArbitraryEncoding{M_1} $ consists of the respective second steps of the emulation of pairwise distributable steps in the source closing all emulations that are not finished in $ N_1 $.
			Then all steps in the sequence are pairwise distributable.
			Then (1)~$ N_1 \red N $ is contained in this sequence, (2)~$ N_1 \red N $ and any step in this sequence are distributable, or (3)~$ N_1 \red N $ is in conflict with one of the steps in this sequence.
			\begin{enumerate}
				\item Assume that $ N_1 \red N $ is contained in $ N_1 \red^\ast \ArbitraryEncoding{M_1} $, \ie a step of the sequence $ N_1 \red^\ast \ArbitraryEncoding{M_1} $ and $ N_1 \red N $ reduce exactly the same capabilities in the same way (picking the same summands).
					Since distributable steps are confluent, then we can reorder the steps in the sequence $ N_1 \red^\ast \ArbitraryEncoding{M_1} $ such that $ N_1 \red N \red^\ast \ArbitraryEncoding{M_1} $.
					Then $ M \red^\ast M_1 = M' $ and $ N \red^\ast \ArbitraryEncoding{M'} $, where $ \ArbitraryEncoding{M} \red^\ast \ArbitraryEncoding{M'} $ contains at most twice as much steps as $ M \red^\ast M' $ and where $ N \red^\ast \ArbitraryEncoding{M'} $ contains no step on label $ \lbl{enc_o} $.
				\item Assume that $ N_1 \red N $ and any step in $ N_1 \red^\ast \ArbitraryEncoding{M_1} $ are distributable.
					Then either emulates a step on a conditional (by reducing the conditional itself) or it is an interaction on label $ \lbl{enc_o} $ to start an emulation.
					In the latter case, we need a second step to finish the emulation.
					We proceed with an induction on the derivation of $ N_1 \red N $ by the rules in Figure~\ref{fig:reduction} to show that $ M_1 \red M' $ and $ N \red^\ast \ArbitraryEncoding{M'} $, where $ \ArbitraryEncoding{M_1} \red^\ast \ArbitraryEncoding{M'} $ contains at most two steps and where $ N \red^\ast \ArbitraryEncoding{M'} $ contains no step on label $ \lbl{enc_o} $.
					\begin{description}
						\item[\RChoice:] Since $ N_1 \red^\ast \ArbitraryEncoding{M_1} $ completes all emulations, $ N_1 = \pa{\p}{\choice{\colorbox{green!20}{\ssinp{\q}{\lbl{enc_o}}}\seq\soutprlbl{\q}{\lab_i}{v_i}\seq\ArbitraryEncoding{\PP_i}}{i \in I}} \Par \pa{\q}{\colorbox{green!20}{\ssout{\p}{\lbl{enc_o}}}\seq\choice{\sinprlbl{\p}{\lab_j'}{x_j}\seq\ArbitraryEncoding{Q_j}}{j \in J}} \Par N' $ and $ N = \pa{\q}{\soutprlbl{\p}{\lab_n}{v_n}\seq\ArbitraryEncoding{\PP_n}} \Par \pa{\q}{\choice{\sinprlbl{\p}{\lab_j'}{x_j}\seq\ArbitraryEncoding{Q_j}}{j \in J}} \Par N' $ for some $ n \in I $.
							Then $ M_1 = \pa{\p}{\choice{\soutprlbl{\q}{\lab_i}{v_i}\seq\PP_i}{i \in I}} \Par \pa{\q}{\choice{\sinprlbl{\p}{\lab_j'}{x_j}\seq Q_j}{j \in J}} \Par M_1' $.
							By safety, there is some $ m \in J $ such that $ \lab_n = \lab_m' $.
							By \RChoice, then $ N \red \pa{\p}{\ArbitraryEncoding{\PP_n}} \Par \pa{\q}{\ArbitraryEncoding{Q_m}\subst{v_n}{x_m}} \Par N' $.
							By Lemma~\ref{lem:niSCBSintoBS}, $ \ArbitraryEncoding{Q_m}\subst{v_n}{x_m} = \ArbitraryEncoding{Q_m\subst{v_n}{x_m}} $.
							Then $ M_1 \red M' $ and $ N \red \ArbitraryEncoding{M'} $, where $ M' = \pa{\p}{\PP_n} \Par \pa{\q}{Q_m\subst{v_n}{x_m}} \Par M_1' $.
							Thus $ M \red^\ast M' $ and $ N \red^\ast \ArbitraryEncoding{M'} $.
						\item[\RCondT:] Then $ N_1 = \pa{\p}{\cond{\true}{\ArbitraryEncoding{P_1}}{\ArbitraryEncoding{P_2}}} \Par N' $ and $ N = \pa{\p}{\ArbitraryEncoding{P_1}} \Par N' $.
							Then $ M_1 = \pa{\p}{\cond{\false}{P_1}{P_2}} \Par M_1' $, $ M' = \pa{\p}{P_1} \Par M_1' $, and $ N \red^\ast \ArbitraryEncoding{M'} $, where $ N \red^\ast \ArbitraryEncoding{M'} $ contains at most two steps.
							By \RCondT, then $ M_1 \red M' $.
						\item[\RCondF:] Then $ N_1 = \pa{\p}{\cond{\false}{\ArbitraryEncoding{P_1}}{\ArbitraryEncoding{P_2}}} \Par N' $ and $ N = \pa{\p}{\ArbitraryEncoding{P_2}} \Par N' $.
							Then $ M_1 = \pa{\p}{\cond{\false}{P_1}{P_2}} \Par M_1' $, $ M' = \pa{\p}{P_2} \Par M_1' $, and $ N \red^\ast \ArbitraryEncoding{M'} $, where $ N \red^\ast \ArbitraryEncoding{M'} $ contains at most two steps.
							By \RCondF, then $ M_1 \red M' $.
						\item[\RCong:] In this case $ N_1 \equiv N_2 \red N_3 \equiv N $.
							By \RCong, then $ N_1 \red N_3 $.
							By the induction hypothesis, $ M_1 \red M' $ and $ N_3 \red^\ast \ArbitraryEncoding{M'} $, where $ \ArbitraryEncoding{M_1} \red^\ast \ArbitraryEncoding{M'} $ contains at most two steps and where $ N_3 \red^\ast \ArbitraryEncoding{M'} $ contains no step on label $ \lbl{enc_o} $.
							Then $ M \red^\ast M_1 \red M' $ and thus $ M \red^\ast M' $.
							By \RCong, $ N \red^\ast \ArbitraryEncoding{M'} $.
					\end{description}
				\item Assume that $ N_1 \red N $ is in conflict with one of the steps in $ N_1 \red^\ast \ArbitraryEncoding{M_1} $.
					Then $ N_1 \red N $ is an interaction that does not reduce the label $ \lbl{enc_o} $.
					Since the first step already picks an output, $ N_1 \red N $ differs from the step, it is in conflict to, by picking another input with the same prefix.
					Then $ N_1 \red N $ and its conflicting step reduce exactly the same participants: $ \pa{\p}{\soutprlbl{\q}{\lab}{v}\seq\ArbitraryEncoding{\PP}} $ and $ \pa{\q}{\choice{\sinprlbl{\p}{\lab_j'}{x_j}\seq\ArbitraryEncoding{Q_j}}{j \in J}} $.
					There are $ n, m \in J $ such that $ \lab_n' = \lab = \lab_m' $, where $ \lab_n' $ is used to obtain $ N $.
					Then $ N $ and the continuation of the conflicting step differ only by replacing $ \ArbitraryEncoding{Q_n}\subst{v}{x_n} $ with $ \ArbitraryEncoding{Q_m}\subst{v}{x_m} $.
					Then there is some $ M' $ such that $ M' $ differs from $ M_1 $ only by replacing $ Q_n\subst{v}{x_n} $ with $ Q_m\subst{v}{x_m} $.
					Then $ M \red^\ast M' $ and $ M \red^\ast M_1 $ have the same length.
					By Lemma~\ref{lem:niSCBSintoBS}, $ \ArbitraryEncoding{Q_n}\subst{v}{x_n} = \ArbitraryEncoding{Q_n\subst{v}{x_n}} $.
					Then $ N_1 \red N \red^\ast \ArbitraryEncoding{M'} $.
					\qedhere
			\end{enumerate}
	\end{description}
\end{proof}

Again, safety ensures that for every output that was picked in the first step (with the fresh label $ \colorbox{green!20}{\lbl{enc_o}} $) there will be a matching input in the second step.
Because of that, we need safety in the case of a single step from $ \ArbitraryEncoding{M} $ that is an interaction.

\begin{theorem}[From \smpst into \mpst]
	\label{thm:smpstIntoMpst}
	$ \arbitraryEncoding $ in Figure~\ref{fig:encodingSmpstIntoMpst} is a good encoding from \smpst into \mpst.
\end{theorem}

\begin{proof}
	By Lemma~\ref{lem:typeSMPintoMP}, $ \arbitraryEncoding $ is an encoding from \smpst into \mpst.
	Compositionality follows from the definition of $ \arbitraryEncoding $ in Figure~\ref{fig:encodingSmpstIntoMpst}.
	By Lemma~\ref{lem:niSMPintoMPT}, $ \arbitraryEncoding $ is name invariant.
	By the Lemmata~\ref{lem:occSMPintoMP} and \ref{lem:ocsSMPintoMP}, $ \arbitraryEncoding $ satisfies operational correspondence.
	Divergence reflection follows from Lemma~\ref{lem:ocsSMPintoMP}, since it proves a bound on the number of steps to emulate a single source term step.
	By Figure~\ref{fig:encodingSmpstIntoMpst}, $ M\hasSuccess $ iff $ \ArbitraryEncoding{M}\hasSuccess $.
	By operational correspondence, then $ \arbitraryEncoding $ is success sensitive.
	Because the parallel operator is translated homomorphically, $ \arbitraryEncoding $ preserves distributability.
	We instantiated $ \asymp $ with $ = $, which is an equivalence and a success respecting reduction bisimulation.
	By Definition~\ref{def:goodEncoding}, then $ \arbitraryEncoding $ is a good encoding from \smpst into \mpst.
\end{proof}

\subsection{Encoding \dmpst into \smpst}
\label{app:encodingDmpstIntoSmpst}

The encoding \dmpst \blueArrow \smpst encodes choice in the same way as \mcbs \blueArrow \scbs (see Figure~\ref{fig:KeysExpressiveness}(2)).
The remaining operators are translated homomorphically except for the creation of the total order on participants, that we take again from \S~\ref{app:totalOrder}.
The encoding function in Figure~\ref{fig:encodingDMPintoSMP} looks the same as the encoding from \mcbs into \scbs in Figure~\ref{fig:encodingMcbsIntoScbs}.
Mind, however, the different source and target languages.

\begin{figure}
	\begin{align*}
		\ArbitraryEncoding{\inact}_{<} &= \inact\\
		\ArbitraryEncoding{\success}_{<} &= \success\\
		\ArbitraryEncoding{\pvar{X}}_{<} &= \pvar{X}\\
		\ArbitraryEncoding{\rec{X} \PP}_{<} &= \rec{X} \ArbitraryEncoding{\PP}_{<}\\
		\ArbitraryEncoding{\left( \choice{\soutprlbl{\q}{\lab_i}{v_i}\seq\PP_i}{i \in I} \right) \OR \left( \choice{\sinprlbl{\q}{\lab_j'}{x_j}\seq Q_j}{j \in J} \right)}_{<}
		&= \begin{cases}
			\begin{array}{l}
				\left( \choice{\soutprlbl{\q}{\lab_i}{v_i}\seq\ArbitraryEncoding{\PP_i}_{<}}{i \in I} \right) \OR\\
				\colorbox{red!10}{\ssout{\q}{\lbl{enc_i}}}\seq\left( \left(\choice{\sinprlbl{\q}{\lab_j'}{x_j}\seq\ArbitraryEncoding{Q_j}_{<}}{j \in J} \right) \OR \colorbox{yellow!30}{\ssinp{\q}{\lbl{reset}}}\seq \choice{\soutprlbl{\q}{\lab_i}{v_i}\seq\ArbitraryEncoding{\PP_i}}{i \in I} \right)
			\end{array} & \text{if } I \neq \emptyset \neq J \wedge \p < \q\\
			\left( \choice{\sinprlbl{\q}{\lab_j'}{x_j}\seq\ArbitraryEncoding{Q_j}_{<}}{j \in J} \right) \OR \colorbox{red!10}{\ssinp{\q}{\lbl{enc_i}}}\seq\choice{\soutprlbl{\q}{\lab_i}{v_i}\seq\ArbitraryEncoding{\PP_i}_{<}}{i \in I} & \text{if } I \neq \emptyset \neq J \wedge \p > \q\\
			\choice{\soutprlbl{\q}{\lab_i}{v_i}\seq\ArbitraryEncoding{\PP_i}_{<}}{i \in I} & \text{if } J = \emptyset \wedge \p < \q\\
			\colorbox{red!10}{\ssinp{\q}{\lbl{enc_i}}}\seq\choice{\soutprlbl{\q}{\lab_i}{v_i}\seq\ArbitraryEncoding{\PP_i}_{<}}{i \in I} & \text{if } J = \emptyset \wedge \p > \q\\
			\colorbox{red!10}{\ssout{\q}{\lbl{enc_i}}}\seq\choice{\sinprlbl{\q}{\lab_j'}{x_j}\seq\ArbitraryEncoding{Q_j}_{<}}{j \in J} & \text{if } I = \emptyset \wedge \p < \q\\
			\left( \choice{\sinprlbl{\q}{\lab_j'}{x_j}\seq\ArbitraryEncoding{Q_j}_{<}}{j \in J} \right) \OR \colorbox{red!10}{\ssinp{\q}{\lbl{enc_i}}}\seq\colorbox{yellow!30}{\ssout{\q}{\lbl{reset}}}\seq\choice{\sinprlbl{\q}{\lab_j'}{x_j}\seq\ArbitraryEncoding{Q_j}_{<}}{j \in J} & \text{if } I = \emptyset \wedge \p > \q
		\end{cases}\\
		\ArbitraryEncoding{\cond{v}{\PP}{Q}}_{<} &= \cond{v}{\ArbitraryEncoding{\PP}_{<}}{\ArbitraryEncoding{Q}_{<}}
	\end{align*}
	where $ \lbl{enc_i} $ and $ \lbl{reset} $ are labels that do not occur in any source term.
	\caption{Encoding from \dmpst into \smpst\label{fig:encodingDMPintoSMP}}
\end{figure}

Also the proofs of results for \dmpst \blueArrow \smpst differ from proofs of results for \mcbs \blueArrow \scbs only in the way observed already in \S~\ref{app:encodingSmpstIntoMpst} in comparison to \S~\ref{app:encodingScbsIntoBs}.
In particular, we have to capture the interleaving of independent actions of different participants.
Since the difference were already explained in \S~\ref{app:encodingSmpstIntoMpst}, we directly come to the main result.

\begin{theorem}[From \dmpst into \smpst]
	\label{thm:dmpstIntoSmpst}
	$ \arbitraryEncoding $ in Figure~\ref{fig:encodingDMPintoSMP} is a good encoding from \dmpst into \smpst.
\end{theorem}

\begin{proof}
	Similar to the proof of Theorem~\ref{thm:mcbsIntoScbs}.
\end{proof}

\subsection{Encoding \dmpst into \mpst}
\label{app:encodingDmpstIntoMpst}

By combining the ideas of \smpst \blueArrow \mpst and \dmpst \blueArrow \smpst in Figure~\ref{fig:encodingDMPintoMP} is a good encoding from \dmpst into \mpst.

\begin{figure}
	\begin{align*}
		\ArbitraryEncoding{\inact}_{<} &= \inact\\
		\ArbitraryEncoding{\success}_{<} &= \success\\
		\ArbitraryEncoding{\pvar{X}}_{<} &= \pvar{X}\\
		\ArbitraryEncoding{\rec{X} \PP}_{<} &= \rec{X} \ArbitraryEncoding{\PP}_{<}\\
		\ArbitraryEncoding{\begin{array}{l}
				\left( \choice{\soutprlbl{\q}{\lab_i}{v_i}\seq\PP_i}{i \in I} \right) \OR\\
				\left( \choice{\sinprlbl{\q}{\lab_j}{x_j}\seq\PP_j}{j \in J} \right)
		\end{array}}_{<}
		&= \begin{cases}
			\begin{array}{l}
				\left( \choice{\colorbox{green!20}{\ssinp{\q}{\lbl{enc_o}}}\seq\soutprlbl{\q}{\lab_i}{v_i}\seq\ArbitraryEncoding{\PP_i}_{<}}{i \in I} \right) \OR\\
				\colorbox{green!20}{\ssinp{\q}{\lbl{enc_o}}}\seq\colorbox{red!10}{\ssout{\q}{\lbl{enc_i}}}\seq\left( \left(\choice{\sinprlbl{\q}{\lab_j}{x_j}\seq\ArbitraryEncoding{\PP_j}_{<}}{j \in J} \right) \OR \colorbox{yellow!30}{\ssinp{\q}{\lbl{reset}}}\seq \choice{\soutprlbl{\q}{\lab_i}{v_i}\seq\ArbitraryEncoding{\PP_i}_{<}}{i \in I} \right)
			\end{array} & \text{if } I \neq \emptyset \neq J \wedge \p < \q\\
			\colorbox{green!20}{\ssout{\q}{\lbl{enc_o}}}\seq\left( \left( \choice{\sinprlbl{\q}{\lab_j}{x_j}\seq\ArbitraryEncoding{\PP_j}_{<}}{j \in J} \right) \OR \colorbox{red!10}{\ssinp{\q}{\lbl{enc_i}}}\seq\choice{\soutprlbl{\q}{\lab_i}{v_i}\seq\ArbitraryEncoding{\PP_i}_{<}}{i \in I} \right) & \text{if } I \neq \emptyset \neq J \wedge \p > \q\\
			\choice{\colorbox{green!20}{\ssinp{\q}{\lbl{enc_o}}}\seq\soutprlbl{\q}{\lab_i}{v_i}\seq\ArbitraryEncoding{\PP_i}_{<}}{i \in I} & \text{if } J = \emptyset \wedge \p < \q\\
			\colorbox{green!20}{\ssout{\q}{\lbl{enc_o}}}\seq\colorbox{red!10}{\ssinp{\q}{\lbl{enc_i}}}\seq\choice{\soutprlbl{\q}{\lab_i}{v_i}\seq\ArbitraryEncoding{\PP_i}_{<}}{i \in I} & \text{if } J = \emptyset \wedge \p > \q\\
			\colorbox{green!20}{\ssinp{\q}{\lbl{enc_o}}}\seq\colorbox{red!10}{\ssout{\q}{\lbl{enc_i}}}\seq\choice{\sinprlbl{\q}{\lab_j}{x_j}\seq\ArbitraryEncoding{\PP_j}_{<}}{j \in J} & \text{if } I = \emptyset \wedge \p < \q\\
			\colorbox{green!20}{\ssout{\q}{\lbl{enc_o}}}\seq\left( \left( \choice{\sinprlbl{\q}{\lab_j}{x_j}\seq\ArbitraryEncoding{\PP_j}_{<}}{j \in J} \right) \OR \colorbox{red!10}{\ssinp{\q}{\lbl{enc_i}}}\seq\colorbox{yellow!30}{\ssout{\q}{\lbl{reset}}}\seq\choice{\sinprlbl{\q}{\lab_j}{x_j}\seq\ArbitraryEncoding{\PP_j}_{<}}{j \in J} \right) & \text{if } I = \emptyset \wedge \p > \q
		\end{cases}\\
		\ArbitraryEncoding{\cond{v}{\PP}{Q}}_{<} &= \cond{v}{\ArbitraryEncoding{\PP}_{<}}{\ArbitraryEncoding{Q}_{<}}
	\end{align*}
	for all $ \mathsf{x} \in \left\{ \mathsf{o}, \mathsf{i} \right\} $, where $ \lbl{enc_i} $, $ \lbl{enc_o} $, and $ \lbl{reset} $ are labels that do not occur in any source term.
	\caption{Encoding from \mcbs into \bs\label{fig:encodingDMPintoMP}}
\end{figure}

\begin{theorem}[From \dmpst into \mpst]
	\label{thm:dmpstIntoMpst}
	$ \arbitraryEncoding $ in Figure~\ref{fig:encodingDMPintoMP} is a good encoding from \dmpst into \mpst.
\end{theorem}

\begin{proof}
	Similar to the proof of Theorem~\ref{thm:mcbsIntoBs}.
\end{proof}

We combine the three encodings \smpst \blueArrow \mpst, \dmpst \blueArrow \smpst, and \dmpst \blueArrow \mpst to show the mutual encodings among the languages \dmpst, \smpst, and \mpst.

\thmMP*

\begin{proof}
	By the Theorems~\ref{thm:smpstIntoMpst}, \ref{thm:dmpstIntoSmpst}, and \ref{thm:dmpstIntoMpst} and the calculus inclusions $ \mpst \subset \smpst \subset \mpst $.
\end{proof}

\subsection{Encoding \mcmpst into \msmpst}
\label{app:encodingMcmpIntoMsmp}

The calculi \mcmpst and \msmpst in the top layer of Figure~\ref{fig:hierarchy} have the same expressive power, although \msmpst does restrict choices with several prefixes from/to the same participant to a separate choice, \ie either only outputs or only inputs but not their mixture.
We observe that combining several different communication partners within a choice is more crucial to the expressive power of choice.

To encode \mcmpst, the \msmpst-calculus has to split the mixed choices per participant.
Therefore, we reuse the splitting of mixed choice in \mcbs \blueArrow \scbs and apply it for every participant in the choice.
The encoding \mcmpst \blueArrow \msmpst is given in Figure~\ref{fig:encodingMCMPintoMSMP} and Figure~\ref{fig:totalOrder}.

\begin{figure}
	\begin{align*}
		\ArbitraryEncoding{\inact}_{<} &= \inact\\
		\ArbitraryEncoding{\success}_{<} &= \success\\
		\ArbitraryEncoding{\pvar{X}}_{<} &= \pvar{X}\\
		\ArbitraryEncoding{\rec{X} \PP}_{<} &= \rec{X} \ArbitraryEncoding{\PP}_{<}\\
		\ArbitraryEncoding{\sum_{\q} \left( \begin{array}{l} \left( \choice{\soutprlbl{\q}{\lab_i}{v_i}\seq\PP_i}{i \in I} \right) \OR\\ \left( \choice{\sinprlbl{\q}{\lab_j'}{x_j}\seq Q_j}{j \in J} \right) \end{array} \right)}_{<}
		&= \sum_{\q} \begin{cases}
			\begin{array}{l}
				\left( \choice{\soutprlbl{\q}{\lab_i}{v_i}\seq\ArbitraryEncoding{\PP_i}_{<}}{i \in I} \right) \OR\\
				\colorbox{red!10}{\ssout{\q}{\lbl{enc_i}}}\seq\left( \left(\choice{\sinprlbl{\q}{\lab_j'}{x_j}\seq\ArbitraryEncoding{Q_j}_{<}}{j \in J} \right) \OR \colorbox{yellow!30}{\ssinp{\q}{\lbl{reset}}}\seq \choice{\soutprlbl{\q}{\lab_i}{v_i}\seq\ArbitraryEncoding{\PP_i}}{i \in I} \right)
			\end{array} & \text{if } I \neq \emptyset \neq J \wedge \p < \q\\
			\left( \choice{\sinprlbl{\q}{\lab_j'}{x_j}\seq\ArbitraryEncoding{Q_j}_{<}}{j \in J} \right) \OR \colorbox{red!10}{\ssinp{\q}{\lbl{enc_i}}}\seq\choice{\soutprlbl{\q}{\lab_i}{v_i}\seq\ArbitraryEncoding{\PP_i}_{<}}{i \in I} & \text{if } I \neq \emptyset \neq J \wedge \p > \q\\
			\choice{\soutprlbl{\q}{\lab_i}{v_i}\seq\ArbitraryEncoding{\PP_i}_{<}}{i \in I} & \text{if } J = \emptyset \wedge \p < \q\\
			\colorbox{red!10}{\ssinp{\q}{\lbl{enc_i}}}\seq\choice{\soutprlbl{\q}{\lab_i}{v_i}\seq\ArbitraryEncoding{\PP_i}_{<}}{i \in I} & \text{if } J = \emptyset \wedge \p > \q\\
			\colorbox{red!10}{\ssout{\q}{\lbl{enc_i}}}\seq\choice{\sinprlbl{\q}{\lab_j'}{x_j}\seq\ArbitraryEncoding{Q_j}_{<}}{j \in J} & \text{if } I = \emptyset \wedge \p < \q\\
			\left( \choice{\sinprlbl{\q}{\lab_j'}{x_j}\seq\ArbitraryEncoding{Q_j}_{<}}{j \in J} \right) \OR \colorbox{red!10}{\ssinp{\q}{\lbl{enc_i}}}\seq\colorbox{yellow!30}{\ssout{\q}{\lbl{reset}}}\seq\choice{\sinprlbl{\q}{\lab_j'}{x_j}\seq\ArbitraryEncoding{Q_j}_{<}}{j \in J} & \text{if } I = \emptyset \wedge \p > \q
		\end{cases}\\
		\ArbitraryEncoding{\cond{v}{\PP}{Q}}_{<} &= \cond{v}{\ArbitraryEncoding{\PP}_{<}}{\ArbitraryEncoding{Q}_{<}}
	\end{align*}
	where $ \lbl{enc_i} $ and $ \lbl{reset} $ are labels that do not occur in any source term.
	\caption{Encoding from \mcmpst into \msmpst\label{fig:encodingMCMPintoMSMP}}
\end{figure}

Since \mcmpst \blueArrow \msmpst is mostly homomorph and translates choice in a very similar way to \mcbs \blueArrow \scbs, we also inherit the properties of \mcbs \blueArrow \scbs.

\begin{theorem}[From \mcmpst into \msmpst]
	\label{thm:mcmpstIntoMsmpst}
	$ \arbitraryEncoding $ in Figure~\ref{fig:encodingMCMPintoMSMP} is a good encoding from \mcmpst into \msmpst.
\end{theorem}

\begin{proof}
	Similar to the proof of Theorem~\ref{thm:mcbsIntoScbs}.
\end{proof}

\thmMCMP*

\begin{proof}
	By Theorem~\ref{thm:mcmpstIntoMsmpst} and since $ \msmpst \subset \mcmpst $.
\end{proof}

\subsection{Encoding \lcnvms into \mcbs}
\label{app:LcnvmsIntoMcbs}

To show that the choice construct introduced by the typed calculus \mcmpst subsumes the choice construct in \cnvms, we strip interleaved sessions and session delegation from \cnvms and limit it to its linear fragment.
Note that also recursion is removed from the calculus this way.
What remains from \cnvms in \lcnvms is its flexible choice construct.

\begin{figure}
	\begin{align*}
		\ArbitraryEncoding{\Gamma \vdash \ResCMVmix{x}{y}{\PP}} &= \ArbitraryEncoding{\Gamma' \vdash \PP}_{x, y}\\
		\ArbitraryEncoding{\Gamma \vdash P \mid Q}_{x, y} &= \ArbitraryEncoding{P}_{x, y} \mid \ArbitraryEncoding{Q}_{x, y}\\
		\ArbitraryEncoding{\inactCMV}_{x, y} &= \inactCMV\\
		\ArbitraryEncoding{\success}_{x, y} &= \success\\
		\ArbitraryEncoding{\ConditionalCMV{v}{P}{Q}}_{x, y} &= \cond{v}{\ArbitraryEncoding{P}_{x, y}}{\ArbitraryEncoding{Q}_{x, y}}\\
		\ArbitraryEncoding{\Gamma \vdash \ChoiceCMVmix{\linCMVmix}{\mathsf{k}}{\left( \PP = \left( \sum_{i \in I} \OutCMVmix{\lab_i}{v_i}{\PP_i} \right) \OR \left( \sum_{j \in J} \InpCMVmix{\lab_j}{z_j}{\PP_j} \right) \right)}}_{x, y}
		&= \begin{cases}
			\pa{\enodei{k}}{\inact} & \text{if } \fv{\PP} = \Set{x, y} \cup \Set{\mathsf{k}}\\
			\pa{\enodei{k}}{\left( \sum_{i \in I} \soutprlbl{\enodei{\overline{k}}}{\lab_i \cdot \colorbox{blue!10}{o}}{v_i}\seq\ArbitraryEncoding{\Gamma_i \vdash \PP_i}_{\mathsf{\overline{k}}} \right)} & \text{if } \fv{\PP} \neq \Set{x, y}\text{, } \mathsf{\overline{k}} \in \Set{x, y} \setminus \Set{\mathsf{k}}\text{,}\\
			\hspace{1.5em} \OR \left( \sum_{j \in J} \ssout{\enodei{\overline{k}}}{\lab_i \cdot \colorbox{blue!10}{i}}\seq\sinprlbl{\enodei{\overline{k}}}{\lab_i}{z}\seq\ArbitraryEncoding{\Gamma_j \vdash \PP_j}_{\mathsf{\overline{k}}} \right) & \phantom{if }\text{and } \PP \text{ is typed as internal in } \Gamma\\
			\pa{\enodei{k}}{\left( \sum_{i \in I} \ssinp{\enodei{\overline{k}}}{\lab_i \cdot \colorbox{blue!10}{i}}\seq\soutprlbl{\enodei{\overline{k}}}{\lab_i}{v_i}\seq\ArbitraryEncoding{\Gamma_i \vdash \PP_i}_{\mathsf{\overline{k}}} \right)} & \text{if } \fv{\PP} \neq \Set{x, y}\text{, } \mathsf{\overline{k}} \in \Set{x, y} \setminus \Set{\mathsf{k}}\text{,}\\
			\hspace{1.5em} \OR \left( \sum_{j \in J} \sinprlbl{\enodei{\overline{k}}}{\lab_i \cdot \colorbox{blue!10}{o}}{z}\seq\ArbitraryEncoding{\Gamma_j \vdash \PP_j}_{\mathsf{\overline{k}}} \right) & \phantom{if }\text{and } \PP \text{ is typed as external in } \Gamma
		\end{cases}\\
	\ArbitraryEncoding{\Gamma \vdash \ChoiceCMVmix{\linCMVmix}{\mathsf{k}}{\left( \PP = \left( \sum_{i \in I} \OutCMVmix{\lab_i}{v_i}{\PP_i} \right) \OR \left( \sum_{j \in J} \InpCMVmix{\lab_j}{z_j}{\PP_j} \right) \right)}}_{\role{k}}
&= \begin{cases}
\left( \sum_{i \in I} \soutprlbl{\enodei{k}}{\lab_i \cdot \colorbox{blue!10}{o}}{v_i}\seq\ArbitraryEncoding{\Gamma_i \vdash \PP_i}_{\mathsf{k}} \right) & \text{if } \PP \text{ is typed as internal in } \Gamma\\
\hspace{1.5em} \OR \left( \sum_{j \in J} \ssout{\enodei{k}}{\lab_i \cdot \colorbox{blue!10}{i}}\seq\sinprlbl{\enodei{k}}{\lab_i}{z}\seq\ArbitraryEncoding{\Gamma_j \vdash \PP_j}_{\mathsf{k}} \right) &\\
\left( \sum_{i \in I} \ssinp{\enodei{k}}{\lab_i \cdot \colorbox{blue!10}{i}}\seq\soutprlbl{\enodei{k}}{\lab_i}{v_i}\seq\ArbitraryEncoding{\Gamma_i \vdash \PP_i}_{\mathsf{k}} \right) & \text{if } \PP \text{ is typed as external in } \Gamma\\
\hspace{1.5em} \OR \left( \sum_{j \in J} \sinprlbl{\enodei{k}}{\lab_i \cdot \colorbox{blue!10}{o}}{z}\seq\ArbitraryEncoding{\Gamma_j \vdash \PP_j}_{\mathsf{k}} \right) &
\end{cases}
	\end{align*}
	\caption{Encoding from \lcnvms into \mcbs\label{fig:encodingLCNVMSintoMCBS}}
\end{figure}

The encoding \lcnvms \blueArrow \mcbs uses the type information of mixed sessions.
A choice typed as internal is translated into an output-guarded choice, whereas external choice is translated into inputs.
Every source step is emulated by one or two steps in the target.

\thmLcnvmsIntoMcbs*

\begin{proof}
	Compositionality follows from the definition of $ \arbitraryEncoding $ in Figure~\ref{fig:encodingLCNVMSintoMCBS}.
	Similarly, name invariance follows directly from the definition of the encoding function, since names are translated to themselves.
	The proofs of operational correspondence require a structural induction.
	Divergence reflection follows operational correspondence, since every emulation of a source step has at most two steps in the target.
	By Figure~\ref{fig:encodingSmpstIntoMpst}, $ M\hasSuccess $ iff $ \ArbitraryEncoding{M}\hasSuccess $.
	By operational correspondence, then $ \arbitraryEncoding $ is success sensitive.
	Because the parallel operator is translated homomorphically, $ \arbitraryEncoding $ preserves distributability.
	By Definition~\ref{def:goodEncoding}, then $ \arbitraryEncoding $ is a good encoding from \lcnvms into \mcbs.
\end{proof}

\subsection{Separation via Leader Election}
\label{app:separationLeaderElection}

We consider \scmpst instead of \mcmpst as source calculus in Theorem~\ref{thm:separationViaLE}, because it is the smallest considered subcalculus of \mcmpst that already has the full abstract expressive power of mixed choice (as in the $ \pi $-calculus).
In \mcmpst and its sub-calculus \msmpst processes within networks are identified by their participant identifier $ \p $.
A process $ \p $ announces itself as leader by sending $ \ssout{\enodei{station}}{\elect} $ to the participant $ \enodei{station} $ that is not participating in the leader election game.
Note that, how exactly the leader is announced after it was elected is not crucial for the following proofs.
All we have to do, is to pick one observable and typed way to announce the leader.
An alternative way without an additional participant such as $ \enodei{station} $ would be for the leader to broadcast its selection to all other participants.

A network $ M_1 \Par \ldots \Par M_n $ is an \emph{electoral system} if in every maximal execution exactly one leader is announced.
Since we do not add the participant $ \enodei{station} $, a participant announces itself as leader by unguarding an output to $ \enodei{station} $ that will not be reduced.

Leader election can be used to distinguish the expressive power of mixed choice from separate choice \eg in the $ \pi $-calculus only if we consider symmetric systems.
Without the requirement on symmetry also asynchronous systems without any choice may be able to solve leader election, by utilizing the asymmetry of the initial network.
In \citep{Palamidessi03} a hypergraph is used to collect the connections between the parts of the considered network, \ie a hypergraph collects shared channels.
Symmetry is then defined by automorphisms on the respective hypergraph.
To simplify the definition of symmetry, we assume a fully connected graph for the interactions in a session not including $ \enodei{station} $.
Note, however, that this decision is not crucial, \ie can be relaxed to not fully connected graphs by using hypergraphs.

Assume a network $ M = M_1 \Par \ldots \Par M_n $ with the participants $ \p_1, \ldots, \p_n $.
An automorphism $ \sigma $ on $ M $ is a permutation on the participants $ \p_1, \ldots, \p_n $.
Since $ \sigma $ is a substitution, we allow to apply $ \sigma $ on terms $ M $, denoted as $ M\sigma $.
The orbit $ \Orbit{\sigma}{n} $ of $ n \in N $ generated by $ \sigma $ is defined as the set of nodes in which the various iterations of $ \sigma $ map $ n $, \ie $ \Orbit{\sigma}{n} = \left\{ n, \sigma(n), \ldots, \sigma^{h - 1}(n) \right\} $, where $ \sigma^i $ represents the composition of $ \sigma $ with itself $ i $ times and $ \sigma^h = \id $.

\begin{definition}[Symmetric System]
	Consider a network $ M = M_1 \Par \ldots \Par M_n $ and let $ \sigma $ be an isomorphism on $ M $.
	$ M $ is \emph{symmetric \wrt $ \sigma $} iff $ M_{\sigma(i)} = M_i\sigma $ for each $ i \in \left\{1, \ldots, n\right\} $.
	$ M $ is \emph{symmetric} if it is symmetric \wrt all the automorphisms on $ M $.
\end{definition}

In the \mcmpst-calculus (just like in the $ \pi $-calculus) we find symmetric electoral systems for many kinds of hypergraphs.
The system $ \mathsf{Election} $ in Example~\ref{ex:leaderElection-process} is a symmetric electoral system with five nodes in the subcalculus \msmpst.
\begin{align*}
	\mathsf{Election} &= M_{\enodei{a}} \Par M_{\enodei{b}} \Par M_{\enodei{c}} \Par M_{\enodei{d}} \Par M_{\enodei{e}}\\
	M_{\enodei{a}} &= \pa{\enodei{a}} \left( \ssout{\enodei{e}}{\leader}. \inact \OR \ssinp{\enodei{b}}{\leader}. \left( \ssout{\enodei{c}}{\leader}. \inact \OR \ssinp{\enodei{d}}{\leader}. \ssout{\enodei{station}}{\elect}. \inact \right) \OR \ssinp{\env}{\delete}. \inact \right)\\
	M_{\enodei{b}} &= \pa{\enodei{b}} \left( \ssout{\enodei{a}}{\leader}. \inact \OR \ssinp{\enodei{c}}{\leader}. \left( \ssout{\enodei{d}}{\leader}. \inact \OR \ssinp{\enodei{e}}{\leader}. \ssout{\enodei{station}}{\elect}. \inact \right) \OR \ssinp{\env}{\delete}. \inact \right)\\
	M_{\enodei{c}} &= \pa{\enodei{c}} \left( \ssout{\enodei{b}}{\leader}. \inact \OR \ssinp{\enodei{d}}{\leader}. \left( \ssout{\enodei{e}}{\leader}. \inact \OR \ssinp{\enodei{a}}{\leader}. \ssout{\enodei{station}}{\elect}. \inact \right) \OR \ssinp{\env}{\delete}. \inact \right)\\
	M_{\enodei{d}} &= \pa{\enodei{d}} \left( \ssout{\enodei{c}}{\leader}. \inact \OR \ssinp{\enodei{e}}{\leader}. \left( \ssout{\enodei{a}}{\leader}. \inact \OR \ssinp{\enodei{b}}{\leader}. \ssout{\enodei{station}}{\elect}. \inact \right) \OR \ssinp{\env}{\delete}. \inact \right)\\
	M_{\enodei{e}} &= \pa{\enodei{e}} \left( \ssout{\enodei{d}}{\leader}. \inact \OR \ssinp{\enodei{a}}{\leader}. \left( \ssout{\enodei{b}}{\leader}. \inact \OR \ssinp{\enodei{c}}{\leader}. \ssout{\enodei{station}}{\elect}. \inact \right) \OR \ssinp{\env}{\delete}. \inact \right)\\
	\sigma &= \left[ \enodei{a} \mapsto \enodei{b}, \enodei{b} \mapsto \enodei{c}, \enodei{c} \mapsto \enodei{d}, \enodei{d} \mapsto \enodei{e}, \enodei{e} \mapsto \enodei{a} \right]\\
	M_{\enodei{b}} &= M_{\enodei{a}}\sigma \quad M_{\enodei{c}} = M_{\enodei{b}}\sigma \quad M_{\enodei{d}} = M_{\enodei{c}}\sigma \quad M_{\enodei{e}} = M_{\enodei{d}}\sigma
\end{align*}
$ \mathsf{Election} $ has ten maximal executions (modulo structural congruence).
These executions can be obtained from the first execution
\begin{align*}
	\mathsf{Election} &\red \pa{\enodei{a}} \left( \ssout{\enodei{c}}{\leader}. \inact \OR \ssinp{\enodei{d}}{\leader}. \ssout{\enodei{station}}{\elect}. \inact \right) \Par M_{\enodei{c}} \Par M_{\enodei{d}} \Par M_{\enodei{e}}\\
	&\red \pa{\enodei{a}} \left( \ssout{\enodei{c}}{\leader}. \inact \OR \ssinp{\enodei{d}}{\leader}. \ssout{\enodei{station}}{\elect}. \inact \right) \Par \pa{\enodei{c}} \left( \ssout{\enodei{e}}{\leader}. \inact \OR \ssinp{\enodei{a}}{\leader}. \ssout{\enodei{station}}{\elect}. \inact \right) \Par M_{\enodei{e}}\\
	&\red \pa{\enodei{c}} \ssout{\enodei{station}}{\elect}. \inact \Par M_{\enodei{e}} \noRed
\end{align*}
by considering the respective alternative steps of symmetric participants.
In each maximal execution exactly one leader is elected; \eg in the execution above $ \enodei{c} $ is elected and announces itself via $ \pa{\enodei{c}} \ssout{\enodei{station}}{\elect}. \inact $.

\begin{lemma}
	The system $ \mathsf{Election} \Par \mathsf{Station} $ in Example~\ref{ex:leaderElection-process} is typed and deadlock-free.
\end{lemma}

\begin{proof}
	In Example~\ref{ex:leaderElection-process} we omit the objects in communication, because they are not relevant for the protocol.
	Assume that in all communications $ \true $ is transmitted and the receiver stores the received value in the variable $ \xx $, where $ \xx $ is not free in $ \mathsf{Election} \Par \mathsf{Station} $.
	By Figure~\ref{fig:typing}, we get $ \types P_{\mathsf{Station}} \as T_{\env} $, $ \types P_{\enodei{a}} \as T_{\enodei{a}} $, $ \types P_{\enodei{b}} \as T_{\enodei{b}} $, $ \types P_{\enodei{c}} \as T_{\enodei{c}} $, $ \types P_{\enodei{d}} \as T_{\enodei{d}} $, and $ \types P_{\enodei{e}} \as T_{\enodei{e}} $, where $ \mathsf{Station} = \pa{\env}{P_{\mathsf{Station}}} $, $ M_{\role{a}} = \pa{\role{a}}{P_{\role{a}}} $, $ M_{\role{b}} = \pa{\role{b}}{P_{\role{b}}} $, $ M_{\role{c}} = \pa{\role{c}}{P_{\role{c}}} $, $ M_{\role{d}} = \pa{\role{d}}{P_{\role{d}}} $, $ M_{\role{e}} = \pa{\role{e}}{P_{\role{e}}} $,
	\begin{align*}
		T_{\env} &= \Tor{\enodei{i} \in \set{\enodei{a}, \enodei{b}, \enodei{c}, \enodei{d},\enodei{e}}}{\tinp{\enodei{i}}{\elect}\Tor{\enodei{i} \in \set{\enodei{a}, \enodei{b}, \enodei{c}, \enodei{d},\enodei{e}}}{\tout{\enodei{i}}{\delete}\tinact}},\\
		T_{\enodei{a}} &= \tout{\enodei{e}}{\leader}\tinact \; \tor \; \tinp{\enodei{b}}{\leader}\left( \tout{\enodei{c}}{\leader}\tinact \; \tor \; \tinp{\enodei{d}}{\leader}\tout{\env}{\elect}\tinact \right) \; \tor \; \tinp{\env}{\delete}\tinact,
	\end{align*}
	$ T_{\enodei{b}} = T_{\enodei{a}}\sigma $, $ T_{\enodei{c}} = T_{\enodei{b}}\sigma $, $ T_{\enodei{d}} = T_{\enodei{c}}\sigma $, and $ T_{\enodei{e}} = T_{\enodei{d}}\sigma $ and the omitted payload type is $ \bool $ in all cases.
	Let $ \LL = \localise{\env} T_{\env}, \localise{\enodei{a}} T_{\enodei{a}}, \localise{\enodei{b}} T_{\enodei{b}}, \localise{\enodei{c}} T_{\enodei{c}}, \localise{\enodei{d}} T_{\enodei{d}}, \localise{\enodei{e}} T_{\enodei{e}} $.
	Since there is no recursion, $ \mathsf{Election} \Par \mathsf{Station} $ has only finitely many maximal executions.
	The reductions of the local types follow closely the reductions of the protocol.
	For instance we have the sequence:
	\begin{align*}
		\LL \red{} & \localise{\env} T_{\env}, \localise{\enodei{a}} \tout{\enodei{c}}{\leader}\tinact \; \tor \; \tinp{\enodei{d}}{\leader}\tout{\env}{\elect}\tinact, \localise{\enodei{b}} \tinact, \localise{\enodei{c}} T_{\enodei{c}}, \localise{\enodei{d}} T_{\enodei{d}}, \localise{\enodei{e}} T_{\enodei{e}}\\
		\red{} & \localise{\env} T_{\env}, \localise{\enodei{a}} \tout{\enodei{c}}{\leader}\tinact \; \tor \; \tinp{\enodei{d}}{\leader}\tout{\env}{\elect}\tinact, \localise{\enodei{b}} \tinact, \localise{\enodei{c}} \tout{\enodei{e}}{\leader}\tinact \; \tor \; \tinp{\enodei{a}}{\leader}\tout{\env}{\elect}\tinact, \localise{\enodei{d}} \tinact, \localise{\enodei{e}} T_{\enodei{e}}\\
		\red{} & \localise{\env} T_{\env}, \localise{\enodei{a}} \tinact, \localise{\enodei{b}} \tinact, \localise{\enodei{c}} \tout{\env}{\elect}\tinact, \localise{\enodei{d}} \tinact, \localise{\enodei{e}} T_{\enodei{e}}\\
		\red{} & \localise{\env} \Tor{\enodei{i} \in \set{\enodei{a}, \enodei{b}, \enodei{c}, \enodei{d},\enodei{e}}}{\tout{\enodei{i}}{\delete}\tinact}, \localise{\enodei{a}} \tinact, \localise{\enodei{b}} \tinact, \localise{\enodei{c}} \tinact, \localise{\enodei{d}} \tinact, \localise{\enodei{e}} T_{\enodei{e}}\\
		\red{} & \localise{\env} \tinact, \localise{\enodei{a}} \tinact, \localise{\enodei{b}} \tinact, \localise{\enodei{c}} \tinact, \localise{\enodei{d}} \tinact, \localise{\enodei{e}} \tinact \noRed
	\end{align*}
	All other maximal sequences of reductions are obtained from the above by symmetry.
	By checking all sequences of reductions of the local types, we obtain $ \safe{\LL} $.
	By Figure~\ref{fig:typing}, then $ \types \mathsf{Election} \Par \mathsf{Station} \as \LL $.
	Similarly by checking all sequences of reductions of local types, we see that $ \LL $ is deadlock-free, \ie $ \df{\LL} $.
	By Theorem~\ref{thm:deadlockfree}, then $ \mathsf{Election} \Par \mathsf{Station} $ is deadlock-free.
\end{proof}

We show that there exists no symmetric electoral system for networks of size five in \scmpst; or more generally no symmetric electoral system for networks of sizes greater one in \scmpst.

A key ingredient to separate the $ \pi $-calculus with mixed choice from the asynchronous $ \pi $-calculus in \citep{Palamidessi03} is a confluence lemma.
It states that in the asynchronous $ \pi $-calculus a step reducing an output and an alternative step reducing an input cannot conflict with each other and thus can be executed in any order.
In the full $ \pi $-calculus this confluence lemma is not valid, because inputs and outputs can be combined within a single choice construct and can thus be in conflict.
We adapt this confluence lemma to reductions instead of labelled steps.
For the \scmpst-calculus we observe that communications with different senders and receivers can also not be in conflict to each other.

\begin{lemma}[Confluence]
	\label{lem:confluence}
	Let $ A $ be a typed term in the \scmpst-calculus.
	Assume that $ A $ can make two steps $ A \red B $ and $ A \red C $ such that in $ A \red B $ an output prefix of participant $ \p $ and an input prefix of participant $ \role{p'} $ is reduced and in $ A \red C $ an output prefix of participant $ \q $ and an input prefix of participant $ \role{q'} $ is reduced with $ \p \neq \q $ and $ \role{p'} \neq \role{q'} $.
	Then there exists $ D $ such that $ B \red D $ and $ C \red D $.
\end{lemma}

\begin{proof}
	Assume the two steps $ A \red B $ and $ A \red C $ as described above.
	By Figure~\ref{fig:reduction}, the step $ A \red B $ implies that $ A $ contains at least two non-empty choices: a choice with an output prefix of participant $ \p $ and a choice with an input prefix of participant $ \role{p'} $ that are modulo structural congruence combined in parallel.
	Similarly, the step $ A \red C $ implies that $ A $ contains at least two non-empty choices: a choice with an output prefix of participant $ \q $ and a choice with an input prefix of participant $ \role{q'} $ that are modulo structural congruence combined in parallel.
	Since $ \p \neq \q $ and $ \role{p'} \neq \role{q'} $, since parallel composed choices have pairwise different participants, and since \scmpst does not allow to combine input and output prefixes in the same choice, the participants $ \p $, $ \role{p'} $, $ \q $, and $ \role{q'} $ are pairwise distinct.
	We conclude that the two steps of $ A $ are distributable, because they are performed by parallel components.
	Then these two steps can be executed in any order as required.
\end{proof}

\begin{wrapfigure}{R}{0.25\textwidth}
	\centering
	\scalebox{0.8}{
	\begin{tikzpicture}[bend angle=20]
		\node (a) at (0, 0.75) {$ A $};
		\node (b) at (1.5, 1.5) {$ B $};
		\node (c) at (1.5, 0) {$ C $};
		\node (d) at (3, 0.75) {$ D $};
		\path[|->] (a) edge (b);
		\path[|->] (a) edge (c);
		\path[|->] (b) edge (d);
		\path[|->] (c) edge (d);
	\end{tikzpicture}
	}
\end{wrapfigure}

The proof of this confluence lemma relies on the observation that the two steps of $ A $ to $ B $ and $ C $ have to reduce distributable parts of $ A $.
Then these two steps are distributable, which in turn allows us to perform them in any order.
This confluence lemma is a direct consequence of restricting choice in the \scmpst-calculus to separate instead of mixed choice.
In \msmpst Lemma~\ref{lem:confluence} does not hold, because the mixed choices allow for situations in that $ \p = \role{q'} $ or $ \q = \role{p'} $ such that the steps $ A \red B $ and $ A \red C $ are in conflict.
With this alternative confluence lemma, we can show that there is no electoral system with more than one node in the \scmpst-calculus.

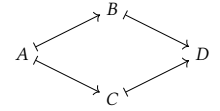
\begin{lemma}[No Electoral System]
	\label{lem:noElectoralSystem}
	Consider a network $ M = M_1 \Par \ldots \Par M_k $ in the \scmpst-calculus with $ k > 1 $.
	Assume an automorphism $ \sigma \neq \id $ with only one orbit, and that $ M $ is symmetric \wrt $ \sigma $.
	Then $ M $ cannot be an electoral system.
\end{lemma}

\begin{proof}
	Assume by contradiction that $ M $ is an electoral system.
	We will show that we can then construct an infinite execution $ E: M \red^\ast M^0 \red^\ast M^1 \red^\ast \ldots $ such that, for each $ j $, $ E_j : M \red^\ast M^j $ does not announce a unique leader and $ M^j $ is still symmetric \wrt $ \sigma $.
	This is a contradiction, because the limit of this sequence is an infinite computation for $ M $ which does not announce exactly one leader.

	The proof is by induction on the current length, denoted by $ h $, of the infinite symmetric execution we have to construct.
	Notice that the assumption of $ \sigma $ generating only one orbit implies that $ \Orbit{\sigma}{i} = \left\{ i, \sigma(i), \ldots, \sigma^{k - 1}(i) \right\} = \left\{ 1, \ldots, k \right\} $, for each $ i \in \left\{ 1, \ldots, k \right\} $.
	We do not require or enforce, that $ M_i $ contains only a single participant.
	However, following the definition of an electoral system, we do require that in every execution every part $ M_i $ can announce at most one leader.
	Moreover, since the parallel composed participants of $ M $ are pair-wise different, no two parts $ M_i $ and $ M_j $ of $ M $ with $ i \neq j $ share the same participants.
	\begin{description}
		\item[Base Case ($ h = 0 $):] Define $ E_0 $ to be the empty execution, \ie $ E_0: M \red^\ast M^0 $ with $ M^0 = M $.
		\item[Induction Step ($ h + 1 $):] Given $ E_h : M \red^\ast M^h = M_1^h \Par \ldots \Par M_k^h $, we construct $ E_{h + 1} : M \red^\ast M^{h + 1} $ as follows.

			If $ M^h $ announces a leader $ \p $, then participant $ \p $ performs $ \ssout{\enodei{station}}{\elect}.Q^h $.
			By symmetry, then some participant $ \sigma(\p) $ with $ \sigma(\p) \neq \p $ performs $ \ssout{\enodei{station}}{\elect}.R^h $, \ie more than one leader is announced.
			This is a contradiction.

			Since $ M $ is an electoral system but $ M^h $ does not yet announces a leader, $ M^h $ has to be able to reduce, \ie there is some $ M' $ such that $ M^h \red M' $.
			This step was performed by one or two of the parts in the network, \ie either $ M_i^h \red M_i' $ and $ M' = M_1^h \Par \ldots \Par M_i' \Par \ldots \Par M_k^h $ or $ M_i^h \Par M_j^h \red M_{i, 1} \Par M_{j, 1} $ and $ M' = M_1^h \Par \ldots \Par M_{i, 1} \Par \ldots \Par M_{j, 1} \Par \ldots \mid M_k^h $ with $ i \neq j $.
			\begin{description}
				\item[$ M_i^h \red M_i' $:] Regardless of whether the step $ M_i^h \red M_i' $ reduces a conditional or performs a communication within part $ i $ of the network, symmetry ensures that the other parts of the network can perform a sequence of steps that leads to a state symmetric to $ M_i' $.
					We choose $ M_i^{h + 1} = M_i' $.
					By symmetry, $ M_{\sigma(i)}^h \red M_{\sigma(i)}^{h + 1}, \ldots, M_{\sigma^{k - 1}(i)}^h \red M_{\sigma^{k - 1}(i)}^{h + 1} $ with $ M_i^{h + 1}\sigma = M_{\sigma(i)}^{h + 1}, \ldots, M_i^{h + 1}\sigma^{k - 1} = M_{\sigma^{k - 1}(i)}^{h + 1} $.
					Since the steps of the different parts of the network are distributable, we obtain $ E_{h + 1} : M \red^\ast M^h \red^\ast M^{h + 1} $, where $ M^{h + 1} = M_1^{h + 1} \Par \ldots \Par M_k^{h + 1} $ and $ M^{h + 1} $ is still symmetric \wrt $ \sigma $.
				\item[$ M_i^h \Par M_j^h \red M_{i, 1} \Par M_{j, 1} $:] Let us denote this step by $ a_1 $.
					A step performed by two processes of the network in the \scmpst-calculus is a communication.
					By Figure~\ref{fig:reduction} and since \scmpst does not contain mixed choices, $ a_1 $ reduces an output-guarded choice of some participant $ \p $ and an input-guarded choice of some participant $ \q $ such that $ \p \neq \q $.
					Without loss of generality, assume that $ \p $ is in part $ M_i^h $ and $ \q $ is in part $ M_j^h $.
					The only other case ($ \p $ is in $ M_j^h $ and $ \q $ is in $ M_i^h $) is similar.
					By symmetry, we find symmetric output-guarded and input-guarded choices in the other parts of the network such that
					\begin{align*}
						a_2: M_{\sigma(i)}^h \Par M_{\sigma(j)}^h & \red M_{\sigma(i), 2} \Par M_{\sigma(j), 2}\\
						& \vdots\\
						a_k: M_{\sigma^{k - 1}(i)}^h \Par M_{\sigma^{k - 1}(j)}^h & \red M_{\sigma^{k - 1}(i), k} \Par M_{\sigma^{k - 1}(j), k}
					\end{align*}
					In the steps $ a_1, \ldots, a_k $ each component of the network is used exactly twice to reduce a choice of the participants $ \sigma^m(\p) $ and $ \sigma^n(\q) $ for some $ m, n \in \left\{ 0, \ldots, k - 1 \right\} $ with $ m \neq n $.
					Since $ \sigma_h $ is an automorphism with only one orbit and since the choice of $ \sigma^m(\p) $ is output-guarded whereas the choice of $ \sigma^n(\q) $ is input-guarded, and since two choices combined in parallel have different participants, $ \sigma^m(\p) \neq \sigma^n(\q) $ for all such cases.
					By repeatedly applying Lemma~\ref{lem:confluence}, then we can perform $ a_1, \ldots, a_k $ in sequence, \ie there are some $ M_1^{h + 1}, \ldots, M_k^{h + 1} $ such that $ E_{h + 1}: M \red^\ast M^h \red^\ast M^{h + 1} = M_1^{h + 1} \Par \ldots \Par M_k^{h + 1} $, where the sequence $ M^h \red^\ast M^{h + 1} $ is obtained from $ a_1, \ldots, a_k $.
					Then $ M^{h + 1}_{\sigma(i)} = M^{h + 1}_i \sigma $, because $ M^h $ is symmetric and in $ M^h \red^\ast M^{h + 1} $ each component of the network is used exactly twice to reduce a choice of participants $ \sigma^m(\p) $ and $ \sigma^n(\q) $ for some $ m, n \in \left\{ 0, \ldots, k - 1 \right\} $ with $ m \neq n $ such that $ M_{\sigma(i)}^h \red^\ast M_{\sigma(i)}^{h + 1}, \ldots, M_{\sigma^{k - 1}(i)}^h \red^\ast M_{\sigma^{k - 1}(i)}^{h + 1} $.
					Thus $ M^{h + 1} $ is still symmetric \wrt $ \sigma $. \qedhere
			\end{description}
	\end{description}
\end{proof}

In the proof we construct a potentially infinite sequence of steps such that the system constantly restores symmetry, \ie whenever a step destroys symmetry we can perform a sequence of steps that restores the symmetry.
Therefore we rely on the assumption of $ \sigma $ generating only one orbit.
This implies that $ \Orbit{\sigma}{i} = \left\{ i, \sigma(i), \ldots, \sigma^{k - 1}(i) \right\} = \left\{ 1, \ldots, k \right\} $, for each $ i \in \left\{ 1, \ldots, k \right\} $.
Because of that, whenever part $ i $ performs a step that destroys symmetry or parts $ i $ and $ j $ together perform a step that destroys symmetry, the respective other parts of the originally symmetric network can perform symmetric steps to restore the symmetry of the network.
Because of the symmetry, the constructed sequence of steps does not elect a unique leader.
Accordingly, the existence of this sequence ensures that $ M $ is not an electoral system.

By the preservation of distributability, encodings preserve the structure of networks; and by name invariance, they also preserve the symmetry of networks.
With operational correspondence, divergence reflection, and barb sensitiveness, any good encoding of $ \mathsf{Election} $ is again a symmetric electoral system of size five, since the combination of these three criteria allows to distinguish between an electoral system and a system that does not elect exactly one leader in every maximal execution.
Since by Lemma~\ref{lem:noElectoralSystem} this is not possible, we can separate the \msmpst-calculus from the \scmpst-calculus by using $ \mathsf{Election} $ from Example~\ref{ex:leaderElection-process} as counterexample.

\thmseparation*

\begin{proof}
	Assume the contrary, \ie there is a good encoding $ \arbitraryEncoding $ from the \msmpst-calculus into the \scmpst-calculus with the renaming policy $ \varphi $.
	Then this encoding translates $ \mathsf{Election} $ in Example~\ref{ex:leaderElection-process}.
	By Definition~\ref{def:distributabilityPreservation} (preservation of distributability),
	\begin{align*}
		\ArbitraryEncoding{\mathsf{Election}} \equiv N_{\varphi(\role{a})} \Par N_{\varphi(\role{b})} \Par N_{\varphi(\role{c})} \Par N_{\varphi(\role{d})} \Par N_{\varphi(\role{e})}
	\end{align*}
	such that $ N_{\varphi(i)} \asymp \ArbitraryEncoding{M_i} $ for all $ i \in \left\{ \role{a}, \ldots, \role{e} \right\} $.
	Remember that $ \mathsf{Election} $ is symmetric.
	For all automorphisms $ \sigma $ we have $ M_{\sigma(i)} = M_i\sigma $ for all $ i \in \left\{ \role{a}, \ldots, \role{e} \right\} $.
	Fix $ \sigma $, \ie let $ \sigma $ be an arbitrary such automorphism, and let $ \sigma' $ be such that $ \varphi(\sigma(n)) = \sigma'{\left( \varphi(n) \right)} $ for all $ n $.
	Then $ \sigma' $ is a permutation (on translated source term names).
	By name invariance, then $ N_{\sigma'(\varphi(i))} = N_{\varphi(\sigma(i))} \asymp \ArbitraryEncoding{M_{\sigma(i)}} = \ArbitraryEncoding{M_i\sigma} \asymp \ArbitraryEncoding{M_i}\sigma' \asymp N_{\varphi(i)}\sigma' $ for all $ i \in \left\{ \role{a}, \ldots, \role{e} \right\} $.
	Since $ \asymp $ is a barb respecting weak reduction bisimulation (see Definition~\ref{def:goodEncoding}), then $ N_{\sigma'(\varphi(i))} \approx N_{\varphi(i)}\sigma' $ for all $ i \in \left\{ \role{a}, \ldots, \role{e} \right\} $, \ie $ \ArbitraryEncoding{\mathsf{Election}} $ is symmetric.
	By the combination of operational correspondence, divergence reflection, and barb sensitiveness, $ \ArbitraryEncoding{\mathsf{Election}} $ is an electoral system, because every maximal execution has to be emulated with the same reachable barbs.
	Since $ \mathsf{Election} $ has no infinite executions, divergence reflection ensures that all executions of $ \ArbitraryEncoding{\mathsf{Election}} $ are finite.
	Then $ \ArbitraryEncoding{\mathsf{Election}} $ is a symmetric electoral system of size five.
	This contradicts Lemma~\ref{lem:noElectoralSystem}.
	We conclude that there is no good encoding from the \msmpst-calculus into the \scmpst-calculus.
\end{proof}

}

\end{document}